\documentclass[12pt,letterpaper,titlepage,openright]{report}

\newcommand{\quotes}[1]{``#1''}%
\def\mam{mAMSB}%
\def\ino{inoAMSB}%
\def\hca{HCAMSB}%
\def\mssm{MSSM}
\def\nino{\tilde{\chi}_{1}^{0}}%
\def\dmchi{\Delta M_{\nino}}%
\def\etm{\slashed E_{T}}%
\def\et{{E_{T}}}%
\def\etcut{{E_{T}^{c}}}%
\def\pt{{p_{T}}}%
\def\invfb{{\textrm {fb}^{-1}}}%
\def\mnot{m_{0}}%
\def\mhf{m_{3/2}}%
\def\tanb{\textrm{tan} \ \beta}%
\def\anot{A_{0}}%
\def\mgut{M_{GUT}}%
\def\mstring{M_{\rm{string}}}%
\def\ohmh{\Omega h^{2}}%
\def\mll{m(l^{+}l^{-})}
\def\xen{XENON}
\def\kah{K\"ahler }
\def\bbar{\bar{b}}%
\def\Dbar{\bar{D}}%
\def\pbar{\bar{p}}%
\def\pos{e^{+}}
\def\sq{{\tilde q}}%
\def\sl{{\tilde l}}%
\def\axino{{\tilde a}}%
\def\hup{H^u}%
\def\hdn{H^d}%
\def\sp{\hat{W}}%
\def\cino{{\tilde W}_{1}}%
\def\nino{{\tilde Z}_{1}}%
\def\gl{{\tilde g}}%
\def\mb{M_{1}}%
\def\mw{M_{2}}%
\def\mgl{M_{3}}%
\def\mg{m_{\tilde g}}%
\def\gl{{\tilde g}}%
\def\sb{{\tilde b}_{1}}%
\def\sbbar{\bar{\tilde b}_{1}}%
\def\st{{\tilde t}_{1}}%
\def\stbar{\bar{\tilde t}_{1}}%
\def\susy{SUSY }%
\def\lsim{\stackrel{<}{\sim}}%
\def\gsim{\stackrel{>}{\sim}}%
\newcommand{\spart}[2]{{\tilde #1}_{#2}}%
\newcommand{\ninos}[1]{{\tilde Z}_{#1}}%
\newcommand{\cinos}[1]{{\tilde W}_{#1}}%

\usepackage[left=1.5in,right=1in,bottom=1in,top=1in]{geometry}
\usepackage{graphicx, color}
\usepackage{subfigure}
\usepackage{setspace}
\usepackage{slashed}
\usepackage{fancyhdr}%
\usepackage{amsmath}
\usepackage{amssymb}
\usepackage{hyperref}
\usepackage{multirow}%
\usepackage{colortbl, pstricks}%
\def\author{Shivakumar Rajagopalan}%
\def\degree{Doctor of Philosophy}%
\def\school{University of Oklahoma}%
\def\title{LHC Phenomenology and Dark Matter Considerations for Various Anomaly Mediated Supersymmetry Breaking Models}%
\def\maketitlepage{%
  \thispagestyle{empty}
  {%
    \centering%
    {%
	\MakeUppercase{\school}\\%
	\MakeUppercase{Graduate College}\\%
	\vfill
      \MakeUppercase{\title}\\%
	\vfill
	\MakeUppercase{A Dissertation}\\
	\MakeUppercase{Submitted to the Graduate Faculty}\\
  	in partial fullfillment of the requirements for the\\ 
	Degree of\\
	\MakeUppercase{\degree}\\%
    }%
	\vfill
      By\\%
	\singlespacing 
      \MakeUppercase{\author}\\%
      Norman, Oklahoma\\%
	2010\\%
	\newpage
  }%
}

\def\membera{Dr. Howard Baer, Chair}%
\def\memberb{Dr. Kimball Milton}%
\def\memberc{Dr. Phillip Gutierrez}%
\def\memberd{Dr. Eddie Baron}%
\def\membere{Dr. Nikola Petrov}%
\def\makesigpage{%
  \thispagestyle{empty}%
  {%
    \singlespacing%
    \centering%
    {%
	\linespread{1.6}%
	\MakeUppercase{\title}\\%
	\vspace{2cm}%
	\MakeUppercase{A Dissertation Approved for the}\\
	\MakeUppercase{Homer L. Dodge Department of Physics and Astronomy}\\%
	\vfill%
	\vspace{2cm}%
	BY%
	\vspace{2cm}%
	\vfill%
    }%
    {\flushright%
    \underline{\hspace{7cm}} \\%
    \membera \\%
    \vfill  
    \underline{\hspace{7cm}} \\%
    \memberb \\%
    \vfill 
    \underline{\hspace{7cm}} \\%
    \memberc \\%
    \vfill 
    \underline{\hspace{7cm}} \\%
    \memberd \\%
    \vfill 
    \underline{\hspace{7cm}} \\%
    \membere \\%
    \vfill 
    \newpage%
    }%
  }%
}%

\def\makecopyright{%
  \thispagestyle{empty}%
  {%
    \centering%
    {%
	\vspace*{20cm}%
	\singlespacing
	\copyright \ Copyright by \MakeUppercase{\author} \ 2010\\%
	All Rights Reserved.%
	\newpage%
    }%
  }%
}%

\def\makededication{%
  \thispagestyle{empty}%
  {%
	\begin{center}
	{\it In loving memory of my father, \\Dr. Raj}.%
	\end{center}
  }%
  \newpage%
}%

\begin{document}%
\doublespacing

\maketitlepage%
\makesigpage%
\makecopyright%
\makededication%

\pagenumbering{roman}%
\setcounter{page}{4}

\chapter*{Acknowledgements}
There are a several people who I would like to thank for helping me in a 
variety of ways while I was busy doing a lot of stuff over the last many 
years.  Without the kind help of people very near and very far away, it 
probably would have been impossible to finish this thesis with less mental 
damage.  So to all of you who helped, thank you kindly. 

And since that probably isn't enough... 

I would first like to thank Howard Baer for being a terrific adviser.  Working
for Howie has helped me to be more practical and has reinforced the importance 
of being independent.  I hold these lessons with the highest regard and will 
carry them with me for many years to come.  

To my family: Mom, Ettan, Chechi, Indu, Ravee Ettan, Diane, Gates, Mason, 
Anjali, and Lexes.  It has been wonderful seeing you when possible and finding 
that, even when so much has changed, those things that bring me peace and 
comfort are always with you.  Although I always wish I could stay longer, 
seeing all of you continually replenishes me and reminds me of what is 
important in life. 

There are also many people at my previous school, Florida State University, 
who stand out in my mind as key figures in my understanding of, well, lots of 
things.  To Laura, I will never forget how you offered your time and how 
incredibly helpful you have always been.  I was lucky to have you as a teacher 
and I hold your friendship with great value.  To my very dear friend Ben 
Thayer, what can I say?  You are a great friend and you were always there, 
though usually 45 minutes late (just kidding... relax).  I have no doubt that 
our old conversations have been reincarnated in this thesis.  Same goes for 
Bryan Field, another very good friend of mine who also helped me very much over 
the years and who is so knowledgeable that he should just go ahead and write an 
encyclopedia. And to Andrew Culham, your friendship had helped me through some 
of the hardest times. Though you did not help with matters of physics, you 
often kept me sane enough to pursue them.

I would like to thank Radovan Derm\' i\v sek and Shanta de Alwis, without whom
these projects would not have been possible, and who were kind of enough to 
read over various sections of this manuscript.  I would also like 
to thank my OU advisory committee members for their help, and to particularly 
thank Kimball Milton for his careful attention to detail when reading the 
manuscript as well.

There are also a few very special thanks that cannot go without saying. To 
Andre Lessa, it is impossible to talk to you without learning something.  You 
are a great source of knowledge and I am happy for your friendship.  I 
especially want to thank you for entertaining the ramblings of a mad man for
the last couple of years.

To Nanda and Diane, I cannot express how grateful I am to both of you for the 
support you have given me always.  You have encouraged me to pursue my dreams 
and I know you have always been there.  For this I am forever indebted to you.  

And finally, I want to give a great `thank you' to my wonderful Tzvetalina.  
You have been so kind and patient and I don't know what I would have done 
without your loving support and ease.  You are a truly tender soul.

\tableofcontents%
%
\newpage
\addcontentsline{toc}{chapter}{List of Tables}
\listoftables
\newpage
\addcontentsline{toc}{chapter}{List of Figures}
\listoffigures
%
\newpage
\addcontentsline{toc}{chapter}{Abstract}
\chapter*{Abstract}
In this thesis we examine three different models in the MSSM context, all of 
which have significant supergravity anomaly contributions to their soft 
masses.  These models are the so-called Minimal, Hypercharged, and Gaugino 
Anomaly Mediated Supersymmetry Breaking models.  We explore some of the string 
theoretical motivations for these models and proceed by understanding how they 
would appear at the Large Hadron Collider (LHC).  Our major results include 
calculating the LHC reach for each model's parameter space and prescribing a 
method for distinguishing the models after the collection of 100 $\invfb$ at 
$\sqrt{s}=14$ TeV.  AMSB models are notorious for predicting too low a dark 
matter relic density.  To counter this argument we explore several proposed 
mechanisms for $non$-$thermal$ dark matter production that act to augment 
abundances from the usual thermal calculations.  Interestingly, we find that 
future direct detection dark matter experiments potentially have a much better 
reach than the LHC for these models.
\newpage%

\pagestyle{fancy}%
\pagenumbering{arabic}%
\setcounter{page}{1}%
\setcounter{chapter}{0}
\linespread{2}%

\chapter{%
  \label{chap:intro}%
  Introduction
}%

\indent%

Particle physics is at an exciting point at the time of writing this thesis.  
The long-awaited start of the Large Hadron Collider (LHC) era has finally begun 
with the first collection of data at the world-record breaking collision energy 
of 7 TeV c.o.m. in March of this year.  After the first two years the 
experiment is planned to run at the full design luminosity accumulating 
ultimately 100-1000 $\invfb$ at a tremendous $\sqrt{s}=$ 14 TeV.  It is 
expected, or at least hoped, that the LHC will shed light on important 
mysteries of the Standard Model (SM) of particle physics by allowing us to 
detect particle states that have yet to be observed at the 
Tevatron.  The most likely of these is the Higgs boson, the remaining piece of 
the SM, which is thought to be responsible for the spontaneous breaking of the 
electroweak symmetry.  The presence of the Higgs boson in the SM is itself a 
strong theoretical motivator that other heavy particle states exist that can 
also be discovered at the LHC.  One very well-motivated class are the 
supersymmetric (\susy) particle states whose masses lie in the TeV range.  The 
discovery of these particles at the LHC would have deep implications for the 
nature of space-time.  Having the potential to discover new physical states 
such as these puts us at a truly unique and exciting time.\\  
\indent%
On the other hand particle physics is also merging increasingly with 
cosmology.  The energy density of the Universe is known to be comprised of Dark 
Energy (71\%), normal baryonic matter (4\%), and a non-luminous form of matter 
known as Dark Matter (DM) which comprises roughly 25\% of the energy density.
If DM is considered to be comprised of particles, it must be massive to account 
for the relic abundance and it must be cold enough to allow for structure 
formation on large scales.  For these reasons, there are no good candidates for 
DM in the SM.  There is, however, a particularly good DM candidate in 
supersymmetric theories in which R-parity is conserved.  The particle is the 
lightest neutralino, and the search for it is an important priority at the LHC 
and in DM experiments around the world.\\
\indent%
The outline of this thesis is as follows.  The remainder of the Introduction is 
intended to provide background for subsequent chapters.  We will first give a 
$very$ brief introduction to the Standard Model (SM) and simultaneously attempt 
to motivate supersymmetry.  Next, we will introduce the idea of the scalar 
superfield before moving on to describe the Minimal Supersymmetric Standard 
Model (MSSM)\footnote{Here, the SM and MSSM are best viewed as the low-energy 
effective theories of a high-energy framework, $i.e.$, $supergravity$.}.  Once 
the fields and interactions of the theory are established, we turn our 
attention to how the scalar superpartners of the SM acquire mass at the weak 
scale.  These quantities are usually given by a combination of low-energy 
constraints ($e.g$, electroweak symmetry breaking parameters, $etc.$) and by 
high-scale physics ($M_{GUT}$, $M_{string}$, $etc.$) that relates to physics at 
the low scale ($M_{weak}$) through renormalization group effects.  For this 
thesis, the latter quantities are, in part, due to anomalies that are present 
in supergravity theories, and a rather technical derivation of these will be 
given Section \ref{sec:grav_soft}.  Theories in which contributions of this 
type are important will be referred to as \quotes{Anomaly-Mediated 
Supersymmetry Breaking (AMSB) models}.  The Introduction concludes with 
descriptions of the computational tools used in this work and should be used 
for reference in the chapters that follow.\\
\indent%
In Chapters \ref{chap:mamsb} - \ref{chap:inoamsb} we will examine three 
different models of the AMSB type.  The first is the prototypical model known 
as the $minimal$ AMSB model, and is \quotes{minimal} in the sense that it has 
the minimum ingredients to be phenomenologically viable.  The second model is 
referred to as \quotes{Hypercharged AMSB} (HCAMSB) because it pairs AMSB mass 
contributions with $U(1)$-contributions at the string scale.  
The third model is actually a class of string theory models with specific 
high-scale boundary conditions and a rather $low$ string scale, referred to 
collectively as \quotes{Gaugino AMSB}, or \quotes{inoAMSB} for short.  Each of 
these chapters gives a full theoretical explanation of the model being 
considered.  They also give a full analysis for LHC physics, that is, we 
describe all renormalization group effects, compute weak-scale parameters, run 
signal and SM-background simulations for $pp$ collisions at $\sqrt{s}=14$ TeV, 
and calculate the reach of the models' parameters for 100 $\invfb$ of 
accumulated data (one year at full design luminosity).\\%
\indent%
It would be greatly insufficient to focus only on collider physics searches 
since future cosmological data will be precise enough to be competitive.  In 
Chapter \ref{chap:dm} we consider the question of dark matter (DM) in AMSB 
models where we take the lightest neutralino to be our DM candidate.  AMSB 
models notoriously yield too little thermally-produced relic abundance to 
account for the measured DM.  However, it is expected that other heavy 
particles present in the early universe could increase the DM abundance in AMSB 
models $non$-$thermally$.  In this chapter we present several such scenarios 
and assume the total DM abundance can be accommodated.  After a review of DM 
theory and direct/indirect detection experiments, we calculate important rates 
for AMSB DM cosmology for all of the models in Chapters \ref{chap:mamsb} - 
\ref{chap:inoamsb}.  We supplement all rates with detailed physical 
explanations.  We also find the reach for each model and compare the results 
with LHC expectations.\\%
\indent
Finally, we conclude with a general overview of the results in Chapter 
\ref{chap:conclusion}.
\section{%
  Standard Model and Supersymmetry
  \label{sec:smandsusy}
}
So far most experimental data in HEP experiments is described by the Standard 
Model of particle physics and this has been the case for 30+ years\footnote
{Important exceptions not described by the SM are $\nu$-oscillations, dark matter, dark energy,
and the baryon asymmetry. }. The basic 
facts of this model are given in this section with the goal in mind of 
extending it to include supersymmetry.\\%
\indent%
The Standard Model is a collection of three identical generations of 
spin-$\frac{1}{2}$ fermion fields, spin-1 vector bosons, and spin-0 scalar 
Higgs fields. The fermions interact through the exchange of spin-1 vector 
bosons that arise through the gauge invariance of both abelian and non-abelian 
interactions.  The interactions of the particles can be trivial or non-trivial 
under each of the $SU(3)_{c}$, $SU(2)_{L}$, and $U(1)_{Y}$ rotations, where 
\quotes{c} stands for the color-charge and \quotes{L} for the weak-isospin, and
\quotes{Y} is the hypercharge of the interacting particles.\\%
\indent%
Those particles that interact through $SU(3)_{c}$ interactions are the 
spin-$\frac{1}{2}$ fermions known as quarks and the interaction is mediated by 
spin-1, color-charged octet of gluon fields.  Fermions that are unaffected by 
$SU(3)_{c}$ interactions are known as leptons.  Left-handed fermions interact 
in very specific pairs (or doublets) and the mediators of the interactions are 
the spin-1 triplet of $W_{\mu}$-bosons.  These interactions do not apply to 
right-handed fermions, and they are the only sources of flavor-change in the SM 
\cite{Amsler:2008zzb} as is described, for example, by 
Cabibbo-Kobayashi-Maskawa mixing for quarks.  And finally, all fermions are 
charged under abelian $U(1)_{Y}$ rotations, and these interactions are mediated 
by the $B_{\mu}$ vector boson.\\%
\indent%
The inclusion of a Higgs particle is necessary to give mass to both the weak sector bosons and matter fields.  The Higgs  
($\Phi = {\phi^{+} \choose \phi^{0}}$) is thought to be a dynamic field with quartic self-interactions, thus its contribution to the Lagrangian is
\begin{equation}
{\cal L}_{Higgs} = |\partial_{\mu}\Phi|^{2} + \mu^{2}\Phi^{\dagger}\Phi
  - \lambda(\Phi^{\dagger}\Phi)^{2}.
\end{equation} 
By allowing $\Phi$ to be a complex doublet that transforms non-trivially under 
the electroweak symmetry, $SU(2)_{L}\times U(1)$, the covariantized action mixes 
the Higgs with weak and hypercharge bosons \cite{Perkins:1972yu}.  When the 
neutral Higgs acquires a VEV, the $SU(2)\times U(1)_{Y}$ symmetry is still a 
good symmetry of the Lagrangian, yet its generators no longer annihilate the 
vacuum.  The symmetry is said to be broken $spontaneously$, and the remnant 
symmetry is the $U(1)_{em}$ of electromagnetism.  In the symmetry breaking  
process, massless degrees of freedom are produced which are subsequently 
transformed into the longitudinal modes of the electroweak bosons (rendering 
them massive).  Also upon acquiring a VEV, the Higgs gives mass to the fermion 
fields through its Yukawa couplings to matter.\\
\indent%
The fundamental particles of the Standard Model are summarized in Table 
\ref{tab:sm} along with the matter and Higgs quantum numbers.  The table tells
how the particles transform under a given symmetry.  For example, left-handed 
fermions form doublets transforming non-trivially under $SU(2)$, right-handed 
fermions are singlets under $SU(2)$, and quarks transform as a triplet of color 
charge under $SU(3)$.  Not shown in the table are two other generations of fermions, identical to the one shown with exactly the same quantum numbers, but with larger masses.  Also, anti-particles have not been included.\\
\begin{table}[t]%
  \begin{center}%
  \begin{tabular}{|ccc|cccc|}%
    \hline\hline %
    &&&&&&\\%
    &&&&
    \multicolumn{3}{c}{Symmetry/Quantum \#s} \vline\\
    &&&&&&\\%
    Fermion 
    & Flavor 
    &&   
    &  $SU(3)_{c}$ 
    &  \ \ \ $SU(2)_{L}$ 
    &  $U(1)_{Y}$ \\ \hline
    &&&&&&\\%
    \multirow{2}{*}{leptons} 
    & ${\nu_{e}\choose e}_{L}$ 
    &&
    &{\bf 1}
    &{\bf 2}
    &-1\\ 
    & $e_{R}$ 
    &&
    &{\bf 1}
    &{\bf 1}
    &-2\\
    &&&&&&\\
    &&&&&&\\    
    \multirow{2}{*}{quarks} 
    & ${u \choose d}_{L}$ 
    &&
    &{\bf 3}
    &{\bf 2}
    &$\frac{1}{3}$\\[3pt]
    & $u_{R}$ 
    &&
    &{\bf 3}
    &{\bf 1} 
    &$\frac{4}{3}$\\[3pt]

    & $d_{R}$ 
    &&
    &{\bf 3}
    &{\bf 1}
    &$\frac{2}{3}$\\[3pt] \hline
    &&&&&&\\    
    Higgs 
    & \multirow{2}{*}{$\Phi = {\phi^{+} \choose \phi^{0}}$} 
    &
    & \multirow{3}{*}{gauge \ $\Biggl\{ $ } 
    & gluon 
    & weak 
    & hypercharge \\
    boson 
    &&&
    & $G_{\mu}^{A}$ 
    & $W_{\mu}^{i}$ 
    & $B_{\mu}$\\ 
    &{\scriptsize ({\bf 1},{\bf 2},1)}&&
    &{\scriptsize $A=1\cdots 8$}
    &{\scriptsize $i=0,+,-$} 
    &\\ 
%
\hline\hline  
  \end{tabular}%
  \caption{%
  Matter, vector boson, and Higgs fields in the Standard Model.
  \label{tab:sm}
}%
  \end{center}%
\end{table}%
\indent%
The SM is completely consistent but there are several reasons why it cannot be the complete description of nature.  A list of some, but not all, of these reasons are given here:
\begin{itemize}%
\item hierarchy problem - radiative corrections to the scalar (Higgs) mass 
       terms are quadratically divergent due to gauge and fermion loops, but unitarity arguments require the mass to be constrained to less than a few hundred GeV \cite{Baer:2006rs}.  If the SM is to be considered an effective field theory below a high-scale $\Lambda$, then the Higgs mass could be subject to excessive and unnatural fine-tuning.
\item Dark Matter (DM) is not yet included in the SM.
\item Gravitational interactions are not present in the SM.
\end{itemize}%
\indent%
It is ideal to have a single framework that can address these issues, and in this thesis supersymmetry is the adopted solution.  A supersymmetry transformation acts on bosonic state to form fermionic states and vice versa.  
The hierarchy problem is famously solved in supersymmetric theories because the 
fermion loop corrections to the Higgs mass are accompanied by bosonic 
corrections.  The new loops are also quadratically divergent but generally 
appear with opposite sign.  In the case of $unbroken$ supersymmetry, where 
fermions and bosons have equal mass, the fermionic and bosonic contributions 
precisely cancel one another to all orders of perturbation theory 
\cite{Baer:2006rs}\cite{Martin:1997ns}\cite{Weinberg:2000cr}.  Even in the 
case of broken supersymmetry, the divergent contributions to the Higgs mass are at most logarithmic and, therefore, not severely destabilizing.\\
\indent%
Supersymmetry also provides a framework for addressing the remaining two issues 
in the list above.  In particular, we will consider $local$ supersymmetry at a 
very high scale that naturally incorporates gravity.  It will be necessary for 
the supersymmetry to be (spontaneously) broken to agree with phenomenology.  
When we define the minimal supersymmetric model and renormalize the model 
parameters at the weak scale, we will encounter natural EW symmetry breaking, 
non-quadratically divergent scalar masses, and supersymmetry provides several 
DM candidates.\\
\indent%
Before developing the superfield formalism in the next section, it is important 
to discuss some technical details regarding why, if supersymmetry is to explain 
the short-comings of the SM, it must be a broken symmetry.  Since the 
supersymmetry generators transform bosons into fermions, they must be fermionic 
and therefore obey $anti$-commutation relations.  When these relations are 
combined with the Poincar\'e algebra, the closed ($graded$) algebra that 
results is known as the super-Poincar\'e algebra.  The squared-momentum 
generator, $(P^{\mu})^{2}$, is a Casimir of the super-Poincar\'e algebra 
\cite{Lykken:1996xt}, and so supersymmetric fermion-boson pairs are expected to 
be degenerate.  If this was a rigid requirement, supersymmetry would already by 
excluded by the fact that no partners of the SM particles have been observed 
with identical mass.  Supersymmetry necessarily has to be a $broken$ symmetry 
to be phenomenologically viable.\\%
\indent%
Viable supersymmetric models must assume breaking in a way that does not re-
introduce quadratic divergences.  In order for this to occur {\it it is 
necessary for the dimensionless couplings of the theory to be unmodified by 
supersymmetry breaking, and that only couplings with positive mass dimension 
are included in the supersymmetry breaking potential}.  Supersymmetry that is 
broken in this way is said to be broken {\it softly}.  This reduces the number 
of additional  terms that can be included in the soft supersymmetry-breaking 
Lagrangian, ${\cal L}_{soft}$, which are: linear (gauge singlet), bilinear 
(including masses), and trilinear ($A$-term) scalar interactions, and bilinear 
gaugino mass terms.
\section{%
  Superfields
  \label{sec:superfields}
}
In this section we briefly review some aspects of the superfield formalism.  We 
will see how boson-fermion pairs are embedded into super-multiplets.  Also in 
this section, key ingredients to supersymmetric theories are defined, and the 
notation to be used in describing the minimal supersymmetric standard model of 
the next section are established.\\
\indent%
Superfields place boson fields in the same multiplet as their fermionic 
partners.  It is convenient to introduce anti-commuting Grassmann variables
arranged as Majorana spinors, $\theta \ (\bar{\theta})$, that multiply the 
fields in order to put them on the same footing.  Fields in a super-multiplet 
that are not multiplied by Grassmann variables are referred to as the 
\quotes{lowest} component, those with one Grassmann variable are the second 
component, $etc$ \footnote{For more discussion on Grassmann variables, see 
\cite{Gates:1983nr}\cite{Wess:1992cp}.}. It is the lowest component that 
determines the type of superfield, $i.e.$, scalar, spinor, vector\footnote{We 
only consider scalar superfields here.  For more information, see \cite{Baer:2006rs}.}, etc.\\
\indent%
A general scalar superfield has many components as seen in
\begin{align}
  \hat\Phi(x,\theta)=&
  S%
  -i\sqrt{2}\bar{\theta}\gamma_{5}\psi%
  -\frac{i}{2}(\bar{\theta}\gamma_{5}\theta)M%
  -\frac{1}{2}(\bar{\theta}\theta)N%
  +\frac{1}{2}(\bar{\theta}\gamma_{5}\gamma_{\mu}\theta)V^{\mu}
  \label{eqn:generalsf}\\%
  &+i(\bar{\theta}\gamma_{5}\theta)[%
    \bar{\theta}(\lambda+\frac{i}{\sqrt{2}}\slashed{\partial}\psi)
    -\frac{1}{4}(\bar{\theta}\gamma_{5}\theta)^{2}[D-\frac{1}{2}\nabla^{2}S]
    \nonumber
  ],\\%
  \nonumber
  \end{align}
but this representation is not irreducible.  Fortunately supersymmetry allows for chiral representations of superfields.  For example, there is a representation where the scalar component and the left-chiral spinor transform into one another without mixing with the corresponding right-handed fields.  A left chiral scalar superfield is of the form
\begin{equation}
  \hat{S}_{L}(x,\theta) = S(\hat x) + i\sqrt{2}\bar{\theta}\psi_{L}(\hat{x})%
  + i\bar{\theta}\theta_{L}F(\hat{x})\\%
\end{equation}
where the lowest (Grassmann) component is a scalar, the second component is the partner fermion, and F is an auxiliary field required to balance off-shell degrees of freedom\footnote{Additionally $\hat{x}_{\mu}=x_{\mu}+i\bar{\theta}\gamma_{5}\gamma_{\mu}\theta$, but we will not focus on technical details of this sort.}.  
Similarly a right chiral scalar superfield is of the form
\begin{equation}%
  \hat{S}_{R}(x,\theta) = S(\hat{x}^{\dag}) 
    - i\sqrt{2}\bar{\theta}\psi_{R}(\hat{x}^{\dag})%
    + i\bar{\theta}\theta_{R}F(\hat{x}^{\dag}),\\%
\end{equation}
but will frequently be recast as the conjugate of a left-chiral scalar field.
In short-hand, chiral scalar superfields can be referred to by their 
components, ($S,\psi_{L},F$).  To be clear, the scalar component is not a 
spinor and does not have helicity.  It contains the annihilation operator of 
the superpartner of a chiral fermion.  Under supersymmetry transformations the 
components of the left-chiral superfield transform into one another:
\begin{align}
  \delta S & = -i\sqrt{2}\bar{\alpha}\psi_{L}\\%
  \delta\psi_{L} & = -\sqrt{2}F\alpha_{L}%
    + \sqrt{2}\slashed{\partial}S\alpha_{R}\\%
  \delta F & = i\sqrt{2}\bar{\alpha}\slashed{\partial}\psi_{L}.%
  \label{eqn:fterm}
\end{align}
\indent%
It is intriguing that the $F$-term of the superfield transforms as a total 
derivative (Equation (\ref{eqn:fterm})) because field combinations that transform 
as total derivatives bring dynamics to the action.  This property of the $F$-
term is true even of products of chiral superfields because, as can be shown,  
they are themselves chiral superfields.  The general polynomial of left-chiral 
superfields is another left-chiral superfield known as the $superpotential$, 
and its $F$-term ($\bar{\theta}\theta_{L}$-component) must be appended to the 
Lagrangian.  In a renormalizable theory, the superpotential is at most a cubic 
polynomial by dimensional analysis.\\ 
\indent%
It can also be shown that the $(\bar{\theta}\gamma_{5}\theta)^{2}$-component, 
or \quotes{$D$-term}, of a general superfield (Equation (\ref{eqn:generalsf})) 
transforms as a total derivative under supersymmetry transformations .  The 
product of {\it chiral} superfields, however, does not have an interesting 
$D$-term because it is already a total derivative, as in 
\begin{equation}
  \hat{S_L} \ni \frac{1}{8}(\bar{\theta}\gamma_{5}\theta)^{2}\nabla^{2}S.
\end{equation}
This term is automatically zero in the action and does not produce any 
dynamics, and is for this reason that the superpotential does not contribute 
$D$-terms.  However, a polynomial of mixed chirality can give important $D$-term 
contributions to the Lagrangian, and this function is referred to as the \kah 
potential.  It is important to note that the \kah potential is at most 
quadratic by renormalizability and real by hermiticity of the Lagrangian 
\cite{Baer:2006rs}.  It is then taken to have the form 
\begin{equation}%
K = \sum_{i,j=1}^{N} A_{ij}\hat{S}_{i}^{\dagger}\hat{S}_{j},
\end{equation}%
the $D$-term of which is also included in the Lagrangian as with the $F$-term 
contributions.

\section{%
  \label{sec:mssm}
  Minimal Supersymmetric Standard Model
}%
We are now ready to build the minimal extension of the Standard Model (SM) that 
incorporates supersymmetry, the Minimal Supersymmetric Standard Model (MSSM).  
That is, we seek the minimal extension of the SM that includes broken 
supersymmetry, and that is both phenomenologically and theoretically safe.   In 
order to do this, it is first assumed that the theory will have the SM gauge 
symmetry group: $SU(3)_{c}\times SU(2)_{L}\times U(1)$.  Thereafter, all gauge 
and matter fields of the SM must be promoted to superfields.  The matter 
superfields must be L-chiral fields as required by the superpotential.  Thus, 
for each $chirality$ of every SM fields, we will choose there will be one $left
$-chiral superfield assigned.\\%
\indent%
Extending the SM Higgs field to a superfield will add a partner fermion, a higgsino ($\tilde h$).  Having only one extra fermion re-introduces $U(1)^{3}_{Y}$ and $U(1)_{Y}SU(2)^{2}_{L}$ gauge anomalies \cite{Terning:2006bq} that are canceled successfully in the SM.  If instead there are two Higgs doublets in the theory, with opposite hypercharges, $Y=1$ and $Y=-1$, their fermionic partners have the correct quantum numbers to satisfy the anomaly cancellation.  Furthermore, a single Higgs doublet is not allowed in the MSSM because the 
lower-component fermions of the SM weak doublets would receive their mass from the conjugate of the Higgs, which is a $right$-chiral superfield.  Interactions of $right$-chiral superfields are forbidden in the superpotential
\cite{Baer:2006rs}, and we are forced to accept at least two Higgs doublets into the MSSM.  We denote by $\hat{H}_{u}$ the Higgs doublet of superfields that is associated with the mass of $Y=1$ fermions, and $\hat{H}_{d}$ associated with mass of $Y=-1$ fermions.  The matter and Higgs superfields of the MSSM are shown in Table 
\ref{tab:mssm}.
\begin{table}%
  \begin{center}%
  \begin{tabular}{|c|ccc|}%
    \hline\hline%
    Field & $SU(3)_{C}$ & $SU(2)_{L}$ & $U(1)_{Y}$\\%
    \hline 
    &&&\\%
    $\hat{L}$=$\hat{\nu}_{eL} \choose \hat{e}_{L}$ & {\bf 1} 
      & {\bf 2} & -1\\[3pt]%
    $\hat{E^{c}}$ & {\bf 1} & {\bf 1} & 2\\[9pt]%
    $\hat{Q}$=$\hat{u_{L}} \choose \hat{d_{L}}$ & {\bf 3} & {\bf 2}%
      & $\frac{1}{3}$\\[3pt]%
    $\hat{U}^{c}$ & {\bf 3}$^{*}$ & {\bf 1} 
      & \hspace{-.3cm}$-\frac{4}{3}$\\[3pt]%
    $\hat{D}^{c}$ & {\bf 3}$^{*}$ & {\bf 1} & $\frac{2}{3}$\\[9pt]%
    $\hat{H}_{u}$=$\hat{h}_{u}^{+} \choose \hat{h}_{u}^{0}$ 
      & {\bf 1} & {\bf 2} & 1\\[3pt]%
    $\hat{H}_{d}$=$\hat{h}_{d}^{-} \choose \hat{h}_{d}^{0}$ & {\bf 1}%
      & {\bf 2}$^{*}$ & -1 \\[6pt]%
    \hline\hline%
  \end{tabular}%
\caption{%
  Matter and Higgs superfields in the MSSM.
  \label{tab:mssm}
}%
\end{center}%
\end{table}%

Next it is necessary to define the superpotential of the theory.  The 
superpotential, denoted by $\sp$, contains $SU(2)\times U(1)$ invariant 
combinations of chiral matter superfields and Higgs fields.  The matter fields
are coupled to Higgs fields through Yukawa coupling matrices, while the
$\mu$-term couples $\hup$ and $\hdn$.  In the MSSM it is:\\%
\begin{equation}
  \sp=\mu\hat{H}_{u}^a\hat{H}_{da} 
  +\sum_{i,j=1,3}[%
    ({\bf f}_{u})_{ij}\hat{Q}_{i}^{a}\hat{H}_{ua}\hat{U}_{j}^{c}
    +({\bf f}_{d})_{ij}\hat{Q}_{i}^{a}\hat{H}_{da}\hat{D}_{j}^{c}
    +({\bf f}_{e})_{ij}\hat{L}_{i}^{a}\hat{H}_{da}\hat{E}_{j}^{c}
  ],%
\end{equation}
where $a$ is an $SU(2)$ index and $c$ stands for conjugation.\\
\indent%
This superpotential is not completely general because, in its construction, 
terms that would lead to baryon (B) and/or lepton (L) number violation have been carefully omitted.  There are renormalizable terms that could be added that are gauge and supersymmetrically invariant, but the presence of these terms would have physical consequences that are highly constrained by experiment (for instance, proton decay).  The omission of these operators is made possible by imposing a discrete symmetry on the superpotential, known as R-parity.  When 
\begin{equation}%
R = (-1)^{3(B-L)+2s}
\end{equation}
is conserved ($s$ is the spin of the state), renormalizable $B$- and 
$L$-violating interactions will be forbidden .  Furthermore, conservation of 
R-parity leads to three important consequences \cite{Terning:2006bq}:
\begin{itemize}%
\item[1.] superpartners will always be produced in $pairs$ at colliders;
\item[2.] the decays of the SM superpartners (including extra higgsinos) produce
	    an odd number of the final state lightest SUSY particle (LSP), which 
	    for our purposes this is a neutralino;
\item[3.] the LSP is absolutely stable and therefore may be a good dark matter 
	    candidate.
\end{itemize}%
We accept R-parity as part of the definition of the MSSM and should stay mindful of these consequences in the coming chapters.\\
\indent%
The final step in constructing the MSSM is to include all gauge-invariant 
{\it soft} supersymmetry-breaking terms into the Lagrangian.  These 
terms are thought to arise from the interactions between MSSM fields and a \quotes{hidden} sector where SUSY is broken (see Section 
\ref{sec:grav_soft}).   These terms raise the masses of supersymmetric partners 
of the SM fields and are needed for electroweak symmetry breaking (see next 
subsection).   The MSSM soft Lagrangian is \cite{Baer:2006rs}
\begin{align}%
{\cal L}_{soft} = 
 	     & -\bigl[\spart{Q}{i}^{\dagger}{\bf m^{2}_{Q}}_{ij}\spart{Q}{j} +
		  \spart{d}{Ri}^{\dagger}{\bf m^{2}_{D}}_{ij}\spart{d}{Rj} +
		  \spart{u}{Ri}^{\dagger}{\bf m^{2}_{U}}_{ij}\spart{u}{Rj} + 
	      \nonumber \\
	     &  \qquad \spart{L}{i}^{\dagger}{\bf m^{2}_{L}}_{ij}\spart{L}{j} +
	 	  \spart{e}{Ri}^{\dagger}{\bf m^{2}_{E}}_{ij}\spart{e}{Rj}
		  + m_{H_{u}}^{2}|H_{u}|^{2} + m_{H_{d}}^{2}|H_{d}|^{2}\bigr] 
	      \nonumber \\
	    & -\frac{1}{2}\bigl[M_{1}\bar{\lambda_{0}}\lambda_{0} 
		  + M_{2}\bar{\lambda_{A}}\lambda_{A}
		  + M_{3}\bar{\gl}_{B}\gl_{B}\bigr] 
	      \nonumber \\
	    & -\frac{1}{2}\bigl[M'_{1}\bar{\lambda_{0}}\gamma_{5}\lambda_{0} 
		  + M'_{2}\bar{\lambda_{A}}\gamma_{5}\lambda_{A}
		  + M'_{3}\bar{\gl}_{B}\gamma_{5}\gl_{B}\bigr] 
	      \nonumber \\
	    &  + \bigl[({\bf a_{u}})_{ij}\epsilon_{ab}
                  \spart{Q}{i}^{a}H_{u}^{b}\spart{u}{Rj}^{\dagger}  
               + ({\bf a_{d}})_{ij}\spart{Q}{i}^{a}H_{da}\spart{d}{Rj}^{\dagger}
		   + ({\bf a_{e}})_{ij}\spart{L}{i}^{a}H_{da}\spart{e}{Rj}^{\dagger}
		   + h.c.\bigr]
	      \nonumber \\
	    &  + \bigl[({\bf c_{u}})_{ij}\epsilon_{ab}
                  \spart{Q}{i}^{a}H_{d}^{*b}\spart{u}{Rj}^{\dagger}  
               + ({\bf c_{d}})_{ij}\spart{Q}{i}^{a}H_{*ua}
                   \spart{d}{Rj}^{\dagger}
		   + ({\bf c_{e}})_{ij}\spart{L}{i}^{a}H_{ua}^{*}
	             \spart{e}{Rj}^{\dagger}  + h.c.\bigr]
	      \nonumber \\
	    & +\bigl[bH_{u}^{a}H_{da} + h.c.\bigr].
  \label{eqn:Lsoft}
\end{align}%
In short, the parameters above are the scalar mass matrices of lines 1 \& 2, 
the gaugino mass terms of lines 3 \& 4, the trilinear $A$-term couplings from 
supersymmetry breaking in line 5, trilinear terms of line 6 \& 7.  It is seen 
that these terms give the MSSM an extremely large number of parameters.  We 
will see that in AMSB-type models the number of parameters will be reduced from 
the 178 in the soft Lagrangian above to just a few.  This is one of the many 
attractive features of the class of models considered in this thesis.
\subsection{Electroweak Symmetry Breaking in the MSSM}%
As in the Standard Model, the electroweak symmetry is spontaneously broken by 
minimizing the potential in the scalar sector.  We have seen that due to 
supersymmetry the scalar sector includes much more than just the Higgs 
particle.  The potential is extended now to include effects of all matter 
scalars along with all possible effects that originate in \susy breaking.  The 
scalar potential is then the sum of the various terms:%
\begin{equation}%
  V_{MSSM} = V_{F} + V_{D} + V_{soft}
\end{equation}%
with%
\begin{align}%
  V_{F} & = \sum_{i}|F_{i}|^{2}%
    = \sum_{i} \ \Bigl|\frac{\partial W}{\partial \hat{S}_{i}}\Bigr| %
    ^{2}_{\hat{S}=S}, \qquad \textrm{and} \\%
  V_{D} & = \frac{1}{2}\sum_{A} 
    \Bigl[\sum_{i}\hat{S}_{i}^{\dagger}gt_{A}S_{i}\Bigr]^{2},%
\end{align}%
and $V_{soft}$ comes from lines 1, 2, 5, 6, and 7 of Equation (\ref{eqn:Lsoft}).
As in the SM, the Higgs VEVs should be responsible for the breaking.  In the
usual manner, gauge symmetry allows the VEV of $H_{u}$ to be rotated to its 
lower neutral component.  Upon minimizing with respect to the other component 
of $H_{u}$, it is necessarily so that $\langle h_{d}^{-} \rangle$ = 0 
\cite{Baer:2006rs}. 
Then only the potential of the neutral Higgs fields needs to minimized in the 
breaking of electroweak symmetry.  In this way, provided no other scalars are 
allowed to develop VEVs, only charge-conserving vacuua can occur in the MSSM.\\%
\indent
It will be important to understand how the electroweak symmetry is broken properly in order to put constraints on parameters in a model.  Schematically the minimization requires the following:%
\begin{itemize}%
  \item[1.] the potential is extremized with respect to both $h_{u}^{0}$ and 
            $h_{d}^{0}$ and their conjugates through its first derivatives, i.e.
	        \begin{center}%
	          $\frac{\partial V}{\partial h_{i}^{0}} 
		      \textrm{ (and conjugates)} = 0; $%
	        \end{center}%
  \item[2.] to ensure that EW-breaking occurs the origin must be a maximum, 
	      i.e., the determinant of the Hessian should be negative there, and 
	      this imposes the condition%
	        \begin{center}%
		    $(B\mu)^{2} > (m_{H_{u}}^{2} + \mu^{2})(m_{H_{d}}^{2} 
		      + \mu^{2});$%
	        \end{center}%
  \item[3.] and finally, in order that the potential is bounded from below the
	      (D-term) quartic terms must be non-zero (positive at infinity) and 
	      this results in the condition%
	        \begin{center}%
		    $2|B\mu| < m_{H_{u}}^{2} + m_{H_{d}}^{2} + 2\mu^{2}.$%
	        \end{center}%
\end{itemize}%
Electroweak symmetry is broken properly when these conditions are met because a
well-defined minimum develops that does not break charge.  It is customary to 
define the ratio of the Higgs VEVs as
\begin{center}%
$\tanb = \bigl(\frac{v_{u}}{v_{d}}\bigr)$,%
\end{center}%
and to recast the important potential minimization conditions as
\begin{align}%
  & B\mu = \frac{(m_{H_{u}}^{2} + m_{H_{d}}^{2} 
    + 2\mu^{2})\textrm{sin}2\beta}{2} \quad \textrm{and}\\%
  & \mu^{2} = \frac{m_{H_{d}}^{2} - m_{H_{u}}^{2}(\tanb)^{2}}
    {\textrm{tan}^{2}\beta - 1} - \frac{M_{Z}^{2}}{2},
    \label{eq:mincond}
\end{align}%
where $M_{Z} = \frac{g^2 + g'^{2}}{2} (v_{u}^{2} + v_{d}^{2})$ is the tree-level
result for the $Z^{0}$ mass when the effects of EW-breaking on gauge bosons are 
considered.\\%
\indent%
In the SM, the required shape of the potential is achieved through the inclusion of a tachyonic scalar that transforms non-trivially under the EW
symmetry and acquires a VEV.  The same is true in the MSSM, with the exception
that the potential is generated naturally at the weak scale through RGE effects
rather than put in by hand.  As SUSY parameters are evolved from the GUT scale down to the weak scale, the large top Yukawa coupling drives the up-type squared
Higgs mass to negative values, thus triggering what is know as radiative 
electroweak symmetry breaking (REWSB).\\
\indent%
The $\mu$ value calculated through Equation \ref{eq:mincond} is purely 
phenomenological.  In actuality, it is difficult to understand why $\mu$ should 
be so small considering it appears in a supersymmetric superpotential term and 
should naturally be of order the (high) supersymmetry breaking scale (perhaps 
of order the Planck scale, $M_{\rm{Pl}}$) \cite{Brignole:1997dp}, but such high values would 
destroy the mechanism of electroweak symmetry breaking.  This is known as the 
\quotes{$\mu$-problem}.  Typically the low $\mu$ value is assumed to arise by 
some other mechanism, most commonly that of Guidice and Masiero 
\cite{Giudice:1988yz}.\\
\indent%
The procedure described above is true in general for minimizing the scalar 
potential, however there are important cases where radiative corrections need 
to be considered.  For example, without radiative corrections, the MSSM would 
already be excluded by Higgs mass bounds.  For the sake of brevity we do not 
dwell on these corrections, but their importance is noted.\\%
\indent%
Now that we have seen the MSSM and its parameters, we can determine the physical
mass eigenstates that are important for any type of phenomenology.  The 
following subsections give a general overview of the contributions to the mass 
matrices at the weak scale in the various sectors.
\subsection{Neutralinos and Charginos}%
The spontaneous electroweak symmetry breaking leads to mixing of fields with the
same electric charge, spin, and color quantum numbers.  The spin-$\frac{1}{2}$,
color-neutral fermions mix and come as wino-bino-higgsino mass eigenstate 
combinations.  Those combinations that are neutral are called {\it neutralinos} 
and those that are charged are {\it charginos}.\\
\indent%
The Lagrangian mass terms for the neutral fields can be written as 
\begin{equation}
  {\cal L}_{neutralino}= -\frac{1}{2}\bar{\Psi}{\cal M}_{neutralino}\Psi
\end{equation}
where 
$\Psi^{T}=({\tilde h}_{u}^{0}, {\tilde h}_{d}^{0}, {\tilde W}^{3}, 
{\tilde b})$
contains the neutral up- and down-type higgsinos, wino, and bino respectively.
The tree-level mass matrix\footnote{The actual mass matrix has 1-loop contributions, the most important of which on the upper-left diagonal.}  for this sector is
\begin{equation}
 {\cal M}_{neutralino} = \left(
  \begin{array}{cccc}
   0   & \mu & -\frac{gv_{u}}{\sqrt{2}} & \frac{g'v_{u}}{\sqrt{2}} \\
   \mu & 0   & \frac{gv_{d}}{\sqrt{2}}  & -\frac{g'v_{d}}{\sqrt{2}}\\
   -\frac{gv_{u}}{\sqrt{2}}    &  \frac{gv_{d}}{\sqrt{2}} & M_{2} & 0 \\
   \frac{g'v_{u}}{\sqrt{2}} & -\frac{g'v_{d}}{\sqrt{2}} & 0 & M_{1}\\
  \end{array} \right).
  \label{eqn:nino_matrix}%
\end{equation}
After diagonalizing the mass matrix we find the mass eigenstates, $\ninos{i}$, in linear combinations of the higgsinos and gauginos, ${i.e.}$,
\begin{equation}
  \ninos{i}= v_{i(1)}{\tilde h}_{u}^{0} + v_{i(2)}{\tilde h}_{d}^{0} 
    + v_{i(3)}{\tilde W}^{3} + v_{i(4)}{\tilde b}
\end{equation}
Note that the neutralino mass matrix is hermitian and will have real 
eigenvalues.\\%
\indent%
Charginos are linear combinations of the charged gauginos, ${\tilde W}^{\pm}$, 
and charged higgsinos, ${\tilde h}_{u}^{-}$ and ${\tilde h}_{d}^{+}$.  They appear in the Lagrangian as 
\begin{equation}
  {\cal L}_{chargino} = -\bar{\Phi}({\cal M}_{charged}P_{L} 
    + {\cal M}_{charged}^{T}P_{R})\Phi ,
\end{equation}
where the negatively charged, two-component field is 
$\Phi^{T} = (\frac{\lambda_{1} + {\it i}\lambda_{2}}{2}, 
  P_{L}\spart{h^{-}}{d}-P_{R}\spart{h^{+}}{u})$, and the tree-level matrix is
\begin{equation}
 {\cal M}_{charged} = \left(
  \begin{array}{cc}
    M_{2} & -g v_{d} \\
    -g v_{u} & -\mu \\
  \end{array} \right).
  \label{eqn:cino_matrix}
\end{equation}
The diagonalization of ${\cal M}_{charged}$ is more involved than in the 
neutral case as it requires two unitary matrices, U and V, and this procedure 
can be found in \cite{Baer:2006rs}.  In the end, there are two mass eigenstates 
for each charge in the combinations 
\begin{align}
\spart{W}{i}^{-} = U_{i(1)}{\tilde W}^{-} + U_{i(2)}{\tilde h}_{u}^{-}\\
\spart{W}{i}^{+} = V_{i(1)}{\tilde W}^{+} + V_{i(2)}{\tilde h}_{d}^{+}.
\end{align}
The physics mass eigenvalues will depend on $\mu$, $M_{2}, M_{W}$, and $\tanb$. 
Note that the mass matrix is tree-level and loop corrections are not shown 
here.\\%
\indent%
In general, there may be a high degree of mixing for any of the neutralino and
chargino mass eigenstates.  It turns out that of the parameters 
$\mu, M_{1}, M_{2}$, and $\tanb$ that contribute to the mixing, AMSB type 
models tend to have a very light $M_{2}$ value.  This results in both the 
lightest neutralino and lightest chargino being {\it wino}-like, a fact that 
will have strong implications for LHC phenomenology (Chapters \ref{chap:mamsb}- 
\ref{chap:inoamsb}) and Dark Matter (Chapter \ref{chap:dm}) as we will see. 
\subsection{Sfermion Masses}%
Unlike the SM, there are four mass contributions to the squarks and sleptons.  
They are: superpotential contributions that rely and Higgs VEV; generational 
soft masses from SUSY breaking; trilinear interactions with Higgs fields; and 
D-term contributions.  The result of these contributions is to mix left and right sfermions of the same flavor with $2\times2$ matrices that are
straight-forwardly diagonalized.  The procedure is not very illuminating for the current discussion, and the reader is referred to \cite{Baer:2006rs} 
\cite{Martin:1997ns} for more discussion.  In the end, the mass eigenvalues 
will depend on $\mu, \tanb$, trilinear couplings, SM EW parameters, and the 
corresponding fermion mass.%
\subsection{Gluino Mass}%
The gluino mass is not tied to the EW symmetry breaking sector, and its 
presence is purely due to supersymmetry breaking and renormalization, and is 
parameterized by $M_{3}$.  The gluino must be a mass eigenstate because it is 
the only color octet fermion and $SU(3)_{C}$ is not a broken symmetry.  In the 
next section we will see how this and other gaugino masses are generated by 
anomalous SUGRA effects.
\section{%
  \label{sec:grav_soft}%
  Supergravity Soft Terms%
}%
The MSSM is considered to be an effective theory containing information about supersymmetry breaking through the soft terms of Equation (\ref{eqn:Lsoft}).  
In this section the origin of the soft terms is outlined.  Most importantly, the explanation of how the supergravity anomaly imparts mass is explained.  The class of models where these contributions dominate (or is comparable to) all other forms of soft mass generation are known as Anomaly Mediated Supersymmetry Breaking Models, and all of the models in this thesis are of this type.\\%
\indent%
It has already been remarked that the generators of supersymmetry 
transformations are spinors, and it follows that the variational parameter of 
supersymmetry is also a spinor.  When supersymmetry is \quotes{localized} and 
combined with the symmetries of general relativity, the resulting theory is 
called {\it supergravity}
\cite{Freedman:1976xh}\cite{Wess:1992cp}\cite{Martin:1997ns}.  Supergravity has 
a supermultiplet of gauge particles: the spin-2 graviton, and its super-partner, 
the spin-$\frac{3}{2}$ gravitino, both of which are massless so long as 
supersymmetry is unbroken.  However, when supersymmetry is broken spontaneously 
by the VEV of the auxiliary component of a \quotes{hidden} supermultiplet X, 
$i.e.$, $<F_{X}>$, a massless goldstone fermion, or $goldstino$, is produced.  
This goldstino becomes the longitudinal mode of the gravitino thereby imparting 
mass, and this mechanism is called the $super$-$Higgs$ mechanism due to its 
obvious parallels to SM Higgs mechanism. The gravitino mass, $\mhf$, is given by
\begin{equation}%
\mhf = \frac{<F_{X}>}{\sqrt{3}M_{\rm{Pl}}},
\end{equation}%
and plays a key role in soft mass generation when supersymmetry breaking is 
communicated to the MSSM particles.  In the case of gravity-mediated supersymmetry breaking, MSSM fields interact with the hidden sector mainly through gravitational interactions.  These interactions induce soft masses for the scalars that are suppressed by powers of $M_{\rm{Pl}}$ and typically of the order \cite{Terning:2006bq}
\begin{equation}%
m_{soft} \sim \frac{<F_{X}>}{M_{\rm{Pl}}}.
\end{equation}%
When the MSSM is coupled to gravity, the effective soft Lagrangian below the 
Planck scale is given as in Equation (\ref{eqn:Lsoft}) with coefficients that 
depend on powers of $m_{\rm{soft}} \sim \mhf$.  However these tree-level effects 
can be suppressed relative to the quantum anomaly contributions to be described 
in the next subsection, and we will see that these contributions will require 
typically higher values of $\mhf$ than is typical for gravity mediated 
scenarios.

\subsection{  
  \label{subsec:amsb}  
  Anomaly Mediated Supersymmetry Breaking%
}%

In this section we examine how soft masses are generated by anomalous supergravity (local susy) effects\footnote{The following description follows closely to the arguments of Kaplunovsky and Louis \cite{Kaplunovsky:1994fg}
and de Alwis \cite{deAlwis:2008aq}.}. %
Low-energy type II-B string theory enforces the following functional as the 
classical Wilsonian action for a generic super Yang-Mills coupled to gravity
(see \cite{Gates:1983nr}\cite{Wess:1992cp} for more discussion):
\begin{align}
  S= &-3\int d^{8}z{\bf E}
        ~exp\{-\frac{1}{3}K(\Phi,\bar{\Phi};Q,\bar{Q}e^{2V})\} 
    \label{eqn:sugra_action} \\
     & \quad + \quad \bigl(\int d^{6}z
       {\cal E}~
       [W(\Phi,Q) + \frac{1}{4}f_{ab}
       {\cal W}^{a\alpha}
       {\cal W}^{b}_{\alpha}] + h.c.\bigr) \nonumber \\
    & \quad + ({\rm higher \ order \ derivatives}) \nonumber,
\end{align}
where K is the \kah potential, W the superpotential, $f_{ab}$ are the gauge 
coupling, and ${\cal W^{\alpha}}$ is the gauge field strength associated with 
the prepotential V.  R is the chiral curvature superfield, {\bf E} is the full
superspace measure, and ${\cal E}$ is the chiral superspace measure.\\
\indent%
The issue of the anomaly arises with the first term in the functional above 
because it is not in the standard form with Einsteinian gravity.  It also has 
the improper form for kinetic metric, given by the second derivative of the 
\kah potential.  What is desired instead is to have the 
fields transformed out of the SUGRA frame (as it appears in 
(\ref{eqn:sugra_action})) into the more physical Einstein-\kah frame that has 
Einsteinian gravity as well as canonical matter normalizations.  The 
supergravity effects are disentangled via {\it Weyl-rescaling} of the metric, 
{\it i.e.}, %
\begin{equation}
  g_{mn}=\hat{g}_{mn}\cdot e^{-\frac{1}{3}K},
  \label{eq:weyl}
\end{equation}
and by rescaling fermion fields by exponentials of $K$ in order that fermions
and bosons belonging to the same supermultiplet are normalized in the same way
\cite{Wess:1992cp}\cite{Kaplunovsky:1994fg}.\\%
\indent%
The torsion and the chirality constraints of supergravity (SUGRA) are invariant 
under Weyl transformations, given by re-weightings with parameter $e^{\tau}$ and 
$e^{\bar{\tau}}$, where $\tau$ is a chiral superfield.  Some of these 
transformations are given below:%
\begin{center}%
  ${\bf E} \rightarrow e^{2(\tau + \bar{\tau})}{\bf E}$,%
    ~~ $\cal E$$\rightarrow$$e^{6\tau}$$\cal E$,\\%
  $V \rightarrow V$,%
    ~ $\nabla_{\alpha}\rightarrow e^{\tau-2\bar{\tau}}(\nabla_{\alpha}+...)$,\\%
    ~ $\Phi\rightarrow\Phi$,%
    ~ $Q \rightarrow Q$,%
    ~ \& ~ $W_{\alpha}\rightarrow e^{-3\tau}W_{\alpha}.$%
\end{center}%
The SUGRA action itself is not invariant under these transformations.  However, 
we can make insertions of the \quotes{Weyl compensator}, $C$, such that the action is invariant:%
\begin{eqnarray}
  S= -3\int d^{6}z{\cal E}~(\frac{-\bar{\nabla}^{2}}{4} + 2R)
        ~C\bar{C}~exp\{-\frac{1}{3}K(\Phi,\bar{\Phi};Q,\bar{Q}e^{2V})\} 
	 \label{eqn:weylsugra}\\
       \quad + \quad \bigl(\int d^{6}z
       {\cal E}~
       [C^{3}W(\Phi,Q) + \frac{1}{4}f_{ab}
       {\cal W}^{a\alpha}
       {\cal W}^{a}_{\alpha}] + h.c.\bigr). \nonumber
\end{eqnarray}
The invariance under Weyl symmetry is established by the following 
transformations for the compensator:
\begin{center}%
$C\rightarrow e^{-2\tau}C$.
\end{center}%
Because it does not appear in the action with derivatives, it is 
not a propagating degree of freedom and the action is completely equivalent to 
the case without the use of a compensator.  This internal symmetry is broken 
when $<C>\ne 0$ (however the actual value does not matter as it can be chosen 
to be anything), but the breakdown is to nothing.\\
\indent%
At the quantum level, however, the measure is not invariant due to the chiral 
anomaly \cite{Amati:1988ft} for matter fermions, $\Psi$:
\begin{equation}
  d[\Psi] \rightarrow d[\Psi]exp\Bigl\{\frac{3c_{a}}{16\pi^2} 
    \int dz^{6}{\cal E}~ 2\tau {\cal W}^{\alpha}{\cal W}_{\alpha} + h.c.,
    \Bigr\}
\end{equation}
with 
\begin{equation}
c_{a} = T(G_{a}) - \sum_{r} T_{a}(r),
\end{equation}
where $T_{a}(r)$ is the trace over the squared-generator matter representation 
a, and $T(G_{a})$ is the trace over the adjoint squared generator.  In order to 
maintain local Weyl invariance, this anomaly must be canceled by the replacement
\begin{equation}
  f_{a}(\Phi) \rightarrow {\tilde f}_{a} = f_{a}(\Phi) 
  - \frac{3c_{a}}{8\pi^2}~ {\rm ln} C.
\end{equation}
This ensures that the original SUGRA action of Equation (\ref{eqn:sugra_action}) is 
equivalent to the Weyl symmetric action Equation (\ref{eqn:weylsugra}).  We can
see that in the gauge $C=1$, the effect is nil and we are in the original SUGRA
frame.  The \kah-Einstein frame corresponds to the choice  \\
\begin{equation}
  {\rm log}~C + {\rm log}~\bar{C} = \frac{1}{6}K|_{\rm harmonic} ~,
  \label{eqn:EKgauge}
\end{equation}
where on the right we take only the piece that is the sum of chiral plus 
anti-chiral parts.\\
\indent%
We must also do matter field redefinitions in order for them to have canonical
normalizations.  To do this we expand the \kah potential in matter fields
\begin{equation}
  K({\Phi,\bar{\Phi},Q,\bar{Q}e^{2V}}) = K_{m}(\Phi,\bar{\Phi}) 
  + Z_{I\bar{J}}\bar{Q}^{\bar{J}}e^{2V}Q^{I} + \dots.
\end{equation}
For simplicity we consider a single matter field multiplet in representation r 
(the case of more fields is easily generalizable) and find that the kinetic 
terms are contained in 
\begin{equation}
\int dz^{8}{\bf E}C\bar{C}e^{-\frac{1}{3}K_{m}}Z_{r}(\Phi,\bar{\Phi})\bar{Q}e^{2V}Q, 
  \quad Z_{r}=Z_{r}^{\dagger}.
\end{equation}
If one does a field transformation of the form $Q\rightarrow e^{\tau_{Z}}Q$,
the functional measure is again not invariant.  This results in the Konishi 
anomaly \cite{Konishi:1983hf}\cite{Shifman:1986zi} with another contribution
to the measure of the form \cite{deAlwis:2008aq}
\begin{equation}
exp\Bigl\{\frac{-T_{a}}{16\pi^2} 
	  \int dz^{6}{\cal E}~ 2\tau_{Z} {\cal W}^{\alpha}{\cal W}^{\alpha} + h.c. \Bigr\}.
\end{equation}
This subsequently implies redefining the gauge coupling function:
\begin{equation}
  H_{a} \equiv f_{a}(\Phi) - \frac{3c_{a}}{8\pi^2}~ {\rm ln} C
    -\frac{T_{a}(r)}{4\pi^{2}}\tau_{Z}.
  \label{eqn:gauge_coupling}
\end{equation}
We then get canonical normalization of the matter kinetic term,
\begin{equation}
  \int dz^{8}{\bf E}C\bar{C} 
    e^{-\frac{1}{3}K_{m}} 
    e^{\tau_{Z}+\bar{\tau}_{Z}}
    Z_{r}(\Phi,\bar{\Phi})\bar{Q}e^{2V}Q,
\end{equation}
by simply making the choice
\begin{equation}
  \tau_{Z}+\bar{\tau}_{Z} = {\rm ln}(C\bar{C} e^{-\frac{1}{3}K_{m}} 
    Z_{r})|_{harmonic}.
\end{equation}
This together with the appropriate gauge coupling redefinition, Equation 
(\ref{eqn:gauge_coupling}), ensures that the action remains locally Weyl 
invariant.\\%
\indent%
The final step is to determine the gauge couplings and gaugino masses as they
relate to the auxiliary components of all fields, $viz.$, compensator, moduli,
and matter fields.  Denote the lowest component of the transformed gauge 
coupling function (Equation (\ref{eqn:gauge_coupling})) by $h_{a} = H_{a}|$, and 
its real part by $h_{aR} = \Re H_{a}|$.  The mass and gauge coupling are related
by
\begin{align}
  \frac{m_{a}}{g_{a}^{2}} 
    & = \Re[F^{i}\partial_{i}f_{a}(\Phi)]| 
      - \frac{b_{a}}{8\pi^{2}}\frac{F^{C}}{C}
      -\frac{T_{a}(r)}{4\pi^2}F^{i}
        \partial_{i}({\rm ln}(e^{1\frac{1}{3}K_{m}}Z))| 
      \label{eqn:amsb_coupling_mass}\\
    & = \Re[F^{i}\partial_{i}f_{a}(\Phi)]| 
      - \frac{c_{a}}{8\pi^{2}}F^{i}\partial_{i}K_{m}
      -\frac{T_{a}(r)}{4\pi^2}F^{i}
        \partial_{i}({\rm ln}Z_{r}))|, \nonumber
\end{align}
where $i$ is for all moduli of the underlying string theory.  In going to the second line we have imposed the Einstein-\kah gauge by using the F-component of Equation (\ref{eqn:EKgauge}).  The gauge coupling on the other hand is given by
$g_{a}^{-2} = h_{aR}$ and in the Einstein-\kah frame it is
\begin{equation}
  g_{a}^{-2} = \Re f(\Phi)| - \frac{c_{a}}{16\pi^{2}}K_{m}| 
    - \frac{T_{a}(r)}{8\pi^2}{\rm ln}Z_{r}|.
\end{equation}
Then, when considering F-type breaking, these expressions give the appropriate 
gaugino masses and gauge couplings after the \kah potential and the original 
gauge coupling function ($f_{a}$) of the theory have been specified.\\
\indent%
Suppose the theory has a cut-off $\Lambda$, and we wish the renormalize the
gaugino masses and the gauge couplings by evaluating them at a scale $\mu$.  This is done by shifting the Wilsonian gauge coupling function by a finite (1-loop) renormalization,
\begin{equation}
  H_{a}(\Phi,\tau,\tau_{Z}) \rightarrow
    H_{a}(\Phi,\tau,\tau_{Z},\mu) = H_{a}(\Phi,\tau,\tau_{Z})
      -\frac{b_{a}}{8\pi^2}{\rm ln}\frac{\Lambda}{\mu},
  \label{eqn:gaugerun}
\end{equation} 
and re-evaluating the expressions (here, the $b_{a}$ values are the $\beta$-
function coefficients).  However, the gauge couplings we have derived until 
this point are not physical because the kinetic terms for the gauge fields are 
not normalized properly.  We again must do a rotation into the proper frame, 
but this time of the gauge fields.  The details are omitted here, but the 
result is only to shift $\Re H_{a}$ by again using a new chiral superfield
transformation parameters, $\tau_{Z}$ (see \cite{ArkaniHamed:1997mj}\cite{deAlwis:2008aq} for details):
\begin{equation}
  \Re H_{a} \rightarrow \Re H_{a} - \frac{T_{a}}{8\pi^2}\Re \tau_{V} =
    \frac{1}{g_{phys}^{2}}.
\end{equation}
\indent%
It is important to stress that this derivation of the soft mass contributions
has made no impact on the scalar sector of the theory.  The 
\quotes{Weyl} anomaly is often referred to as the \quotes{rescaling} anomaly 
referring to the fact that many authors have renormalized
gauge couplings with an implicit compensator field associated with the mass
scale of the theory, $i.e.$, $\beta$-function running appears usually as
${\rm ln}(\frac{|C|\Lambda}{\mu})$ instead of as it is seen in Equation 
(\ref{eqn:gaugerun}) (without $|C|$).  It is this fact that has led authors to
derive soft contributions for the $scalars$ of the theory in addition to 
gaugino masses \cite{Randall:1998uk}.  However, it has been argued strongly in 
\cite{deAlwis:2008aq} that the problem with the usual derivation is that it is 
based on conformal invariance rather than strictly on Weyl invariance alone.  
It is ultimately argued in that paper that the Weyl anomaly does not contribute 
at all to scalar masses.\\%
\indent%
In this thesis, a somewhat impartial approach is taken regarding the puzzles 
surrounding the anomaly.  The models that are examined in the following 
chapters will have anomaly contributions as prescribed in the \quotes{usual} 
derivations (mAMSB and HCAMSB) that include scalar masses (resulting from
anomaly rescaling) that evolve as
\begin{equation}
  m_{i}^{2} \propto \frac{d \gamma_{i}}{d \ {\rm log \mu}} \mhf^{2},
\end{equation}
where $\gamma_{i}$ is the anomalous dimension.  We also consider cases that
contain anomaly soft contributions to gauginos as advocated by de Alwis 
\cite{deAlwis:2008aq}, as it is in the inoAMSB class.  In all cases considered 
here, upon supersymmetry breaking, the relation (\ref{eqn:amsb_coupling_mass}) 
implies soft gaugino masses of the form\footnote{This is derived explicitly for 
the \ino case in Section \ref{sec:inoamsb_setup}.}
\begin{equation}
M_{a} = \frac{b_{a} g_{a}^{2}}{16 \pi^{2}}\mhf, \ \ a = 1,2,3, 
\end{equation}
with $b_{a}=(\frac{33}{5}, 1, -3)$.  In any case, we will see that the various 
models will be distinguishable at the LHC, at least at the 100$\invfb$ level 
with $\sqrt{s}=$ 14 TeV.
\section{%
  \label{sec:comptools}
  Computational Tools%
}%
\subsection{%
  \label{subsec:rge}
  Renormalization Group Equations%
}%
As described earlier in this Introduction, inputs at the high-scale (the
GUT or string scales for instance) are required as boundary conditions for
evolution of soft \susy breaking parameters.  These parameters then are used in 
determining physical mass parameters at the TeV scale by matching the MSSM 
renormalization group equations (RGEs) with low energy boundary conditions, 
i.e., weak-scale gauge and third generation Yukawa couplings.\\
\indent%
All of the RGE parameter solutions in this research are obtained through an 
up-down iterative matching procedure using the Isasugra subprogram of Isajet
v7.79\footnote{Isajet is publicly available code and can be found at 
\href{http://www.nhn.ou.edu/~isajet/}{http://www.nhn.ou.edu/$\sim$isajet/} .}
\cite{Paige:2003mg}.  Isasugra implements a full two-loop RGE running of the 
MSSM parameters in the $\overline{DR}$-scheme using Runge-Kutta integration.  
The iteration starts with high-precision weak-scale gauge and Yukawa couplings, 
runs up to the the high scale where the user parameters are used as inputs, and 
then returns to the weak scale.  At the end of the iteration, MSSM masses are 
recalculated and the RGE-improved 1-loop corrected Higgs potential is 
minimized.  The $\beta$-functions are also re-evaluated at each threshold 
crossing during each iteration.  The iterations terminate at the prescribed 
level of convergence, which is that all RGE solutions except $\mu$ and $B\mu$ 
are within 0.3\% of the last iteration.  Because of their rapid variation at 
the weak-scale, the latter are required to have convergence at the 5\% level.\\%
\indent
Isajet v7.79 has the mAMSB and HCAMSB models coded into it.  No extra coding was
necessary for inoAMSB as the high-scale inputs are non-universal gaugino masses
in the ratio 6.6:1:-3 as usual for AMSB, while all scalar and $A$-parameters 
inputs are highly suppressed, i.e., zero.  For convenience, the parameters of 
each of the three models described in this work are listed in Table 
\ref{tab:modpars}.\\%
\begin{table}
  \begin{center}
    \begin{tabular}{|ll|}%
	\hline\hline%
	&\\%
      ~~Model  & ~~Parameters\\
	&\\%
	\hline
	&\\%
      mAMSB:   & $\mnot, \mhf, \tanb, sign(\mu)$\\[3pt]%
      HCAMSB:  & $\alpha, \mhf, \tanb, sign(\mu)$\\[3pt]%
      inoAMSB: & $\mhf, \tanb,  sign(\mu),$\\%
               & $(M_{1},M_{2},M_{3})=(6.6,1,-3)\times\mhf$\\
      &\\%
    \hline\hline%
    \end{tabular}
    \caption{%
      \label{tab:modpars}
    }%
  \end{center}
\end{table}
\subsection{%
  \label{subsec:eventsim}
  Event Simulation
}%
Isajet 7.79 - 7.80 is used for the simulation of signal and background events at the LHC. A toy detector simulation is employed with calorimeter cell size
$\Delta\eta\times\Delta\phi=0.05\times 0.05$ and $-5<\eta<5$. The hadronic calorimeter (HCAL) energy resolution is taken to be $80\%/\sqrt{E}+3\%$ for $|\eta|<2.6$ and forward calorimeter (FCAL) is $100\%/\sqrt{E}+5\%$ for 
$|\eta|>2.6$. The electromagnetic (ECAL) energy resolution is assumed to be $3\%/\sqrt{E}+0.5\%$. We use the UA1-like jet finding algorithm GETJET with jet cone size $R=0.4$ and require that $E_T(jet)>50$ GeV and $|\eta (jet)|<3.0$. Leptons are considered isolated if they have $p_T(e\ or\ \mu)>20$ GeV and $|\eta|<2.5$ with visible activity within a cone of $\Delta R<0.2$ of 
$\Sigma E_T^{cells}<5$ GeV. The strict isolation criterion helps reduce
multi-lepton backgrounds from heavy quark ($c\bar c$ and $b\bar{b}$) 
production.\\
\indent%
We identify a hadronic cluster with $E_T>50$ GeV and $|\eta(j)|<1.5$ as a 
$b$-jet if it contains a $B$ hadron with $p_T(B)>15$ GeV and $|\eta (B)|<3$ within a cone of $\Delta R<0.5$ about the jet axis. We adopt a $b$-jet tagging efficiency of 60\%, and assume that light quark and gluon jets can be mis-tagged as $b$-jets with a probability $1/150$ for $E_T<100$ GeV, $1/50$ for $E_T>250$ GeV, with a linear interpolation for $100$ GeV$<E_T<$ 250 GeV
\cite{Kadala:2008uy}.\\%
\indent%
Isajet is capable of generating events for a wide variety of models.  Once the parameters of the theory are defined, RGE evolution determines weak scale masses
and $2\rightarrow 2$ processes are weighted by their cross sections and 
generated.\\
\indent%
In addition, background events are generated using Isajet for
QCD jet production (jet-types include $g$, $u$, $d$, $s$, $c$ and $b$
quarks) over five $p_T$ ranges as, for example, in Tables \ref{tab:hcaC2} and
\ref{tab:ino_c2cuts}. 
Additional jets are generated via parton showering from the initial and final 
state hard scattering subprocesses.  Also generated are backgrounds in the 
$W+jets$, $Z+jets$, $t\bar{t}(173.1)$ and $WW,\ WZ,\ ZZ$ channels. The $W+jets$ 
and $Z+jets$ backgrounds use exact matrix elements for one parton emission, but 
rely on the parton shower for subsequent emissions.
\subsection{Dark Matter}%
For all the models considered in this work, to calculate the relic density of 
neutralinos and direct detection rates we used, respectively, the IsaRED and 
IsaRES subroutines of the IsaTools package found in Isajet \cite{Paige:2003mg}. 
To calculate indirect detection rates following from neutralino scattering, 
annihilation, and co-annihilations, the DarkSUSY\footnote{Publicly available at 
\href{http://www.physto.se/~edsjo/darksusy/}
{http://www.physto.se/$\sim$edsjo/darksusy/} .} \cite{Gondolo:2004sc} package 
was used.  DarkSUSY, by default, uses the Isasugra subprogram of Isajet v7.69 
\cite{Paige:2003mg} to calculate the SUSY spectrum, with exception to Higgs 
masses for which it uses FeynHiggs \cite{Heinemeyer:1998yj}.  In order to 
perform the calculations for our particular models, we transplanted the Isajet 
v7.79 code into the DarkSUSY package.

\chapter{%
  \label{chap:mamsb}
  Minimal AMSB
}%

With the origin of the anomaly contributions to the particle spectrum understood
we can now look at models with high scale inputs that are connected to the TeV
scale through renormalization group running.  The first model we encounter has 
been analyzed extensively in the literature 
\cite{Barr:2002ex}\cite{Paige:1999ui} and will serve as the model of comparison 
for the other AMSB models to be considered.  The model is known as the 
$Minimal$ Anomaly Mediated Supersymmetry Breaking Model (mAMSB) and its most
important aspects, relevant for LHC and DM considerations, are highlighted in 
this chapter.\\%
\indent%
The \mam model has the attractive feature that it depends on only a few 
GUT-scale input parameters:%
\begin{equation}%
\mnot, \mhf, \tanb, \ {\rm and \ sign}(\mu),
\label{mamsb_pspace}
\end{equation}%
each of which have been mentioned in Chapter \ref{chap:intro} except for
$\mnot$, which will be discussed shortly.  As discussed in Section 
\ref{sec:grav_soft}, the AMSB contributions to the gaugino masses are given by  
\begin{align}%
     M_{1}  & =   \frac{33}{5} \ \frac{g_{1}^{2}}{16\pi^{2}}\mhf^{2} 
	 \label{eqn:amsb_m1}\\%
     M_{2}  & = \ \ \ \  \frac{g_{2}^{2}}{16\pi^{2}}\mhf^{2} \\%
     M_{3}  & =   -3\frac{g_{3}^{2}}{16\pi^{2}}\mhf^{2},%
     \label{eqn:amsb_m3}
\end{align}%
where $g_{i}$ are the running MSSM gauge couplings.  Furthermore, the scalar 
masses and trilinear parameters are given by
\begin{align}%
  m_{{\tilde f}}^{2}  & =   -\frac{1}{4}\{\frac{d\gamma}{dg}\beta_{g}
    +\frac{d\gamma}{df}\beta_{f}\}\mhf^{2} \ \ {\rm and}\\
  A_{f}  & =   \frac{\beta_{f}}{f}\frac{\mhf}{16\pi^{2}},%
\end{align}%
where $\beta_{g}$ and $\beta_{f}$ are, respectively, the gauge coupling and 
Yukawa coupling $\beta$-functions, and their correspond 
$anomalous \ dimensions$ are denoted by $\gamma$.  Note that the above soft 
terms are parametrically tied to supersymmetry-breaking through $\mhf$ (Section 
\ref{sec:grav_soft}).  Also note that all of the above equations are valid at 
$any$ scale.\\
\indent
There are several phenomenologically important points to be made about the AMSB 
contributions.
\begin{itemize}%
\item[$i.$] Anomaly contributions to scalars are the same for particles with 
		the same quantum numbers, while first and second generation Yukawa 
		couplings are negligible.  In the case that AMSB dominates the 
		soft contributions, flavor-violation is safely avoided.  In Figure
		\ref{fig:soft_amsb}, the running of the soft parameters 
		$\mb, \mw, M_{3}$, the third generation squark doublet mass 
		($\sqrt{m^{2}_{\spart{Q}{3}}}$), the right-handed sbottom mass 
		($\sqrt{m^{2}_{\spart{b}{R}}}$), and the bottom trilinear parameter 
		($A_{b}$) are shown.  There is implicit dependence on $A_{b}$ in 
		both of the squark mass parameters.  
  	      \begin{figure}[h!]%
		  \begin{center}%
		    \includegraphics[width=.8\textwidth]{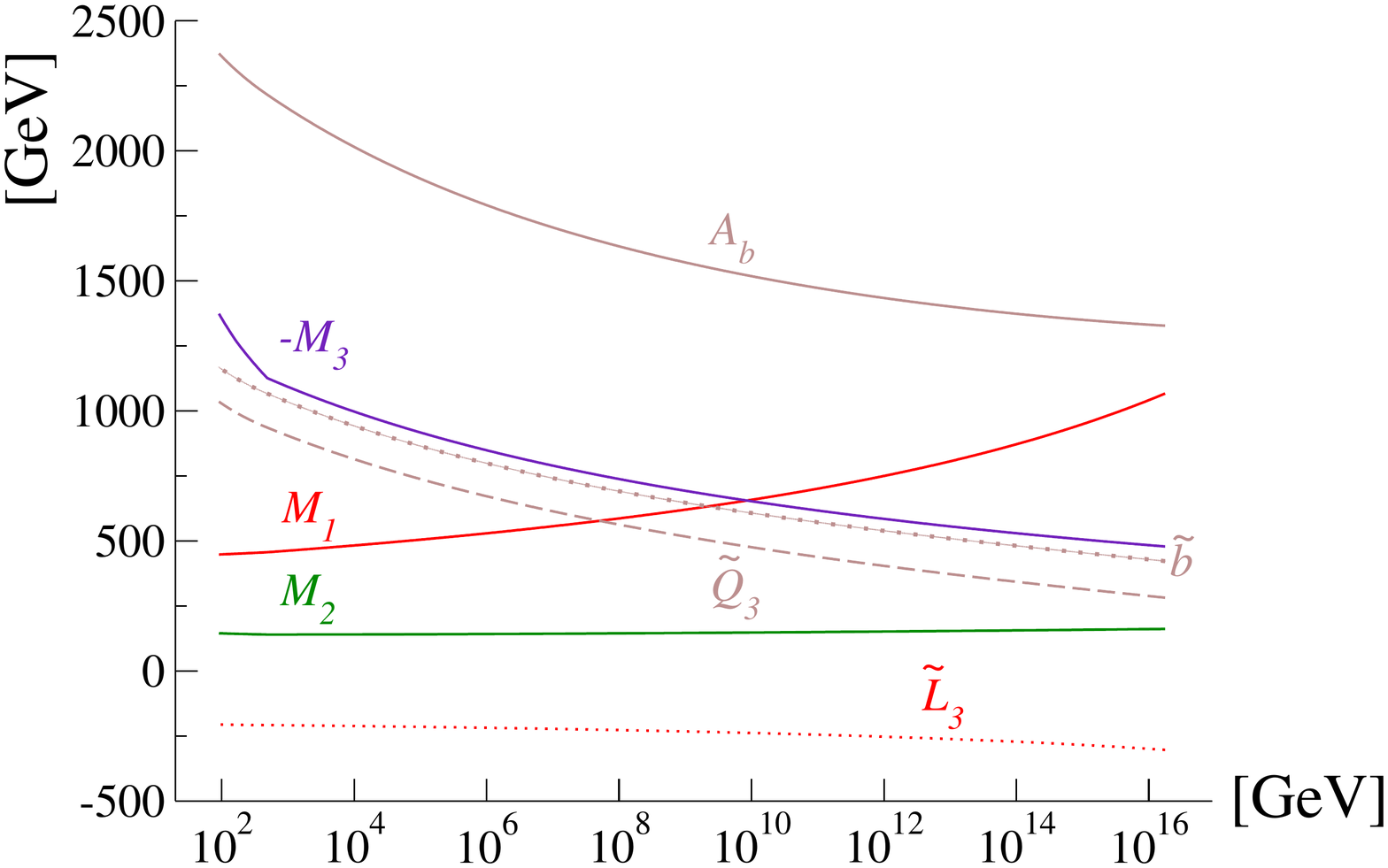}%
		    \caption{%
			Running of soft parameters to the weak scale in the mAMSB 
			scenario for the point ($\mnot$, $\mhf$, $\tanb$) = 
			(0 GeV, 50 TeV, 10) with $\mu > 0$.
		      \label{fig:soft_amsb}%
		    }%
		  \end{center}%
		\end{figure}%
\item[$ii.$] The anomaly contributions to scalars are partially determined by 
		 anomalous dimensions, which are negative for sleptons.  After RGE 
		 running to the weak scale, sleptons remain with negative masses 
		 (see Figure $\spart{L}{3}$ in \ref{fig:soft_amsb}).  The $minimal$ 
		 solution to this problem is to introduce the ad hoc parameter, 
		 $\mnot$, in \ref{mamsb_pspace} at the $GUT$-scale to 
		 prevent tachyonic sleptons at the weak scale.  This is achieved 
		 simply by adding $\mnot^{2}$ to all scalar squared-masses and the $
		 \mnot$ value is taken as a free parameter of the \mam model.
\item[$iii.$] Scalars with different helicities are nearly (left/right)
		 degenerate at the weak scale in the \mam model.  This is despite 
		 possible splitting at the $GUT$ scale.  This is demonstrated in
		 Figure \ref{fig:ltrt_amsb} where, for example, selectrons and up-
		 squarks masses are run from the $GUT$ scale to the weak scale for 
		 a mAMSB point with $\mnot$=300 GeV, $\mhf=$ 50 TeV, $\tanb=10$, 
		 and $\mu > 0$.  While left and right sparticle masses are split at 
		 the high scale, the evolution (accidentally) drives them closer at
		 lower scales.
		 \begin{figure}[h]%
		   \begin{center}%
		     \includegraphics[width=.8\textwidth]{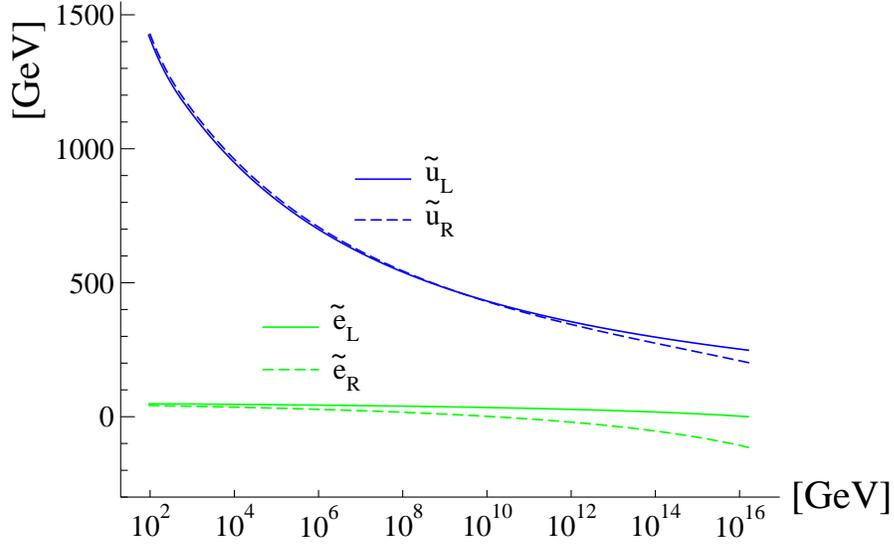}%
		     \caption{%
		       Demonstration of near-degeneracy of left and right 
			 superparticles at weak scale in the mAMSB scenario for the 
			 point ($\mnot$, $\mhf$, $\tanb$) = 
			 (300 GeV, 50 TeV, 10) with $\mu > 0$.
		       \label{fig:ltrt_amsb}%
		     }%
		   \end{center}%
		 \end{figure}%
\item[$iv.$] Requiring weak-scale gaugino masses places $\mhf$ at much
		  higher values than in other supersymmetry-breaking scenarios.  
		  With $\mhf \sim O(10-100)$ TeV range, gravitinos have a short 
		  enough lifetime to avoid the gravitino problem 
		 \cite{Kohri:2005wn}\cite{Kawasaki:2008qe}\cite{Pradler:2006hh}.
\item[$v.$] After evaluating gaugino masses at the weak scale (Equations 
(\ref{eqn:amsb_m1} - \ref{eqn:amsb_m3}) ), $|M_{2}|$ is 
		invariably the lightest of them (see Figure \ref{fig:soft_amsb}).  
		This, along with the potential minimization conditions, affects the 
		diagonalization of the neutralino and chargino mass matrices of 
		Equations (\ref{eqn:nino_matrix}) and (\ref{eqn:cino_matrix}) to 
		produce $wino$-like lightest mass-eigenstates in both cases. Thus 
		$\nino$ and $\cino$ have nearly degenerate masses in AMSB type 
		models.
\end{itemize}%

This last point is particularly important because it potentially allows for a 
discrimination between AMSB and other types of supersymmetry-breaking 
effects at the LHC.  Denote the mass difference between the lightest chargino 
and the lightest neutralino, $m_{\cino} - m_{\nino}$, by $\dmchi$.  As $\dmchi$ 
decreases, the chargino decay width decreases and its lifetime becomes longer. 
Thus, there is a possibility of detecting $\cino$s as highly ionizing tracks 
(HITS) in LHC detectors. \\
\indent%
$\dmchi \sim O({\rm MeV})$ splittings are rather small but $\cino$ can usually 
decay to e or $\mu$ through the three-body process $\cino \rightarrow e\nu 
\nino$.  When the gap opens beyond the pion mass ($\sim$ 140 MeV) hadronic 
decays of the chargino are allowed \cite{Chen:1999yf}\cite{Kuhn:1990ad}.
Further enlarging the mass gap leads to channels with 2$\pi$ and 3$\pi$ final 
states.  The $\cino$ width would be smallest when it decays to a single lepton 
and obviously grows as other channels open.  Chargino detection then falls 
into two categories \cite{Barr:2002ex}:\\%
\begin{itemize}%
  \item $\dmchi < m_{\pi^{+}}$ : $\pi$ decay modes are unavailable and $\cino$ 
	  can leave a track far out in the muon chambers!%
  \item $m_{\pi^{+}} < \dmchi < 200$ MeV : high $p_{T}$ charginos decay in the
        inner detector region.  The decays $\cino \rightarrow \pi^{\pm}\nino~$ 
        and $\cino \rightarrow  l\nu\nino~$ are accessible with $l$ and $\pi$
 	  emitted softly.  This regime has a clear signal.  The charginos appear
	  as highly ionizing tracks in the inner detector and decays to SM 	
	  particles.  But the SM particles are too soft for energy deposits in
    	  the Ecal or the Hcal and appear to contribute to the missing energy 	
	  already taken away by the $\nino$.  This leads to an observable HIT 
	  that terminates, or a stub, and large amounts of $\etm$.%
\end{itemize}%
We will come back to this in Chapters \ref{chap:hcamsb} and \ref{chap:inoamsb}
when we discuss the LHC phenomenology of AMSB models at greater depth.

\chapter{%
  Hypercharged AMSB
  \label{chap:hcamsb}
}%

\section{%
  Introduction to the HCAMSB Model
  \label{sec:hcaintro}
}%
Hypercharged anomaly-mediation is composed of two mechanisms that induce 
masses for visible sector matter fields.  The soft masses come from the anomaly 
mediation already discussed and an additional $U(1)$ mediation, and the latter 
depends intricately on a particular D-brane setup.  While this form of 
mediation is not general in the way that anomaly contributions are, it is an 
interesting pathway to understanding how D-brane constructions can impact the 
visible sector.  In the next section the description of how the $U(1)$ gives 
masses to MSSM fields is given.  Then the $U(1)$ will be paired with AMSB to 
give the full HCAMSB model.  Following the next section the phenomenology of 
the model will be discussed.\\ %
\section{%
  Geometrical Setup with D-branes
  \label{sec:hcamod}
}%
There should be some comments made about D-branes first because the model 
relies on them. D-branes are extended objects on which strings can terminate 
with `D'irichlet boundary conditions.  For this discussion we consider only 
type IIB string theory with Dp-branes\footnote{
Dp-branes have d-p-1 Dirichlet and p+1 Neumann boundary conditions.
} %
that have $p = 1 - 9$, odd.  These branes fill the usual 3+1 space-time, but 
can have superfluous dimensions that must be wrapped by internal cycles to make 
them effectively invisible.  A D7 brane, for example, requires a 4-cycle 
wrapped within the internal geometry, and D5 requires a 2-cycle, etc.\\%
\indent
D-branes have interesting features that are useful for constructing realistic
field theories.  
Most importantly for this discussion are the following two 
properties: 1.) chiral matter can exist as open strings at the intersection of 
two D-branes, and 2.) interactions with local curved geometry lead to bound 
states of D-branes known as \quotes{fractional branes} when branes of different 
dimensions are involved.  In this model the bound states of branes occur at 
singularities in the Calabi-Yau (CY) manifold.  There are two CY singularities; 
the brane located at one of the singularities will be the visible brane, and 
the other will be hidden.  In addition, the two branes will share properties 
that allow for the U(1) mediation.\\%
\subsection{%
  \label{subsec:u1med}
  U(1) Mediation of \susy Breaking%
}%
The main idea behind $U(1)$ mediation is that, given a proper geometrical setup, 
a brane can communicate \susy breaking to another brane despite there not being 
any open string modes connecting them.  An F-type breaking occurring on one 
brane can be mediated to another through bulk closed-string modes that have 
special couplings to gauge fields on the branes.
\indent%
To understand the mechanism, first consider a single D5 brane separated within 
the CY manifold with a U(1) symmetry and associated gauge field $A$.  Dp-branes 
are themselves sources of generalized gauge fields $C_{p+1}$, for p=1, 3, \& 5 
in IIB theories \cite{Douglas:2006es} and exist in the bulk.  $C_{p+1}$ have 
induced linear Chern-Simons couplings to the brane gauge fields.  In 
particular, $C_{4}$ couples to $A$ through
\begin{align}
{\cal L}_{CS} = C_{4}\wedge dA + A \wedge F_{5},
\label{lagrangian}
\end{align}
where $F_{5}$ is the five-form field strength of $C_{4}$.  $F_{5}$ has the 
special property of self-duality in 10 dimensions, that is, $F_{5} = *F_{5}$.  
The equations of motion for this Lagrangian are%
\begin{align}
dF_{5}=dA\wedge \delta_{brane}, \qquad \textrm{and} \qquad 
*F_{5}=dC_{4}+A\wedge\delta_{brane.}%
\end{align}
\indent 
Now consider a 2-cycle $\alpha$ wrapped by the D5 brane, and the four-cycle 
$\beta$ dual to $\alpha$.  When the extra dimensions are reduced, $C_{4}$ leads 
to a massless two-form for each two-cycle wrapped by the brane, and each 
4-cycle leads to a massless scalar.  For the cycles $\alpha$ and $\beta$ this 
means that we have%
\begin{align}
C & =  \int_{\alpha} C_{4} \qquad \textrm{(2-form)}\\%
\varphi & = \int_{\beta} C_{4} \qquad \textrm{(scalar)},%
\end{align}
and we assume that there are no other cycles around.  $C$ and $\varphi$ are 
related via the self-duality of $F_{5}$.  A unique basis can be chosen for the 
expansion of $F_{5}$ consisting of a 2-form $\sigma$ and its dual 4-form $\rho$ 
that satisfy the following properties:%
\begin{align}%
  \int_{\alpha} \sigma = 1, \qquad \int_{\beta} \rho = 1, \quad \textrm{and}   
    \quad \int_{CY} \sigma \wedge \rho = 1.%
\end{align}%
These forms are related by Hodge duality:
\begin{align}%
  *_{6} \ \rho = \mu^{2}\sigma, \qquad \int_{CY}\sigma\wedge *_{6} \ \sigma =
    \frac{1}{\mu^{2}},
\end{align}%
\noindent where $\mu \sim$ string scale and characterizes the size of the 
compactification.  We can expand $F_{5}$ in this basis and KK reduce it:
\begin{center}
$F_{5} = dC\wedge \sigma + (d\phi + A)\wedge \rho + ... $.
\end{center}
From this self-duality of $F_{5}$ is satisfied provided that
\begin{align}
*_{4}dC =  \mu^{2}(d\phi +A),
\label{equ:stuckrel}
\end{align}
where $\mu^{2}$ appears in the $\sigma$ and $\rho$ duality relations.  This 
equation is a solution to the equations of motion of the action dual to the 4D 
version of Equation (\ref{lagrangian}) + $C$-kinetic term.  This results in a 
low energy mass term for A.\\
\indent
Now when we consider the actual setup which is similar to that already 
considered but consists of two D5 branes, one visible (V) and one hidden (H).  
There are $U(1)$ gauge bosons on the branes denoted now by $A_{V}$ and $A_{H}$, 
and each D5 brane wraps its own two- and four-cycles (and associated 2- and 4- 
forms), denoted by $\alpha_{V}$, $\alpha_{H}$, $\beta_{V}$, and $\beta_{H}$.  
Actually we choose the CY geometry such that these cycles are topologically the
same.  For instance, if $\alpha_{V}$ and $\alpha_{H}$ are topologically the 
same two-cycle they can be continuously transformed into each other.  If we 
follow the same procedure outlined above, we again arrive at Equation 
\ref{equ:stuckrel} but with $A \rightarrow A_{V} + A_{H}$, and this leads to a 
mass for the combination $A_{V} + A_{H}$.  With string scale compactifications, 
this combination is quite heavy, and is lifted from the low-energy spectrum.  
However, the remaining light combination, $A_{V} - A_{H}$, does survive to low 
energies as a light vector boson \cite{Dermisek:2007qi}.  In this model, this 
combination is identified with the ordinary hypercharge boson, and the effects 
of supersymmetry breaking are imparted to the superpartner of $A_{H}$.  Thus, 
the bino acquires extra soft contributions that will alter the usual \mam 
contributions in interesting ways to be described in the next section.\\%
%
%
%
\section{%
  \label{sec:spect}
  Spectrum and Parameter Space
}%
\begin{figure}[t]
  \begin{center}%
    \includegraphics[width=.75\textwidth]{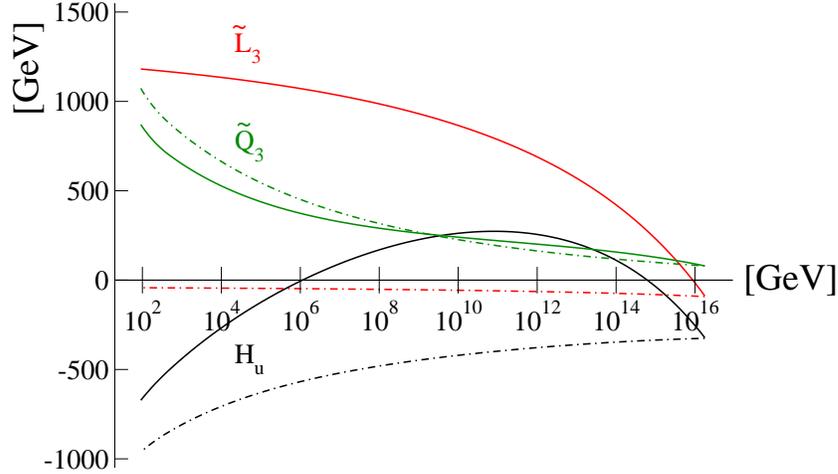}%
    \caption{%
      \label{fig:scalarrun}%
      HCAMSB (solid) and pure AMSB (dotted) mass evolution from GUT scale to 
	the weak scale for third-generation sleptons, third-generation squarks, 
	and the up-type Higgs.%
    }%
  \end{center}%
\end{figure}

\begin{figure}
  \begin{center}%
    \includegraphics[angle=-90,scale=.3,trim= 0cm 0cm 3cm 0cm,clip,
      width=.75\textwidth]{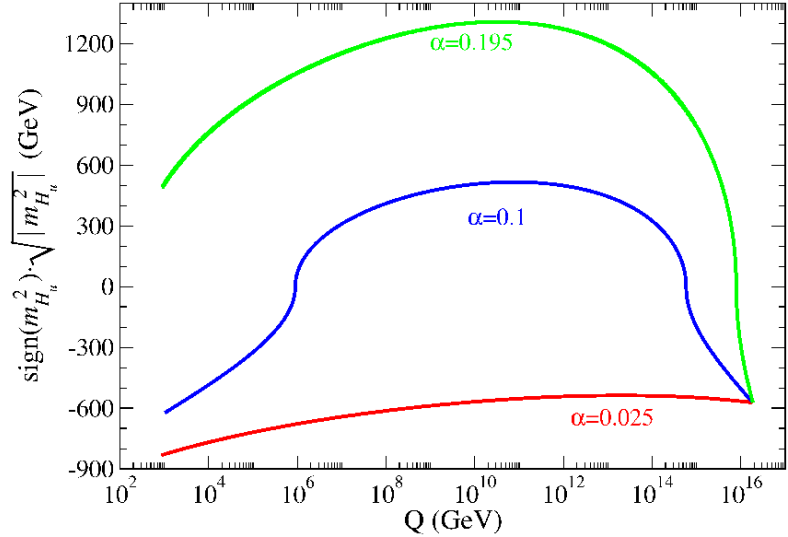}%
    \caption{$m_{H_{u}}^{2}$ evolution for the three values of $\alpha$: 0.025,
      0.01, and 0.195.
      \label{fig:higgsrun}%
    }%
  \end{center}%
\end{figure}

\begin{figure}
  \begin{center}%
    \includegraphics[width=.75\textwidth]{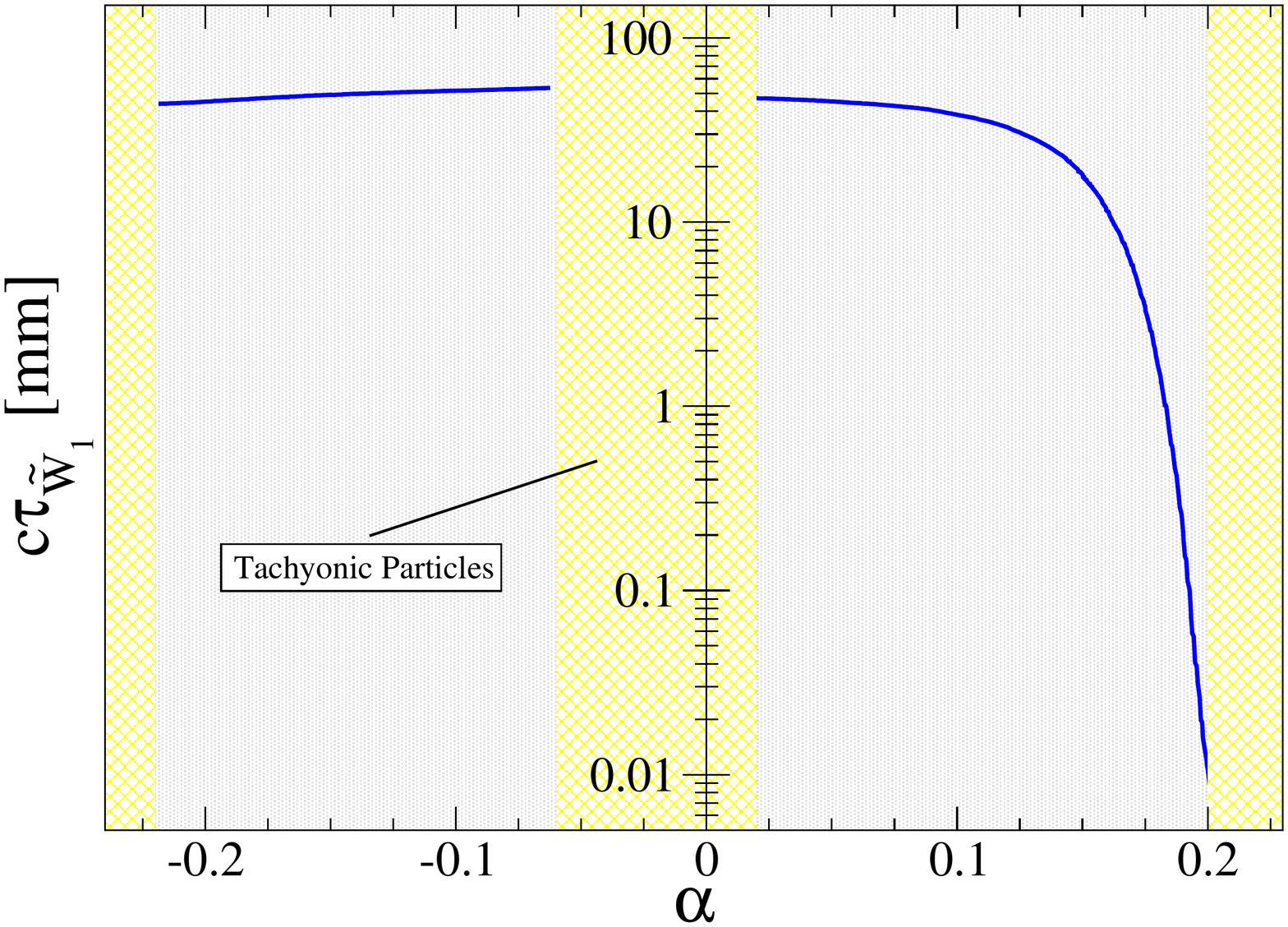}%
    \caption{$\spart{W}{1}$ lifetime as a function of $\alpha$.
      \label{fig:w1lifetime}%
    }%
  \end{center}%
\end{figure}

In this section we examine the mass spectrum and understand how it evolves from 
GUT-scale running to the weak scale.  Because we have the Feynman rules
for the MSSM, all rates can be calculated as long as the masses are given.
This includes rates of rare processes that can be sensitive to the choice of 
model parameters.  These measurements are used to place constraints
on the parameter space of the theory.  We will also constrain the parameter 
space as much as possible from other theoretical considerations such as the 
requirement of proper electroweak symmetry breaking.\\%
\indent
The soft mass contribution RGEs are%
\begin{align}
  M_{1} & =  {\tilde M}_{1} 
                + \frac{33}{5}\frac{g_{1}^{2}}{16\pi^{2}}\mhf^{2}\\[3pt]%
  M_{a} & =  \frac{b_{a}g_{a}^{2}}{16\pi^{2}}\mhf^{2} \quad a=2,3\\[3pt]%
  m_{i}^{2} & =  \frac{1}{4}\bigr\{%
    \frac{\partial \gamma}{\partial g}\beta_{g} 
    + \frac{\partial \gamma}{\partial f}\beta_{f}\bigl\}\mhf^{2}\\[3pt]%
  A_{f} & =  \frac{\beta_{f}}{f}\mhf.%
\end{align}%
The difference between these and mAMSB renormalizations is that the equation 
for $M_{1}$ has an extra ${\tilde M}_{1}$ input at the high scale (GUT scale 
chosen for convenience) that accounts for the hypercharge contribution.  Winos 
and gluinos only receive negligible two-loop contributions from the bino
\cite{Dermisek:2007qi}. %
To compare the $U(1)$ effects with those of AMSB we rewrite ${\tilde M}_{1}$
proportional to $\mhf$ as
\begin{center}%
  ${\tilde M}_{1}$ = $\alpha~\mhf$\\%
\end{center}%
so that the bino soft mass reads as%
\begin{equation}%
  M_{1} =  (\alpha + \frac{33}{5}\frac{g_{1}^{2}}{16\pi^{2}})\mhf.
  \label{eqn:hca_bino}%
\end{equation}%
The parameter space of the HCAMSB model is then
\begin{center}%
  $\alpha, \mhf, \tanb,$ and $sign(\mu)$,\\%
\end{center}%
which resembles the mAMSB space except that $\alpha$ will be the parameter that 
helps to avoid tachyonic sleptons at the weak scale.\\%
\begin{figure}
  \begin{center}%
    \subfigure[]{\includegraphics[width=0.7\textwidth]{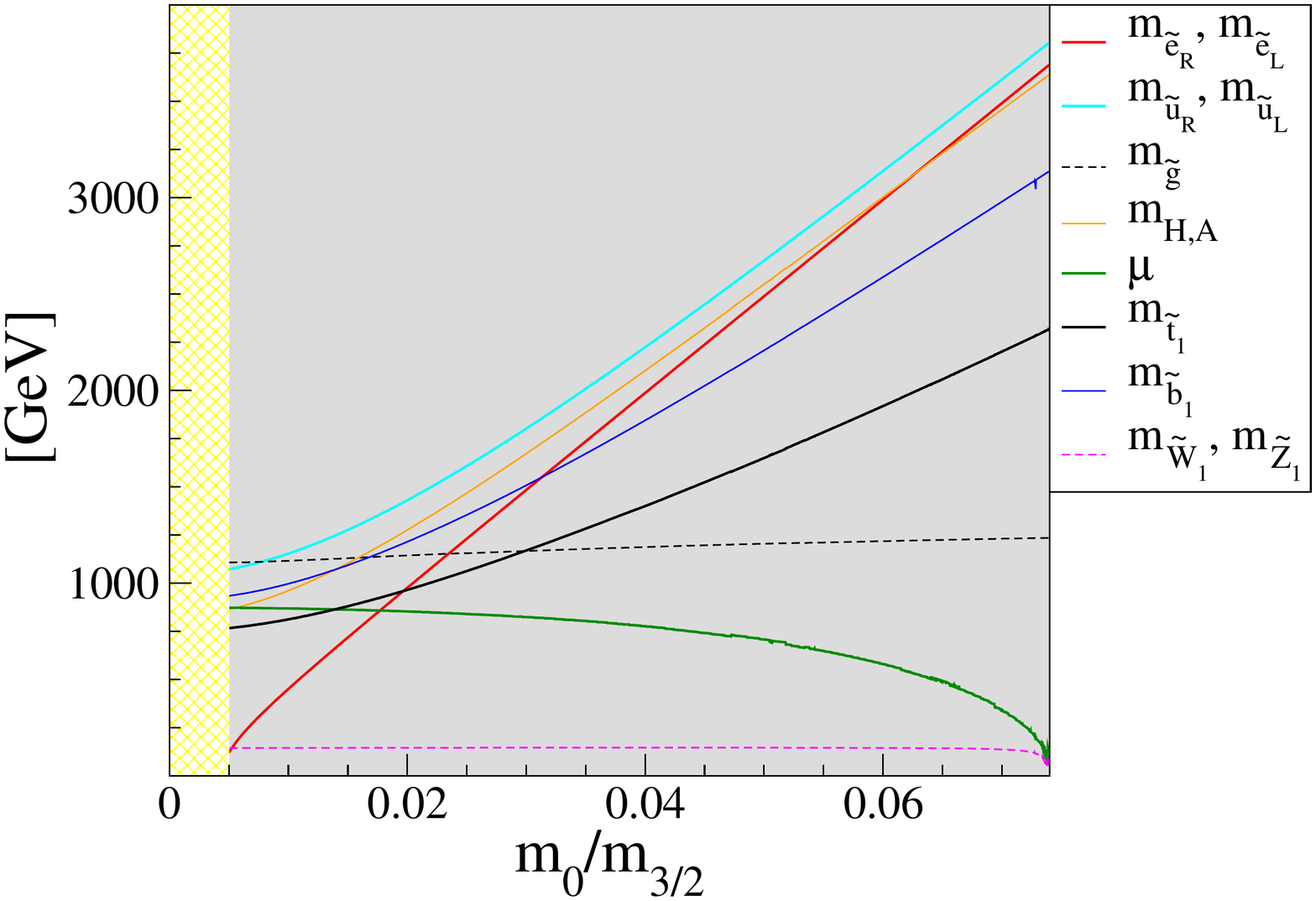}}
    \subfigure[]{\includegraphics[width=0.7\textwidth]{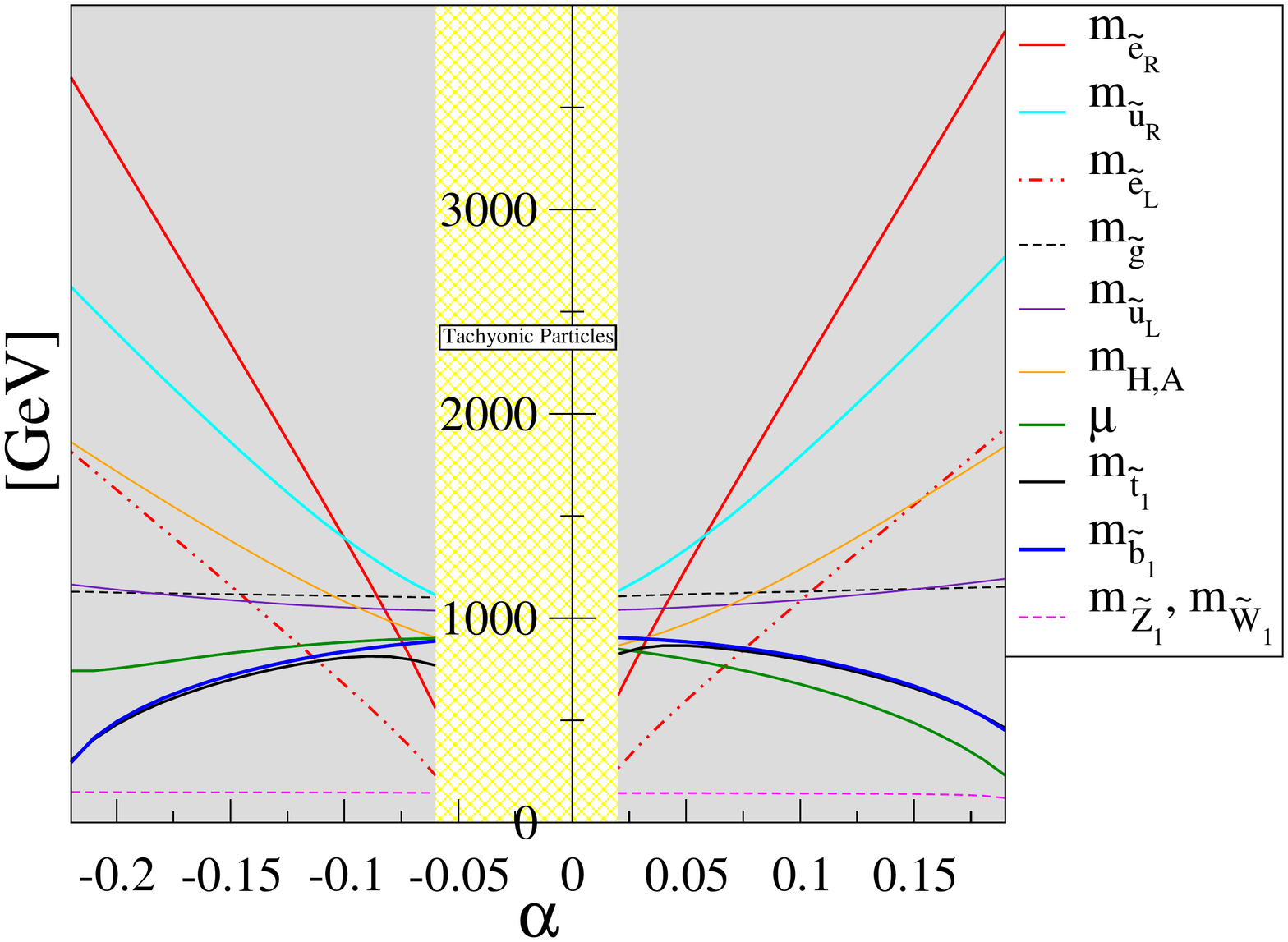}}
    \caption{%
      \label{fig:spectra}%
      Mass spectrum for \mam (left) and \hca (right) with model parameters 
      $\mhf=50$ TeV, $\tanb =10$, and $\mu >0$.%
    }%
  \end{center}%
\end{figure}%
\indent
Figure \ref{fig:spectra} shows the particle mass spectrum at the weak scale for
both the \mam and \hca models against $\mnot$ and $\alpha$ respectively.  In 
order to compare the effect of $\mnot$ with that of $\alpha$ (= 
$\frac{{\tilde M}_{1}}{\mhf}$) we consider its ratio to $\mhf$, while fixing 
$\mhf = 50$ TeV and $\tanb=$ 10.  We will fully explore the impact of varying 
the other parameters later.\\%
\indent
These plots have a few qualitative similarities.  The first feature is that 
each has a yellow-shaded region corresponding to RGE solutions with tachyonic 
sleptons.  These regions are forbidden because the scalar potential should 
not be minimized by charged scalars, as this would lead to breakdown of 
electric charge conservation.  These regions are where they would be 
expected; for \mam this is near $\mnot \sim 0$, and for \hca it is around 
$\alpha \sim 0$ where the bino contribution is small and pure AMSB is 
recovered.  In general, as each parameter increases the general trend is for 
masses to increase (although here are important exceptions in each case).  Each 
has nearly-degenerate $\spart{Z}{1}$ and $\cino$ and their masses remain 
relatively flat for all $\mnot$ and $\alpha$.  This is also the case for the 
$\gl$.  Conversely $\mu$ is seen to decrease with increasing values of the 
parameters.  The upper edges themselves are due to improper breaking of the  
electroweak symmetry which is signaled by $\mu^{2} < 0$.\\%
\indent
There are also notable differences between the spectra that eventually lead to 
distinct phenomenology.  In the \mam case, there is a near-degeneracy between 
left and right particles of the same flavor due to the nearly-equal 
$\beta$-functions and that $\mnot^{2}$ is universally added to all scalar 
squared masses.  \hca on the contrary exhibits a left-right split spectrum. For 
example, \mam has $m_{{\tilde e}_{R}}$ $\simeq$ $m_{{\tilde e}_{L}}$ but these 
values can be seen to differ by over 0.3 TeV in the \hca case for all 
$\alpha$.\\%
\indent
The plot also shows that stop and sbottom masses actually decrease in \hca for
larger hypercharge contributions unlike the other scalars of the theory.  Within
\mam all scalars increase with $\mnot^{2}$ because this contribution is simply
added to all high-scale, scalar, squared soft mass values .\\%
\indent
The parameters $\mu$, $\mb$, and $\mw$ determine the composition of the 
neutralinos and are different between \mam and \hca models.  These parameters 
mix to form the eigenstates of the neutralino mass matrix, ${\tilde Z}_{i}$. 
Because the values of these parameters and their relative ordering determine 
the composition of the LSP, they are also responsible for its interaction 
properties.  For example, Table \ref{tab:ninocomp} shows the ordering of 
neutralino mass parameters and the main components of the neutralino mass 
eigenstates for the benchmark point for \mam and Point 1 for \hca (the 
selection of representative points will discussed in the next section).  Since 
$\mw$ has the smallest value in each case the LSP is wino-like with a small 
mixture of bino and higgsino components.  However, $\ninos{2}$ is bino-like for 
\mam and mainly higgsino for \hca.  We can see already that when $\ninos{2}$ is 
produced in collisions that its decays will be heavily model-dependent and will 
lead to final states with either strong bino or higgsino couplings.  This will 
be crucial is distinguishing between the two models in the LHC section.\\%
\begin{table}%
\begin{center}%
  \begin{tabular}{|c|c|c|}%
    \hline%
    &&\\%
    \multirow{2}[4]{2cm}{Neutralino Eigenstates} & \mam & \hca \\[6pt]%
    \cline{2-3}%
    &&\\%
    & $\mu > \mb > \mw$ & $\mb > \mu > \mw$ \\[6pt]%
    \hline %
    &&\\[3pt]%
    ${\tilde Z}_{1}$ & 99.09 \% wino; $\sim$ 1 \% bino-higgsino%
      & 99.04 \% wino; $\sim$ 1 \% bino-higgsino\\[6pt]%
    ${\tilde Z}_{2}$ & 99.35 \% bino; $<$ .7 \% higgsino-wino%
      & 70.30 \% higgsino; 29.09 \% bino\\[6pt]
    ${\tilde Z}_{3}$ & 99.72 \% higgsino;  $<$ .30 \% bino-wino%
      & 99.74 \% higgsino; $<$ .30 \% bino-wino\\[6pt]
    ${\tilde Z}_{4}$ & 98.73 \% higgsino; $<$ .30 \% wino-bino%
      & 70.88 \% bino; 28.93 \% higgsino\\[6pt]%
    &&\\[1pt]%
    \hline%
  \end{tabular}%
  \caption{%
    \label{tab:ninocomp}
    Compositions of the neutralino mass eigenstates.  Both models have 
    wino-like LSP, but heavier states differ due to order of $\mu$, $\mb$, and 
    $\mw$.
  }%
\end{center}%
\end{table}%
\indent%
We should now turn our attention to understanding the sources of the spectrum 
patterns described in the previous few paragraphs.  In order to understand 
mass parameters at the weak-scale we need to examine how they evolve from the 
GUT scale.  It is seen from the RGEs that the AMSB contribution to scalar 
masses are determined by anomalous dimensions, $\gamma_{i}$.  Because 
$\gamma_{i}$  are positive for squarks and negative for sleptons, the masses of 
particles begin at the GUT scale with their respective signs.  Each scalar 
receives a contribution from the hypercharge mediation of the form 
\cite{Dermisek:2007qi}
\begin{equation}%
  \delta m_{i}^{2}(Q) = -\frac{3}{10\pi^{2}}g_{1}^{2}Y_{i}^{2}M_{1}^{2}
    log \Bigl(\frac{Q}{M_{GUT}} \Bigr)
\end{equation}
where $Y_{i}$ is the hypercharge of the $i^{th}$ scalar.  This contribution 
serves to uplift masses as the scale $Q$ decreases.  This accounts for 
the large left-light splitting between the scalars in figure \ref{fig:spectra}, 
since right-handed particles have a greater hypercharge in general. As $Q$ 
approaches the weak-scale the masses become larger and Yukawa terms tend to 
dominate the RGEs which leads to suppression of masses.  Figure 
\ref{fig:scalarrun} shows these effects for hypercharge anomaly mediation and 
for pure AMSB.  In the case of pure AMSB, the sleptons begin with negative mass 
parameters and are tachyonic at the weak scale as expected.  Sleptons have 
large hypercharge absolute values, and the figure shows that the hypercharge 
contribution lifts slepton masses to positive values for the HCAMSB case.\\%
\indent
Pure hypercharge mediation however has the opposite effect on the third 
generation squarks.  As we evolve to the weak scale, the left-handed stops may 
become tachyonic because they have relatively small hypercharge values and the 
same large Yukawa couplings of top quarks.  But the stops also receive AMSB 
contributions that leave them with positive masses in the HCAMSB scenario.  The 
anomaly and hypercharge mediation mechanisms are complementary in the respect 
that each helps to avoid the tachyonic particles that are present when only the 
other mechanism is present.\\%
\indent
We can understand the extra suppression with increasing $\alpha$ for third 
generation squarks by taking a closer look at the RGEs.  Both $\spart{t}{L}$
and $\spart{b}{L}$ belong to the same $SU(2)$ doublet, $Q_{3}$.  The running of
the doublet mass includes the terms \cite{Baer:2006rs}%
\begin{align}
  \frac{d  m_{Q_{3}}^{2}}{d (log \ Q)} \ni Y_{t}^{2}X_{t} + Y_{b}^{2}X_{b}
\end{align}
where $X_{t,b}$ is defined to be
\begin{align}
  X_{t} & = m_{Q_{3}}^{2} + m_{\spart{t}{R}}^{2} + m_{H_{u}}^{2} + A_{t}^{2}\\%
  X_{b} & = m_{Q_{3}}^{2} + m_{\spart{b}{R}}^{2} + m_{H_{d}}^{2} + A_{t}^{2}.%
\end{align}
Again, right-handed particles have larger hypercharge than $SU(2)$ doublets, and
the values $m_{\spart{t}{R}}^{2}$ and $m_{\spart{b}{R}}^{2}$ steepen the running
slope near the weak scale.  Then it is evident that for larger values of 
$\alpha$ in HCAMSB, the third generation doublet receives extra suppression from
Yukawa effects relative to other generations.  This suppression affects 
$\spart{t}{L}$- and $\spart{b}{L}$-production rates at the LHC for the \hca 
model as will be discussed in the next section.\\%
\indent
We can also see which RGE effect leads to $\mu$-suppression with increasing 
$\alpha$.  The tree-level scalar minimization given by Equation 
\ref{eq:mincond} shows that $\mu$ goes as $-m_{H_{u}}^{2}$, and increasing 
$m_{H_{u}}^{2}$ implies decreasing $\mu$.  The $m_{H_{u}}^{2}$ RGE 
includes the terms
\begin{equation}
  \frac{d m_{H_{u}}^{2}}{d \ log(Q)} \ni -\frac{3}{5}g_{i}^{2}M_{i}^{2}
    + 3Y_{t}^{2}X_{t}.
\end{equation}
Again the large $M_{1}$ value uplifts $m_{H_{u}}^{2}$ in the early running from
$Q = M_{GUT}$, and at low $Q$ Yukawa effects again dominate over the 
hypercharge effects.  This is shown in Figures \ref{fig:scalarrun} and 
\ref{fig:higgsrun}.  In Figure \ref{fig:higgsrun} three values of $\alpha$ have
been chosen and the 1-loop RGEs have been used in the evolution.  As $\alpha$
increases the Higgs mass increases at the weak scale.  For the largest $\alpha$
shown the Higgs mass is positive which could imply negative $\mu^{2}$ and
therefore no electroweak symmetry breaking.  However, when large 1-loop 
corrections are added to the scalar potential $\mu^{2}$ is once again positive.
In Figure \ref{fig:higgsrun}, $\alpha \sim 0.195$ is the upper edge of 
parameter space beyond which EW-breaking does not occur.\\%
\indent
As already mentioned, $\mu$ is an important parameter in the neutralino sector.
Towards higher values of $\alpha$, $\mu$ decreases as mentioned above and moves
nearer to the value of $M_{2}$ (over all of the parameter space $\spart{Z}{1}$ 
is {\it mainly} wino and $M_{\spart{Z}{1}} \simeq M_{2}$).  Here the 
$\spart{Z}{1}$ mass state is a mixture of wino and higgsino.  Because of this,
the mass splitting $\dmchi$ between $\spart{W}{1}$ and $\spart{Z}{1}$ will
increase leading to a shorter lifetime of the former.\\%
\indent
To understand this last point, note that in the limit of $\mu$, $M_{1} >> M_{2}
$ the mass eigenstates $\spart{Z}{1}$ and $\spart{W}{1}^{\pm}$ form an 
$SU(2)$ triplet with common mass $M_{2}$.  The symmetry is broken by 
gaugino-higgsino mixing which leads to mass splitting between the neutralino 
and the chargino.  As we saw in Chapter \ref{chap:mamsb}, $\dmchi$ is important 
in the detection of charginos.  We now see that the detection of the chargino 
can depend on the value of $\alpha$, and its lifetime is shown in Figure
\ref{fig:w1lifetime}.  As can be seen, larger $\alpha$ leads to shorter 
$\spart{W}{1}$ lifetimes.\\%
\indent
All of these results can be summarized as in Table \ref{tab:cases} for the \mam 
point with $(\mnot,\mhf,\tanb)$ = (300 GeV, 50 TeV, 10), and for similar 
parameter choices for two \hca points, with a low and high value of $\alpha$, 
0.025 and 0.195, defined respectively as points HCAMSB1 and HCAMSB2.  By 
comparing the model lines we see that the table reflects all of the \hca 
features: L-R splitting, wino-like $\spart{Z}{1}$/$\spart{W}{1}$, $m_{Q_{3}}$- 
and $\mu$-suppression and $\dmchi$ increase with increasing $\alpha$, etc.  \\
\begin{table}[t]
  \begin{center}
    \begin{tabular}{|lccc|}
      \hline\hline
	&&&\\%
      parameter & mAMSB & HCAMSB1 & HCAMSB2 \\
	&&&\\%
      \hline
	$\alpha$    & --- & 0.025 & 0.195  \\
	$\mnot$       & 300 & --- & --- \\
	$\mhf$   & $50\ {\rm TeV}$ & $50\ {\rm TeV}$ & $50\ {\rm TeV}$ \\
	$\tan\beta$ & 10 & 10 & 10 \\
	$M_1$       & 460.3   & 997.7 & 4710.5 \\
	$M_2$       & 140.0   & 139.5 & 137.5 \\
	$\mu$       & 872.8 & 841.8 & 178.8 \\
	$m_{\spart{g}{}}$   & 1109.2 & 1107.6 & 1154.2 \\
	$m_{\spart{u}{L}}$ & 1078.2 & 1041.3 & 1199.1 \\
	$m_{\spart{u}{R}}$ & 1086.2   & 1160.3 & 2826.3 \\
	$m_{\spart{t}{1}}$& 774.9 & 840.9 & 427.7 \\
	$m_{\spart{t}{2}}$& 985.3 & 983.3 & 2332.5 \\
	$m_{\spart{b}{1}}$ & 944.4 & 902.6 & 409.0  \\
	$m_{\spart{b}{2}}$ & 1076.7 & 1065.7 & 1650.7 \\
	$m_{\spart{e}{L}}$ & 226.9 & 326.3  & 1973.1 \\
	$m_{\spart{e}{R}}$ & 204.6 & 732.3  & 3964.9 \\
	$m_{\spart{W}{2}}$ & 879.2 & 849.4 & 233.1 \\
	$m_{\spart{W}{1}}$ & 143.9 & 143.5 & 107.1 \\
	$m_{\spart{Z}{4}}$ & 878.7 & 993.7 & 4727.2 \\ 
	$m_{\spart{Z}{3}}$ & 875.3 & 845.5 & 228.7 \\ 
	$m_{\spart{Z}{2}}$ & 451.1 & 839.2 & 188.6 \\ 
	$m_{\spart{Z}{1}}$ & 143.7 & 143.3 & 105.0 \\ 
	$m_A$       & 878.1 & 879.6 & 1875.1 \\
	$m_h$       & 113.8 & 113.4 & 112.1 \\ 
	\hline\hline
    \end{tabular}%
    \caption{Parameters and masses in~GeV units for three case study points 
	mAMSB, HCAMSB1 and HCAMSB2 using Isajet 7.79 with $m_t=172.6$ GeV and 
      $\mu 	>0$.%
    }%
    \label{tab:cases}
  \end{center}
\end{table}

\section{%
  \label{sec:hca_constraints}
  HCAMSB Model Constraints%
}
\begin{figure}[t]%
  \begin{center}%
    \includegraphics[width=1\textwidth]{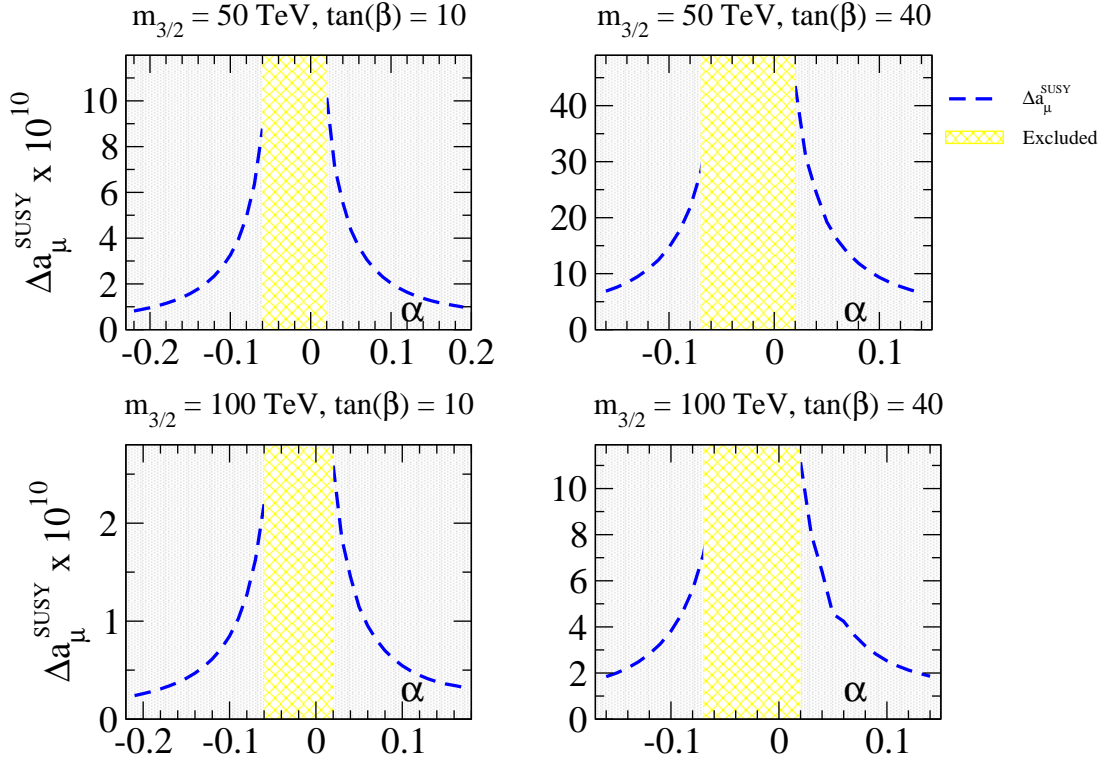}%
    \caption{The \susy contribution to $(g-2)_{\mu}$ as a function of $\alpha$.
      \label{fig:g_2}%
    }%
  \end{center}%
\end{figure}

We now examine the allowed parameter space of the \hca model and begin to infer
some sparticle mass ranges observable at the LHC. There are both theoretical
and experimental constraints for the model parameter space.  On the theory side 
we have the requirement of proper electroweak symmetry breaking which implies 
that only $m_{H_{u}^{2}}$ can become less than zero because the scalar 
potential cannot be minimized by fields with conserved charges.  If other mass 
parameters (including $\mu$) become negative and/or $H_{u}$ does not, then the 
corresponding region of parameter space is prohibited.  Experimental constraints
come in two varieties: direct and indirect.  The direct constraints considered
are the LEP2 mass limits on $m_{\spart{W}{1}}$ or $m_{h}$.  Indirect 
constraints come from measurements that would be sensitive to supersymmetric 
particles appearing in loops.  The indirect constraints considered here are the 
branching ratio from inclusive radiative B-meson decays, 
$BF(b \rightarrow s \gamma)$ \cite{Misiak:2006zs}, and $(g-2)_{\mu}$ 
measurements.  Consideration of cosmological constraints are postponed until 
Chapter \ref{chap:dm} where Dark Matter in AMSB models is discussed.\\%
\begin{figure}[t!]
  \begin{center}%
    \includegraphics[width=1\textwidth]{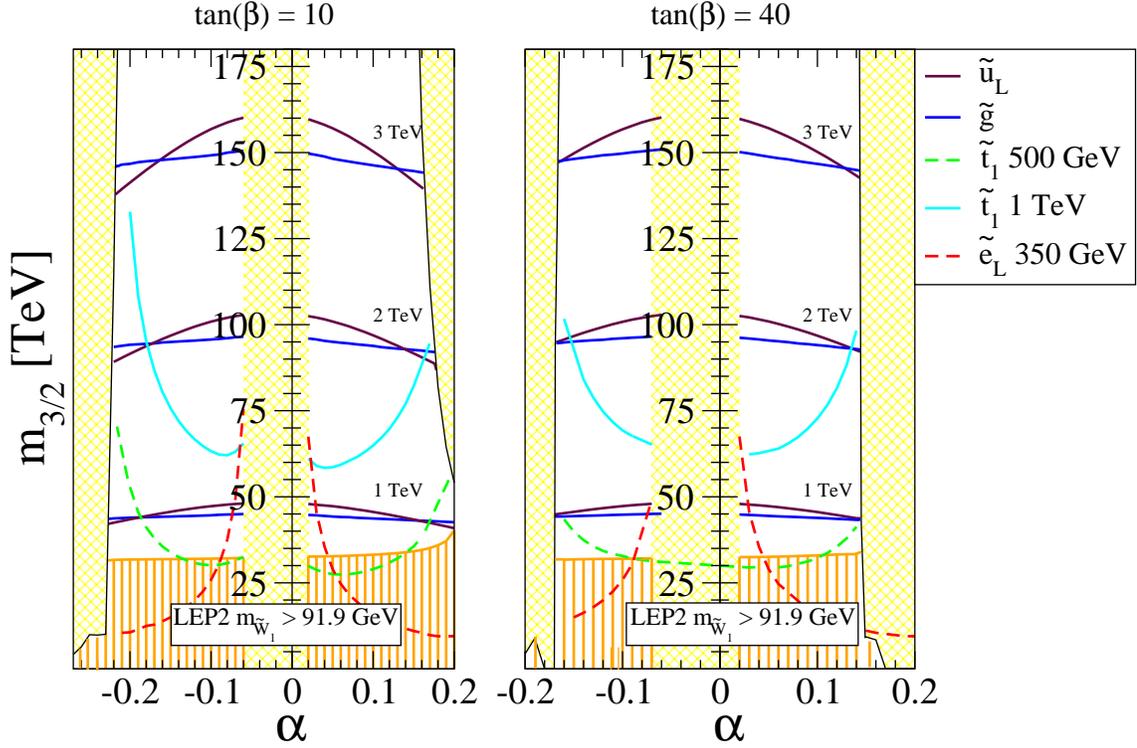}%
    \caption{Allowed $\mhf$-$\alpha$ parameter space for $\tanb = 10$ and $40$ 
	with $\mu > 0$ and $m_{t} = 172.6$ GeV.  The orange vertical lines at 
	lower values of $\mhf$ represent the LEP2 excluded region where 
	$m_{\cino} < 91.9$ GeV.  The yellow regions at the far-right, far-left, 
	and center are regions where electroweak symmetry is improperly broken.
	The white regions are acceptable, and constant-mass contours for 
	$\spart{u}{1},\spart{g}{},\spart{t}{1}$, and $\spart{e}{L}$ are shown.
    }%
    \label{fig:reachpar}%
  \end{center}%
\end{figure}%
\indent%
We begin exploring the parameter space by plotting various masses in the 
$\mhf-\alpha$ plane in Figure \ref{fig:reachpar}.  The plots in the figure have
excluded regions for improper EW symmetry breaking (yellow-thatched region) due
to tachyonic sleptons at $\alpha \sim 0$ and $\mu^{2} < 0$ at the extreme 
$\alpha$ values.  The orange region at lower $\mhf$ values is where $m_{\cino}$ 
is below the LEP2 limit of 91.9 GeV \cite{LEPSUSYWG:2002} in the search for 
nearly degenerate $\cino$s and $\nino$s.\footnote{%
The LEP2 limit on the Higgs mass is $m_{H_{(SM)}} > 114.4 GeV$.  While this 
limit is possibly constraining there is an estimated $\pm 3$ error in the 
calculation of $m_{h}$.  Since $m_{h} \gtrsim 111 GeV$ over the entire allowed 
parameter space we do not show this constraint in Figure \ref{fig:reachpar}.
}  \\%
\indent
The plot shows that the region with $\mhf \lesssim 30$ TeV are excluded by the 
LEP2 chargino limit.  For $\mhf \sim 30$ TeV, we have for the gluino $\mg \sim$ 
730 GeV which is beyond the reach of any reasonable search at the Tevatron, so 
the discovery potential for this model must be investigated for the LHC. There 
is not an upper-bound on the gravitino mass, but the plot extends up to 
$\mhf \sim$ 150 and 160 TeV for the 3 TeV contours in $\mg$ and 
$m_{\spart{u}{L}}$ respectively.  These are somewhat above the reach of 2.1 TeV 
for squarks and 2.8 TeV for gluinos \cite{Barr:2002ex} predicted in the case of 
\mam for the LHC.  We will explore the reach of this model in Section 
\ref{sec:hcalhc}.  \\%
\begin{figure}[t]
  \begin{center}%
    \includegraphics[width=1\textwidth]{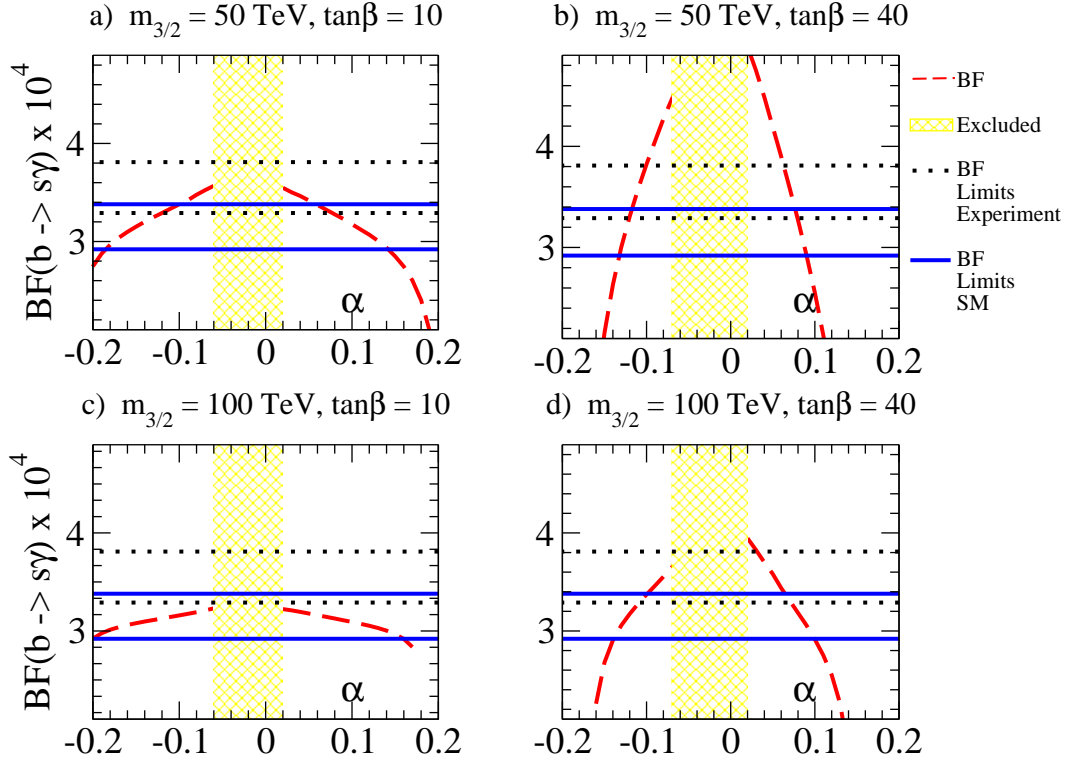}%
    \caption{The branching fraction for $b \rightarrow s \gamma$ vs. $\alpha$ 
	for combinations of $\mhf =$ 50 and 100 TeV and $\tanb =$ 10 and 40.
      \label{fig:bsgamma}%
    }%
  \end{center}%
\end{figure}%
\indent
We also check whether there are regions of parameter space that agree with 
indirect measurements.  Figure \ref{fig:bsgamma} shows the branching fraction
of $b\rightarrow s\gamma$ as a function of $\alpha$ for four pairings of the
$\mhf$ and $\tanb$ parameters to span the space.  The dashed red line 
represent the \hca values calculated with using the Isatools subroutine 
ISABSG \cite{Baer:1996kv}.  The blue line is the theory result for the SM at
order $\alpha_{s}^{2}$ \cite{Misiak:2006zs} with the range of values 
$BF(b\rightarrow s\gamma) = 3.15 \pm 0.23 \times 10^{-4}$.  The black-dotted
lines are the combined experimental branching fraction results of CLEO, BELLE, 
and BABAR \cite{Barberio:2006bi} and have values in the range 
$BF(b\rightarrow s\gamma) = 3.55 \pm 0.26 \times 10^{-4}$.  The plot shows that
there are regions of near-agreement in the parameter space of each case.  
Exceptions include the low $\alpha$ region in frame b) where the BF is too big, 
and very high $\alpha$ values in frames a), c), and d) where BF is too small.
\indent
Finally, we close this section with contribution to $\frac{(g-2)_{\mu}}{2}$, 
denoted as $\Delta a_{\mu}^{\susy}$, in the \hca model.  Figure \ref{fig:g_2} 
shows $\Delta a_{\mu}^{\susy}$ calculated with ISAAMU from Isatools 
\cite{Baer:2001kn}, with four plots with high and low values of $\mhf$ and 
$\tanb$.  In each of the four cases, low $\alpha$ leads to relatively light 
$\spart{\mu}{L}$ and $\spart{\nu}{L}$ that appear in loop corrections to the 
photon vertex.  This appears in the plots as larger corrections at low $\alpha$
to SM predictions.  Parameter values with 
$\Delta a_{\mu}^{\susy} \gtrsim 60 \times 10^{-10}$ or $< 0$ are disfavored
\cite{Baer:2006rs}.

\section{%
  \label{sec:hcasig}%
  HCAMSB Cascade Decays Patterns%
}%
\begin{figure}%
  \begin{center}%
    \includegraphics[width=.75\textwidth]{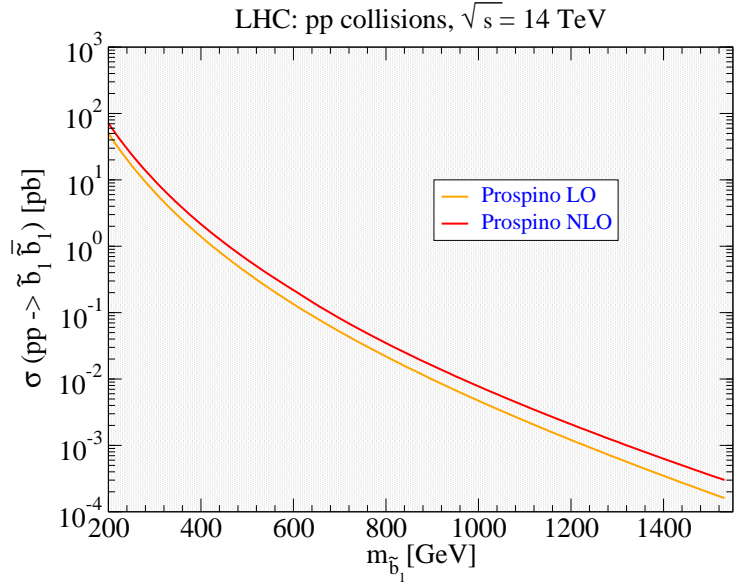}%
    \caption{Total $\spart{b}{1}\bar{\tilde b}_{1}$ production cross section.
      \label{fig:sbottomxs}%
    }%
  \end{center}%
\end{figure}

\begin{figure}%
  \begin{center}
    \subfigure[]{\includegraphics[width=0.7\textwidth]{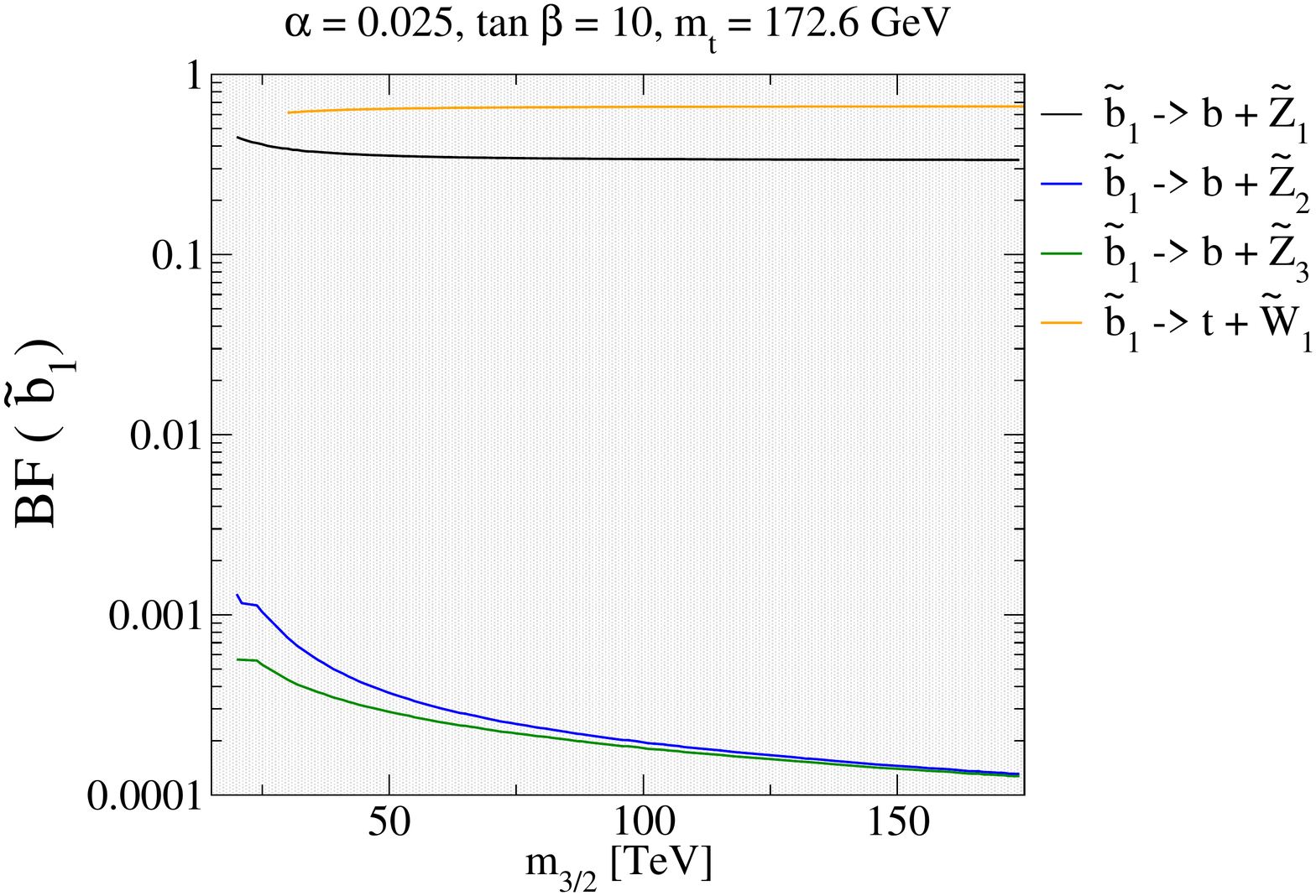}}\\%
    \subfigure[]{\includegraphics[width=0.7\textwidth]{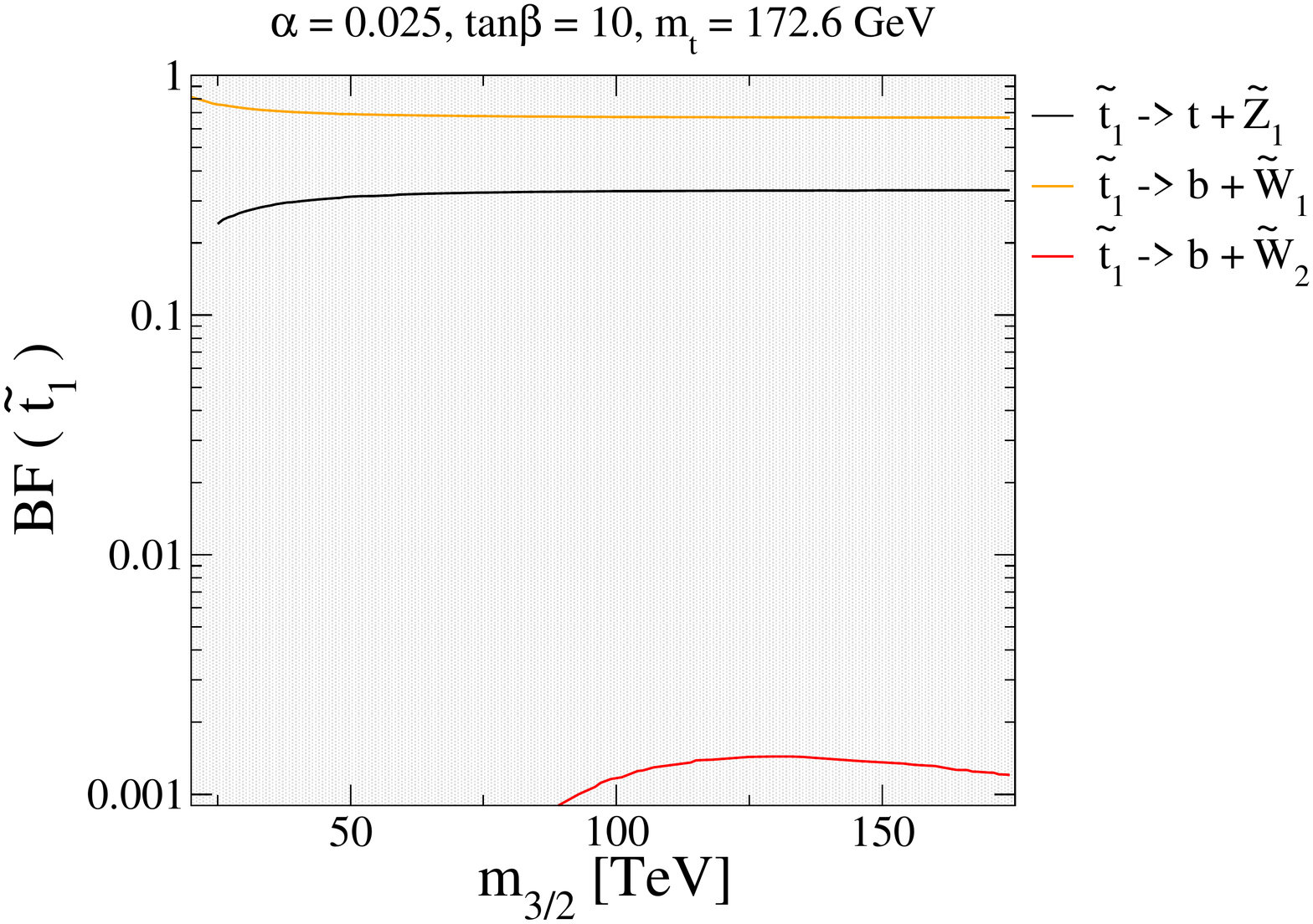}}%
    \caption{Branching fraction for $\sb$ (a) and $\st$ (b) vs. $\mhf$ 
	       for model parameters $\alpha=0.025$, $\tanb = 10$, and 
             $sgn(\mu) > 0$.
      \label{fig:bfm32}%
    }%
  \end{center}%
\end{figure}

All of the signatures at the LHC for the HCAMSB model will be shown in this 
section.  That is, the physical outcome of those regions of parameter space 
that are not already excluded will be explored.  We will also rely on the 
findings of section \ref{sec:spect} to understand the production rates when 
necessary.\\%
\indent
We begin by examining the largest LHC production cross sections in order to 
understand what are the \hca signatures.  Note that all of the following 
analysis is done for the 14 TeV pp beams.  Table \ref{tab:hcarates} shows for 
the three model lines the total \susy production cross section and the 
percentages for the pair-production for gluinos, squarks, EW-inos, sleptons, 
and light stops.  It is seen that the EW-ino pairs dominate over all other 
production rates while squarks and gluinos are produced in lower, but still 
significant, amounts.\\%
\indent
The dominant EW-ino cross sections come from the reactions
$pp \rightarrow \spart{W}{1}^{+}\spart{W}{1}^{-}X$ and
$pp \rightarrow \spart{W}{1}^{\pm}\spart{Z}{1}X$.  However, the $\cino$ decays
and the $\nino$ do not lead to calorimeter signals that can serve as triggers 
for LHC detectors.  So we must instead look at squark and gluino production 
mechanisms. \\%
\begin{table}[h!]%
  \begin{center}%
    \begin{tabular}{|lccc|}%
	\hline \hline%
	&&&\\%
	& \mam & \hca 1 & \hca 2 \\[6pt]%
	\hline
	&&&\\[0pt]%
	$\sigma\ [{\rm fb}]$ & $7.7\times 10^3$ & $7.4\times 10^3$ & $1.8\times
        10^4$ \\[3pt]%
	$\spart{g}{} ,\spart{q}{}\ pairs$ & 15.0\% & 15.5\% & 14.3\% \\[3pt]%
	$EW$-$ino\ pairs$ & 79.7\% & 81.9\% & 85\% \\[3pt]%
	$slep.\ pairs$ & 3.7\% & 0.8\% & -- \\[3pt]%
	$\spart{t}{1}{\tilde{\bar{t}}}_{1}$ & 0.4\% & 0.2\% & 5.5\% \\[3pt]%
	&&&\\%
	\hline
	&&&\\%
	$BF(\spart{Z}{2}\to\spart{Z}{1} Z)$ & $0.01\%$ & 7.7\% & 22.3\% \\%
	&&&\\%
	\hline \hline%
    \end{tabular}%
    \caption{\hca rates for the LHC.
      \label{tab:hcarates}
    }%
  \end{center}%
\end{table}%
\indent%
We first consider that $\mhf$ =  30 TeV, the lowest value not excluded by 
experiment, and qualitatively discuss the emergence of the final states.  
Because of the absence of third-generation partons in the initial state, squark 
and gluino rates are determined by \susy QCD and only depend on their 
respective masses\footnote{For production/decay rates see Appendices of Ref. 
\cite{Baer:2006rs}.}.  Since $m_{\spart{q}{L}}$, $m_{\spart{q}{R}}$ and $\mg$ 
have similar values for low $\alpha$, the final states $\gl\gl,\gl\sq$, and 
$\sq \sq$ are produced at similar rates.  In general, R-squarks are heavier 
than their L-partners due to the U(1) contribution.  Already for low $\alpha$, 
$m_{\spart{u}{R}}$ and $m_{\spart{c}{R}}$ $> \mg$ (similarly to $\mhf$ = 50 TeV 
in Figure \ref{fig:spectra}).  The subsequent decays of these squarks enhance 
the production of gluinos through $\spart{u}{R} \rightarrow u \gl$ and 
$\spart{c}{R} \rightarrow c \gl$.  Gluinos finally decay in quark-squark pairs, 
and they have the highest rates into $b\sbbar + h.c.$ and $t\stbar + h.c.$ 
and subdominant rates into other $q\sq_{L}$ pairs.\\%
\indent%
Conversely, as $\alpha$ increases, right-handed sparticle masses become larger 
to the point that eventually they cannot be produced in collisions.  At higher
values of $\alpha$, left-sparticles become heavier than ${\tilde g}$, while 
$\spart{t}{1}$ and $\spart{b}{1}$ are significantly lighter and again can be
found in the main quark-squark decay modes of gluinos.  At the highest $\alpha$
values, gluinos decay only into quark-squark pairs involving $\st$s or 
$\sb$s: $\gl\rightarrow t\spart{t}{1}$ or $b\spart{b}{1}$.\\%
\indent%
We also find for high $\alpha$ that, in addition to 
$\gl\rightarrow\st/\sb + X$, direct production of $\sb\sbbar + h.c.$ and 
$\st\stbar + h.c.$ pairs dominate over $\gl\gl, \gl\spart{u}{L}$, and 
$\gl\spart{c}{L}$.  The  $\spart{t}{1}$ and $\spart{b}{1}$ mass eigenstates are 
mainly L-squarks at high $\alpha$, and appear approximately in a weak doublet.  
Thus they have nearly the same mass, $m_{\spart{t}{1}} \approx$ 
$m_{\spart{b}{1}}$, and their production rates are nearly identical.  Figure 
\ref{fig:sbottomxs} shows the direct production cross section of sbottom pairs 
using the program Prospino2.1 \cite{Beenakker:1996ed}, and it is understood 
that pair-production cross section of stops is nearly equal.\\%
\indent%
To recap the findings of the previous three paragraphs we see that light stop
and sbottom squark, as well as top and bottom quarks are produced in the 
following ways:\\%
\begin{center}%
  \begin{tabular}{cl}%
    low $\alpha$:  & $\bullet$ gluino production and main decays to quark-
			    squark pairs;\\
    high $\alpha$: & $\bullet$ gluino production and subsequent decay purely 
				     to quark-squark pairs;\\%
	   		 & $\bullet$ direct stop and sbottom pair production.\\%
    \end{tabular}%
  \end{center}%
\noindent
Obviously the production of stops and sbottoms is important in \hca 
phenomenology.  Then to proceed we need to examine the decay patterns of these 
particles to arrive at the final state.\\%
\begin{figure}[th!]%
  \begin{center}
    \includegraphics[width=.75\textwidth]{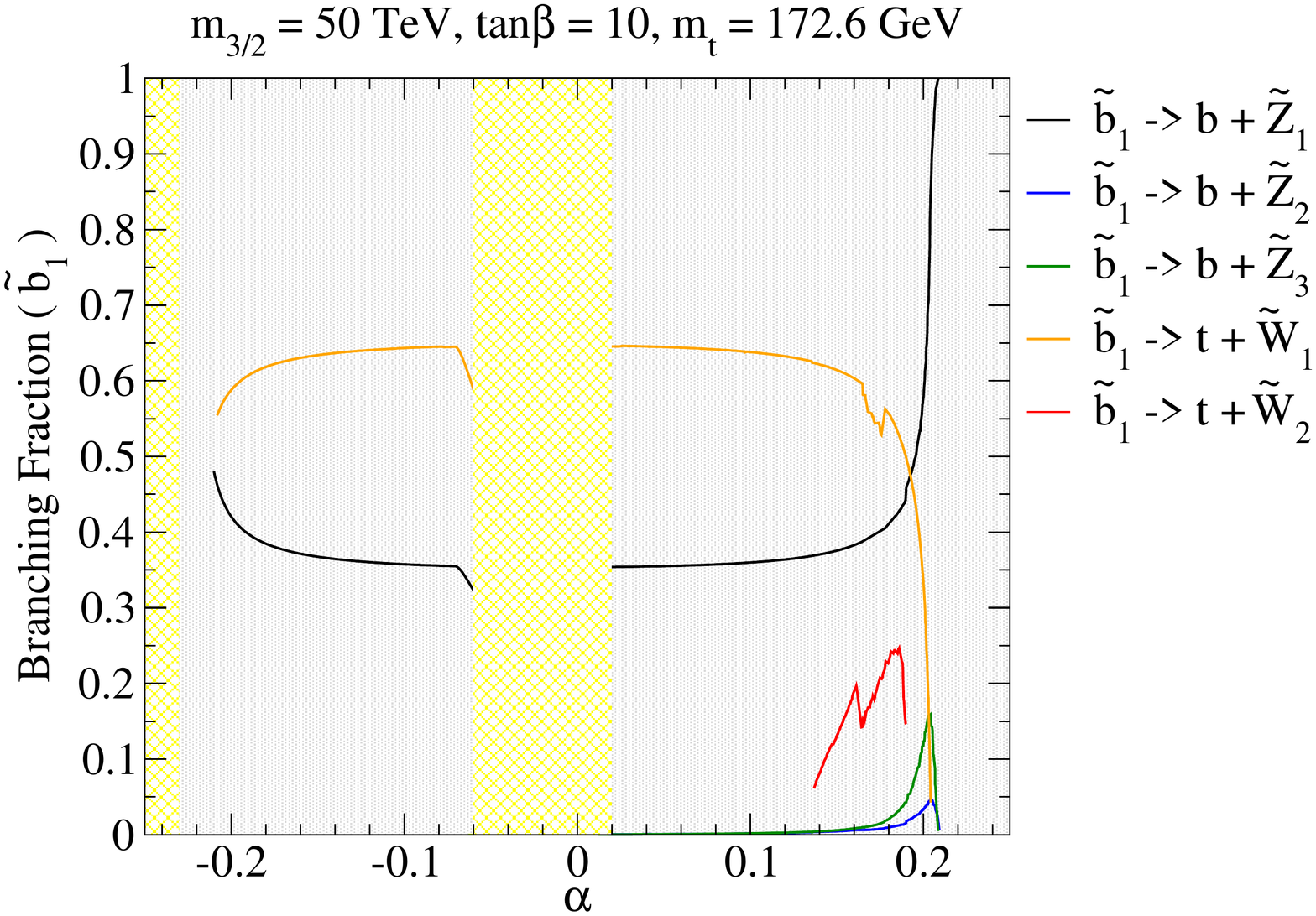}%
    \caption{Branching fraction of $\sb$ for model parameters $\mhf$ = 50 TeV,
	$\tanb = 10$, and $sgn(\mu) > 0$.
      \label{fig:b1bfs}%
    }%
  \end{center}%
\end{figure}%
\begin{figure}[h!]%
  \begin{center}
    \includegraphics[width=.75\textwidth]{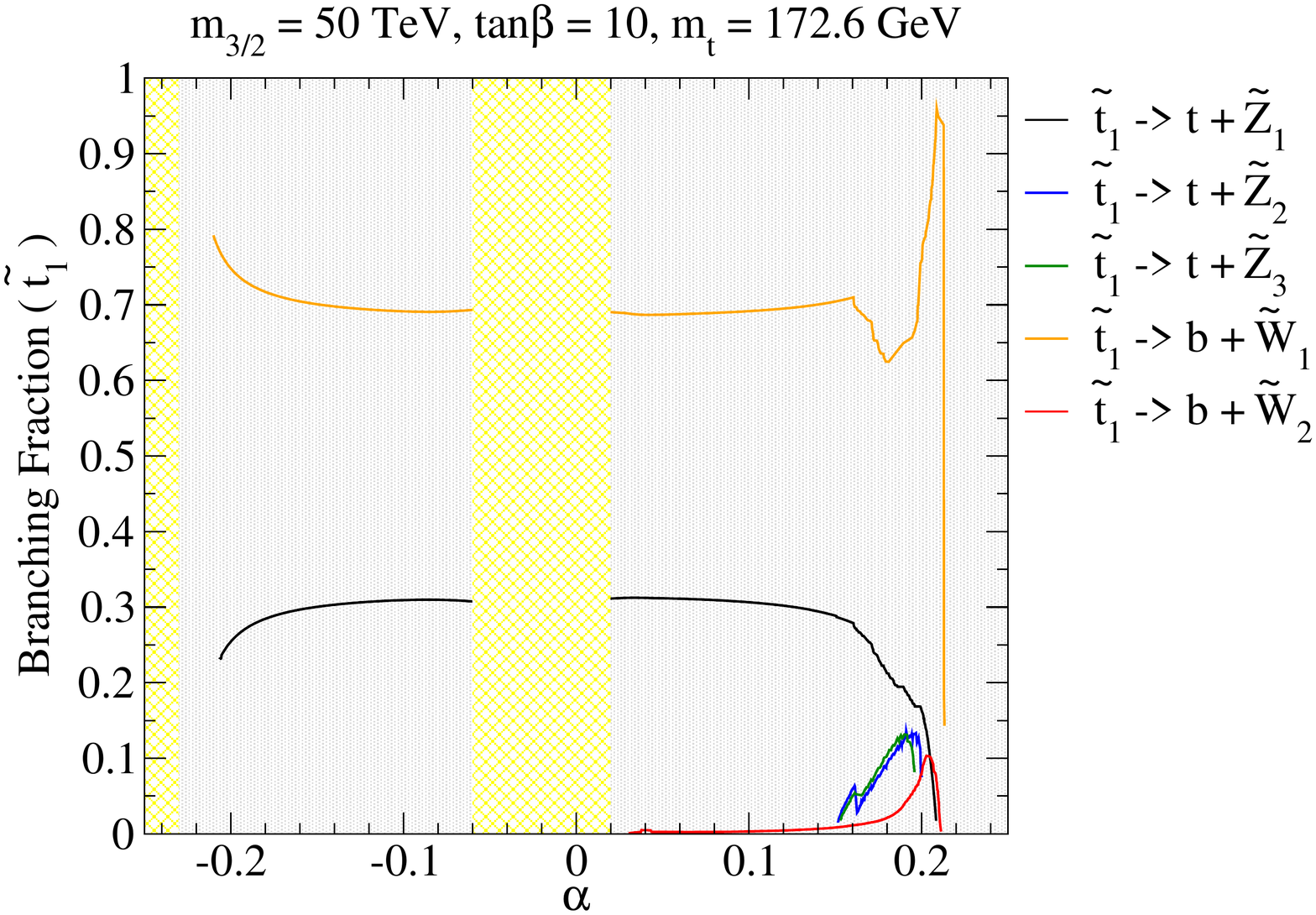}%
    \caption{Branching fraction of $\st$ for model parameters $\mhf$ = 50 TeV,
	$\tanb = 10$, and $sgn(\mu) > 0$.
      \label{fig:t1bfs}%
    }%
  \end{center}%
\end{figure}%
\indent
We now move from $\mhf =$ 30 TeV up to 50 TeV to match the benchmark points and 
we find that the features of the previous paragraphs are unchanged.  To 
proceed, the branching fractions of the lightest sbottoms and stops are plotted 
as a function of $\alpha$ in Figures \ref{fig:b1bfs} and \ref{fig:t1bfs}.  The 
features of low $\alpha$ are simple: both light sbottoms and stops decay to 
lightest charginos with about 67\% and to lightest neutralinos with about 33\%, 
and those make up the entire branching fractions.  As $|\alpha|$ increases, 
$\mu$ decreases and approaches $M_{2}$, and again $\nino$ makes the transition 
from wino-like to a wino-higgsino mix.  $\spart{Z}{2}$ and $\spart{Z}{3}$ are 
mainly higgsino (as in Table \ref{tab:ninocomp}) and become lighter with 
decreasing $\mu$, so that the decay channels 
$\sb\rightarrow b\ninos{2}, b\ninos{3}, t\cinos{2}$ and
$\st\rightarrow t\ninos{2}, t\ninos{3}, b\cinos{2}$ open.  Finally, at the 
largest values of $|\alpha|$, the sbottoms and stops decrease in mass and the 
decay modes close.  In particular, it can be seen in Figure \ref{fig:t1bfs} that
all two-body modes for the light stop are closed, but its decays proceed via 
the $\st\rightarrow bl\nu\nino$ and $c\nino$.\\%
\indent
We also check what effects increasing $\mhf$ has on the branching fractions.  
Figure \ref{fig:bfm32} shows the branching fractions verses $\mhf$ for sbottoms 
and stops at low $|\alpha|$.  It is seen in both cases that out to very large 
values of $\mhf$, two-body decays modes with $\cino$ and $\nino$ are 
dominant while all others are subdominant.  Explicitly, these modes are
$\st\rightarrow b\cino, t\nino$ and $\sb\rightarrow b\nino,t\cino$.\\%
\indent%
Then, to summarize the \hca signatures, we have seen that gluino and 
squark production always lead to cascade decays to third-generation quarks and 
squarks.  From these products should emerge high multiplicities of $b-jets$ 
as well isolated leptons emerging from the $t\rightarrow bW$ decays.  There 
should also be significant $\etm$ due to the presence of $\nino$s and $\nu$s 
escaping the detectors.  In addition to these, there is also the possibility of
detecting highly-ionizing tracks as is usually for AMSB models.  This is due to
the wino-like nature ($M_{2} << M_{1},\mu$) of $\cino$ and $\nino$ states and 
to their near mass degeneracy that leads to a narrower width for the former.  
Because $\cino\rightarrow \pi^{+}\nino$ has too soft a pion to leave a 
calorimeter signal (Chapter \ref{chap:mamsb}), the 
HITs should be seen to abruptly terminate without a calorimeter signal.
\section{%
  \label{sec:hcalhc}
  HCAMSB and the LHC
}%
In this section we analyze events using the Isajet 7.79 \cite{Paige:2003mg} 
event generator (for specific details of the simulation, see section
\ref{sec:comptools} of the Introduction).  As we saw in section 
\ref{sec:hca_constraints}, the 
lowest gluino mass is too large to be detected at the Tevatron; we then must 
understand how \hca would appear at the LHC.  It is expected that the LHC 
phenomenology will have similarities to mAMSB on general grounds.  So part of 
the aim of this section will be to see in which ways this theory will be 
distinguishable from mAMSB at the LHC.\\%
\indent%
The discussion in this section is split into two subsections.  In the first we
flesh out the final states of cascade decays and optimize the cuts used to 
observe them.  In the subsection that follows we analyze the 
reach for the LHC.

\subsection{%
  \label{subsec:finalstates}%
  Final State Analysis%
}%
\begin{figure}[t]%
  \begin{center}%
    \subfigure[]{\includegraphics[scale=.5,trim= 0cm 0cm 5cm 17cm,clip]
	{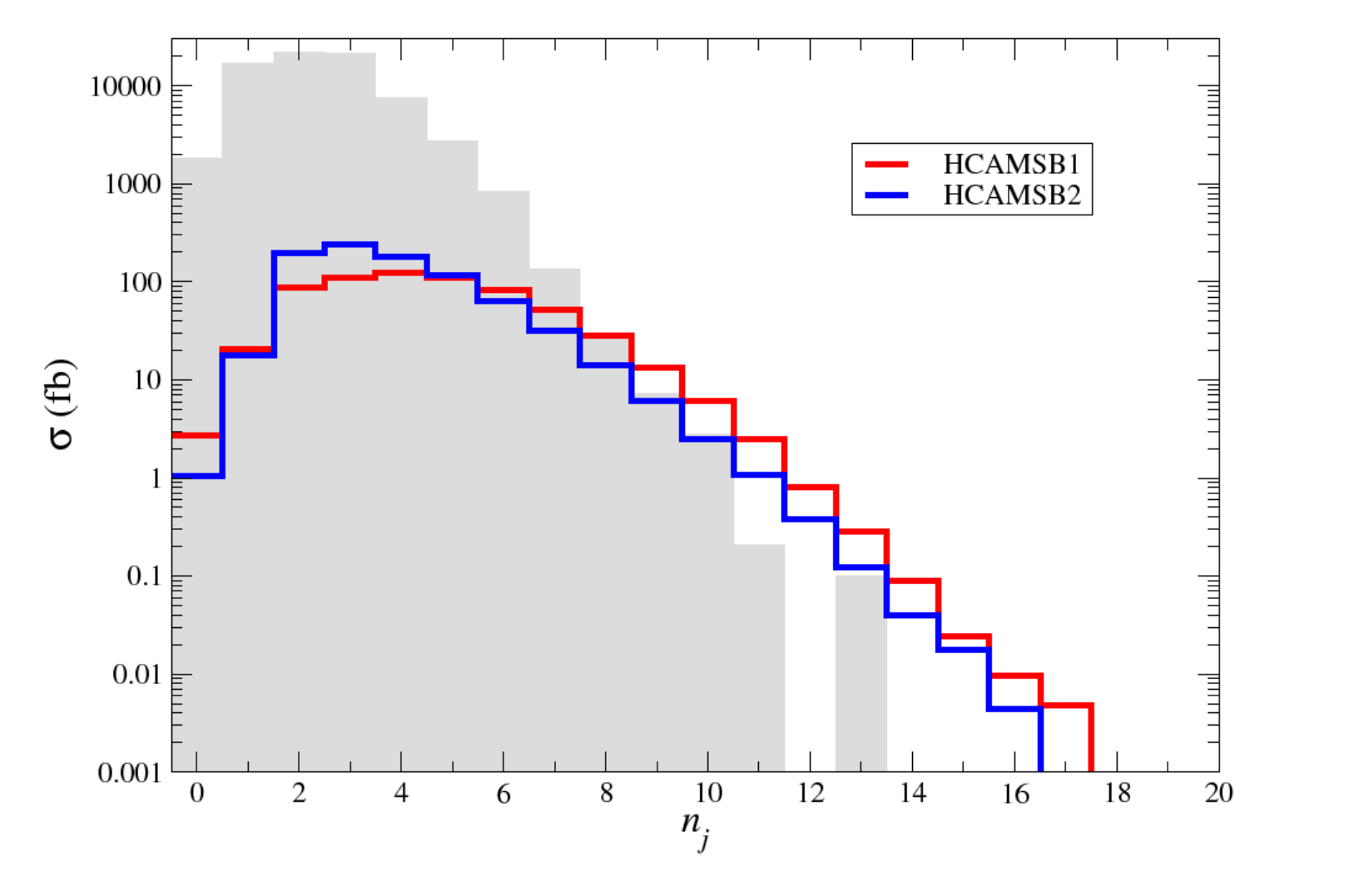}}%
    \subfigure[]{\includegraphics[scale=.5,trim= 0cm 0cm 5cm 17cm,clip]
 	{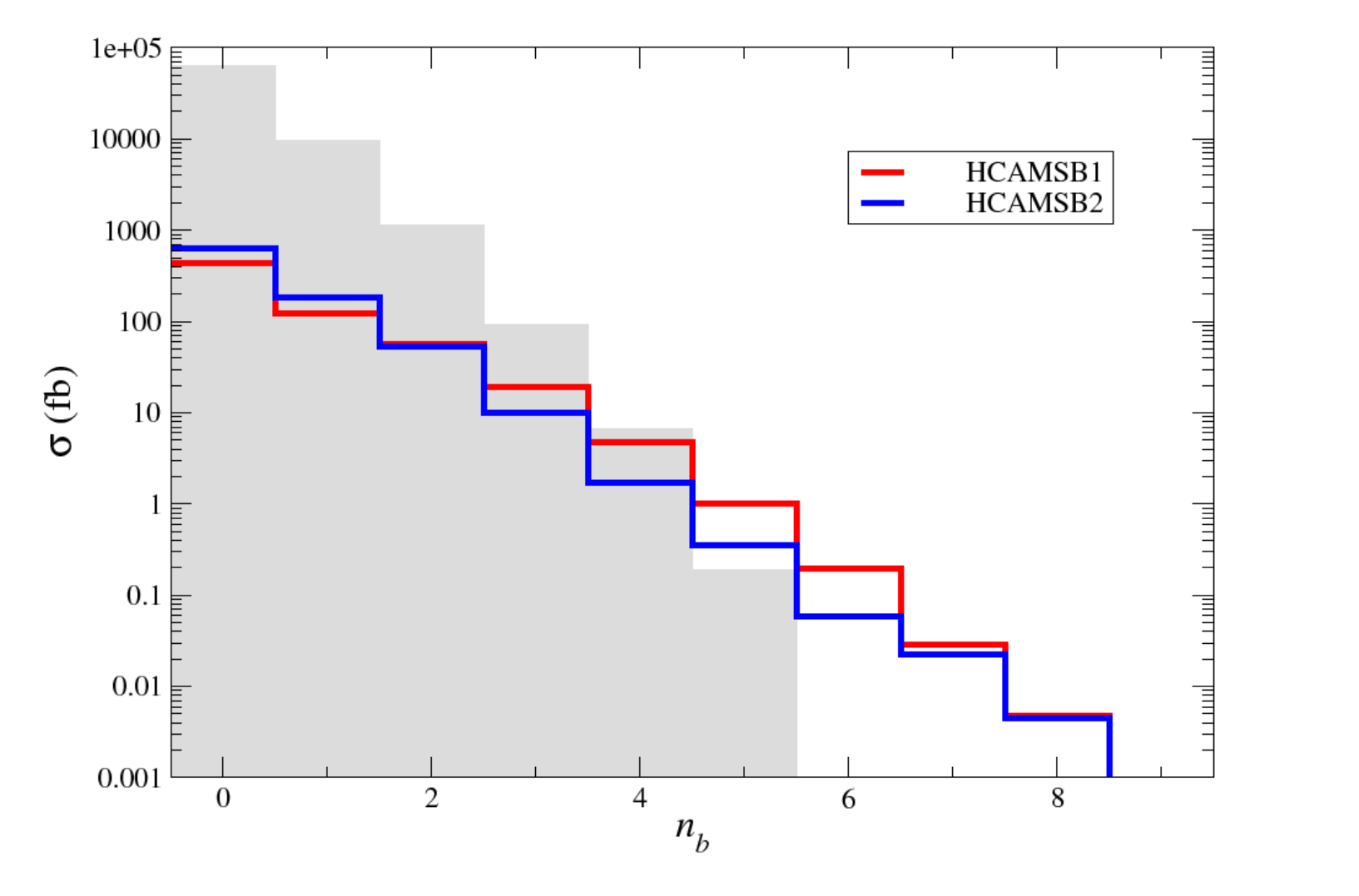}}\\%
    \subfigure[]{\includegraphics[scale=.5,trim= 0cm 0cm 5cm 17cm,clip]
	{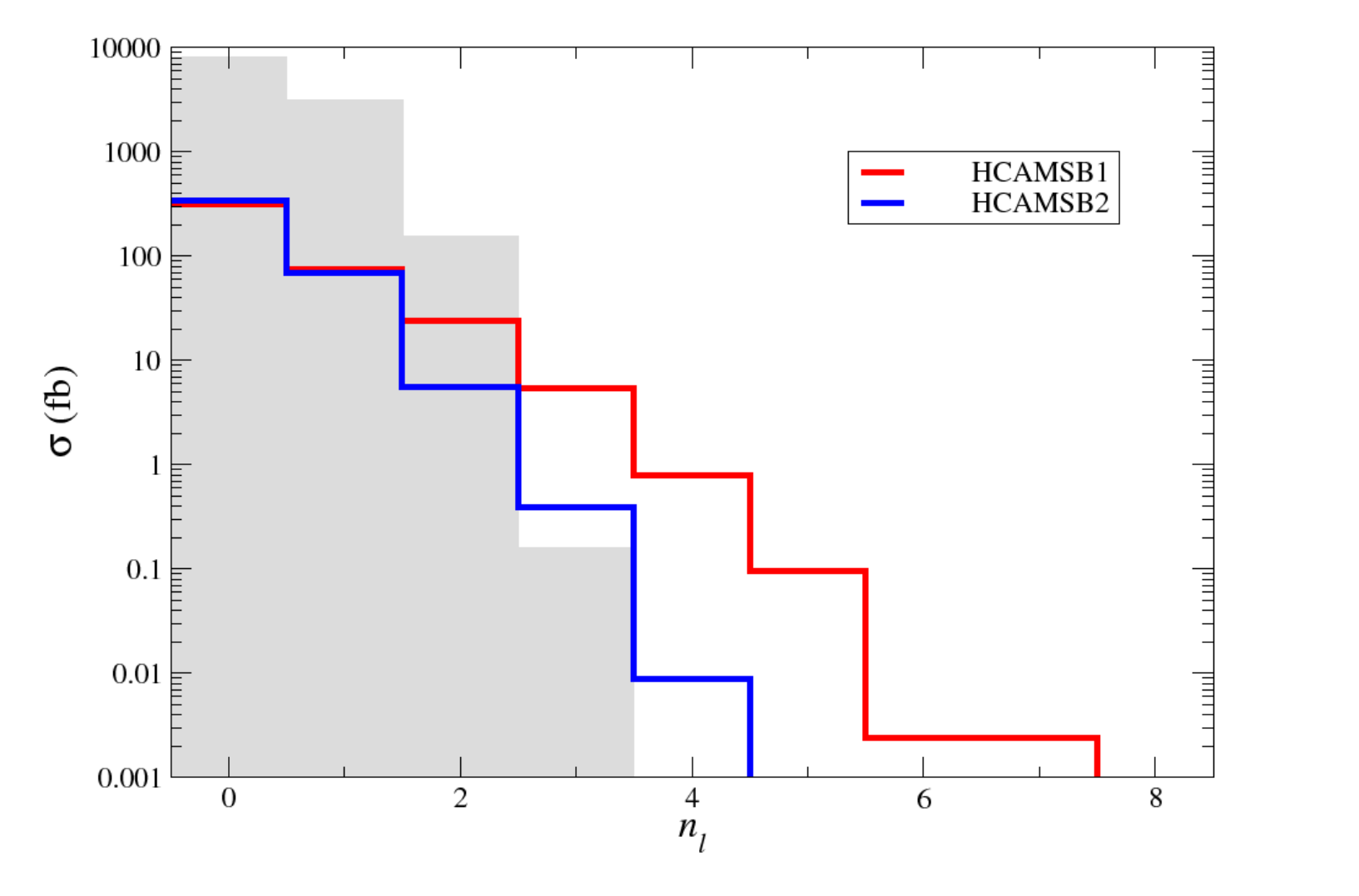}}%
    \caption{Jet, b-jet, and lepton multiplicity distributions for \hca points 
      1 (red) \& 2 (blue), and SM background (gray), all after C1 cuts.
      \label{fig:hcamult}%
    }%
  \end{center}%
\end{figure}

\begin{figure}[t]%
  \begin{center}
    \subfigure[]{\includegraphics[scale=.5,trim= 0cm 0cm 5cm 17cm,clip]
      {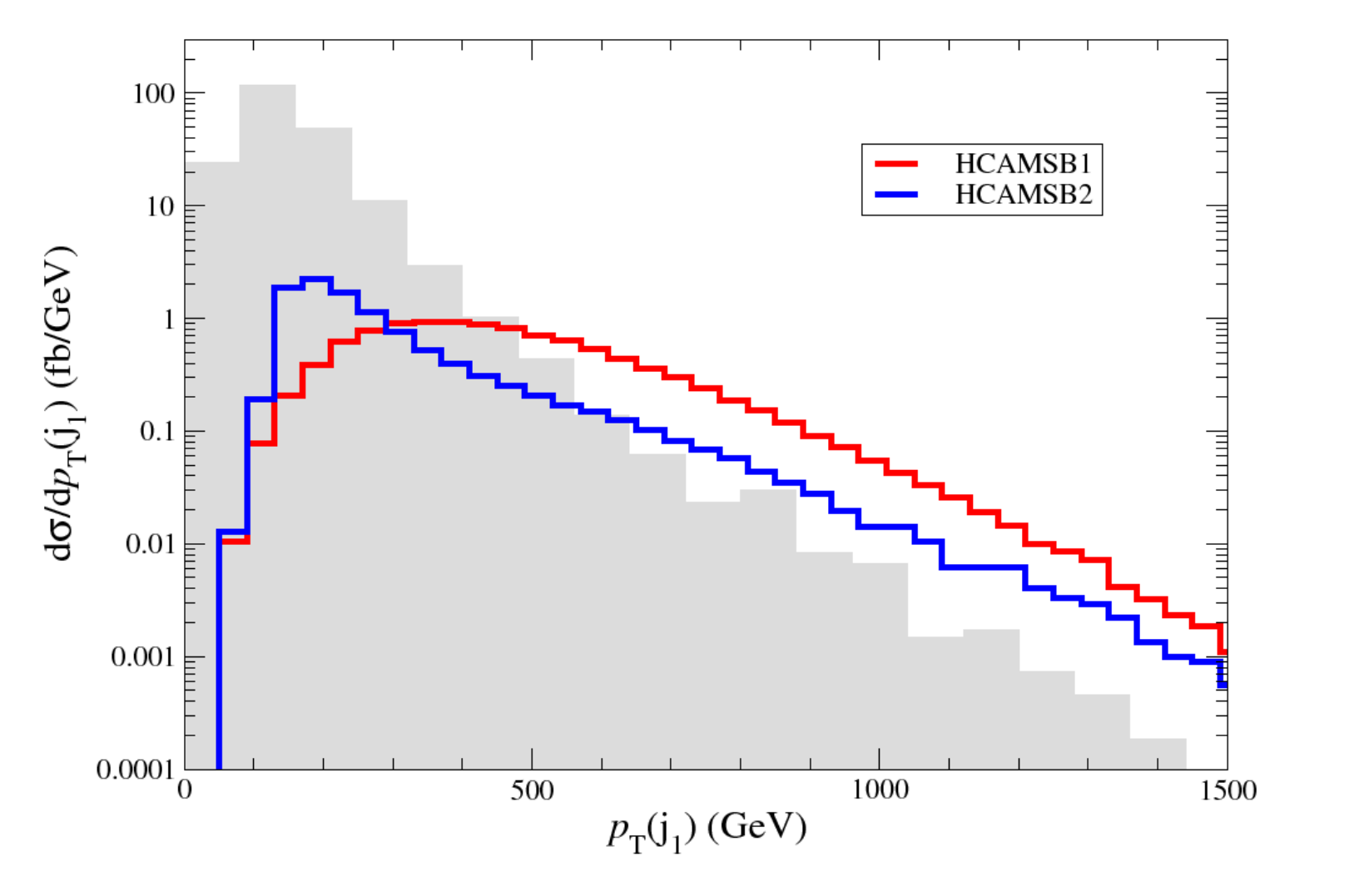}}%
    \subfigure[]{\includegraphics[scale=.5,trim= 0cm 0cm 0cm 17cm,clip]  
      {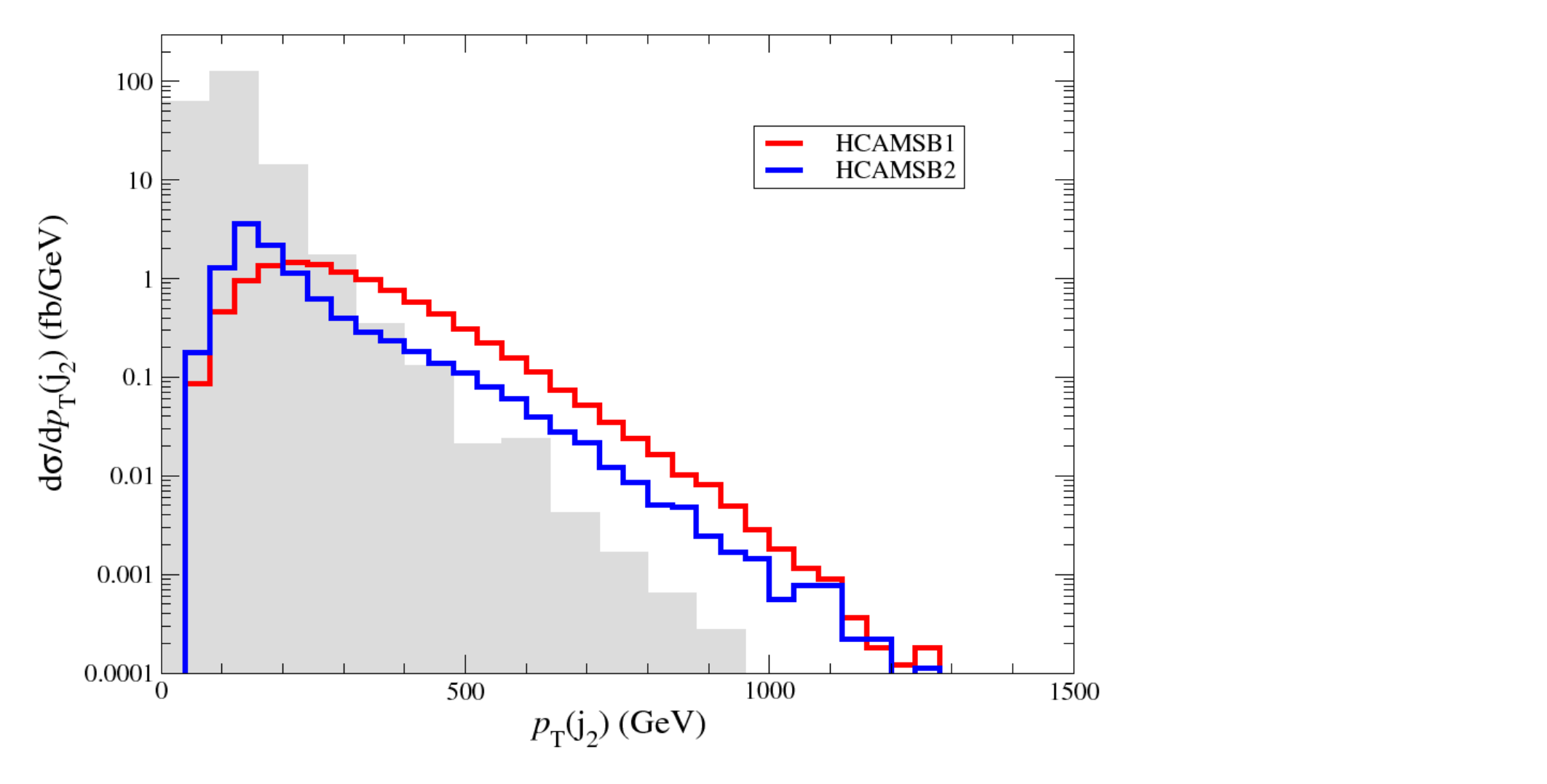}}\\%
    \subfigure[]{\includegraphics[scale=.5,trim= 0cm 0cm 5cm 17cm,clip]
      {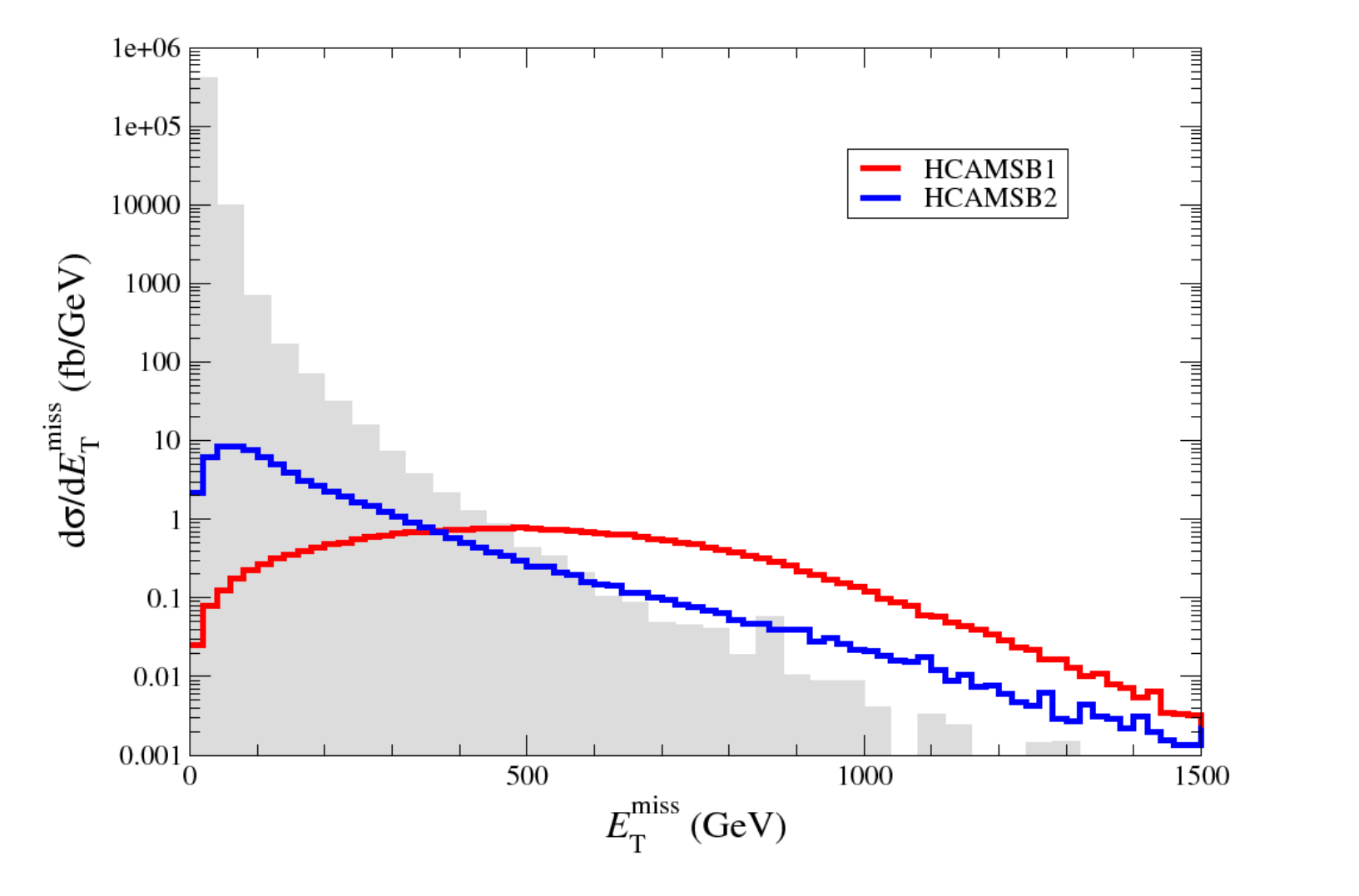}}%
    \subfigure[]{\includegraphics[scale=.5,trim= 0cm 0cm 0cm 17cm,clip]
      {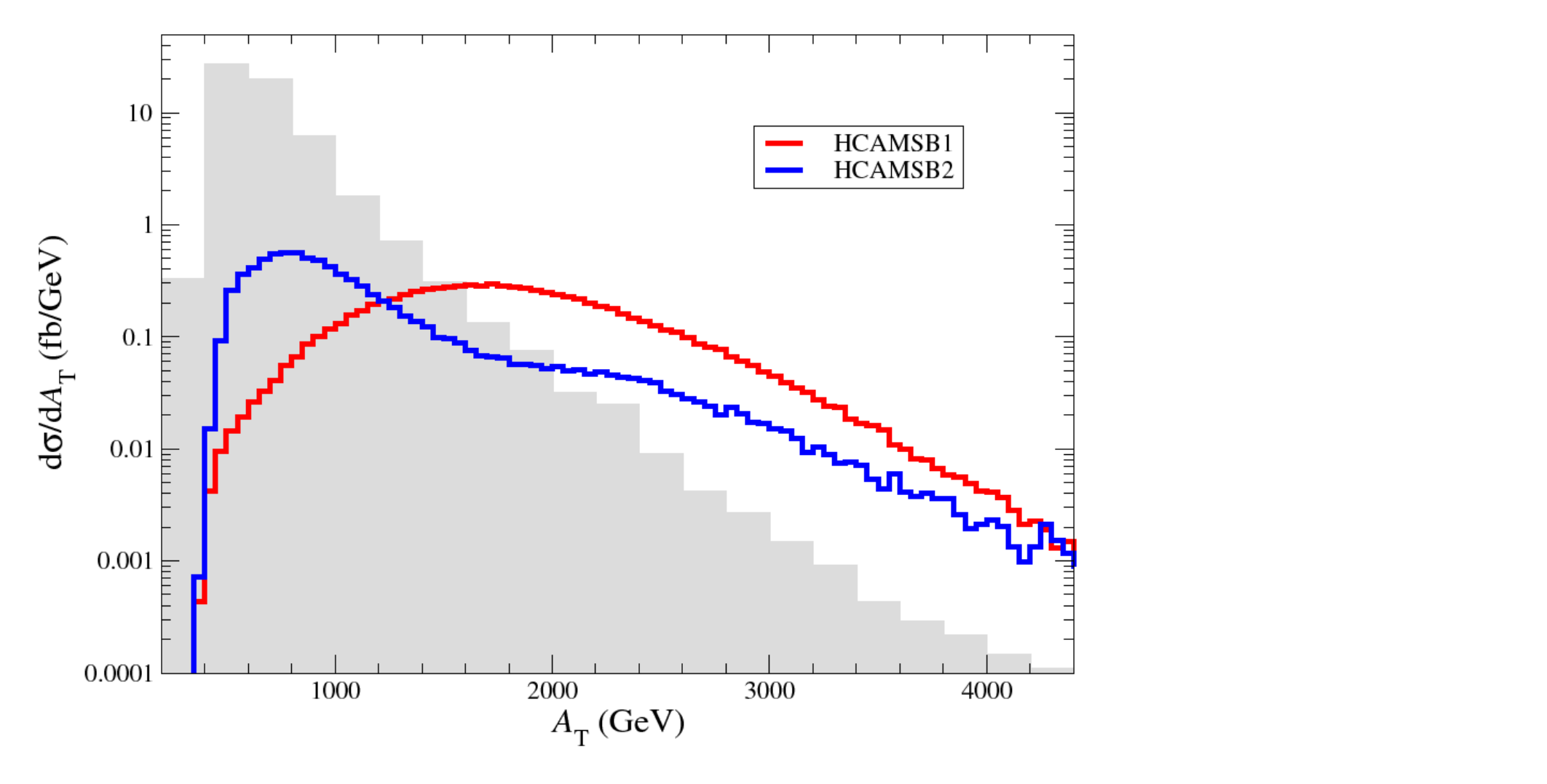}}\\%
    \caption{Kinematic distribution for \hca points 1 (red) \& 2 (blue) and SM  
      background (gray) after C1 cuts.
      \label{fig:hcakin}%
    }%
  \end{center}%
\end{figure}

We generate 2M events for the points HCAMSB1 and HCAMSB2 from Table 
\ref{tab:cases} and compare with SM backgrounds.  QCD jets are generated 
with Isajet and include the following jet types: $g, u, d, s, c$, and
$b$.  Additional jets are produced in parton showering and other parton-level
processes considered.  Other backgrounds include the following: $W+jets$, 
$Z+jets$, $t\bar{t}$(172.6 GeV), and $WW, WZ$, and $ZZ$ vector boson production.
Both $W+jets$ and $Z+jets$ have exact matrix elements for one parton emission
and use parton showering for subsequent emissions.\\%
\indent%
We would like to apply cuts to these backgrounds without overly diminishing the 
signal (a \quotes{good} signal to be quantified in the next subsection on the
LHC reach).  Initially we choose the following cuts labeled C1:\\%
\begin{itemize}%
  \item $n(jets) \geq 4$,%
  \item $\etm > max$(100 GeV, 0.2 $M_{eff}$),%
  \item $\et(j_{1},j_{2},j_{3},j_{4}) >$ 100, 50, 50, 50 GeV,%
  \item Transverse sphericity: $S_{T} > 0.2$,%
\end{itemize}%
\noindent where $M_{eff} = \etm + \et(j_{1}) + \et(j_{2}) + \et(j_{3}) + 
\et(j_{4})$.  After applying the C1 cuts we arrive at Figures \ref{fig:hcamult} 
and \ref{fig:hcakin}.  Figure \ref{fig:hcamult} shows three multiplicity 
distributions: (a) number of jets, (b) number of b-jets, and (c) number of 
leptons.  The kinematic distributions in Figure \ref{fig:hcakin} are for (a) 
hardest jet $p_{T}$, (b) second hardest jet $p_{T}$, (c) $\etm$, and (d) the 
augmented effective mass, $A_{T}$ (this is the effective mass as earlier, but 
also includes $\sum \et(leptons)$ and the $\et$ of all jets).  These 
distributions are used to improve on the cuts and each is discussed in the 
following.\\%
\noindent {\bf Figure \ref{fig:hcamult}}:
\begin{itemize}
  \item[] (a) Jet multiplicity (after relaxing $n(jets) \geq 4$) after C1 
 	    cuts   
          -- signals for \hca points 1 \& 2 do not exceed background until very 
	    high jet multiplicities.  For instance the HCAMSB1 distribution does 
	    not exceed background until $n(jet) \sim 9$.  The selection of 
	    $n(jet) \geq 4$ should be beneficial in this case.
  \item[] (b) $b-jet$ multiplicity (after relaxing $n(jets) \geq 4$) after C1 
          cuts  -- signal cross section appears harder than background due to 
	    the appearance of extra $b$ and $t$ quarks from cascade decays as 
	    discussed in Section \ref{sec:hcasig}.  The signals exceed background 
	    around $n(b-jet) \sim 5$.  We conservatively choose events with at 
	    least a single $b-jet$ to cut down on background.
  \item[] (c) Lepton multiplicity after C1 cuts -- we can see a much harder
	    signal distribution of leptons than for background due to the 
	    presence of leptons from cascade decays as in the discussion of 
	    Section \ref{sec:hcasig}.  The HCAMSB1 signal appears even stronger 
	    because sleptons are lighter at lower $\alpha$.  Signal exceeds 
	    backgrounds around $n(l) \sim 3$, and HCAMSB1 has sufficient strength 
	    to be visible here with only a few fb$^{-1}$ of LHC data 
	    \cite{Baer:2008kc}\cite{Baer:2008ey}.
\end{itemize}
\noindent {\bf Figure \ref{fig:hcakin}}:
\begin{itemize}
  \item[] (a) $\et$ of hardest jet after C1 cuts ($\et(j_{1})$ cut relaxed) -- 
  	    the HCAMSB2 signal peaks for $\et(jet_{1}) \sim 150$ GeV because of 
	    $\sb\sbbar$ production followed by $\sb\rightarrow b\nino$.  Signal 
	    exceeds background around 450 GeV for HCAMSB1 and 550 GeV for HCAMSB2.
  \item[] (b) $\et$ of second hardest jet after C1 cuts ($\et(j_{2})$ cut 
          relaxed) -- Similar to (a), but signal exceeds background at 
	    $\et(jet_{2}) \sim$ 350 GeV for HCAMSB1 and $\sim$ 450 GeV for 
	    HCAMSB2.
  \item[] (c) $\etm$ after C1 cuts -- the HCAMSB2 distribution is softer than 
    	    for HCAMSB1 because the $\etm$ originates from third-generation 
	    squarks, $e.g. \ \sb \rightarrow b\nino$, whereas in the latter case 
	    harder $\nino$s are produced from TeV-scale squarks and gluinos.
  \item[] (d) Augmented effective mass -- $A_{T} =$ $\etm + \sum\et(jets)$
    	    $+ \sum\et(isolated \ leptons)$.  HCAMSB1 has a smooth distribution 
	    and exceeds background at $A_{T}\sim 1600$ GeV.  The HCAMSB2 
	    distribution interestingly has two components: a soft component with 
	    peak at $A_{T}\sim 750$ GeV due to third-generation squark pair 
	    production, and a hard component with peak at $A_{T}\sim 2000$ GeV 	
	    due to $\gl$ and $\sq_{L}$
    production.%
\end{itemize}

\begin{figure}[h]%
  \begin{center}
    \includegraphics[width=.75\textwidth,trim= 0cm 0cm 5cm 17cm,
      clip]{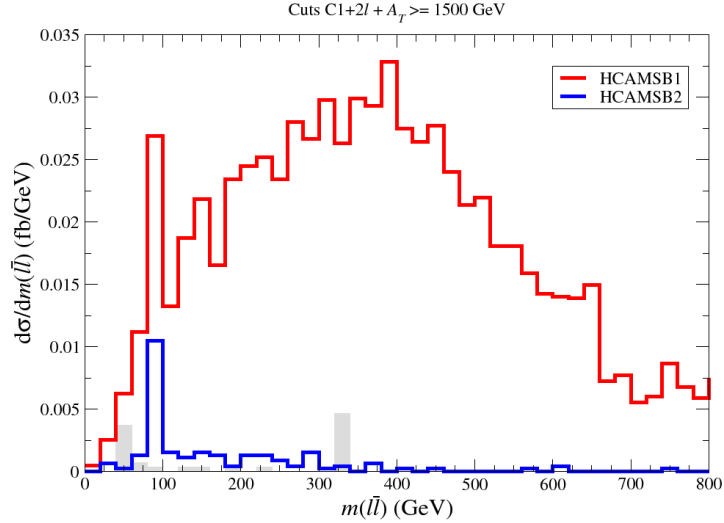}\\%
    \caption{$m(l^{+}l^{-})$ distribution for \hca points 1 (red) \& 2 (blue) 
    and SM background (gray) after C1 cuts.
      \label{fig:hcamll}%
    }%
  \end{center}%
\end{figure}

\indent
And finally, in addition to the distributions just discussed, we consider Figure 
\ref{fig:hcamll}, which shows the oppositely-signed (OS), dilepton invariant 
mass distribution for HCAMSB points 1 \& 2 and the background after C1 cuts and 
an extra cut of $A_{T} > 1500$ GeV.  This type of distribution is useful in SUSY
studies because of the appearance of kinematic mass edges due to 
$\ninos{2}\rightarrow\spart{l}{}^{\pm}l^{\mp}$ or
$\ninos{2}\rightarrow\l^{\pm}l^{\mp}\nino$ decays.  For \mam models the 
bino-like $\ninos{2}$ produce a mass edge since it efficiently decays 
to $\spart{l}{R}^{\pm}l^{\mp}$.  In the \hca model however, $\ninos{2}$ is 
higgsino-like and rather heavy (see Table \ref{tab:ninocomp} for a comparison).
In this case the decay channels are $\ninos{2}\rightarrow\cino^{\pm}W^{\mp}$,
$\nino h$, and $\nino Z$.  Most importantly is the last decay channel which 
should always be open (except when $\mu \rightarrow 0$, when the $\alpha$ is at
its very highest value) and occurs with a branching fraction at the tens of 
percent level.  The decay to $\nino Z$ of the bino-like $\ninos{2}$ of \mam is 
highly suppressed due to the structure of $Z\nino\ninos{2}$ coupling
\cite{Baer:2006rs}, where Z couples to neutralinos only through their higgsino 
component.  Thus we expect to be able to distinguish
between \hca and \mam models in the OS dilepton, invariant mass distribution
because the former has a Z-peak structure and the latter does not.  This is 
clear from Figure \ref{fig:hcamll}, which shows a clear Z-peak structure in \hca
points 1 \& 2, and a diminished SM background.

\subsection{%
  \label{subsec:hcareach}%
  LHC Reach%
}%
\begin{figure}%
  \begin{center}
    \subfigure[]{\includegraphics[width=.75\textwidth,trim= 0cm 0cm 0cm 
      0cm,clip]{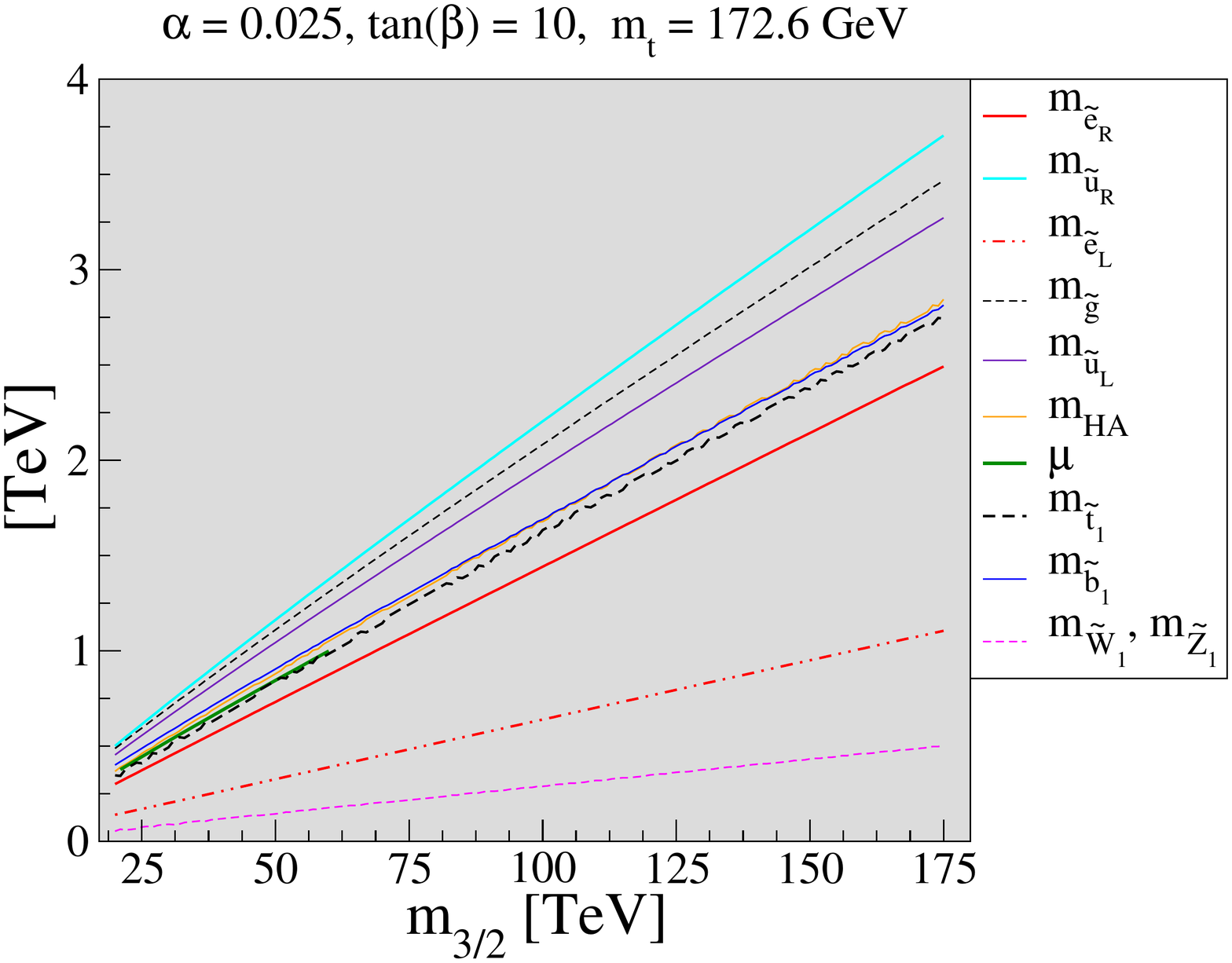}}\\%
    \subfigure[]{\includegraphics[width=.75\textwidth,trim= 0cm 0cm 0cm 
      0cm,clip]{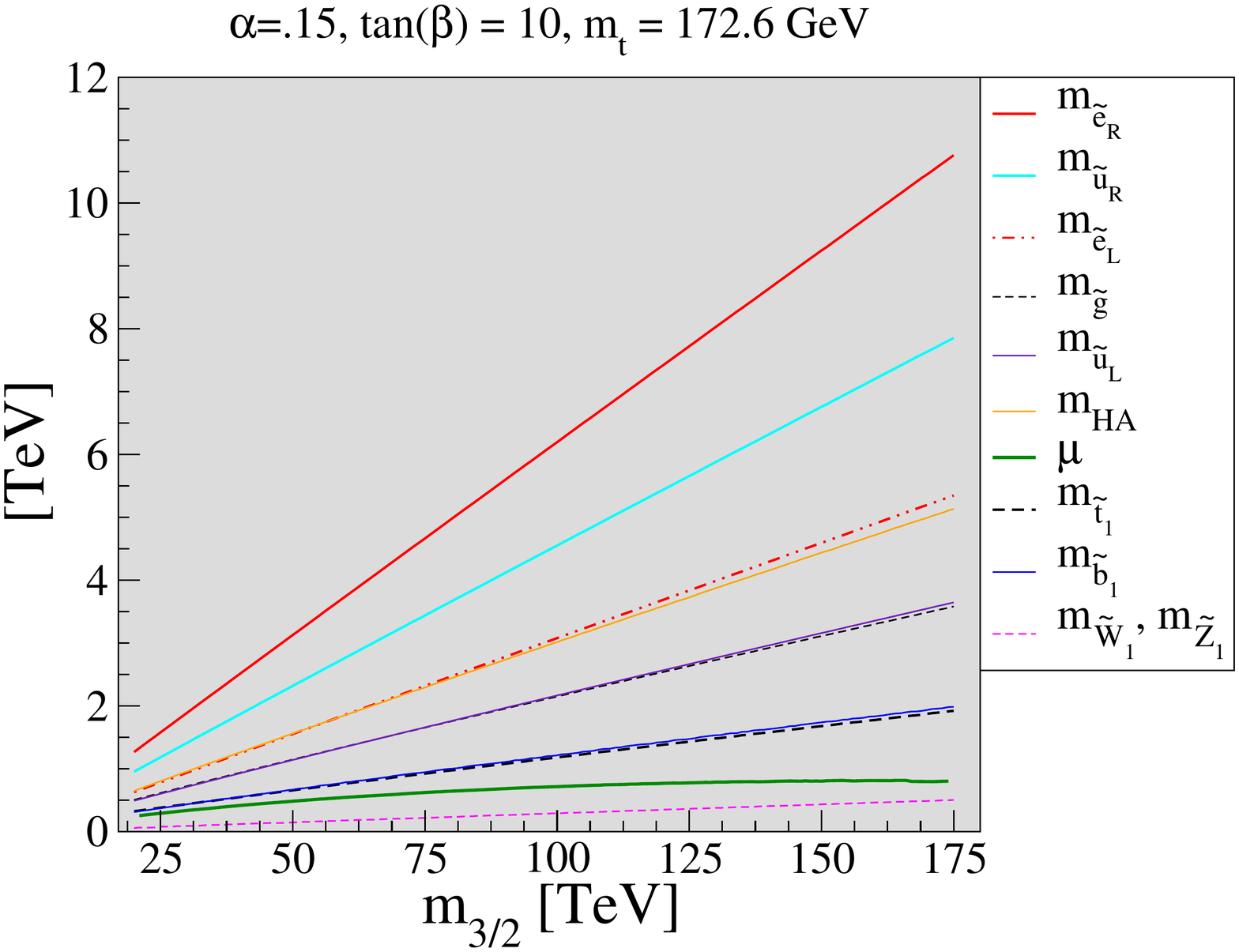}}%
    \caption{Spectrum versus $\mhf$ for \hca points 1 \& 2.
    }%
      \label{fig:hcaspecm32}%
  \end{center}%
\end{figure}

In this section we would like to understand the range of parameters accessible
at the LHC.  To this end, we vary over the range $\mhf = 30 - 200$ TeV and we 
adopt the two following model lines (again, low and high $\alpha$):
\begin{itemize}
\item[] HCAMSB1: $\alpha=0.025,\tanb=10,\mu>0$, 
\item[] HCAMSB2: $\alpha=0.15,\tanb=10,\mu>0$.
\end{itemize}
We take a somewhat lower value of $\alpha$ in HCAMSB2 than previously because
$\mhf$ could only be extended up to $\sim 60$ TeV before EW symmetry is 
improperly broken (see Figure \ref{fig:reachpar}).  The spectrum is shown for 
the two points above while varying the $\mhf$ parameter in Figure 
\ref{fig:hcaspecm32}.  The relative ordering of particle masses are unchanged.

\begin{figure}[t!h!]%
  \begin{center}
    \includegraphics[scale=.3,trim= 0cm 0cm 0cm 0cm,clip]{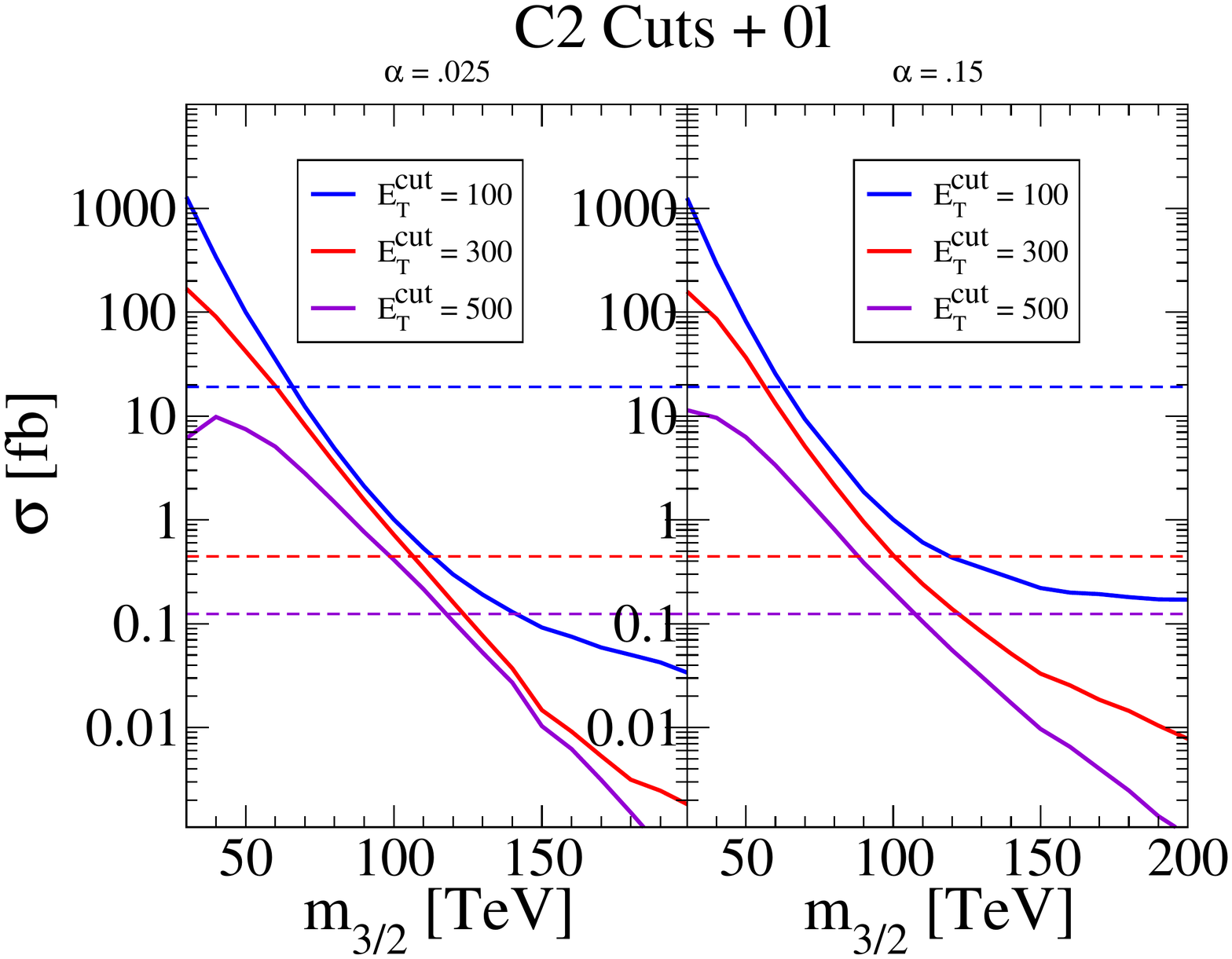}%
    \includegraphics[scale=.3,trim= 0cm 0cm 0cm 0cm,clip]{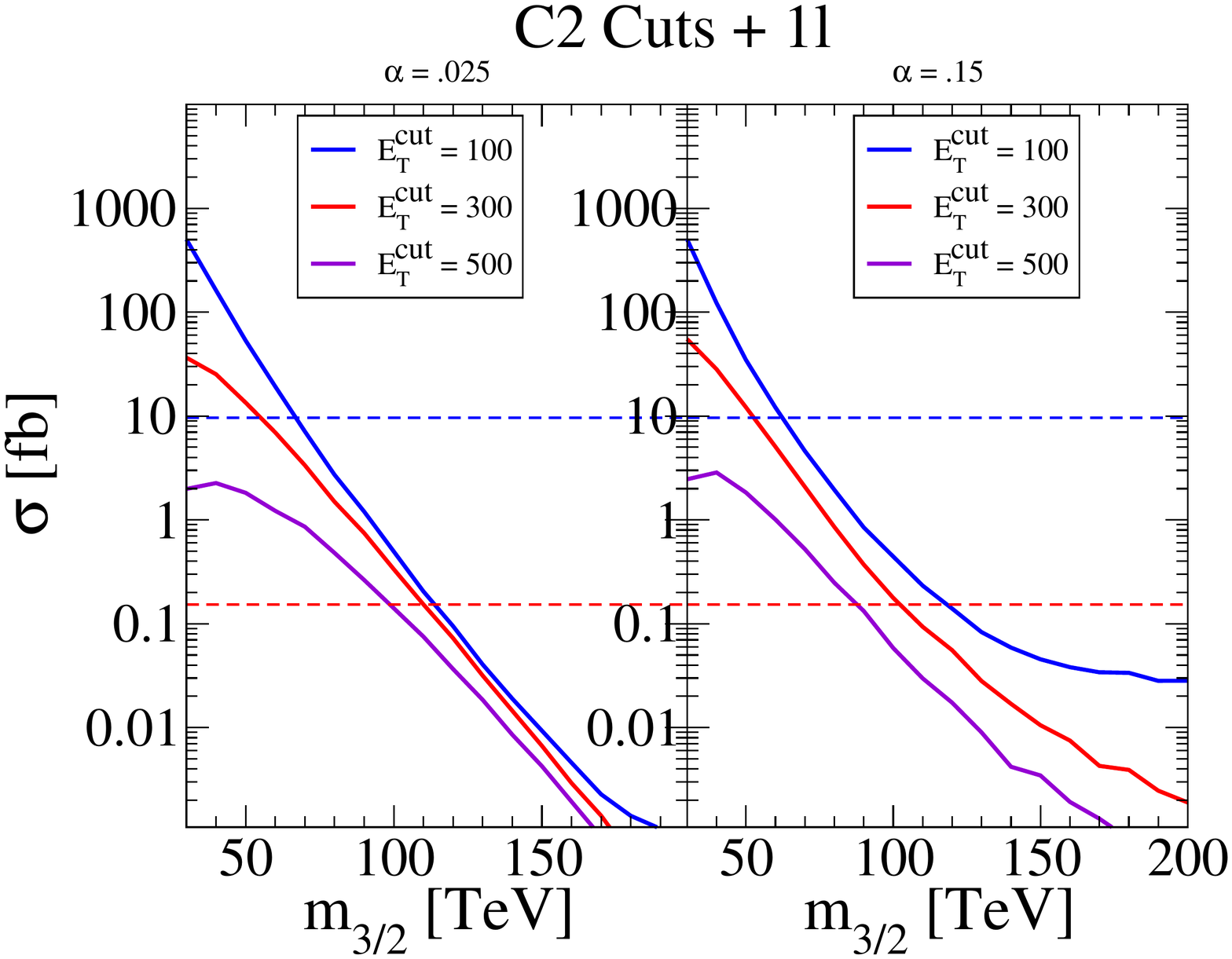}\\%
    \includegraphics[scale=.3,trim= 0cm 0cm 0cm 0cm,clip]{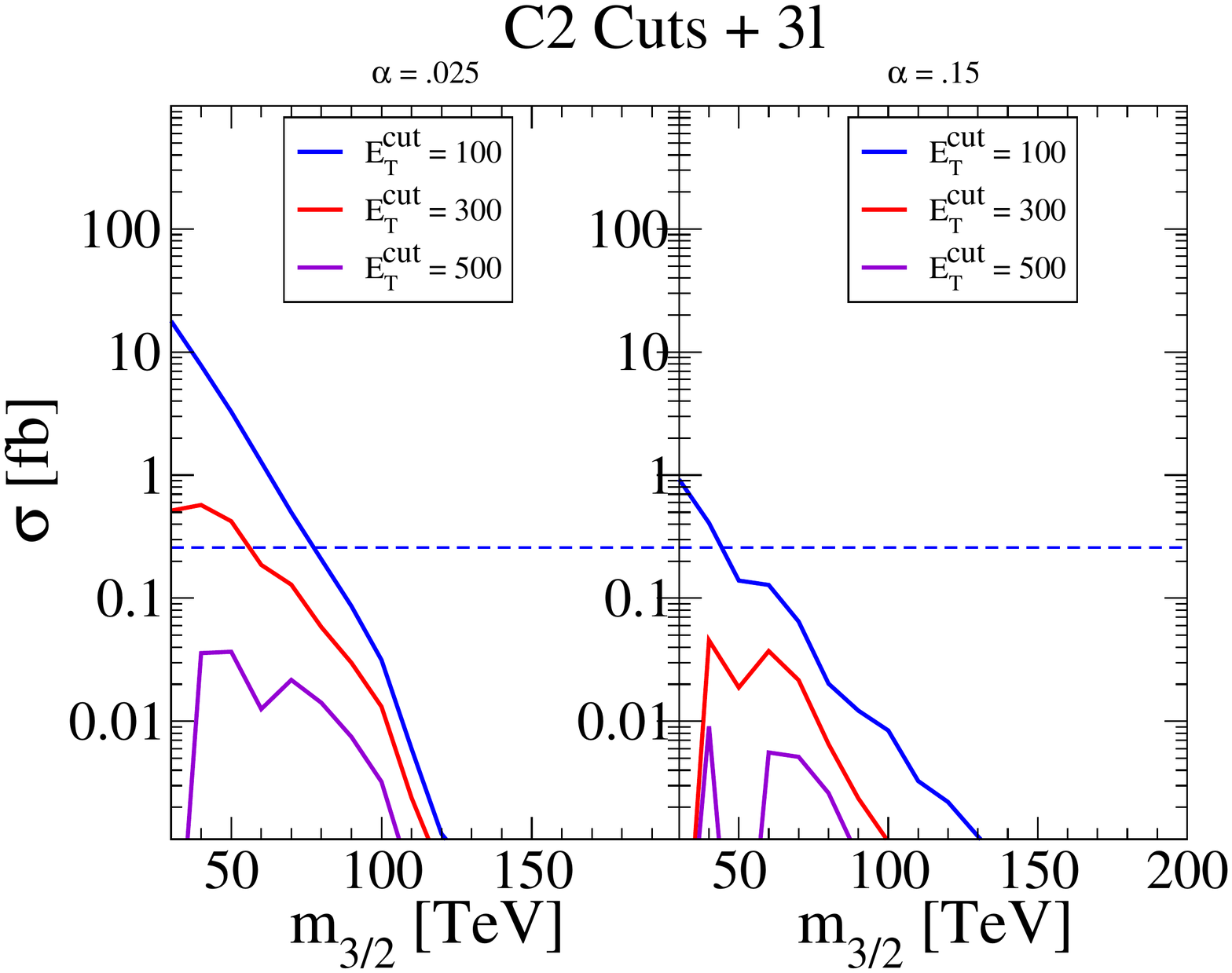}%
    \includegraphics[scale=.3,trim= 0cm 0cm 0cm 0cm,clip]{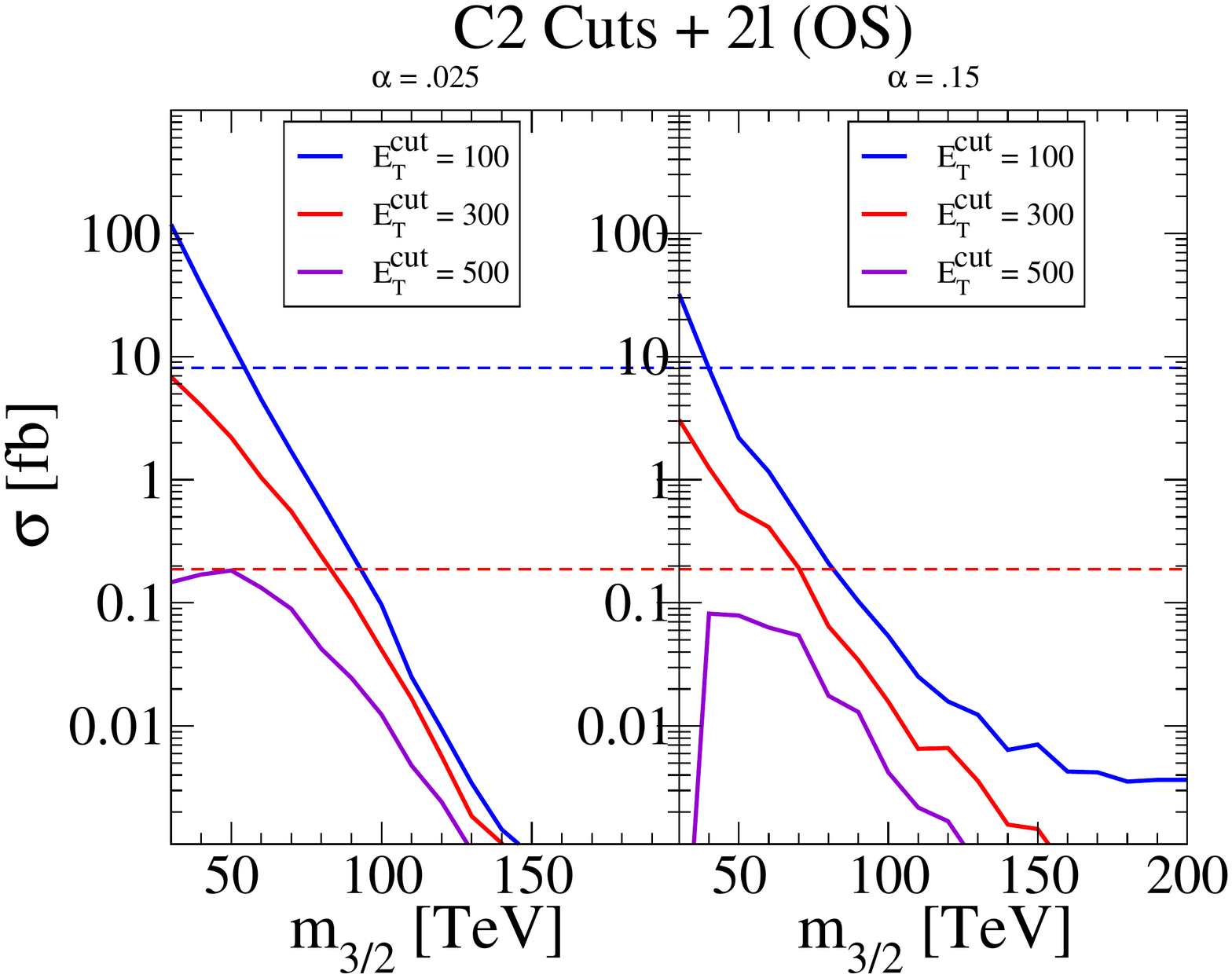}\\%
    \includegraphics[scale=.3,trim= 0cm 0cm 0cm 0cm,clip]{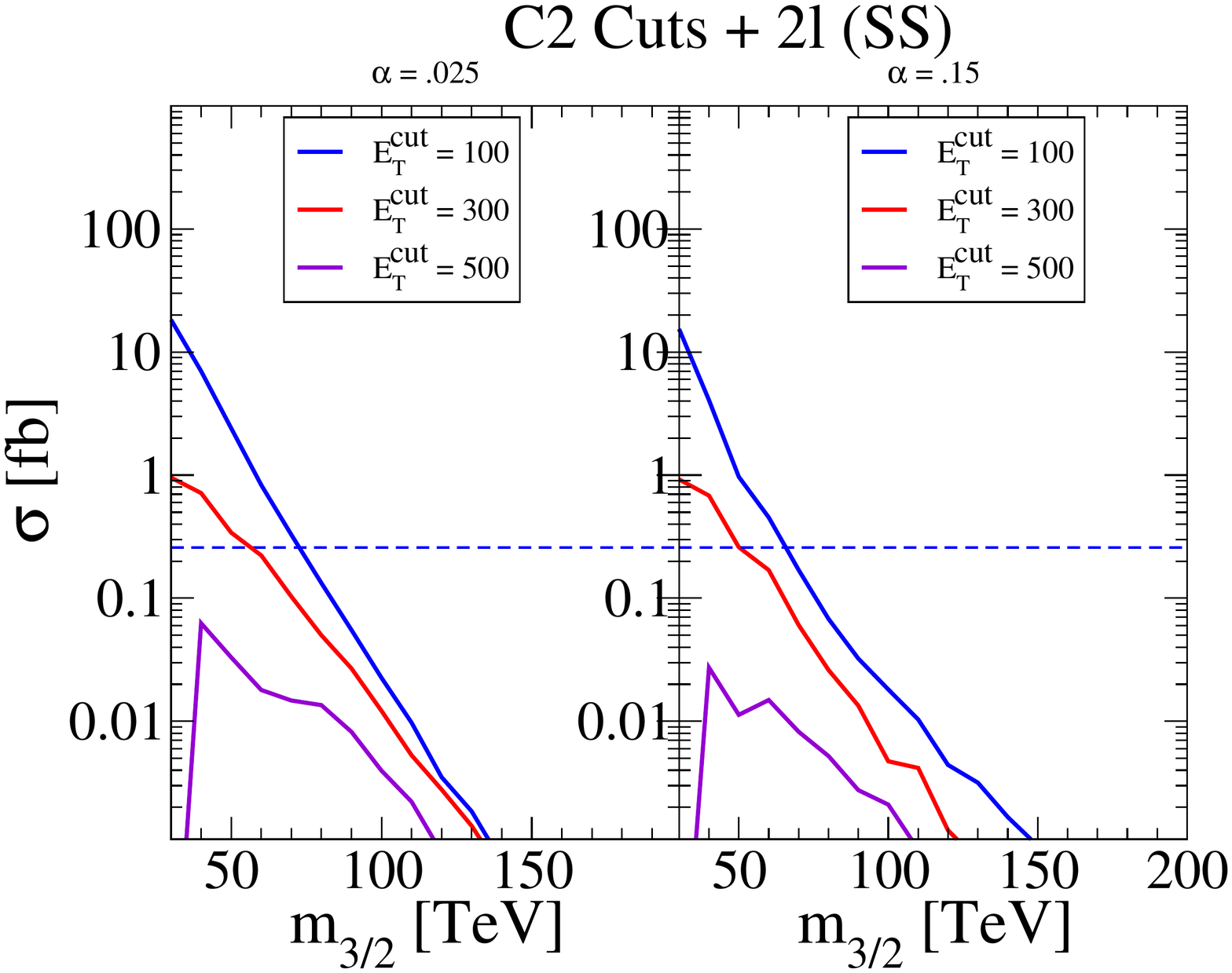}%
    \caption{Cross section versus $\mhf$ for \hca points 1 \& 2, after C2 cuts 
      and in various lepton channels.
    }%
      \label{fig:hcareachm32}%
  \end{center}%
\end{figure}

The distributions of the last section allow us to improve on the background 
cuts.  Instead of C1 cuts we will now use the following set of cuts that we 
label C2:%
\begin{itemize}%
  \item $n(jets) \geq 2$,%
  \item Transverse sphericity: $S_{T} > 0.2$,%
  \item $n(b-jets) \geq 1$,
  \item $\et(j_{1}), \et(j_{2}), \etm  > E_{T}^{cut}$,%
\end{itemize}%
\noindent $E_{T}^{cut}$ will be variable minimum transverse energy value.  
Parameter space points with lower sparticle masses will benefit from a low 
$E_{T}^{cut}$, while points with high sparticle mass but lower production rates
benefit from higher $E_{T}^{cut}$.\\%
\indent
We apply the C2 cuts to the following lepton multiplicity channels: 0lep, 1lep,
OS dilep, SS dilep\footnote{%
We do not require \quotes{same flavor} in the dilepton channels.
}%
, 3lep, and 4lep.  As a supplement to the above cuts, in the 
0lep channel, it is required that the $\etm$-(nearest jet $\et$) transverse 
opening angle is constrained to $30^{\circ} < 
\Delta\phi(\etm,\et(j_{nearest})) < 90^{\circ}$.  Furthermore, all isolated
leptons are required to have a minimum $\pt$ of 20 GeV, and events with single
leptons are required to have transverse mass $M_{T}(l,\etm) \geq 100 GeV$ to 
reject leptons from W decays. \\%
\begin{table}[t]%
  \begin{center}%
    \begin{tabular}{|lccccc|}%
      \hline\hline%
	&&&&&\\%
      Process & $0\ell$ & $1\ell$ & $OS$ & $SS$ & $3\ell$ \\[3pt]%
	&&&&&\\%
      \hline
	&&&&&\\%
	QCD($p_T$: 0.05-0.10 TeV) & -- & -- & -- & -- & -- \\[3pt]%
	QCD($p_T$: 0.10-0.20 TeV) & -- & -- & -- & -- & -- \\[3pt]%
	QCD($p_T$: 0.20-0.40 TeV) & 73.5 & -- & -- & -- & -- \\[3pt]%
	QCD($p_T$: 0.40-1.00 TeV) & 42.6 & 26.5 & 37.3 & -- & -- \\[3pt]%
	QCD($p_T$: 1.00-2.40 TeV) & 0.8 & 0.6 & 0.3 & 0.015 & -- \\[3pt]%
	$t\bar{t}$ & 1253.2 & 341.2 & 224.9 & $ 0.25 $ & $ 0.25 $ \\[3pt]%
	$W+jets; W\to e,\mu,\tau$ & 60.6 & 5.6  & 2.8 & $ -- $ & $ -- $ \\[3pt]%
	$Z+jets; Z\to \tau\bar{\tau},\ \nu s$ & 61.4 & 0.0 & 0.77 & $ -- $ 
	  & $ -- $ \\[3pt]%
	$WW,ZZ,WZ$ & 0.11 & --  & $ -- $ & $ -- $ & $ -- $ \\[3pt]%
	&&&&&\\%
	\hline%
	&&&&&\\%
	$Summed\ SM\ BG$ & 1492.3 & 374.1 & 266.1 & $ 0.26 $ & $ 0.25 $ \\[3pt]%
	&&&&&\\%
	\hline
	&&&&&\\%
	HCAMSB1 & 100.1 & 53.2 & 13.1 & 2.4 & 3.3 \\[3pt]%
	HCAMSB2 & 223.5 & 58.7 & 4.6 & 1.7 & 0.35 \\[3pt]%
	&&&&&\\%
	\hline\hline%
    \end{tabular}%
    \caption{Estimated SM background cross sections (plus two HCAMSB benchmark 
      points) in fb for various multi-lepton plus jets $+\etm$ topologies after 
      cuts C2 with $E_T^c=100$ GeV.
    }%
    \label{tab:hcaC2}%
  \end{center}%
\end{table}%
\indent
The results of C2 cuts on SM background and the two model lines is shown in 
Table \ref{tab:hcaC2} for $\etcut = 100$ and for 2M events.  Backgrounds cannot
be detected for several multi-lepton channels.  We consider an observable 
signal to be one that satisfies the following criteria for an assumed 
integrated luminosity:
\begin{itemize}%
\item[{\it i.}]   $S/B > 0.2$,
\item[{\it ii.}]  Signal has at least 5 events, and
\item[{\it iii.}] $S > 5\sqrt{B}$,
\end{itemize}%
Where $S$ and $B$ are the number of events for signal and background 
respectively.  The first requirement is imposed to prevent the use of data that
has a small signal on top of a large background, and that requires the use 
of extremely precise backgrounds \cite{Baer:2009dn}.  The last requirement 
means that we require the signal to have a $5\sigma$ statistical significance 
above the background.\\%
\indent%
Using the criteria above and assuming 100 $\invfb$ of data, we plot signal and
background versus $\mhf$ for C2 cuts with each of $\etcut = 100, 300,$ and 
$500$ GeV and in each of the lepton multiplicity channels.  Figure 
\ref{fig:hcareachm32} shows the $5\sigma/5$ event SM background values with 
dotted horizontal lines, while the signal is represented by solid lines.  
Signal values above the horizontal lines qualify as observable.  The value of 
$\mhf$ where the solid line meets its corresponding (same color) dotted line is 
the value of $\mhf$ that is accessible at the LHC.  A \susy signal is not 
observable where the solid line appears below the horizontal line.\\%
\indent
The summary of the results from the plots in Figure \ref{fig:hcareachm32} is 
given in Table \ref{tab:hcareachsummary}.  In the table, the upper entries are 
the $\mhf$ reach values for point 1 ($\alpha=0.025$) and the lower entries are 
the same for point 2 ($\alpha=0.15$), and each row corresponds to a different $
\etcut$ value.  The highest gravitino mass probed for $\etcut=100$ GeV is 80 
TeV and is found in the $3l$ channel.  However, we can see by going to the 
harder cut, $\etcut=500$ GeV, in the 0l channel, the reach is extended up to 
115 TeV for $\mhf$ for HCAMSB1.  This value of $\mhf$ corresponds to a gluino 
mass of $\mg\sim$ 2.4 TeV.  Similarly, the best reach for HCAMSB2 is in the 
$0l$ channel with $\etcut=500$ GeV, with the 105 TeV gravitino mass being 
probed, corresponding to a gluino mass of $\sim$ 2.2 TeV.
\begin{table}%
  \begin{center}%
    \begin{tabular}{|lccccc|}%
	\hline\hline%
	&&&&&\\%
	$E_T^c$ (GeV) & $0\ell$ & $1\ell$ & $OS$ & $SS$ & $3\ell$ \\[3pt]%
	&&&&&\\%
	\hline%
	&&&&&\\%
	100 & $65/60$ & $65/65$ &  $55/ 40$ & $70/ 65$ & $80/ 45$ \\[3pt]%
	300 & $105/100$ & $110/105$ &  $85/70$ & $-/-$ & $-/-$ \\[3pt]%
	500 & $115/ 105$ & $-/-$ &  $-/ -$ & $-/ -$ & $-/ -$ \\[3pt]%
	&&&&&\\%
	\hline\hline%
    \end{tabular}%
    \caption{Estimated reach of 100 fb$^{-1}$ LHC for $m_{3/2}$ (TeV) in two 
      HCAMSB model lines: $\alpha =0.025$ (upper entry) and $\alpha =0.15$ 
	(lower entry), in various signal channels.
    }%
    \label{tab:hcareachsummary}%
  \end{center}%
\end{table}%

\section{%
  \label{sec:hcadiscussion}
  Summary
}%
In this chapter we have given a complete analysis of the \hca model relevant to
physics at the LHC.  We calculated the spectrum from which we were able to 
further calculate branching fractions and production cross sections, and to 
simulate LHC collider events for comparison again SM backgrounds.  The final 
result of our analysis is that after 100 $\invfb$ of data collection running at 
14 TeV, the gravitino mass reach is $\mhf\sim 115$ TeV ($\mg\sim 2.4$ TeV) for 
low values of $\alpha$, and $\mhf\sim 105$ TeV ($\mg\sim 2.2$ TeV) for large 
$\alpha$.  Since the spectra is so similar for $\mu > 0$ and $\mu < 0$, we do 
not expect changes in the reach due to the change in the $sgn(\mu)$.  If 
instead $\tanb$ was moved to higher values, we do not expect there to be 
differences in the $0l$ and $1l$ channels where we have maximal reach.  
Although it is possible there would be differences in the multi-lepton channels 
due to enhanced -ino decays to $\tau$s and $b$s at large $\tanb$.  In 
references \cite{Baer:2000bs}\cite{Barr:2002ex}, the \mam reach was shown to be 
around $\mg\sim 2.75$ TeV for low values of $\mnot$, which is somewhat larger 
than for \hca.  This is due to the fact that squark masses tend to cluster 
around a common mass scale $\mnot$ in \mam, while in the \hca case right and 
left squark states are highly split.\\%
\indent
The \mam and \hca models appear to have similar signatures at the LHC.  Both 
would have multi-jet, multi-lepton, and $\etm$ events.  In addition to these, 
long chargino tracks of $O(1-10)$ cm are possible.  However, there are major
differences between these two models that are summarized below.
\begin{itemize}%
  \item \hca has a highly left-right split scalar spectrum due to the left-
	  right asymmetry in hypercharge assignments.  For \mam there is instead 
	  a near-equality between left and right mass parameters.  The lightest 
	  stau, $\spart{\tau}{1}$, is mainly left-handed, while in \mam it is 
	  mainly right-handed.  While it is possible that the left-right mixing 
	  could be determined at the LHC, this task would be easily done at a 
	  linear collider with polarized beams \cite{Baer:1996vd}.
  \item We saw that high rate production of $b$s and $t$s was due to light 
	  $\st$ and $\sb$ states relative to the gluino.  In particular, with the 
    	  \hca model we expect high multiplicities of final state b-quarks, 
	  whereas for \mam models much few are expected (this also depends on the 
	  chosen $\tanb$).
  \item We also see that the ordering of $\mu > M_{1} > M_{2}$ 
	  implied a bino-like $\ninos{2}$ for mAMSB, while for HCAMSB, with
	  $M_{1} > \mu > M_{2}$, it is higgsino-like.  This 
	  crucially changes the decays of $\ninos{2}$ and leads to an important
	  distinction between \mam and \hca models: the $m(ll)$ distribution for
	  OS, dilepton events has a $Z\rightarrow l^{+}l^{-}$ peak in the latter,
	  while the former would have a smooth distribution with no peak (except
	  at large $\tanb$ where there is greater mixing in the neutralino 
	  sector).  Thus, HITs (see Chapter \ref{chap:mamsb}) from charginos and 
	  a Z-peak from $\ninos{2}$ decays could yield a promising signal for 
	  \hca and LHC.
\end{itemize}%

\chapter{%
  inoAMSB
  \label{chap:inoamsb}
}%

\section{Introduction to inoAMSB Models}%
The combination of anomaly mediated (Section \ref{sec:grav_soft}) and gaugino 
mediated \cite{Kaplan:1999ac} soft \susy breaking parameters is suggested by 
two classes of models that arise in type IIB string theories compactified on a 
Calabi-Yao Orientifold (CYO).  These model types differ in the number of \kah 
moduli and they are:
\begin{itemize}
\item Single \kah modulus (SKM models) -- KKLT type with uplift of soft terms
  and cosmological constant coming from one-loop effects, and generally favors
  smaller CY compactification volumes;
\item Large Volume Scenario (LVS models) -- require at least two moduli and, as
  the name implies, favors larger CY compactification volumes.
\end{itemize}
The F-terms of moduli are responsible for \susy breaking.  In both cases the 
moduli are stabilized using a combination of fluxes and non-perturbative 
effects while the interactions with the MSSM are gravitational.  The models 
also share two crucial features: gauginos receive their masses only from Weyl 
anomalies (string contributions are suppressed) and the soft scalar and 
trilinear parameters are naturally suppressed as usual in gaugino mediated 
scenarios (see \cite{Kaplan:1999ac}\cite{Baer:2006rs}).\\%
\indent%
Having just analyzed the \hca model, it should be possible to study other 
models in a similar fashion.  In this chapter we do just that for a class of 
string models whose soft parameters arise in a pattern that we call the 
\quotes{Gaugino Anomaly Mediated \susy Breaking} pattern, or more succinctly, 
\quotes{inoAMSB} \cite{Baer:2010uy}.  The string theory origin for this model 
is quite different than for the case of HCAMSB, but we will see some similarities 
in the spectra that will lead to collider signatures already familiar from the 
last chapter.  Furthermore, we will more or less follow the same procedure for 
arriving at the collider analysis level.  Schematically this procedure is as 
follows:%
\begin{itemize}%
  \item define the model high-scale boundary conditions,
  \item discuss the evolution of theory parameters from high-scale to TeV 
	  scale,
  \item explore the allowed parameter space subject to experimental constraints,
  \item choose the generally representative points for collider analysis and 
	  examine branching fractions and cross sections to determine possible 
	  signatures,
  \item and finally, determine the parameter reach for an experiment to produce
	  a statistically significant signal.%
\end{itemize}%
\section{%
  Setup of the \ino Models
  \label{sec:inoamsb_setup}
}%

Before moving on to the details of the models we should first consider a few
preliminaries.  A supergravity model is defined in terms of three functions: 
a superpotential, a \kah potential, and the gauge kinetic function.  They each 
have implicit $\Phi$ dependence and can be expanded in a series of \mssm 
superfields ($C^{\alpha}$) as follows:
\begin{align}%
  W & = \hat{W}(\Phi)+\mu(\Phi)H_{d}H_{u}
	  +\frac{1}{6}Y_{\alpha\beta\gamma}C^{\alpha}C^{\beta}C^{\gamma}+\dots\\
  K & = \hat{K}(\Phi,\bar{\Phi}) 
	  + \tilde{K}_{\alpha\beta}(\Phi,\bar{\Phi})C^{\alpha}C^{\beta} 
	  + [Z(\Phi,\bar{\Phi})H_{d}H_{u} + h.c.] + \dots,\\
  f_{a} & = f_{a}(\Phi).
\end{align}%
The role of these moduli fields is to break supersymmetry.  This happens once 
they are stabilized and acquire VEVs such that at least one of them has a
non-zero F-term.\\%
\indent%
Each of the first terms in the above expansions have implicit dependence on the 
Calabi-Yau Orientifold (CYO):
\begin{align}%
  \hat{K} & = -2\rm{ln}({\cal V} + \frac{\xi}{2}) 
    - ln(i\int\Omega\wedge\bar{\Omega}(U,\bar{U})) -ln(S+\bar{S}),\\
  \hat{W} & = \int G_{3}\wedge\Omega + \sum_{i} A_{a}e^{-a_{i}T^{i}}.
\end{align}%
Most notably, the first term in $\hat{K}$ contains the CYO volume, $\cal V$, and
a stringy correction $\hat \xi$ that depends on the Euler character of the CYO 
and the real part of the dilaton superfield S.  The last term in $\hat{W}$ is 
non-perturbative and is responsible for stabilizing the \kah modulus at the 
minimum of the potential as in KKLT models.  $\Omega$ are 3-forms on the CYO 
and are functions of $U_r$, the complex structure moduli.\\%
\indent%
The standard SUGRA potential is given by
\begin{align}%
V(\Phi) = F^{A}F^{\bar{B}}K_{A\bar{B}} - 3|\mhf(\Phi)|^{2},
\end{align}%
where the SUSY breaking F-terms at the minimum are given by 
$F^{A}=e^{K/2}K^{A\bar{B}}D_{\bar{B}}W$, and the covariant derivatives are 
given by $D_{A}=\partial_{A}+K_{A}$, where subscripts  indicate derivatives 
with respect to fields.  Also, $\Phi$ stands collectively for all moduli that 
describe the internal geometry of the CYO plus the axion-dilaton.  The minimum 
of the potential, $V_{0}$, gives the cosmological constant (CC) of the theory.
The CC is known to be small, and taking it to be close to zero implies that
at the minimum of the potential $F^{A}\lsim \mhf$.\\%
\subsection{%
  Single \kah Modulus Scenario (SKM)
}%
This model has IIB string theory compactified on a CYO, has only one \kah 
modulus, T, and the \mssm fields come from open string fluctuations on a stack 
of D3 branes.  Other moduli and the axio-dilaton are stabilized by internal 
fluxes and non-perturbative effects, as in KKLT . \\%
\indent%
At the classical level the CC would be small in magnitude but negative and the
soft masses would be highly suppressed.  However, because SUGRA theories have 
quadratic divergences at the quantum level (unlike broken global \susy), there 
are corrections to these quantities that are dependent on the string scale 
cutoff $\Lambda$.  Such contributions can serve to uplift the CC to small and 
positive values and can generate soft \susy breaking masses proportional to 
$\frac{\Lambda}{4\pi}\mhf$.  The cutoff is taken to be the string scale and can 
be between $\sim 10^{14}$ GeV and $\sim M_{GUT}$.
\subsection{%
  Large Volume Scenario (LVS)
}%
In this class of models \cite{Balasubramanian:2005zx},  IIB string theory is 
again compactified on a CYO. However, now more than one \kah modulus is 
considered, $T_{i}$ $(i = 1,\dots, h_{11} $). In particular, in the simplest 
situation there is a large modulus, $\tau_{b}$, and small moduli, 
($\tau_s$ , $\tau_a$), controlling the overall size of the CYO and the volume 
of two small 4-cycles respectively. The total volume is then given by
\begin{align}%
{\cal V} & = \tau_{b}^{3/2} - \tau_{s}^{3/2} - \tau_{a}^{3/2}.
\end{align}%
This is referred to as a \quotes{Swiss Cheese} model. Again the MSSM may be 
located on D3 branes at a singularity. Alternatively, it could be placed on a 
stack of D7 branes wrapping a four cycle (taken to be the one labelled by the 
index \quotes{a}). In this case, it has been argued \cite{Blumenhagen:2007sm} 
that the necessity of having chiral fermions on this brane prevents this cycle 
from being stabilized by non-perturbative effects and it shrinks below the 
string scale. Effectively, this means that the physics is the same as in the D3 
brane case.\\%
\indent%
Extremizing the potential leads to an exponentially large volume
\cite{Balasubramanian:2005zx} $V\sim e^{a\tau_{s}} , \tau_{s}\sim \hat{\xi}$.  
It turns out that the suppression of FCNC effects lead to $V \sim 10^{5}l_{P}$ 
\cite{deAlwis:2009fn} (where $l_{P}$ is the Planck length), so the string scale 
is $M_{string}\sim M_{P} /\sqrt{V}\sim 10^{15.5}$ GeV. The minimum of the 
potential (CC) is given by $V_{0}\sim-\frac{m_{3/2}^{2}
M_P^2}{\ln m_{3/2}\mathcal{V}}$.  This minimum can be uplifted to zero when $S$ 
and $U_r$ acquire (squared) $F$-terms of the order $\frac{m_{3/2}^{2}M_P^2}{\ln 
m_{3/2}\mathcal{V}}$.  Classical contributions to the scalar and slepton masses 
are also of this same order. With the above lower bound on the volume, this
means that even for $m_{3/2}\sim 100$ TeV,  the classical soft terms are $\lsim 
100$ GeV. Of course if one wants to avoid fine-tuning of the flux 
superpotential, it would be necessary to take even larger values of ${\cal V}$ 
corresponding to a string scale of $10^{12}$ GeV. In this case the classical 
soft terms are completely negligible (for $m_{3/2}\sim100$ TeV) but  the 
(classical) $\mu$-term is also strongly suppressed. \\%
\indent%
In the rest of this section the holomorphic variable associated with the large 
modulus $\tau_b$ will be called $T$. 
\subsection{%
  Gaugino Masses - Weyl Anomaly Effects
}%
For a generic version of supergravity, the gaugino masses satisfy the following 
relation at the classical level: 
\begin{equation}
  \frac{M_{a}}{g_{a}^{2}}=\frac{1}{2}F^{A}\partial_{A}f_{a}(\Phi).
  \label{classicalgaugino}
\end{equation}
In the single \kah modulus model the MSSM resides on D3 branes at a 
singularity. In the LVS case, we may either have the MSSM on D3 branes at a 
singularity, or we may have it on a stack of D7 branes wrapping a four cycle 
which shrinks below the string scale. In both cases  the classical gauge 
coupling function is effectively of the form 
\begin{equation}
  f_a=S\label{eq:f}
\end{equation}
where $S$ is the axio-dilaton (for more details see \cite{deAlwis:2009fn} 
section 3.1 and references therein). The important point is that it is 
independent of the modulus $T$.  So at a classical minimum, where the SUSY 
breaking is expected to be in the $T$ modulus direction, the string theoretic 
contribution to the gaugino mass is highly suppressed.\\%
\indent%
However, there is an additional contribution to the gaugino mass due to the 
(super) Weyl anomaly as discussed in Section \ref{sec:grav_soft}. This comes from 
the expression for the effective gauge coupling superfield that has been 
derived by Kaplunovsky and Louis \cite{Kaplunovsky:1994fg} (KL)%
\footnote{As explained in \cite{deAlwis:2008aq}, the usual formulae for AMSB
need modification in the light of \cite{Kaplunovsky:1994fg}.%
}.  For the gaugino masses, the relevant contribution comes from taking
the $F$-term of 
\begin{equation*}
  H_{a}(\Phi,\tau,\tau_{Z})=f_{a}(\Phi)-\frac{3c_{a}}{8\pi^{2}}{\rm ln} C
  -\frac{T_{a}(r)}{4\pi^{2}}\tau_{Z}.
  \tag{\ref{eqn:gauge_coupling}}
  \label{eq:H}
\end{equation*}
Here, the first term on the RHS is the classical term; the second comes from 
the anomaly associated with rotating to the Einstein-\kah frame.  
$c_{a}=T(G_{a})-\sum_rT_{a}(r)$ is the anomaly coefficient and the last term 
comes from the anomaly associated with the transformation to canonical kinetic
terms for the MSSM fields. Also note that we have ignored the gauge kinetic 
term normalization anomaly \cite{ArkaniHamed:1997mj,deAlwis:2008aq} which is a 
higher order effect. The chiral superfields $\phi, \phi_r$ that generate these 
transformations are given by, 
\begin{eqnarray}%
  {\rm ln}C+{\rm ln}\bar{C} & = & \frac{1}{3}K|_{\rm harmonic} ,\\
  \tau_{Z}+\bar{\tau}_{Z} & = & \ln\det\tilde{K}_{\alpha\bar{\beta}}^{(r)} .
\end{eqnarray}%
The instruction on the RHS of the first equation is to take the sum of the 
chiral and anti-chiral (i.e. harmonic) part of the expression.  After 
projecting the appropriate $F$ terms we arrive at the following expression: 
\begin{equation}%
  \frac{2M_{a}}{g_{a}^{2}}=F^{A}\partial_{A}f_{a}-\frac{c_{i}}{8\pi^{2}}
  F^{A}K_{A}-\sum_{r}\frac{T_{i}(r)}{4\pi^{2}}F^{A}\partial_{A}\ln\det\tilde{K}
  _{\alpha\bar{\beta}}^{(r)} \label{amsbgaugino}.
\end{equation}%
The first (classical) term is greatly suppressed relative to $\mhf$ since the 
$T$-modulus does not contribute to the classical gauge coupling function as 
discussed earlier (see paragraph after (\ref{classicalgaugino})). The dominant 
contribution therefore comes from the last two (Weyl anomaly) contributions. It 
turns out that (after using the formulae $F^{T}=-(T+\bar{T})m_{3/2}$, 
$K_{T}=-3/(T+\bar{T})$ and 
$\tilde{K}_{\alpha\bar{\beta}}=k_{\alpha\beta}/(T+\bar{T})$ which are valid up 
to volume suppressed corrections), this yields%
\footnote{Note that we expect the Weyl anomaly expressions for the gaugino 
masses given below to be valid only because of the particular (extended 
no-scale) features of this class of string theory models. It so happens that
these are exactly the same as the expressions given in what is usually
called AMSB: but that is an accident due entirely to the fact that
in these extended no-scale models the relationship $F^{A}K_{A}\simeq3m_{3/2}$
is true.%
},%
\begin{equation}
  M_{a}=\frac{b_{a}g_{a}^{2}}{16\pi^{2}}m_{3/2},\label{eq:weylgaugino}
\end{equation} 
where $b_a=-3T(G_{a})+\sum_rT_{a}(r)$ is the beta function coefficient. 
\subsection{%
  Scalar Masses, Trilinear Couplings, $\mu$ and $B\mu$ terms
}%
Here are summarized the results from this class of string theory models for the 
values of the  soft parameters at the UV scale, {\it i.e.} 
$\Lambda\sim M_{string}\sim M_P/\sqrt{{\cal V}}$. These values should be the 
initial conditions for the RG evolution of these parameters. In the LVS case, 
it was estimated \cite{deAlwis:2009fn} that the lower bound on the CYO volume 
was ${\cal V}>10^5$. Also, typical values of $h_{21}\sim O(10^2)$ are chosen 
for the number of complex structure moduli.  The gravitino mass is chosen to be 
$\mhf\sim |W|M_P/{\cal V}\sim 50\ {\rm TeV}$.  Such a large value of $\mhf$ 
allows us to avoid the SUGRA gravitino problem, which leads to a disruption of 
Big Bang nucleosynthesis if $m_{3/2}\lsim 5$ TeV and $T_R\gsim 10^5$ GeV 
\cite{Kohri:2005wn}\cite{Kawasaki:2008qe}\cite{Pradler:2006hh}.\\%
\indent%
Unlike the gaugino masses, scalar masses and trilinear soft terms do not 
acquire corrections from the Weyl anomaly. They are essentially given at the UV 
scale by their classical string theory value plus one loop string/effective 
field theory corrections. In the $h_{11}=1$ case, the classical soft terms are 
essentially zero while in the LVS case %
\begin{equation}
  m_{0}\sim O\left(\frac{m_{3/2}}{\sqrt{\ln m_{3/2}{\cal V}}}\right), \ \ %
  \mu\sim\frac{B\mu}{\mu}\lsim \sqrt{h_{21}}m_{0},\ \ A_0\ll m_{0}. %
  \label{eq:lvsclassical} 
\end{equation}
As discussed in \cite{deAlwis:2009fn} the estimate of the $\mu,B\mu$ term comes 
from the generic case of assuming that the uplift of the negative CC is 
distributed amongst the complex structure moduli as well as the dilaton. It was 
estimated in that reference that the $F$-term for a complex structure modulus 
is $|F_m|\lsim m_{3/2}/\sqrt{h_{21}(\ln m_{3/2})\cal{V}}$. Since the expression 
for the $\mu$ term involves a sum over $h_{21}$ terms this gives the  estimate 
in (\ref{eq:lvsclassical}).  The parametric dependence on the volume factor in 
the upper bounds  for the $\mu,B\mu/\mu$ terms goes as 
$\sqrt{h_{21}}/\sqrt{\cal V}$ and clearly favors the ``small" values CYO 
volume  i.e. ${\cal V}\sim 10^5$ and the large  values of the number of complex 
structure moduli ($\sim 10^2$). However it  should be stressed that these 
estimates are on a different footing than the rigorous calculation of $m_0$ and 
the gaugino masses,  and indeed it is possible that these  LVS models  may have 
a $\mu$-problem.\\%
\indent%
After adding quantum corrections at the UV scale, both cases give similar 
values for the soft terms.  As an example, this is illustrated for two values 
for the CYO volume:   
\begin{itemize}%
\item ${\cal V}\sim 10^5, M_{string}\sim\Lambda\sim10^{-2.5}M_P\sim 10^{15.5}\ 
      {\rm GeV}$.  Then,  
        \begin{equation}%
          \mu \sim\frac{B\mu}{\mu}\lsim 250\ {\rm GeV},\, m_{0}\sim 25\ 
	    {\rm GeV},    \,A_0\ll m_0 .
        \label{GUT}
        \end{equation}%
\item ${\cal V}\sim 10^{12},\ M_{string}\sim\Lambda\sim10^{-6}M_P\sim 10^{12}
      \ {\rm GeV}$. Then, 
	  \begin{equation}%
          \mu \sim\frac{B\mu}{\mu}\lsim 10^{-1}\ {\rm GeV}, \, m_{0}\sim 10^{-2}
	    \ {\rm GeV},\, A_0\ll m_0 .\label{INT}
	  \end{equation}%
\end{itemize}%
The second very large volume case can be accessed only in the LVS model. Our 
favored case however is the first one with a volume around ${\cal V}\sim 10^5$ 
in Planck units. It should be noted that even with this value, it is only 
expected that stringy corrections to the numerical estimates of the above 
values of the soft masses etc. will be of order $\alpha'/R_{CYO}^2\sim 1/{\cal
{V}}^{1/3}\lsim 10\%$.\\%
\indent%
The first case is at the lower bound for the volume.  This gives the largest 
allowable string scale. This is still somewhat below the apparent unification 
scale,  but it is close enough that (allowing for undetermined $O(1)$ factors) 
the GUT scale may be used as the point at which to impose the boundary 
conditions. This is useful for the purpose of comparing with other models of 
SUSY mediation where it is conventional to use the GUT scale.\\%
\indent%
The second case above corresponds to choosing generic values of the flux 
superpotential, while the first needs a fine tuned set of fluxes to get 
$|W|\sim 10^{-8}$, in order to have a gravitino mass of  $\sim 10^2$ TeV, 
though in type IIB string theory general arguments show that there exist a 
large number of solutions which allow this. The most significant problem with 
the second case (apart from the fact that there is no hope of getting a GUT 
scenario) is  the extremely  low upper bound on the $\mu$ term. In other words, 
there is a serious $\mu$- problem. The first case also may have a $\mu$ term 
problem, but again since these estimates are accurate only to $O(1)$ numbers, 
it is possible to envisage that the problem can be resolved within the context 
of this model.\\%
\indent%
In any case, the string theory input is used to suggest a class of 
phenomenological models. Given that in both the GUT scale model and the 
intermediate scale model, the soft scalar mass and $A$ term are suppressed well 
below the weak scale and assumed to be zero for these at the UV scale, while 
the gaugino masses at this scale are given by%
\begin{align}%
M_{a}=\frac{b_{a}g_{a}^2}{16\pi^2}\mhf
\end{align}%
as was the case in the last subsection.\\%
\indent%
The case when the input scalar mass $m_0$ is non-negligible will also be 
discussed.  This would be the case for instance in the SKM model with smaller 
volumes and/or larger values of $h_{21}$,  and also in the case of LVS with the 
volume at the lower bound but with larger values of $h_{21}$.%
\section{%
  Spectrum, Parameter Space, and Constraints
}%
\begin{figure}[t]%
  \begin{center}
    \includegraphics[width=.75\textwidth,trim=0cm 0cm 9cm 15cm,clip]
      {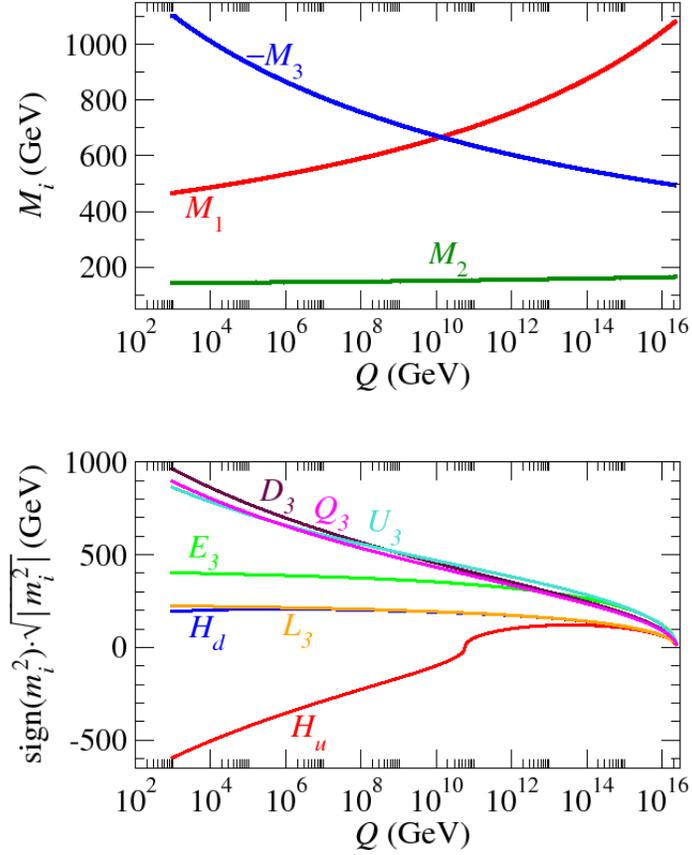}%
    \caption{inoAMSB soft \susy breaking parameters as a function of energy 
      scale Q for model parameters $\mhf$=50 TeV, $\tanb$=10, $\mu>0$, and 
	$\mstring$=$\mgut$.
    }%
      \label{inoamsb:running}%
  \end{center}%
\end{figure}

\begin{figure}[h!]%
  \begin{center}
    \subfigure[]{\includegraphics[width=0.7\textwidth]{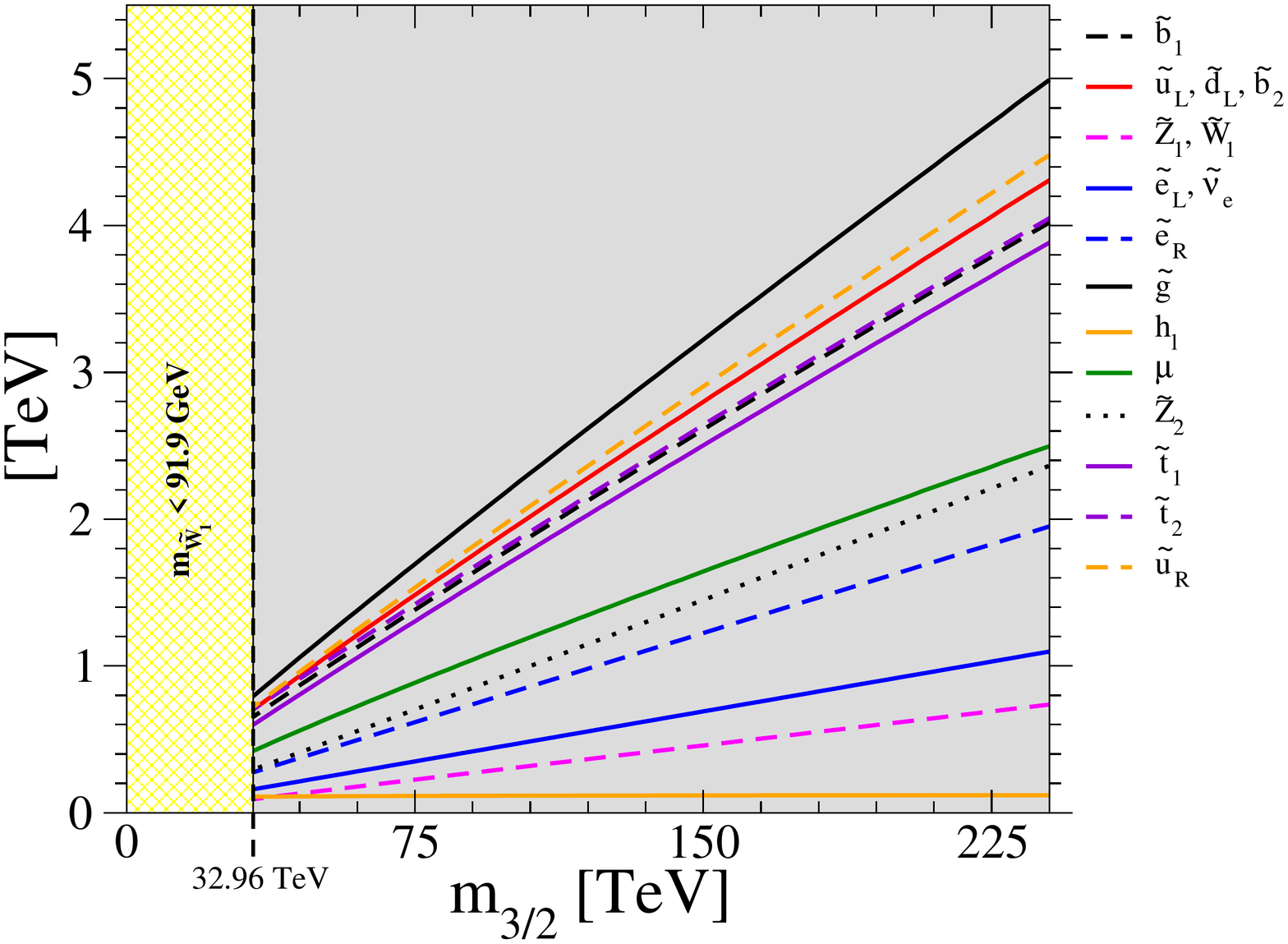}}\\%
    \subfigure[]{\includegraphics[width=0.7\textwidth]{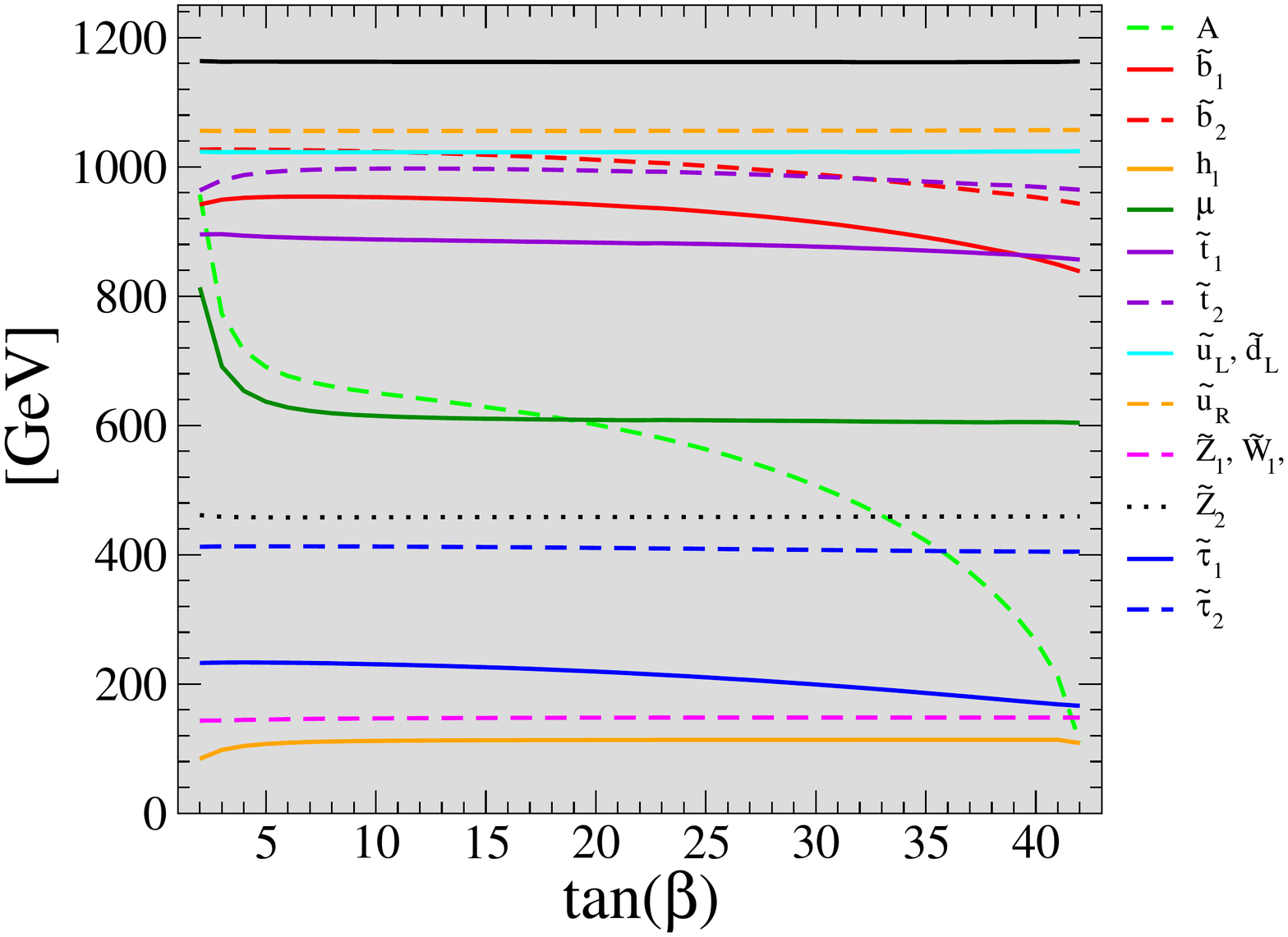}}%
    \caption{Sparticle masses: (a) as function of $\mhf$, with $\tanb=10$, 
	$\mu>0$, and $\mstring=\mgut$;  (b) as a function of $\tanb$, with $\mhf=$
	50 TeV, $\mu>0$, and $\mstring=\mgut$.
    }%
      \label{fig:ino_spectrum}%
  \end{center}%
\end{figure}

\begin{figure}[bh!]%
  \begin{center}
    \includegraphics[width=.6\textwidth,angle=-90,trim= 0cm 0cm 11cm 15cm,clip]
      {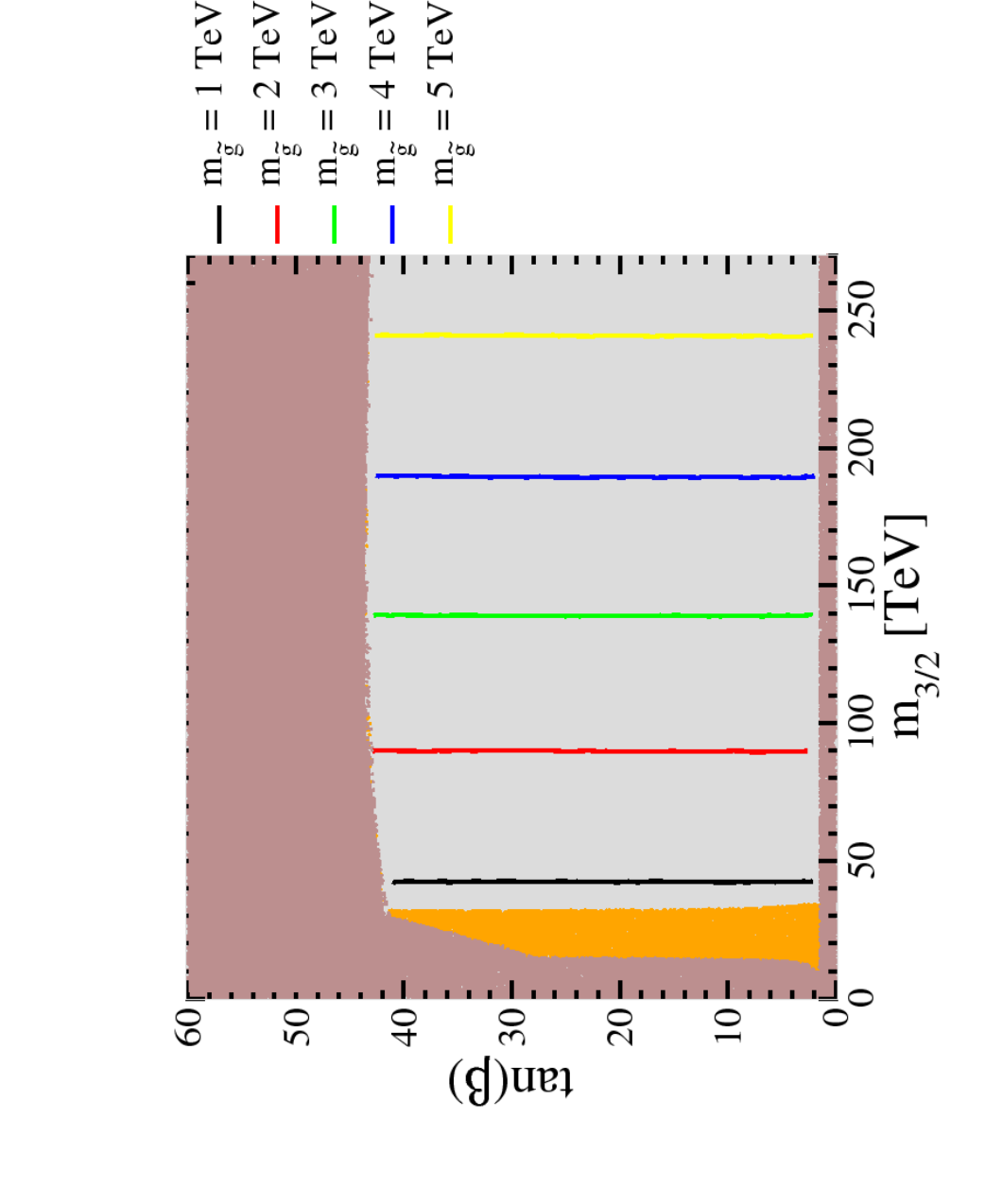}%
    \caption{$\tanb-\mhf$ parameter space, with $\mu>0$ and $\mstring=\mgut$.
	Also shown are gluino mass contours in the parameter plane.
    }%
    \label{inoamsb_pspace}%
  \end{center}%
\end{figure}

\begin{figure}[t!]%
  \begin{center}
    \includegraphics[width=.75\textwidth,trim= 0cm 0cm 0cm 0cm,clip]
      {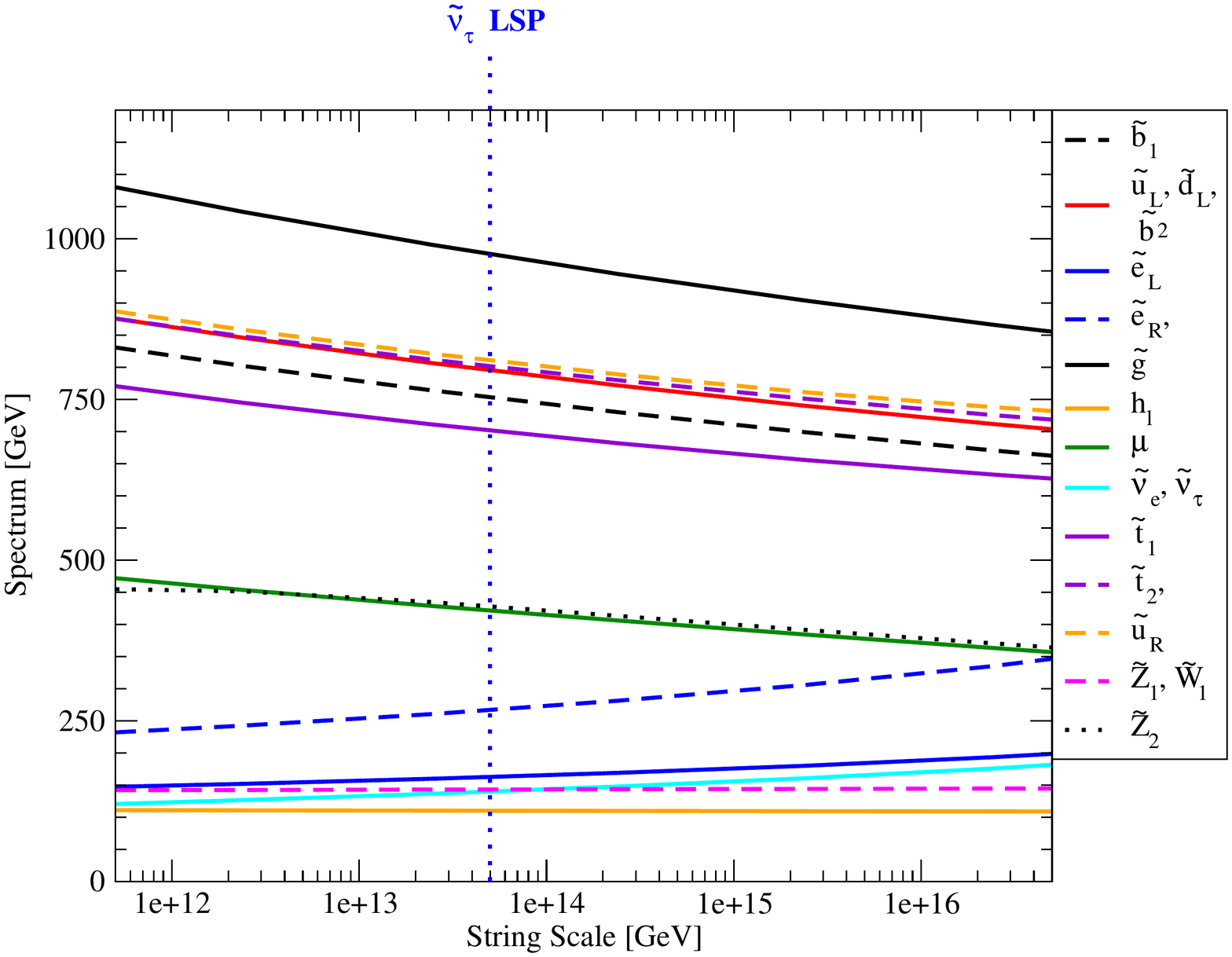}%
    \caption{\ino spectrum versus string scale for $\mhf$ = 50 TeV, $\tanb=10$,
	and $\mu > 0$.
    }%
    \label{fig:ino_spectrum_mstr}%
  \end{center}%
\end{figure}

\begin{figure}[t!]%
  \begin{center}
    \includegraphics[width=.75\textwidth,trim= 0cm 0cm 0cm 0cm,clip]
	{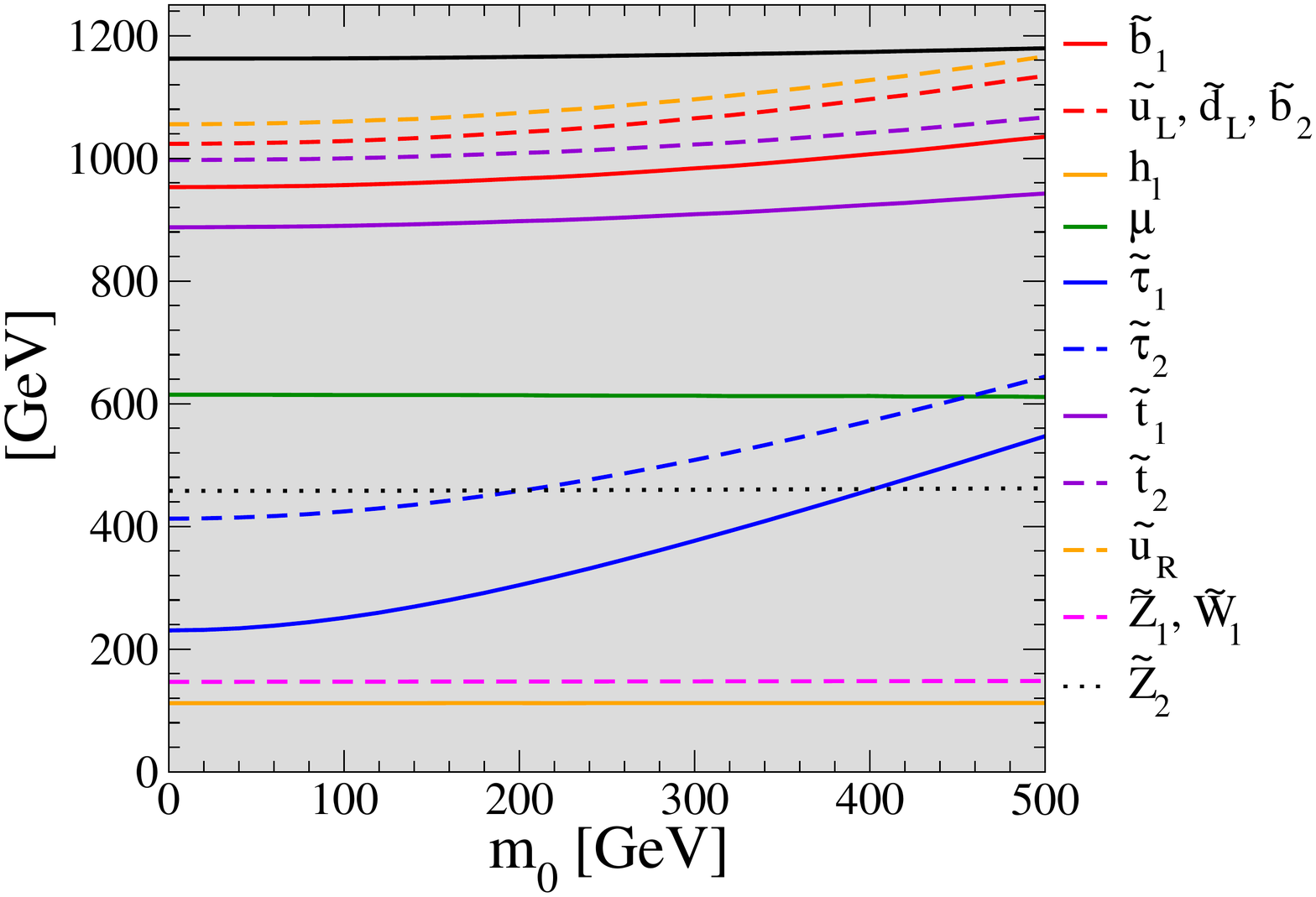}%
    \caption{\ino spectrum versus classical common soft mass, $\mnot$.  
      The parameters for this plot are $\mhf=50$ TeV, $\tanb=10$, and $\mu>0$.
    }%
    \label{fig:ino_spectrum_m0}%
  \end{center}%
\end{figure}

The parameter space for this model is%
\begin{center}%
$\mhf, \tanb$, and $sgn(\mu)$,%
\end{center}%
and, given the discussion of the previous section the boundary conditions 
\begin{equation}%
\mnot=\anot=0
\end{equation}%
are imposed along with AMSB masses for gauginos at $M_{string/GUT}$.  Note that 
even when $\mnot$ and $\anot$ are negligible at the GUT, scalars will be 
massive at the weak scale due to uplift from gauginos in the renormalization 
group running.\\%
\indent%
To illustrate the running of soft masses from the GUT scale to the weak scale,
Figure \ref{inoamsb:running} shows gaugino masses (top) and scalars (bottom) 
for the point $\mhf$ = 50 TeV, $\tanb = 10$, and $\mu > 0$ for 
$\mstring=\mgut$.  Gauginos begin with the usual ratio of ($\mb:\mw:\mgl$) =
(6.6 : 1 : -4.5), but at the weak scale the 1 and 3 masses switch in ordering 
leaving $|\mgl| >> \mb > \mw$.  The lower value of $\mw$ indicates that the 
lightest neutralino will be wino-like.  Scalars begin with negligible masses
as required by the boundary conditions.  However the mass parameters are 
uplifted by radiative self-couplings.  In particular, the large value of $\mb$
is responsible for the uplift near the $\mgut$.  In the early running the 
right-slepton, $E_{3}$, moves to the highest values because it has the largest
hypercharge ($Y$=2).  Because the hypercharges for the left-sleptons are lower 
than for the right-sleptons, they appear to have lower masses at the weak scale,
which leaves us with a left-right split spectrum for sleptons.  This effect is 
similar to what was seen in the case of HCAMSB, although the splitting for that 
model was more severe because of the much heavier $\mb$.  Squarks receive extra
uplift from QCD effects leaving them in the TeV range, while sleptons are 
generally lower: approximately between 200-400 GeV.  Also, since $\nino$ is 
wino-like and $\mw < \sqrt {m_{{\tilde L}_{3}}^{2}}$, charged LSPs are not of 
concern. \\%
\indent%
At the weak scale, physical masses and mixings are computed as usual and 1-loop
corrections are added.  The resulting spectrum for two \ino points with 
$\mhf$ = 50 and 100 TeV is listed in columns 3 and 4 of Table 
\ref{tab:inospectrum}. In column 1 of the same table is the \mam point with
$\mnot = 300$ GeV in column 2 is an \hca point with $\alpha = 0.025$ for 
comparison.  By inspection it is seen that the features of the previous
paragraph are represented in the table: large left-right splitting of sleptons,
generally heavier squarks, and a $\nino$ with a mass close to $\mw$ reflecting
that it is wino-like.  By comparison, \hca has a much larger left-right 
splitting in the sleptons (reflective of the heavier $\mb$ in that case), 
whereas \mam has a much smaller splitting (nearly-degenerate with 
$m_{\spart{e}{R}} < m_{\spart{e}{L}}$).\\
\begin{table}[th!]%
  \begin{center}%
    \begin{tabular}{|lcccc|}%
      \hline\hline%
	&&&&\\
      parameter & mAMSB & HCAMSB1 & inoAMSB1 & inoAMSB2 \\
	&&&&\\
	\hline%
	&&&&\\
	$\alpha$    & --- & 0.025 & --- & ---  \\
	$m_0$       & 300 & --- & --- & --- \\
	$m_{3/2}$   & $50\ {\rm TeV}$ & $50\ {\rm TeV}$ & $50\ {\rm TeV}$ & $100\ 
        {\rm TeV}$ \\
	$\tan\beta$ & 10 & 10 & 10 & 10\\
	$M_1$       & 460.3   & 997.7 & 465.5 & 956.1 \\
	$M_2$       & 140.0   & 139.5 & 143.8 & 287.9 \\
	$\mu$       & 872.8 & 841.8 & 607.8 & 1127.5 \\
	$mg$   & 1109.2 & 1107.6 & 1151.0 & 2186.1 \\
	$m_{\spart{u}{L}}$ & 1078.2 & 1041.3 & 1011.7 & 1908.7 \\
	$m_{\spart{u}{R}}$ & 1086.2   & 1160.3 & 1045.1 & 1975.7 \\
	$m_{\spart{t}{1}}$& 774.9 & 840.9 & 878.8 & 1691.8 \\
	$m_{\spart{t}{2}}$& 985.3 & 983.3 & 988.4 & 1814.8 \\
	$m_{\spart{b}{1}}$ & 944.4 & 902.6 & 943.9 & 1779.5  \\
	$m_{\spart{b}{2}}$ & 1076.7 & 1065.7 & 1013.7 & 1908.3 \\
	$m_{\spart{e}{L}}$ & 226.9 & 326.3  & 233.7 & 457.8 \\
	$m_{\spart{e}{R}}$ & 204.6 & 732.3  & 408.6 & 809.5 \\
	$m_{\cinos{2}}$ & 879.2 & 849.4 & 621.2 & 1129.8 \\
	$m_{\cinos{1}}$ & 143.9 & 143.5 & 145.4 & 299.7 \\
	$m_{\ninos{4}}$ & 878.7 & 993.7 & 624.7 & 1143.2 \\ 
	$m_{\ninos{3}}$ & 875.3 & 845.5 & 614.4 & 1135.8 \\ 
	$m_{\ninos{2}}$ & 451.1 & 839.2 & 452.6 & 936.8 \\ 
	$m_{\ninos{1}}$ & 143.7 & 143.3 & 145.1 & 299.4 \\ 
	$m_A$       & 878.1 & 879.6 & 642.9 & 1208.9 \\
	$m_h$       & 113.8 & 113.4 & 112.0 & 116.0 \\ 
	&&&&\\
	\hline\hline%
    \end{tabular}%
    \caption{Masses and parameters in~GeV units
      for four case study points mAMSB1, HCAMSB1, inoAMSB1 and inoAMSB2
      using Isajet 7.80 with $m_t=172.6$ GeV and $\mu >0$. 
      Also listed are the total tree level sparticle production cross section   
      in fb at the LHC.
    }%
    \label{tab:inospectrum}%
  \end{center}%
\end{table}%
\indent%
The \ino parameter space is simpler than either the \mam or \hca case because it
has one less model parameter.  We begin by recognizing that once a point is 
chosen, the increasing $\mhf$ value increases all masses but leaves the 
relative hierarchy unchanged.  This is demonstrated in Figure 
\ref{fig:ino_spectrum} (a) where the spectrum is plotted versus $\mhf$ for 
$\tanb = 10$ and $\mu > 0$.  Everywhere in the parameter space of the \ino model
the hierarchy is generally in the order $\mg > m_{\tilde q}$ $> |\mu|$
$ > m_{\spart{e}{R}}$ $> m_{\spart{e}{L}, \spart{\nu}{L}}$ $> m_{\nino,\cino}$.
We then expect squark pairs to be produced in LHC events either directly or 
through the decay of pair-produced gluinos.  Furthermore, because squarks are 
heavier than sleptons we can expect squark cascade decays to two hard jets + 
isolated leptons through $\sq\rightarrow$ $q\ninos{i}\rightarrow$ 
$q\sl^{\pm}l^{\mp}$.\\%
\indent%
In frame (b) of Figure \ref{fig:ino_spectrum}, taking $\mhf$ = 50 TeV, an
example of the spectrum versus $\tanb$ is given.  Some interesting effects 
occur in the region of high $\tanb$.  There are noticeable dips in the masses 
for ${\tilde b}$ and ${\tilde \tau}$ eigenstates since larger $\tanb$ results 
in larger Yukawa couplings for these particles, and thus greater suppression of 
masses when approaching the TeV scale.  The $m_{A}$ parameter decreases most
significantly in the plot.  Because of the increased down-type Yukawa couplings
for large $\tanb$, $m_{H_{d}}$ is pushed towards negative values.  Since 
the EW-breaking minimization conditions give 
$m_{A}^{2}\sim m_{H_{d}}^{2} - m_{H_{u}}^{2}$, the pseudoscalar too decreases 
for larger $\tanb$.  Negative values of $m_{H_{d}}^{2}$ signal improper 
breaking and this occurs around $\tanb\sim42$.  And finally, $m_{A}$ cannot 
fall below the LEP2 bounds on the Higgs.  In the plot $m_{h}\sim 111$ GeV which 
we consider near the acceptable edge due to the $\pm3$ GeV error in the theory 
calculation.  As $\tanb$ increases, the LEP2 bound is violated, and shortly 
thereafter the EW symmetry is not properly broken.  It can also been seen that
at the lowest $\tanb$ values, the lightest Higgs is too light.\\%
\indent%
With the features of Figure \ref{fig:ino_spectrum} in mind, we look at the 
entire $\tanb-\mhf$ parameter space in Figure \ref{inoamsb_pspace}.  The orange 
region is specifically excluded by LEP2 bounds on the chargino mass, and all 
$\mhf$ values between this region and zero are likewise excluded.  The brown 
regions are also excluded as in Figure \ref{fig:ino_spectrum} (b): the high 
$\tanb$ regions has improper EW-breaking and the lower region (just above 
$\tanb=0$) is where the light Higgs is too light.  Gluino contours are also 
shown in the plane, and it is seen that the major (quasi-linear) dependence is 
on $\mhf$.  A gluino with the higher masses shown will not be accessible at the 
LHC.\\%
\indent%
Until now we have made the assumption that $\mstring=\mgut$, but this is not
necessarily so as pointed out earlier.  In fact, we should expect that 
$\mstring$ will be somewhat lower than the GUT scale in order to have enough
FCNC suppression.  Figure \ref{fig:ino_spectrum_mstr} is a plot of the spectrum 
as a function of the string scale, and the latter is taken as low as $10^{11}$ 
GeV.  As $\mstring$ is lowered the spectrum is seen to spread out, and there are
cases of mass re-ordering.  The most important feature of this plot is that 
around $5\times10^{13}$ GeV, the LSP is the tau-sneutrino.  There are severe 
limits on stable sneutrino dark matter \cite{sneu_dm}.  Then, in order to avoid 
these limits and to have a $\nino$ LSP, we do not consider such low 
$\mstring$.\\%
\indent%
Finally, we also have been assuming that $\mnot=0$.  However this is 
not completely realistic, as they are only expected to be suppressed.  The 
spectrum is shown again in Figure \ref{fig:ino_spectrum_m0}, this time as a 
function of $\mnot$, the classical mass added universally to all scalars.  We 
adopt values $\mhf=50$ TeV and $\tanb =10$ for this plot.  As $\mnot$ increases 
beyond zero, it is seen that the spectra change little so long as 
$m_0\lsim 100$ GeV, and also the mass orderings remain intact. For larger 
values of $m_0$, the left- and right-slepton masses begin to increase, with 
first $m_{\spart{e}{R}}$ surpassing $m_{\ninos{2}}$, and later even 
$m_{\spart{e}{L}}$ surpasses $m_{\ninos{2}}$.  At these high values of $\mnot$, 
decay modes such as $\ninos{2}\rightarrow \l^\pm\sl^\mp$ would become 
kinematically closed, thus greatly altering the collider signatures.  However, 
generically in this class of models, we would not expect such large additional 
contributions to scalar masses.\\%
\subsection{\emph{BF($b\rightarrow s\gamma$)} 
             and \emph{(g-{\bf 2})$_{\mu}$/{\bf 2}}}
\begin{figure}%
  \begin{center}
    \subfigure[]{\includegraphics[width=.65\textwidth,angle=-90,trim= 1cm .9cm 
	11cm 15cm,clip]{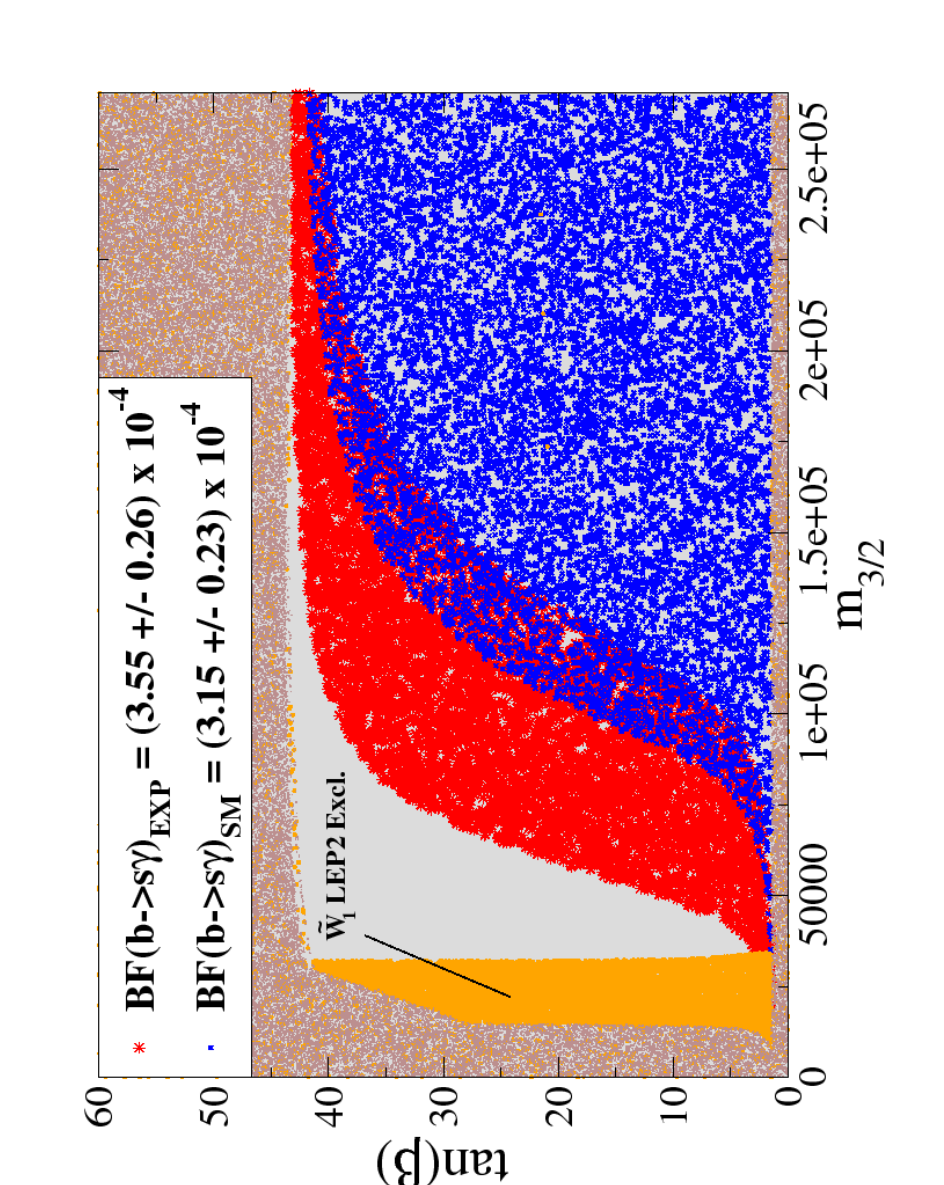}}\\%
    \subfigure[]{\includegraphics[width=.65\textwidth,angle=-90,trim= 1cm .9cm 
	11cm 15cm,clip]{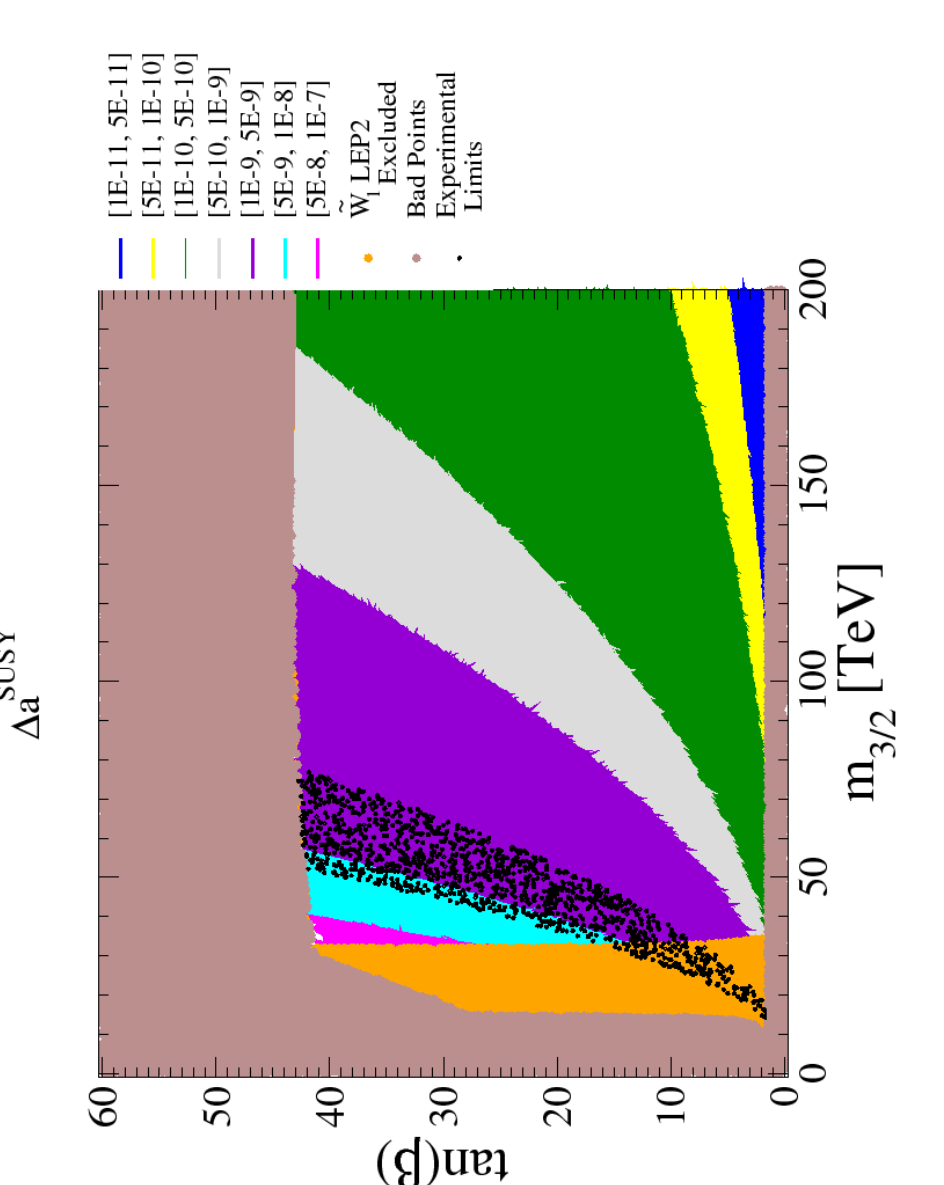}}%
    \caption{(a) $BF(b\rightarrow s\gamma)$ and (b) \susy contribution to 
	$(g-2)_{\mu}$ in the \ino parameter plane with $\mstring=\mgut$.
    }%
      \label{fig:ino_bsgamu}%
  \end{center}%
\end{figure}

Along with experimental constraints on the inoAMSB models from LEP2 limits on 
$m_h$ and $m_{\cino}$, there also exist indirect limits on model parameter 
space from comparing measured values of $BF(b\to s\gamma )$ and 
$\Delta a_\mu\equiv(g-2)_\mu /2$ against SUSY model predictions.  Figure 
\ref{fig:ino_bsgamu} (a) shows regions of the branching fraction for 
$BF(b\rightarrow s\gamma )$ in the inoAMSB model versus $m_{3/2}$ and $\tanb$ 
variation, calculated using the Isatools subroutine ISABSG\cite{Baer:1996kv}. 
The red-shaded region corresponds to branching fraction values within the SM 
theoretically predicted region 
$BF(b\rightarrow s\gamma )_{SM}=(3.15\pm 0.23)\times 10^{-4}$, by a recent 
evaluation by Misiak\cite{Misiak:2006zs}). The blue-shaded region corresponds 
to branching fraction values within the experimentally allowed region
\cite{Barberio:2006bi}: here, the branching fraction $BF(b\to s\gamma )$ has 
been measured by the CLEO, Belle and BABAR collaborations; a combined analysis
\cite{Barberio:2006bi} finds the branching fraction to be 
$BF(b\to s\gamma )=(3.55\pm 0.26)\times 10^{-4}$.  The gray shaded region gives 
too large a value of $BF(b\rightarrow s\gamma)$. This region occurs for low 
$\mhf$, where rather light $\spart{t}{1}$ and $\cino$ lead to large branching 
fractions, or large $\tanb$, where also the SUSY loop contributions are enhanced
\cite{Baer:1998xx}.\\%
\indent%
Figure \ref{fig:ino_bsgamu} (b) shows the plot of the SUSY contribution to 
$\Delta a_\mu$: $\Delta a_\mu^{SUSY}$ (using ISAAMU from Isatools
\cite{Baer:2001kn}). The contribution is large when $m_{3/2}$ is small;
in this case, rather light $\spart{\mu}{L}$ and $\spart{\nu}{\mu L}$ masses lead
to large deviations from the SM prediction.  The SUSY contributions to 
$\Delta a_\mu^{SUSY}$ also increase with $\tanb$. It is well-known that there 
is a discrepancy between the SM predictions for $\Delta a_\mu$, where $\tau$ 
decay data, used to estimate the hadronic vacuum polarization contribution to 
$\Delta a_\mu$, gives rough accord with the SM, while use of 
$e^+e^-\rightarrow hadrons$ data at very low energy leads to a roughly $3\sigma$
discrepancy.  The measured $\Delta a_\mu$ anomaly, given as 
$(4.3\pm 1.6)\times 10^{-9}$ by the Muon $g-2$ Collaboration\cite{Brown:2001}, 
is shown by the black dotted region.
\section{%
  \ino at the LHC
}%
\subsection{Sparticle Production}%
\begin{figure}[t]%
  \begin{center}
    \includegraphics[scale=.5,trim= 0cm 0cm 0cm 0cm,clip]{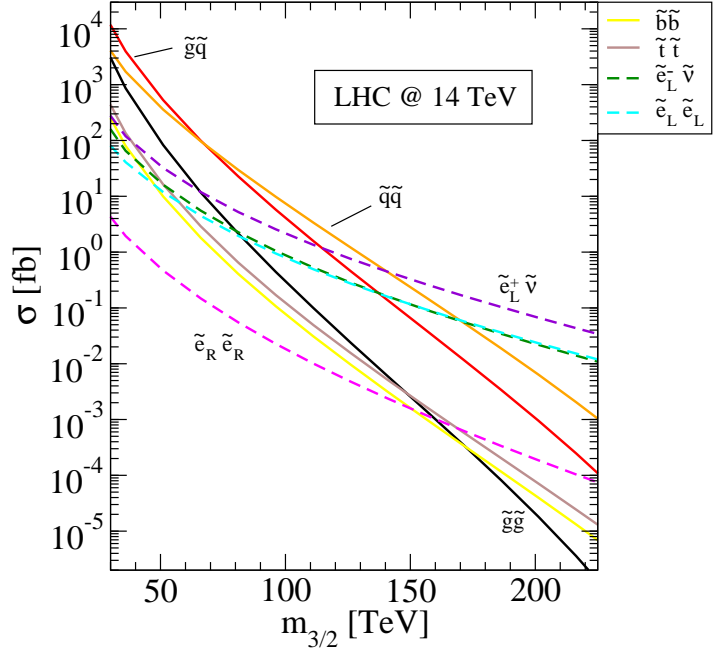}%
    \caption{Production cross sections for various pairs versus $\mhf$ for
	$\tanb$ = 10, and $\mu > 0.$
    }%
      \label{fig:ino_xsection_m32}%
  \end{center}%
\end{figure}

\begin{table}[t]%
  \begin{center}%
    \begin{tabular}{|lcccc|}%
	\hline\hline%
	&&&&\\%
       & mAMSB & HCAMSB1 & inoAMSB1 & inoAMSB2 \\
	&&&&\\%
	\hline%
	&&&&\\%
	$\sigma\ [{\rm fb}]$ & $7.7\times 10^3$ & $7.4\times 10^3$ & $7.5\times 
        10^3$ & $439$ \\
	$\gl ,\spart{q}{}\ {\rm pairs}$ & 15.0\% & 15.5\% & 19.1\% & 3\% \\[3pt]%
	${\rm EW-ino\  pairs}$ & 79.7\% & 81.9\% & 75.6\% & 93\% \\[3pt]%
	${\rm slep.}\ {\rm pairs}$ & 3.7\% & 0.8\% & 3.1\% & 3\% \\[3pt]%
	$\spart{t}{1}\bar{\tilde{t}}_{1}$ & 0.4\% & 0.2\% & 0.1\% & 0\% \\[3pt]%
	&&&&\\%
	\hline\hline%
    \end{tabular}%
    \caption{Cross sections for $pp\rightarrow SUSY$ at 14 TeV.
    }%
    \label{tab:ino_xsects}%
  \end{center}%
\end{table}%
Table \ref{tab:ino_xsects} shows the distributions in the cross section of 
various sparticle pair-production channels.  EW-ino pairs dominate over all 
other forms of production.  As in the case of HCAMSB, because $\nino$ is stable 
and $\cino^{\pm}$ decays to soft $\pi^{\pm}$ and $\nino$, the processes 
$pp\rightarrow\cino^{\pm}\cino^{\mp}$, $\cino^{\pm}\nino$ do not produce 
sufficient visible energy to meet detector trigger requirements.  We must 
instead rely on detecting produced squark and gluino pairs.\\%
\indent%
We will will be mainly interested in $pp\rightarrow \gl\gl,\sq\gl,\sq\sq$ 
because of their significant cross sections and visible final states.  Figure 
\ref{tab:ino_xsects} shows that there are significant cross sections for slepton 
production as well, but those cross sections are still lower than QCD pairs and 
the LHC reach in $\mhf$ is much less.  Therefore, for this analysis we focus 
strictly on QCD pair-production.\\%
\indent%
Since sparticle masses depend mainly on $\mhf$, the plot of multiple 
pair-production cross sections is shown as a function of this parameter in 
Figure \ref{fig:ino_xsection_m32}.  We see that for $\mhf\lsim$ 65 TeV $\gl\sq$ 
and $\sq\sq$ rates are comparable, but for higher $\mhf$ that level squark 
pairs dominate.  Rates for $\gl\gl$ can be significant, but they are never 
dominant.
\subsection{Sparticle Decay}%
We saw that the \ino RGE running leads to a hierarchy of masses in which all 
squark masses are close in value and less than the gluino mass.  Thus, produced
gluinos decay nearly equally to squark species through $\gl\rightarrow q\sq$.
Left squarks decay mainly through $\spart{q}{L}\rightarrow q + wino$, and the
calculation shows $\spart{q}{L}\rightarrow q'\cino$ at 67\% and  
 $\spart{q}{L}\rightarrow q\nino$ at 33\% over the entire parameter space.  
Right squarks on the other hand decay to $q+bino$, and therefore the decays
$\spart{q}{R}\rightarrow q\ninos{2}$ occur at 97\% over all the parameter 
values.\\%
\indent%
The RGE evolution also led to a large left-right splitting in the sleptons, and
this should have noticeable effects.  Left sleptons decay mainly to 
$wino+lepton$ and the calculation shows $\sl_{L}\rightarrow l\nino$ at 33\%
and $\sl_{L}\rightarrow \nu_{lL}\cino$ at 67\% over all parameters.  Sneutrinos
decay invisibly to $\nu_{l}\nino$ at 33\%, but has 66\% visible branching to
$l\cino$.  If right-sleptons were heavy enough, they would decay to 
$bino+lepton$, but since they are too light they decay via $\sl_{R}\rightarrow$
$e\nino$ at 78\%, as well as through three-body modes $\sl^{-}_{R}\rightarrow$
$l^{-}\tau^{+}\spart{\tau}{1}^{-}$ and $l^{-}\tau^{-}\spart{\tau}{1}^{+}$ at 
13\% and 7\% respectively.
\subsection{\ino LHC Events}%
\begin{figure}[h]%
  \begin{center}
    \includegraphics[width=1.2\textwidth,trim=.55cm 0cm 9cm 23.cm,clip] 
      {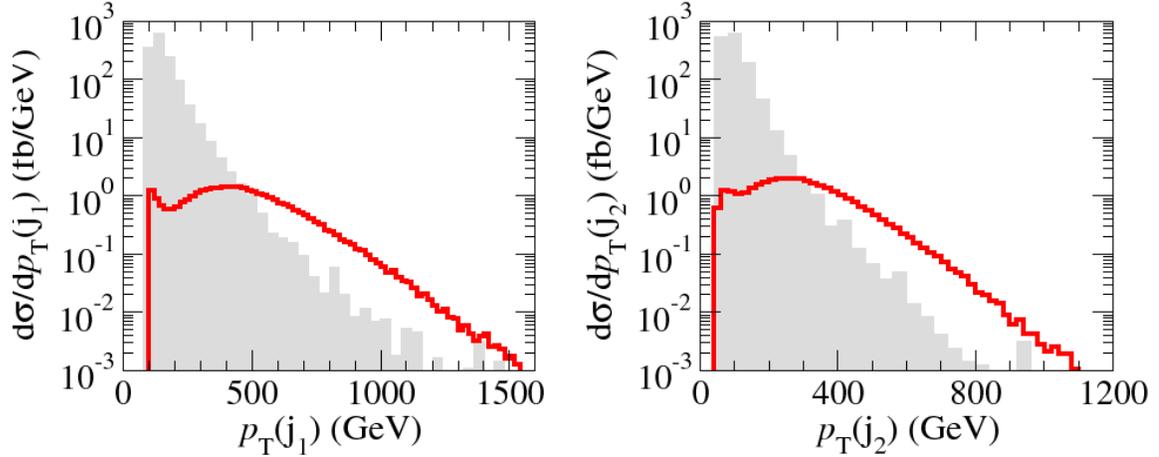}%
    \caption{$\pt$ distributions for hardest (left) and second hardest (right)
	jets for inoAMSB1 and SM backgrounds after C1 cuts for 14 TeV pp 
	collisions.
    }%
    \label{fig:ino_ptjs}%
  \end{center}%
\end{figure}

\begin{figure}[h]%
  \begin{center}
    \includegraphics[width=1.2\textwidth,trim= .55cm 0cm 9cm 23.cm,clip]
      {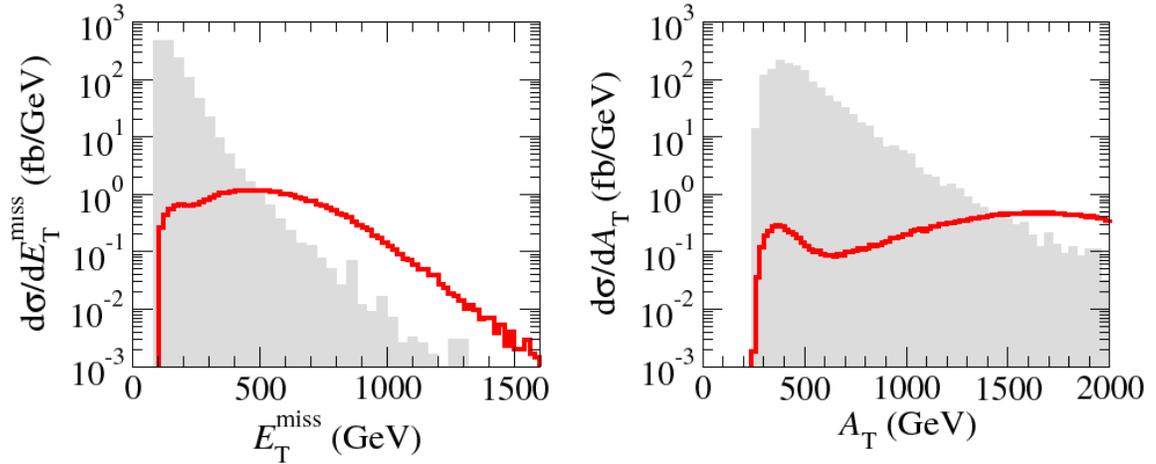}%
    \caption{$\et$ and $A_{T}$ distributions for inoAMSB1 and SM backgrounds 
	after C1 cuts for 14 TeV pp collisions.
    }%
    \label{fig:ino_etat}%
  \end{center}%
\end{figure}

\begin{figure}[h]%
  \begin{center}
    \includegraphics[width=1.2\textwidth,trim= .55cm 0cm 9cm 23.cm,clip]
      {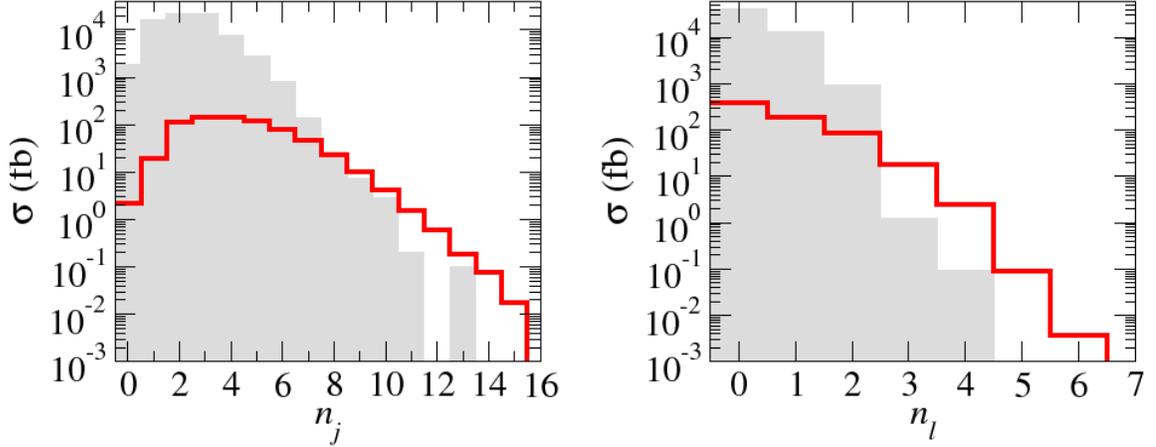}%
    \caption{Jet multiplicity (left) and lepton multiplicity (right) 
	distributions for inoAMSB1 and SM backgrounds after C1 cuts for 14 TeV pp 
	collisions.
    }%
    \label{fig:ino_njnl}%
  \end{center}%
\end{figure}

Isajet \cite{Paige:2003mg} was used for the LHC event simulation, the details 
of which can be found in Section \ref{sec:comptools}.  Two-million events for 
\ino point 1 of Table \ref{tab:inospectrum} were generated in addition to the 
same Standard Model backgrounds discussed in the Chapter \ref{chap:hcamsb}, 
which are $W+jets, \ \ Z+jets$\footnote{Vector boson + jets uses exact matrix 
element for a single parton emission and parton showering for subsequent 
emissions.}, $t\bar{t}, \ \ WW, \ \ WZ$, and $ZZ$.\\%
\indent%
We first apply a rudimentary set of cuts on signal and background that we will
label {\bf C1}.  This first round of cuts will help to understand the 
underlying properties of LHC events and will give us information on how to make
better cuts.  The C1 cuts are as follows:
\begin{itemize}%
  \item $n(jets) \geq 2$
  \item $\etm > max(100 \ \textrm{GeV}, 0.2M_{eff})$
  \item $\et(j_{1}, j_{2})>100,50 \ \textrm{GeV}$
  \item transverse sphericity $S_{T}>0.2$,
\end{itemize}%
where $M_{eff}=\etm+\et(j_{1})+\et(j_{2})+\et(j_{3})+\et(j_{4})$.\\%
\indent%
Since the sparticle production is dominated by $\gl\sq$ and $\sq\sq$ we expect 
to have two very hard jets in each event through the squark decays to EW-ino + 
quark.  This is seen in Figure \ref{fig:ino_ptjs} where the first and second 
hardest $\pt$ distributions are seen to emerge from background around 450 GeV 
and 250-300 GeV respectively.\\%
\indent%
Figure \ref{fig:ino_etat} shows the $\etm$ and $A_{T}(=\etm+\sum^{jets}\et+
\sum^{isolated\atop leptons}\et$) distributions for inoAMSB1 with SM 
backgrounds.  The $\etm$ distribution emerges from background around 500 GeV 
and appears rather hard because squarks have a significant invisible branching 
to $\nino$s.  The $A_{T}$ distribution has a peak around 400 GeV that is
buried under background, and a broad hard peak at higher $A_{T}$.  The softer 
peak originates in chargino, neutralino, and slepton pair production, while the 
hard peak is due to squark and gluino pair production and emerges from 
background around 1400 GeV.
We also examine jet (left) and lepton (right) multiplicities in Figure 
\ref{fig:ino_njnl}.  We expect to see hard dijet events from squark decays, but 
the jet distribution shows a broad peak between $n_{j}=$ 2--5 as well as much 
higher values.  These extra jets occur because of cascade decays and ISR and 
because in this first round of cuts $\etm>$ 50 GeV which is rather low compared 
to when the hardest jets exceed background in Figure \ref{fig:ino_ptjs}.  The 
leptons are seen to stand out above background already at 3 leptons, and so we
see already that with the minimal cuts a signal with 3 leptons, at least 2 jets,
and $\etm$ should appear well above background.\\%
\indent%
Again, as was the case for \mam and \hca already encountered, HITs (see 
Section \ref{chap:mamsb}) from long-lived charginos is an important signal for 
models with wino-like $\nino$s, particularly AMSB models.  If the anomaly 
mediation supersymmetry breaking pattern is the correct one, we expect HITs to 
play an important role in the discovery of \susy at the LHC.  After an AMSB 
discovery, deciphering further whether a model is correct would be the next step.  
After discussing the reach in the next subsection, it will be shown that if 
mAMSB, HCAMSB, and \ino are accessible at the LHC, it would be possible to 
unambiguously distinguish between all three models!  This is a major result of 
this work.\\%
\subsection{%
  \label{subsec:ino_reach}%
  LHC Reach}%
\begin{figure}[t]%
  \begin{center}
  \includegraphics[width=1\textwidth,trim= 0cm 0cm 9cm 20cm,clip]{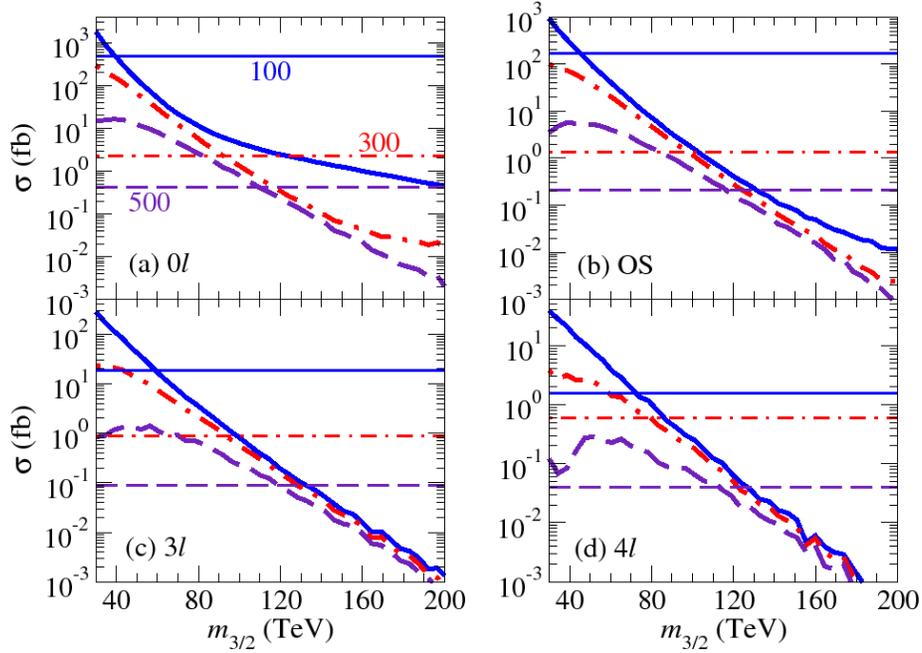}
  \caption{Cross section for multi-jet plus $\etm$ events with {\it a}). 
  $n(\ell) =0$,  {\it b}). OS isolated dileptons {\it c}). isolated $3\ell$s 
  and {\it d}). isolated $4\ell$s at the LHC after cuts $C2$ listed in 
  the text with $E_T^c=100$ GeV (blue solid), $E_T^c=300$ GeV (red dot-dashed) 
  and $E_T^c=500$ GeV (purple dashes), versus $m_{3/2}$, from the inoAMSB model 
  line points with $\tan\beta =10$ and $\mu >0$. We also list the 100 fb$^{-1}$ 
  $5\sigma$, 5 event, $S>0.1\ BG$ limit with the horizontal lines.  
  }%
  \label{fig:ino_reach}
  \end{center}
\end{figure}

Now we would like to find what are the maximum \ino parameter values 
accessible at the LHC after one year of data collection (100 $\invfb$) at 
$\sqrt{s} =$ 14
TeV.  This amounts to finding the reach in the $\mhf$ parameter because
it makes the most significant contribution to sparticle mass as seen in Figure 
\ref{fig:ino_spectrum}.\\%
\indent%
Taking what was learned in the last section by the C1 cuts on signal and 
background, we adopt the following new set of cuts labeled {\bf C2}:
\begin{itemize}%
\item $n(jets) \geq 2$%
\item $S_{T}>0.2$%
\item $\et(j_{1}), \ \et(j_{2}), \ \etm \ > \ \etcut$ 
\end{itemize}%
where $\etcut$ is variable in order to maximize the reach.  Parameter space 
points with lower masses benefit from lower $\etcut$.  Heavier particles 
have lower cross sections but also have higher energy release per event and
thus are more visible for higher $\etcut$.  Additionally, we apply extra cuts
for multilepton channels.  For the $0l$ channel we apply a cut on the 
transverse opening angle between $\etm$ and the nearest jet, $30^{\circ}<$
$\Delta\phi(\etm,\et (j^{near})<90^{\circ})$.  For all isolated leptons a
minimum $\pt$ of 20 GeV is required.  For the sake of brevity we will consider
$0l, 2l(OS), 3l, \textrm{and} \ 4l$ channels because they provide the best 
reach.\\%
\indent%
After applying C2 cuts with $\etcut=100$ GeV on SM backgrounds (2M events each) 
and the inoAMSB1 point,  the total cross sections are shown in Table 
\ref{tab:ino_c2cuts}.  The hard C2 cuts prevent some backgrounds from producing
noticeable cross sections indicated in the table by dashes.  For a signal to be
considered observable, we require that  i.) $S/B > 0.2$, ii.) Signal has at 
least 5 events, and iii.) $S > 5\sqrt{B}$ (5$\sigma$ significance), where $S$ 
and $B$ are respectively the signal and the background numbers of events for
a given luminosity.\\%
\begin{table}[t]%
  \begin{center}%
    \begin{tabular}{|l|ccccc|}%
	\hline\hline%
	&&&&&\\%
	Process & $0\ell$ & $OS$ & $SS$ & $3\ell$ & $4\ell$ \\
	&&&&&\\%
	\hline%
	&&&&&\\%
	QCD($p_T$: 0.05-0.10 TeV) & -- & -- & -- & -- & -- \\[3pt]%
	QCD($p_T$: 0.10-0.20 TeV) & 755.1 & -- & -- & -- & -- \\[3pt]%
	QCD($p_T$: 0.20-0.40 TeV) & 803.8 & 621.1 & 109.6 & 36.5 & --  \\[3pt]%
	QCD($p_T$: 0.40-1.00 TeV) & 209.8 & 304.7 & 72.6 & 29.0 & 2.6 \\[3pt]%
	QCD($p_T$: 1.00-2.40 TeV) & 2.2   & 5.3 & 1.7 & 1.5 & 0.2 \\[3pt]%
	$t\bar{t}$ & 1721.4 & 732.6 & 273.8 & $ 113.3 $ & $ 6.6$ \\[3pt]%
	$W+jets; W\to e,\mu,\tau$ & 527.4 & 22.6  & 8.4 & $ 1.3 $ 
	  & $ -- $ \\[3pt]%
	$Z+jets; Z\to \tau\bar{\tau},\ \nu s$ & 752.9 & 11.1 & 1.3 & $0.2$ 
	  & $ -- $ \\[3pt]%
	$WW,ZZ,WZ$ & 3.4 & 0.3 & $0.25$ & $ -- $ & $ -- $ \\[3pt]%
	&&&&&\\%
	&&&&&\\%
	$Summed\ SM\ BG$ & 4776.1 & 1697.8 & 467.7 & $181.9$ & $9.4$ \\
	&&&&&\\%
	inoAMSB1 & 112.7 & 85.7 & 27.6 & 36.0 & 7.5 \\
	&&&&&\\%
	\hline\hline%
    \end{tabular}
    \caption{Estimated SM background cross sections and the inoAMSB1 
    benchmark point in fb for various multi-lepton plus jets $+\etm$ 
    topologies after cuts C2 with $\etcut=100$ GeV.
    }%
    \label{tab:ino_c2cuts}%
  \end{center}%
\end{table}%
\indent%
Applying these criteria, the model's $\mhf$ reach after 100 $\invfb$ of data
collection is computed for each of the lepton channels and for each of 
$\etcut=100,300$ and 500 GeV.  Figure \ref{fig:ino_reach} shows the plots of the
cross section for the background (horizontal) and signal (decreasing slopes) as 
a function of $\mhf$. In the figure solid blue represents cuts with $\etcut=100$
GeV, red dash-dotted represents 300 GeV, and purple dashed represents 500 GeV.  
Where the signal touches its corresponding (5$\sigma$) background value on the 
plot is the where signal is no longer significant enough to be observable.  
These $\mhf$ values are listed in Table \ref{tab:ino_reach} (along with the SS 
dilepton reach).  It is seen there that the best reach at 100 $\invfb$ for the 
inoAMSB model (with $\tanb=10$ and $\mu>0$) is for multi-jet + $\etm$ 
+ {\bf 3l} which shows significant cross section up to $\mhf\sim 118$ TeV 
($\mg \sim 2.6$ TeV) when  $\etcut=$ 500 GeV.
\begin{table}
  \begin{center}
    \begin{tabular}{|c|ccccc|}
 	\hline\hline%
	&&&&&\\%
	$E_T^c$ (GeV) & $0\ell$ & $OS$ & $SS$ & $3\ell$ & $4\ell$  \\
	&&&&&\\%
	\hline
	&&&&&\\%
	100 & $40$ & $57$ &  62 & $60$ & $75$ \\[3pt]%
	300 & $93$ & $95$ &  85 & $98$ & $80$ \\[3pt]%
	500 & $110$ & $115$ & 105 & $118$ & $110$ \\[3pt]%
	&&&&&\\%
	\hline\hline%
    \end{tabular}
    \caption{Estimated reach of 100 fb$^{-1}$ LHC for $m_{3/2}$ (TeV) in 
      the inoAMSB  model line in various signal channels.
     }%
    \label{tab:ino_reach}
  \end{center}
\end{table}
\subsection{%
  HITs + multi-peak $m(l^{+}l^{-})$ Distribution%
  \label{subsec:ino_signature}%
}%
\begin{figure}[h!]%
  \begin{center}
    \includegraphics[width=1\textwidth,trim= 0cm 0cm 0cm 3.5cm,clip]{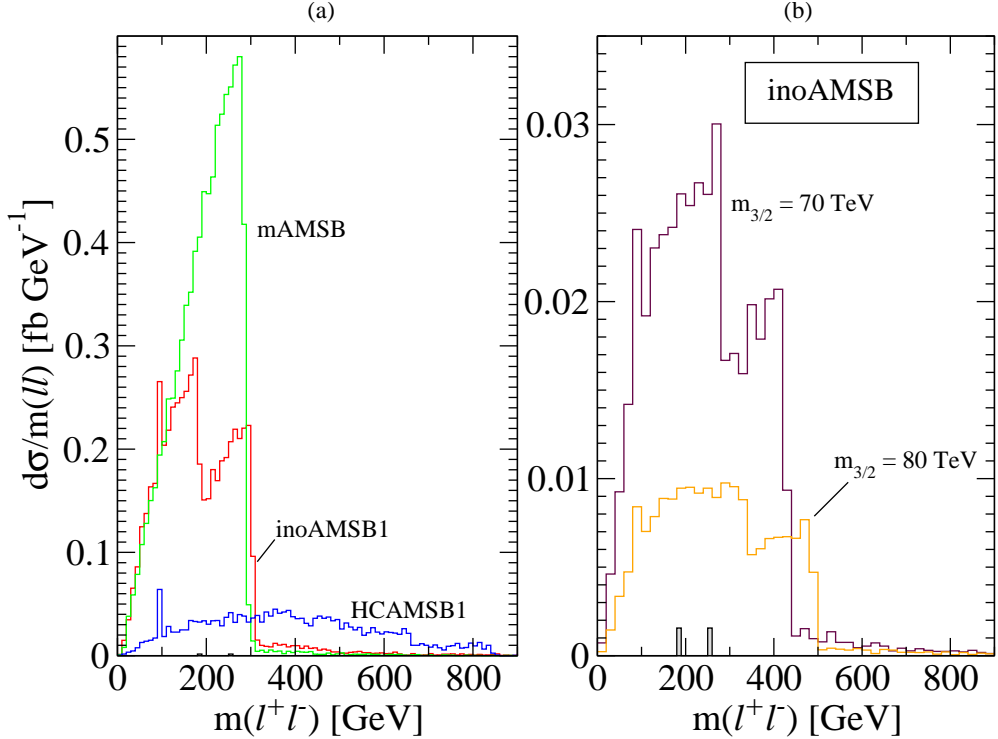}%
    \caption{Dilepton invariant mass distribution. (a) comparison of \ino
	to \mam and \hca.  C1 cuts are used with $\etm>300$ GeV and 
	$A_{T}>900 GeV$.  (b) the same distribution except $\mhf=70$ and 80 TeV 
	values.  The double-edge is still visible at these high $\mhf$ values.
    }%
    \label{fig:ino_mll}%
  \end{center}%
\end{figure}

We next examine the dilepton invariant mass distribution for cascade decay 
events for $\geq 2$ high $\pt$ jets, large $\etm$ and $SF/OS$ dileptons.  It is 
known that the $\mll$ distribution is capable of having a \quotes{kinematic 
mass edge} structure through the decay $\ninos{2}\rightarrow\sl^{\pm}l^{\mp}$ 
or through $\ninos{2}\rightarrow l^{\pm}l^{\mp}\nino$\cite{Paige:1999ui}.  
Because the \ino model predicts L/R splitting in the slepton masses, so it is 
expected that {\it two} edges can occur from the first type of decay through 
$\ninos{2}\rightarrow\sl_{L}^{\pm}l^{\mp}$ and 
$\ninos{2}\rightarrow\sl_{R}^{\pm}l^{\mp}$.\\%
\indent%
It is also expected that a peak can occur in the $\mll$ distribution around 
$M_{Z}$ because of the decays $\ninos{3}\rightarrow Z\nino$, 
$\ninos{4}\rightarrow Z\nino$ and $\cinos{2}\rightarrow Z\cino$.  In the case
of point inoAMSB1, these branching occur for 25\%, 6\%, and 29\% 
respectively.\\%
\indent%
Figure \ref{fig:ino_mll} frame (a) shows the $\mll$ distribution for inoAMSB1
(red line) along with \mam (green) and \hca (blue) points.  The cuts are C1 
with $\etm > 300$ GeV and $A_{T} >$ 900 GeV, and the SM backgrounds are 
completely suppressed.  The peak at $M_{Z}$ is clear and there is a $\ninos{2}$ 
decay double-edge structure calculated to be at values
\begin{align}%
\mll \leq m_{\ninos{2}}\sqrt{1-\frac{m_{\sl}^{2}}{m_{\ninos{2}}^{2}}}
\sqrt{1-\frac{m_{\nino}^{2}}{m_{\sl}^{2}}} = \textrm{182 GeV and 304 GeV.}%
\end{align}%
The former is for $\ninos{2}$ to $\sl_{R}$ decays where as the latter is for the
decays to left-sleptons.\\%
\indent
Unlike inoAMSB, the \mam plot exhibits only a single edge because \mam
points predict nearly-degenerate L and R sleptons.  \mam also lacks a Z-peak 
because the second heaviest neutralino, $\ninos{2}$, is bino-like and thus 
has suppressed couplings to $\nino Z$.  \hca does not have an edge
because here $\ninos{2}$ is higgsino-like and has different decay products: 
$\ninos{2}\rightarrow \cino^{\pm} W^{\mp}$, $\nino h$, and $\nino Z$.  Because 
the last  decay channel should always be open, \hca has a smooth $\mll$ 
distribution with a visible Z-resonance.\\%
\indent%
We conclude that \ino, \mam, and \hca models can be distinguished at the LHC.  
First, AMSB models in general are identified by a wino-like $\nino$ with near 
degeneracy to the lightest chargino.  This would be seen in gluino and squark
cascade decay events including a long-lived chargino with a terminating track 
(HIT signal).  Then the different $\mll$ distributions could be used to 
distinguish between the \ino, \hca, or \mam cases.  These results are 
summarized in Table \ref{tab:ino_signatures} for the reader's convenience.
\begin{table}%
  \begin{center}
    \begin{tabular}{|ccccc|}%
	\hline\hline%
	&&&&\\%
      Model   & HITs & \# of edges & $\ninos{2}$ 
	  & $m_{\sl_{L}} \stackrel{?}{\approx} m_{\sl_{R}}$\\%
	&&&&\\%
	\hline%
	&&&&\\%
	inoAMSB & Yes  & 2           & bino        & No \\[3pt]%
      mAMSB   & Yes  & 1           & bino        & Yes \\[3pt]%
      HCAMSB  & Yes  & 0           & higgsino    & No \\[3pt]%
	&&&&\\%
      \hline\hline%
    \end{tabular}
    \caption{%
	The three AMSB models can be distinguished in the dilepton invariant mass
	distribution by the number of edges.  The structures depend crucially on
	the type of $\ninos{2}$ and the slepton masses predicted by the models.
      \label{tab:ino_signatures}
    }%
  \end{center}
\end{table}
\section{%
  Summary
}%
In this chapter, we have examined the phenomenology of supersymmetric models 
with the boundary conditions $m_0\sim A_0\sim 0$ at $M_{string}$, while gaugino 
masses assume the form as given in AMSB.  We have labeled this class of 
boundary conditions gaugino-AMSB boundary conditions, or \ino for short. Such 
boundary conditions can arise in type IIB string models with flux 
compactifications. They are very compelling in that off-diagonal flavor 
violating and also $CP$ violating terms are highly suppressed, as in the case 
of no-scale supergravity or gaugino-mediated SUSY breaking models.  However, 
since gaugino masses assume the AMSB form at $M_{GUT}$, the large $U(1)_Y$ 
gaugino mass $M_1$ pulls slepton masses to large enough values through 
renormalization group evolution that one avoids charged LSPs (as in NS or 
inoMSB model) or tachyonic sleptons (as in pure AMSB models).\\%
\indent%
The expected sparticle mass spectrum is very distinctive. Like mAMSB and 
HCAMSB, it is expected that lightest neutralino ($\nino$) will be wino-like, and
a quasi-stable chargino ($\cino$) would leave observable highly ionizing tracks 
in collider detectors. The spectrum is unlike mAMSB in that a large mass 
splitting is expected between left- and right- sleptons. The case when the 
string scale, $M_{string}$, is much lower than $M_{GUT}$ was also 
investigated.  In this case, the entire spectrum becomes somewhat spread-out, 
and if $M_s\lsim 10^{14}$ GeV, then the left-sneutrino becomes the LSP, which 
is excluded by double beta decay experiments \cite{sneu_dm}.\\%
\indent%
We also saw in detail some aspects of LHC collider signatures.  Since 
$m_{\sq}<m_{\gl}$ in inoAMSB models, we expect dominant $\sq\sq$ and $\sq\gl$ 
production at LHC, followed by 2-body $\sq$ and $\gl$ decays. This leads to
collider events containing at least two very high $p_T$ jets plus $\etm$ as is
indicative of squark pair production.\\%
\indent%
While squark and gluino cascade decay events should be easily seen at the LHC 
(provided $\mhf\lsim 110$ TeV), the signal events should all contain visible 
HITs, which would point to a model with $m_{\cino}\simeq m_{\nino}$, as occurs 
in anomaly-mediation where $M_2<M_1,\ M_3$ at the weak scale.  We find a LHC 
reach, given 100 fb$^{-1}$ of integrated luminosity, out to $m_{3/2}\sim 118$ 
TeV, corresponding to a reach in $m_{\gl}$ of about 2.6 TeV.  We note here that 
if a signal is found at the outer edges of the reach limit, the signal will 
consist of typically just a few (5-10) events (due to hard cuts) over a small 
background. The signal events should include the characteristic presence of 
HITs, which should be absent in the background. As data accumulates, signals 
should also appear in the complementary channels, thus building confidence in a 
discovery.\\%
\indent%
We also find that the invariant mass distribution of SF/OS dilepton pairs 
should have a distinctive two-bump structure that is indicative of neutralino 
decays through both left- and right- sleptons with a large slepton mass 
splitting. This distribution would help distinguish inoAMSB models from HCAMSB, 
where a continuum plus a $Z$-peak distribution is expected, or from mAMSB, 
where the two mass edges (present only if $m_0$ is small enough that 
$m_{\spart{l}{L}}$ and $m_{\spart{l}{R}}$ are lighter than $m_{\ninos{2}}$) 
would be very close together, and probably not resolvable.

\chapter{%
  \label{chap:dm}
  Dark Matter in AMSB Models
}%

\section{Introduction to Dark Matter in AMSB Models}%
Now that we know what to expect for mAMSB, HCAMSB, and \ino models at the LHC, 
we can now turn our attention to using astrophysical methods to observe them.  
In particular we are interested in {\it i}) whether these models can account 
for the observed cold dark matter (CDM) abundance
\begin{equation}%
\ohmh = 0.1123 \pm 0.0035 \ \ 68\% CL
\label{equ:wmap7}%
\end{equation}%
according to WMAP7 \cite{Komatsu:2010fb} measurements, where 
$\Omega = \frac{\rho}{\rho_{c}}$, and $\rho_{c}=1.88\times 10^{-29} \ h^{2}$ 
${\rm g \ cm^{-3}}$ is the critical closure density; {\it ii}) if current or 
future experiments can either confirm or rule out regions of their parameter 
spaces;  and {\it iii}) how cosmological data compares with LHC data.\\%
\indent%
At first sight there appear to be problems with AMSB models arising from the 
fact that the LSPs annihilate and co-annihilate (with the lightest chargino)
too efficiently \cite{Chen:1996ap} and leave the relic density of thermally 
produced neutralinos 1--2 orders of magnitude below the experimental value in 
Equation \ref{equ:wmap7}.  Indeed, this is a major issue with AMSB models that 
has led many to consider other model types to be more interesting.  However, 
it is argued here that when one looks back into the history of the Universe it 
is important not only to consider thermal production of LSPs in quasi-thermal 
equilibrium, but it is also important to consider the impact of 
{\it non-thermal} production of LSPs: production of LSPs through the decay of 
heavy particles that may have existed early on.\\
\indent%
In this chapter we will explore several possibilities to increase the CDM 
abundance(s) that are strongly motivated in theory.  String theories require 
the presence of heavy moduli fields (and other heavy scalars) and local \susy 
theories require gravitinos, both of which decay through gravitational 
interactions and can add to $\ohmh$. We also consider here the case where the 
strong CP problem is solved by invoking the axion ($a$) solution.  When this 
solution is combined with the \susy framework, the axion's supersymmetric 
partner, the axino ($\tilde{a}$), appears in the spectrum 
\cite{Rajagopal:1990yx} and its mass is tied (model-dependently) to the \susy 
breaking sector.  This means that the axino can be heavy enough that its decay 
chains terminate with LSP production or it can even serve as the LSP itself and 
contribute, along with the axion, to the CDM abundance.\\
\indent%
Non-thermal production mechanisms must not only produce enough DM to account
for WMAP, but must also be careful not to disrupt Big Bang Nucleosynthesis 
(BBN) which includes measurements of light element abundances, baryon-entropy 
ratio, and the neutron-proton ratio.  With these requirements being met, the 
AMSB models' parameters can be further constrained and the reaches will be 
shown for direct and indirect detection experiments.\\%
\indent%
This chapter is organized as follows.  In Section \ref{sec:ntpamsb}, four 
non-thermal production mechanisms of dark matter will be described.  The first 
two result in an increase of {\it wino}-neutralino DM as a result of moduli 
decay and (thermally-produced) gravitino decay.  The remaining mechanisms
will invoke the Peccei-Quinn (PQ) solution to the strong CP problem.  The third
mechanism will increase the wino DM through heavy axino decays and there will 
be an additional component of axion ({\it a}) DM from the breaking of the PQ 
symmetry.  In the fourth mechanism we take the axino ($\tilde{a}$) to be the 
LSP such that it is produced both thermally and in {\it neutralino} decays.  
This last case will have a mixture of axion and axino CDM.\\%
\indent%
Dark matter may be observable in future experiments through direct 
detection (DD) where the relic DM particles scatter from target nuclei in the 
experimental apparatus.  It is also possible that it will be observed through 
indirect detection (ID) where DM annihilate into SM particles somewhere else in 
the galaxy and we see its products.  A brief description of CDM experiments is 
given Section \ref{sec:dmexp}.\\%
\indent%
Sections \ref{sec:ddrates} and \ref{sec:idrates} contain the results of this 
research.  In Section \ref{sec:ddrates}, the calculated DD cross sections for 
AMSB wino DM are given.  In Section \ref{sec:idrates}, the ID rates for mAMSB, 
HCAMSB, and \ino models are also given.  The discussions in these sections 
describe the interactions that lead to the calculated rates and explore the 
current experimental exclusions and future experimental reaches of the 
parameter space.  Finally, we close this chapter in Section 
\ref{sec:dmsummary} with a summary and conclusion.\\%

\section{%
  \label{sec:ntpamsb}
  Thermal and Non-thermal Production of Dark Matter in AMSB Models
}%
As described in the Introduction of this chapter, dark matter (DM) production 
in the early universe can happen both thermally and non-thermally.  In this 
section we explore four cases where the abundance of DM is composed of thermal 
and non-thermal components.  In addition, we will see in these cases the DM can 
be wino (as for our AMSB models), axion, axino, or mixtures of these.
\subsection{Neutralino CDM via Moduli Decay}%
Shortly after the introduction of AMSB models, Moroi and Randall proposed a 
solution to the AMSB dark matter problem based on augmented neutralino 
production via the decays  of moduli fields in the early universe
\cite{Moroi:1999zb}. The idea here is that string theory is replete with 
additional moduli fields: neutral scalar fields with gravitational couplings to 
matter. In generic supergravity theories, the moduli fields are expected to 
have masses comparable to $m_{3/2}$.  When the Hubble expansion rate becomes 
comparable to the moduli mass $m_\phi$, then an effective potential will turn 
on, and the moduli field(s) will oscillate about their minima, producing 
massive excitations, which will then decay to all allowed modes: {\it e.g.} 
gauge boson pairs, higgs boson pairs, gravitino pairs, $\cdots$. The neutralino 
production rate via moduli decay has been estimated in Ref. \cite{Moroi:1999zb}.
It is noted in Ref. \cite{Acharya:2009zt} that the abundance-- given by
\begin{equation}%
  \Omega_{\nino}^{mod.}h^2\sim 0.1\times\left(\frac{m_{\nino}}{100\ 
    {\rm GeV}}\right)
    \left(\frac{10.75}{g_*}\right)^{1/4}\left(\frac{\sigma_0}{\langle\sigma v  
    \rangle}\right)
    \left(\frac{100\ {\rm TeV}}{m_\phi}\right)^{3/2}
\end{equation}%
with $\sigma_0=3\times 10^{-24}$ cm$^3$/sec -- yields nearly the measured dark 
matter abundance for wino-like neutralino annihilation cross sections and 
$m_{\phi}\sim 100$ TeV.\footnote{In inoAMSB models, we expect moduli with SUSY 
breaking scale masses, $m_\phi\sim m_{3/2}/\sqrt{V}\ll m_{3/2}$, where $V$ is 
the (large) volume of the compactified manifold: $V\sim 10^5$ in Planck units.
In this case, the mechanism would not so easily apply.}  These authors dub this 
the ``non-thermal WIMP miracle''.\\%
\indent%
A necessary condition for augmented neutralino production via scalar field 
decay is that the re-heat temperature of radiation $T_R$ induced by moduli 
decays is bounded by $T_R\gsim 5$ MeV (in order to sustain Big Bang 
Nucleosynthesis (BBN) as we know it), and $T_R< T_{fo}$, where $T_{fo}$ is the 
freeze-out temperature for thermal neutralino production 
$T_{fo}\sim m_{\nino}/20$.  If $T_R$ exceeds $T_{fo}$, then the decay-produced 
neutralinos will thermalize, and the abundance will be given by the thermal 
calculation as usual.\\%
\indent%
This ``low re-heat'' neutralino production mechanism has been investigated 
extensively by Gondolo, Gelmini, et. al. 
\cite{Gelmini:2006pw}\cite{Gelmini:2006pq}\cite{Gelmini:2006mr}. The low 
re-heat neutralino abundance calculation depends on the input value of $T_R$ and 
the ratio $b/m_\phi$, where $b$ is the average number of neutralinos produced 
in moduli decay, and $m_\phi$ is the scalar field mass.  They note that 
theories with an under-abundance of thermally produced neutralino CDM with 
$\Omega_{\nino}^{TP}\gsim 10^{-5}\left(\frac{100\ {\rm GeV}}{m_{\nino}}\right)$ 
can always be brought into accord with the measured DM abundance for at least 
one and sometimes two values of $T_R$.\footnote{References 
\cite{Gelmini:2006mr}\cite{Gelmini:2006pq}\cite{Gelmini:2006pw}
also shows that an overabundance of thermally produced neutralino CDM can also 
be brought into accord with the measured abundance via dilution of the 
neutralino number density by entropy injection from the $\phi$ field decay. 
Since this case does not attain in AMSB models (unless $m_{\nino}\gsim 1300$ GeV), we will neglect it here.}\\%
\indent%
While the low $T_R\sim 10-1000$ MeV scenario with DM generation via scalar 
field decay is compelling, we note here that it is also consistent with some 
baryogenesis mechanisms: {\it e.g.} Affleck-Dine baryogenesis wherein a large 
baryon asymmetry is generated early on, only to be diluted to observable 
levels via moduli decay \cite{Kawasaki:2007yy}, or a scenario wherein the 
baryon asymmetry is actually generated by the moduli decay 
\cite{Kitano:2008tk}.%
\subsection{Neutralino CDM via Gravitino Decay}%
\begin{figure}[htbp]%
  \begin{center}%
    \includegraphics[angle=0,width=0.75\textwidth]{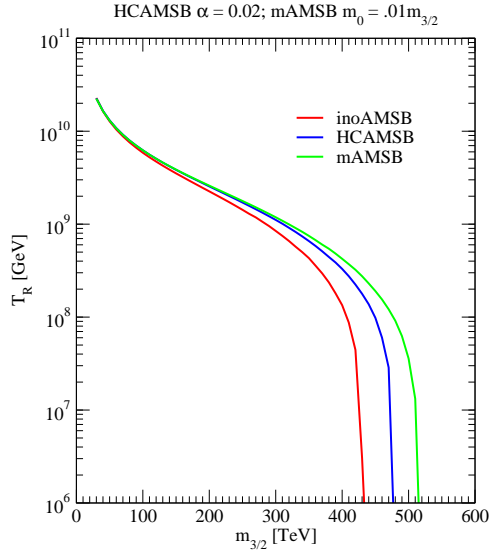}%
    \vspace{-.5cm}%
    \caption{%
      Plot of allowed region of $T_R\ vs.\ m_{3/2}$ plane allowed for wino-like 
      neutralino DM from thermal production plus thermally produced gravitino   
      decay.  For HCAMSB $\alpha=0.02$ and for mAMSB $\mnot=0.01 \ \mhf$.
    }%
    \label{fig:tr_m32}%
  \end{center}%
\end{figure}

An alternative possibility for augmenting the production of wino-like 
neutralinos in AMSB models is via gravitino production and decay in the early 
universe. While gravitinos would not be in thermal equilibrium during or after 
re-heat, they still can be produced thermally via radiation off ordinary
sparticle scattering reactions in the early universe. The relic density of 
thermally produced gravitinos as calculated in Ref's \cite{Pradler:2006qh}
\cite{Bolz:2000fu} is given by
\begin{equation}%
  \Omega_{\tilde{G}}^{TP}h^2=\sum_{i=1}^{3}\omega_ig_i^2\left(1
  +\frac{M_i^2}{3m_{3/2}^2}\right)\log\left(\frac{k_i}{g_i}\right)
  \left(\frac{m_{3/2}}{100\ {\rm GeV}}\right)\left(\frac{T_R}{10^{10}\ 
  {\rm GeV}}\right),
\end{equation}%
where $g_i$ and $M_i$ are the gauge couplings and gaugino masses evaluated at 
scale $Q=T_R$, and $\omega_i=(0.018,0.044,0.117)$ and $k_i=(1.266,1.312,1.271)$.
Each gravitino ultimately cascade decays down to the wino-like $\nino$ state, 
so the neutralino relic density is given by
\begin{equation}%
  \Omega_{\nino}h^2 =\Omega_{\nino}^{TP}h^2 
  +\frac{m_{\nino}}{m_{3/2}}\Omega_{\tilde{G}}^{TP}h^2 .
\label{eq:gravo}
\end{equation}%
A plot of the value of $T_R$ and $m_{3/2}$ which is required to yield 
$\Omega_{\nino}h^2=0.11$ 
from Equation \ref{eq:gravo} is shown in Figure \ref{fig:tr_m32} for mAMSB 
($m_0=0.01m_{3/2}$), HCAMSB ($\alpha=0.02$) and inoAMSB using $\tan\beta =10$ 
and $\mu >0$.  The region above the  $\Omega_{\nino}h^2=0.11$ curves would 
yield too much dark matter, while the region below the curves yields too little.
We should consider the curves shown in Figure \ref{fig:tr_m32} as only 
indicative of the simplest scenario for wino production via gravitino decay. 
Three other effects can substantially change the above picture from what is 
presented in Equation \ref{eq:gravo}. 
\begin{itemize}
  \item On the one hand, if moduli fields $\phi_m$ exist with mass 
        $m_{\phi_m}>2m_{3/2}$, then gravitinos can also be produced via moduli 
        production and decay \cite{Kohri:2005ru}\cite{Asaka:2006bv}. The exact 
        abundance of these moduli-produced gravitinos is very model dependent, 
        and depends on the moduli and gravitino mass and branching fractions.
  \item A second case arises if we consider gravitino production via inflaton 
        decay at the end of inflation \cite{Endo:2007sz}.  This production 
        mechanism depends on unknown properties of the inflaton: {\it e.g.} its 
        mass and branching fractions, and the re-heat temperature generated by 
        inflaton decay.  These latter quantities are very model dependent.
  \item Additional entropy production generated via the inflaton, moduli and 
        gravitino decays may also dilute the above relic abundance in Equation 
        \ref{eq:gravo}.
\end{itemize}
We will bear in mind that these possibilities permit much lower or much higher 
values of $T_R$ and $m_{3/2}$ than those shown by the $\Omega_{\nino}h^2=0.1$ 
contour of Figure \ref{fig:tr_m32}.
\subsection{%
  Neutralino CDM from Heavy Axino Decays Mixed with Axion CDM
}%
A third mechanism for increasing the wino-like relic abundance is presented in 
Ref. \cite{Choi:2008zq}, in the context of the PQMSSM. If we adopt the Peccei-
Quinn (PQ) solution to the strong $CP$ problem within the context of 
supersymmetric models, then it is appropriate to work with the PQ-augmented 
MSSM, which contains in addition to the usual MSSM states, the  axion $a$, the 
$R$-parity even saxion field $s$, and the spin-${1\over 2}$ $R$-parity odd 
axino $\axino$. The axino can serve as the lightest SUSY particle if it is 
lighter than the lightest $R$-odd MSSM particle. The $a$ and $\axino$ have 
couplings to matter which are suppressed by the value of the PQ breaking scale 
$f_a$, usually considered to be in the range 
$10^9\ {\rm GeV}\lsim f_a\lsim 10^{12}$ GeV\cite{axion_review}.\\%
\indent%
In Ref. \cite{Choi:2008zq}, it is assumed that $m_{\axino}>m_{\nino}$, where 
$\nino$ is the LSP. In the AMSB scenarios considered here, we will assume 
$T_R\lsim 10^{10}$ GeV, so as to avoid overproduction of dark matter via 
gravitinos.  With these low values of $T_R$, we are also below the axino 
decoupling temperature $T_{\axino -dcp}=10^{11}\ {\rm GeV}\left(
\frac{f_a}{10^{12}\ {\rm GeV}}\right)^2\left(\frac{0.1}{\alpha_s}\right)^3$, 
so the axinos are never considered at thermal equilibrium
\cite{Rajagopal:1990yx}.  However, axinos can still be produced thermally via 
radiation off usual MSSM scattering processes at high temperatures. The 
calculation of the thermally produced axino abundance, from the hard thermal 
loop approximation, yields \cite{Brandenburg:2004du}
\begin{equation}%
  \Omega_{\axino}^{TP}=h^2\simeq 5.5 g_s^6\ln\left(\frac{1.211}{g_s}\right)
    \left(\frac{10^{11}\ {\rm GeV}}{f_a/N}\right)^2
    \left(\frac{m_{\axino}}{0.1\ {\rm GeV}}\right)
    \left(\frac{T_R}{10^4\ {\rm GeV}}\right)
  \label{eq:Oh2_TP}
\end{equation}%
where $g_s$ is the strong coupling evaluated at $Q=T_R$ and $N$ is the
model dependent color anomaly of the PQ symmetry, of order 1.
Since these axinos are assumed quite heavy, they will decay to $g\gl$ or 
$\ninos{i}\gamma$ modes, which further decay until the stable LSP state, 
assumed here to be the neutral wino, is reached.\\%
\indent%
If the temperature of radiation due to axino decay ($T_D$) exceeds the 
neutralino freeze-out temperature $T_{fo}$, then the thermal wino abundance is 
unaffected by axino decay.  If $T_D<T_{fo}$, then the axino decay will 
{\it add} to the neutralino abundance. However, this situation breaks up into 
two possibilities: {\it i}). a case wherein the axinos can dominate the energy 
density of the universe, wherein extra entropy production from heavy axino 
decay may dilute the thermal abundance of the wino-like LSPs, and {\it ii}). a 
case where they don't.  In addition, if the yield of winos from axino decay is 
high enough, then additional annihilation of winos after axino decay may occur; 
this case is handled by explicit solution of the Boltzmann equation for the 
wino number density. Along with a component of wino-like neutralino CDM, there 
will of course be some component of vacuum mis-alignment produced axion CDM: 
thus, in this scenario, we expect a WIMP/axion mixture of CDM.
\subsection{%
  Mixed Axion/Axino CDM%
}%
\begin{figure}[htbp]
  \begin{center}
    \includegraphics[angle=0,width=0.75\textwidth]{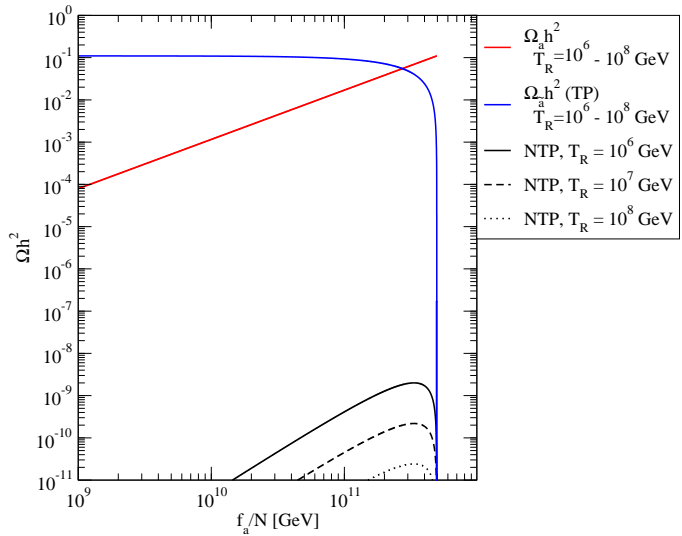}
    \vspace{-.5cm}
    \caption{Abundance of TP and  NTP axino DM and vacuum-misalignment
      production of axion CDM versus $f_a/N$, for various values of $T_R$.
     }
    \label{fig:oh2_ata}
  \end{center}
\end{figure}

\begin{figure}[htbp]
  \begin{center}
    \includegraphics[angle=-90,width=0.75\textwidth]{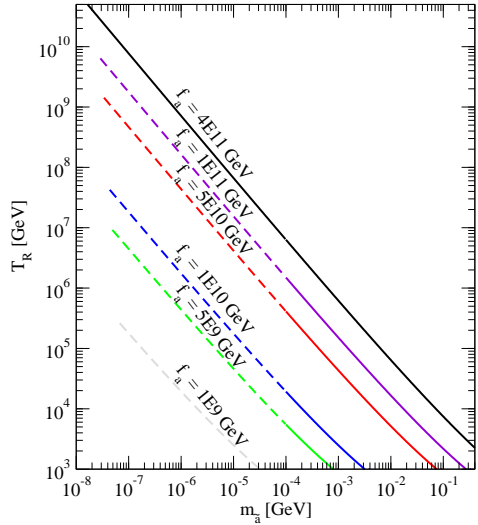}
    \vspace{-.5cm}
    \caption{%
      Plot of $T_R$ needed to ensure $\Omega_{a\axino}h^2=0.1$ for HCAMSB 
      benchmark Pt. 1, versus $m_{\axino}$ for various values of the PQ 
      breaking scale $f_a$.  The dashed curves yield mainly warm axino DM, 
      while solid curves yield mainly cold mixed axion/axino DM. 
    }%
    \label{fig:trvsmax}
  \end{center}
\end{figure}

In this case, we again consider the PQMSSM, as in the last subsection. But now, 
we consider a light axino with $m_{\axino}<m_{\nino}$, so that $\axino$ is the 
stable LSP \cite{Covi:1999ty}.  Here, the thermally produced wino-like 
neutralinos will decay via $\nino\to \axino\gamma$, so we will obtain a very 
slight dark matter abundance from neutralino decay:
$\Omega_{\axino}^{NTP}=\frac{m_{\axino}}{m_{\nino}}\Omega_{\nino}h^2$, since 
each thermally produced neutralino gives rise to one non-thermally produced 
(NTP) axino. We will also produce axinos thermally via Equation 
\ref{eq:Oh2_TP}. Finally, we will also produce axion CDM via the vacuum mis-
alignment mechanism\cite{absik}:
$\Omega_a h^2\simeq \frac{1}{4}\left(\frac{f_a/N}{10^{12} \ 
{\rm GeV}}\right)^{7/6}\theta_i^2$ (we will take here the initial mis-alignment 
angle $\theta_i\simeq 1$). The entire CDM abundance is then the sum 
\begin{equation}%
  \Omega_{a\axino}h^2=\Omega_{\axino}^{NTP}h^2
    +\Omega_{\axino}^{TP}h^2+\Omega_{a}h^2 .
\end{equation}%
In this case, the TP axinos constitute CDM as long as $m_{\axino}\gsim 0.1$ 
MeV. The NTP axinos constitute warm DM for $m_{\axino}\lsim 1$ GeV
\cite{Jedamzik:2005sx}, but since their abundance is tiny, this fact is largely 
irrelevant. The entire CDM abundance then depends on the parameters 
$f_a$, $m_{\axino}$ and $T_R$; it also depends extremely weakly on 
$\Omega_{\nino}h^2$, since this is usually small in AMSB models.\\%
\indent%
As an example, we plot in Fig. \ref{fig:oh2_ata} the three components of mixed 
axion/axino DM abundance from HCAMSB benchmark point 1 in Reference 
\cite{Baer:2009wz}: $\alpha =0.025$, $m_{3/2}=50$ TeV, $\tan\beta =10$ and 
$\mu >0$. The neutralino thermal DM abundance would be 
$\Omega_{\nino}h^2=0.0015$ if the $\nino$ was stable.  We require instead 
$\Omega_{a\axino}h^2 =0.11$, and plot the three components of 
$\Omega_{a\axino}h^2$ versus $f_a/N$, for three values of $T_R=10^6$, $10^7$ 
and $10^8$ GeV. The value of $m_{\axino}$ is determined by the constraint 
$\Omega_{a\axino}h^2=0.11$. We see that at low values of $f_a/N$, the NTP axino 
abundance is indeed tiny. Also the axion abundance is tiny since the assumed 
initial axion field strength is low. The TP axino abundance dominates. As 
$f_a/N$ increases, the axion abundance increases, taking an ever greater share 
of the measured DM abundance. The TP axino abundance drops with increasing 
$f_a/N$, since the effective axino coupling constant is decreasing. Around 
$f_a/N\sim 3\times 10^{11}$ GeV, the axion abundance becomes dominant. It is in 
this range that ADMX\cite{admx} would stand a good chance of measuring an axion 
signal using their microwave cavity experiment.\\%
In Figure \ref{fig:trvsmax}, we again require $\Omega_{a\axino}h^2=0.11$ for 
HCAMSB benchmark point 1, but this time plot the value of $T_R$ which is needed 
versus $m_{\axino}$, for various values of $f_a/N$. The plots terminate at high 
$T_R$ in order to avoid reaching the axion decoupling temperature $T_{a-dcp}$. 
Dashed curves indicate regions where over 50\% of the DM is warm, instead of 
cold. Solid curves yield the bulk of DM as being cold.\\%
\indent%
We see that for very light axino masses, and large values of $f_a$, the value 
of $T_R$ easily reaches beyond $10^6$ GeV, while maintaining the bulk of dark 
matter as cold.  Such high values of $T_R$ are good enough to sustain
baryogenesis via non-thermal leptogenesis\cite{ntlepto}, although thermal 
leptogenesis requires $T_R\gsim 10^{10}$ GeV\cite{Buchmuller:2002rq}. Since 
$f_a$ is quite large, we would expect that the dominant portion of DM is 
composed of relic axions, rather than axinos; as such, detection of the relic 
axions may be possible at ADMX\cite{admx}.  While Figures \ref{fig:oh2_ata} and 
\ref{fig:trvsmax} were created for the HCAMSB model, quite similar results are 
obtained for the mAMSB or inoAMSB models.
\section{%
  \label{sec:dmexp}%
  Dark Matter Detection
}%
In this section a brief description of Dark Matter experiments are given. These 
break down into two categories: direct detection (DD) and indirect detection 
(ID) and each are described separately.
\subsection{Direct Detection}%
Direct detections experiments are those that are intended to measure properties
of Weakly Interacting Massive Particles (WIMPs) via their direct elastic 
scatterings off of nuclei.  This is in contrast to the ID experiments (described
in the next subsection) where WIMP interactions lead to different final state 
particles, and these products are observed and analyzed.\\%
\indent%
DD experiments are considered to be extremely important in WIMP searches for a 
number of reasons.  Although a neutralino mass should in principle be 
re-constructable through LHC data, its interactions are much more difficult to 
decipher as a result of its inertness in the detector.  If there are WIMPS with 
weak interactions, DD experiments should be able to probe the couplings of these
particles.  Furthermore, $\etm$ signatures in a collider experiment do not give
us information on the stability of particles on cosmological time scales, but 
only on collider time scales.  Thus only DD experiments can tell us whether a
particle produced in collisions is \quotes{absolutely} stable.  And finally, DD 
experiments do not have the serious systematic uncertainties that ID 
experiments have.  This is because the WIMP properties are measured directly, 
and because the backgrounds to these experiments are well understood and, in 
principle, can be controlled.  The largest uncertainties in direct detection 
experiments comes from the observed local density of WIMPs in the vicinity of 
the earth and their velocity distributions.  Increased local density of DM 
through clumping, etc. would have an impact on the rate of interactions in DD 
experiments.\\
\indent%
Dark matter must be cold and the calculations of wimp-nucleon scattering cross 
sections must be made in the zero momentum transfer limit.  In \susy 
theories, the interactions of neutralinos with quarks in the proton are 
dominated by t-channel CP even Higgs exchange\footnote{s-channel squark 
exchange is also possible, but suppressed.}\cite{Hooper:2008sn}.  
This is because the relevant vertex is gaugino-higgsino-Higgs
\cite{Baer:2006rs}.  Thus, purely gaugino or purely higgsino neutralinos will 
have suppressed couplings to nuclei\cite{Hooper:2008sn}.\\%
\indent%
The total elastic scattering cross section for WIMP-nucleon has contributions 
from a part that is spin-independent and from one that is spin-dependent.  
Because the WIMP de Broglie wavelength is expected to be of the order the 
dimension of heavy atoms, the WIMP will scatter from the coherent composite of 
nucleons.  Thus the spin-independent cross section will be enhanced by factors 
of $A^{2}$, where A is the atomic number of the target nucleus.  The spin-
dependent contribution to the cross section only occurs through spin 
interactions and scales with $J(J+1)$.\\%
\indent%
The best limits on the spin-dependent cross section come from the Cryogenic 
Dark Matter Search (CDMS: Ge/Si detectors) experiment which exclude WIMPs with 
weak scale masses and cross sections at the level of $5\times10^{-8}$ pb 
level.  We will compare our model to the spin-independent limits set by CDMS 
and other experiments in the text.  We will also compare spin-dependent cross 
sections as well, but we should note that the constraints placed by spin-
dependent DD experiments however are not nearly as strong as for spin-
independent cross sections.\\%
\indent%
The results of the XENON experiments (liquid Xenon detectors) are expected to 
surpass limits set by its competitors due to its size.  XENON has already taken 
DM searches from the 10 to 100 kg-scale detectors.  Impressively, future 
XENON experiments will be at the ton-scale with SI reaches in the 
$10^{-10}-10^{-11}$ pb range. 
\subsection{Indirect Detection}%
Indirect detection (ID) observations are complementary to direct detection (DD) 
of DM, and there is a wide variety of ID programs that seek annihilation 
products of DM particles from displaced sources.  Any locus of strong 
gravitation potential is capable of squeezing WIMPs close enough to interact 
highly with one another.  Examples include the galactic center, dwarf 
spheroidal galaxies, the Sun, the core of the Earth, and inhomogeneities in 
WIMP halo profiles.  However, different sources will provide different 
information about DM depending on the strength and distance of the source, the 
distribution of DM around the source, and on the mass and coupling of the 
WIMPs. A few general descriptions of potential ID signals are described in this 
subsection\footnote{see \cite{Baer:2003bp}\cite{Hooper:2008sn} for reviews.}
\begin{itemize}%
\item[] {\bf Gamma Rays}
\end{itemize}%
The galactic center of the Milky Way and dwarf spheroidal galaxies in or near 
the Milky Way are the most capable of gathering DM because of their large 
gravitation pull.  But these sources are relatively far away and so evidence of
WIMP interactions must arrive at Earth unattenuated and undeflected by large 
magnetic field variations in the galaxy.  Because gamma rays do not carry 
charge and travel in straight lines they retain their spectral information and 
hence are excellent probes of interactions at far distances.\\%
\indent%
The gamma ray fluxes produced by annihilations of DM for any structure and at 
any distance was performed in \cite{Bergstrom:2001jj} and analyzed in
\cite{Taylor:2002zd}.  The expression for the flux of gamma rays is
\begin{equation}
  \Phi_{\gamma}(E_{\gamma}) \approx 2.8 \times 10^{-12} \rm{cm^{-2}s^{-1}}
    \frac{dN_{\gamma}}{dE_{\gamma}}\biggl(
    \frac{<\sigma v>}{3\times10^{-26}cm^{2}/s}\biggr)
    \biggl(\frac{1 \ \rm{TeV}}{m_{\chi}}\biggr)^{2}
    J(\Delta\Omega,\psi)\Delta\Omega,
\end{equation}
where $\Delta\Omega$ is the solid angle observed and the interaction cross 
section appears explicitly and $\frac{dN_{\gamma}}{dE_{\gamma}}$ is the 
spectrum.  $J(\Delta\Omega,\psi)$ depends only on the 
DM distribution and is averaged over the solid angle to give
\begin{equation}%
  J(\psi) = \frac{1}{8.5 \rm{kpc}}\biggl(\frac{1}{.3 \rm{GeV/cm^{3}}}\biggr)^2
    \int_{los}\rho^{2}(r(l,\psi))dl,
\end{equation}%
where the DM density, $\rho$, appears quadratically, $\psi$ is the angle from 
the galactic center, and the integral takes place over the line of sight 
(los).\\%
\indent%
The spectrum for standard astrophysical sources is generally expected to have 
power law behavior \cite{Bertone:2007ki}, i.e. 
$\frac{dN_{\gamma}}{dE_{\gamma}} \propto E_{\gamma}^{-\alpha}$.  The HESS
collaboration has found a high energy source of gamma rays near the galactic 
center \cite{Aharonian:2004wa} that exhibits the power law behavior with 
$\alpha = 2.2 \pm 0.09 \pm 0.05$ above the threshold 165 GeV.  It is difficult 
to pick out a signal of DM annihilations from this background, but it is still 
an interesting target for Fermi-LAT that explores $E_{\gamma} > 1$ GeV, which 
is well below the high HESS energy threshold.%
\begin{itemize}%
\item[] {\bf Antimatter}
\end{itemize}%
\noindent Charged matter including positrons ($e^{+}$), antiprotons 
($\bar{p}$), and antideuterons ($\bar{D}$) are also interesting probes of DM 
and are produced in WIMP pair-annihilations in the halo.  In the case of $wino$
DM pairs will annihilate most efficiently to combinations of Higgs and gauge 
bosons, whose further decays lead to antimatter \cite{Hooper:2008sn}.  Of 
these products, cosmic positrons are the most interesting.  Because positrons 
lose the majority their energy over kiloparsec lengths, they only probe the 
local DM (in sharp contrast to gamma rays).  Since DM densities are better 
understood locally than, say, at the galactic center, positron measurements are
subject to fewer systematic uncertainties than other antimatter species.\\%
\indent%
Cosmic ray protons interacting with our galaxy's interstellar gas would provide 
the largest background sources of $\Dbar$ and $\pbar$ fluxes.  These types of 
reactions rarely produce low-energy particles, much less nucleons with energies 
small enough to match the binding energy of $\Dbar$ (2.2 MeV) 
\cite{Donato:1999gy}.  $\Dbar$ production is indeed very rare and the 
background of these particles is highly suppressed at low energies, with 
estimated flux of $\Phi_{\Dbar}\sim 5\times 10^{-13}$ 
[GeV cm$^{2}$ s sr]$^{-1}$.  For the future GAPS\cite{Mori:2001dv} 
experiment, sensitive to fluxes above $1.5 \times 10^{-11}$ 
[GeV cm$^{2}$ s sr]$^{-1}$, the background is essentially absent 
\cite{Cui:2010ud}.  An anomalous amount of detected low-energy $\Dbar$ could be 
a signature of neutralino annihilations in the halo, because the hadronization 
of WIMP hadronic annihilation channels ($\nino \nino \rightarrow q\bar{q}$) are 
confined by color to occur in the rest frame of the halo.  Vector boson and 
Higgs channels can also inject significant amounts of $\Dbar$ for light Higgs 
and lower values of WIMP mass \cite{Cui:2010ud}, especially near the $\Dbar$ 
threshold.\\
\indent%
Antiproton production is not rare, in contrast to $\Dbar$, but it also would 
not be a clean signal of WIMP annihilation at low energies.  This is because, 
of the two, $\pbar$ is more susceptible to ionization losses, synchrotron 
radiation, and solar modulation, allowing it to eventually populate the lower 
energies \cite{Donato:1999gy}.
\begin{itemize}%
\item[] {\bf Neutrinos}
\end{itemize}%
WIMPs are thought to become captured by elastically scattering from hydrogen 
and helium in the Sun and dissipating energy.  High-energy neutrinos may be 
produced in WIMP annihilations in the core of the Sun.  The $\nu_{\mu}$s can 
then undergo $\nu_{\mu} + q \rightarrow q' + \mu$ scattering to produce muons 
in terrestrial ice or water.  Thus, the detection of muons on Earth from the 
WIMP annihilation to neutrinos depends critically on the Sun capture rate of 
WIMPs, which further depends on the WIMP's nucleon cross section, its density 
near the Sun, and its mass.\\%
\indent%
The the capture rate depends on the effective WIMP-nucleus cross section which 
is%
\begin{equation}%
  \sigma_{eff} = \sigma_{H}^{SD} + \sigma_{H}^{SI} + 0.07\sigma_{He}^{SI},
\end{equation}%
where the factor 0.07 reflects the relative abundance if He to H along with 
other dynamical factors \cite{Hooper:2008sn}.  An important point here is that 
while spin-independent WIMP-nucleus cross sections tend to be highly 
constrained by DD experiments, SD cross sections are not so constrained and can 
be conceivable much higher. Thus ID experiments such as ICECUBE should be able 
to probe much lower values of the spin-dependent cross section than DD 
experiments such as COUPP.
\section{%
  \label{sec:ddrates}
  Direct Detection Rates for the AMSB Models
}%
\begin{figure}[t]
  \begin{center}
    \includegraphics[width=1.0\textwidth]{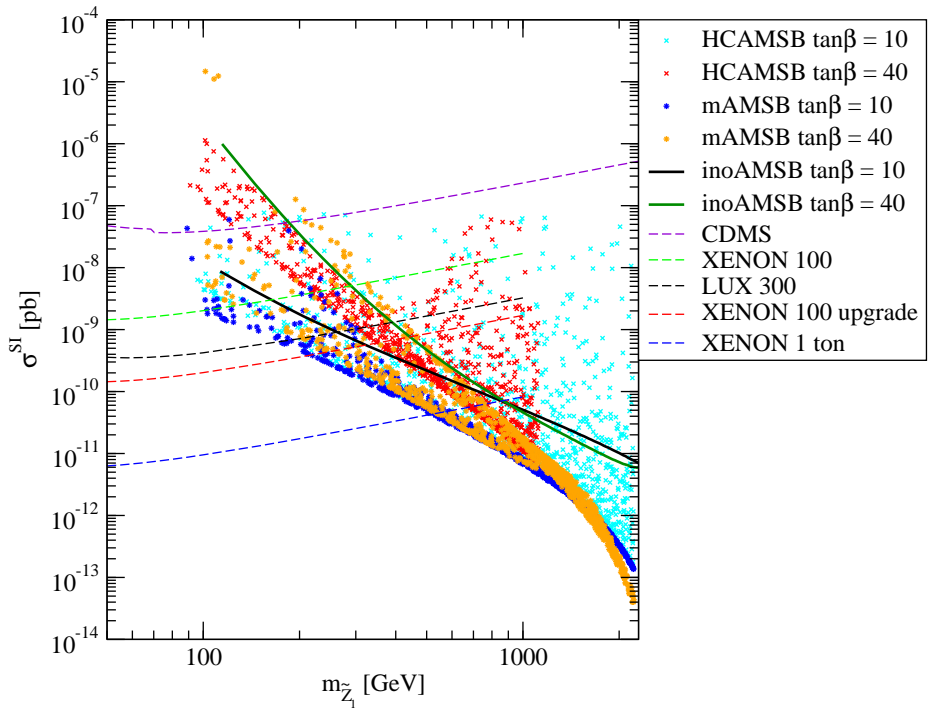}
    \caption{%
      Spin-independent $\nino-p$ scattering cross section versus $m_{\nino}$ 
	for mAMSB, HCAMSB and inoAMSB models for $\tanb =10$ and 40 and 
	$\mu >0$. The parameters $m_{3/2}$ and also $m_0$ (for mAMSB) and  
	$\alpha$ (for HCAMSB) have been scanned over.  We also show the CDMS 	
	limit and projected Xenon and LUX sensitivities.
    }%
    \label{fig:si_xs}
  \end{center}
\end{figure}

The ability to directly detect WIMPs depends not only on the mass, 
interactions, and the local density of the candidate, but also on the local 
velocity distribution.  In the standard cosmological scenario WIMPs are 
thermally produced in an expanding universe, they freeze out, and then fall 
into gravitational wells.  It is generally assumed, at present, that WIMPs 
conform to a Maxwell-Boltzmann velocity distribution, $i.e.$, 
$\ f(v)\sim v^{2}e^{-v^{2}/v_{0}^{2}}$, where $v_{0}^{2} \sim 220$ km/s is the 
Sun's velocity about the galactic center.  The same distribution holds for our 
cases, even despite the fact that $non-thermal$ production of the DM candidates 
(moduli, gravitino, axino decays) can distort the initial velocity 
distributions of the WIMPs.  This is because the original velocity distributions
are red-shifted away, and the Maxwell-Boltzmann distributions arise as usual 
from gravitational in-fall.  Thus, the calculations or WIMP-target 
scattering cross sections for direct detection can be carried out as usual.\\%
\indent%
As described in Section \ref{sec:dmexp}, WIMP scattering cross sections from 
target nuclei are described through spin-independent (SI) and spin-dependent 
(SD) parts.  Because SI cross sections receive enhancement from the mass of 
heavy targets DD will have better constraints of SI interactions.  The SI 
WIMP-nucleon cross sections are calculated using  the Isatools subroutine 
IsaReS \cite{Baer:2003jb}.\\%
\indent%
In Figure \ref{fig:si_xs} we scan over all parameters of the mAMSB, HCASMB, and
\ino parameter spaces, with a low and high value of $\tanb$ and $\mu > 0$, and 
plot the $\nino-p$ scattering cross section versus $m_{\nino}$.  Only points 
with $m_{\cino} > 91.9$ GeV are retained, as required by LEP2 
\cite{LEPSUSYWG:2002}.  The \ino points appear as lines because they have one 
less parameter than either of the mAMSB or HCAMSB models.\\%
\indent%
Several crucial features emerge from this plot.  
\begin{itemize}%
\item First, we note that for a given value of $m_{\nino}$, the value of 
$\sigma (\nino p)$ is bounded from below, unlike the case of the mSUGRA model.  
That means that wino-WIMP dark matter can be either detected or excluded for a 
given $m_{\nino}$ value. 
\item Second, we note that the cross section values generally fall in the range 
that is detectable at present or future DD experiments. The purple contour, for 
instance, exhibits the CDMS reach based on 2004-2009 data, and already excludes 
some points, especially those at large $\tanb$. We also show the reach of 
Xenon-100, LUX, Xenon-100 upgrade, and Xenon 1 ton\cite{dm_limits}.  These 
experiments should be able to either discover or exclude AMSB models with 
$m_{\nino}$ values below $\sim 90,\ 100,\ 200$ and 500 GeV respectively.  These 
WIMP masses correspond to values of $m_{\gl}\sim 690,\ 770,\ 1540$ and 3850 
GeV, respectively! The latter reach far exceeds the 100 fb$^{-1}$ of integrated 
luminosity reach of LHC for $m_{\gl}$, which were shown in Chapters 
\ref{chap:hcamsb} and \ref{chap:inoamsb} to be $m_{\gl}\sim 2.2-2.4$ for \hca 
and $m_{\gl}<2.6$ TeV for \ino respectively.  For inoAMSB models, where the 
minimal value of $\sigma^{SI}(\nino p)$ exceeds that of mAMSB or HCAMSB for a 
given $m_{\nino}$ value, the Xenon 1 ton reach is to $m_{\nino}\sim 800$ GeV, 
corresponding to a reach in $m_{\gl}$ of 6200 GeV!%
\end{itemize}%

\begin{figure}[t]
  \begin{center}
    \includegraphics[width=1.0\textwidth]{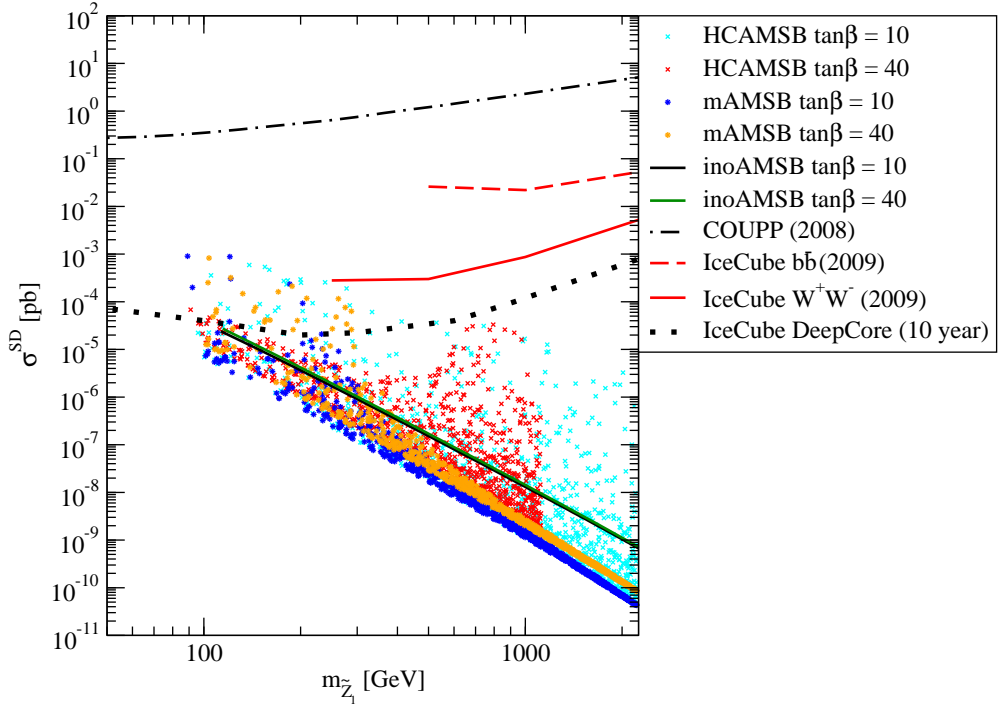}
    \caption{%
	Spin-dependent $\nino-p$ scattering cross section versus $m_{\nino}$ for 
	mAMSB,  HCAMSB and inoAMSB models for $\tanb =10$ and 40 and 
	$\mu >0$. The parameters $m_{3/2}$ and also $m_0$ (for mAMSB) and 
	$\alpha$ (for HCAMSB) have been scanned over.  We also show the COUPP and 
	IceCube limits in $\sigma^{SD}(\nino p)$.
    }%
    \label{fig:sd_xs}
  \end{center}
\end{figure}

In Figure \ref{fig:sd_xs} we scan the parameter spaces again and plot the 
spin-dependent DD cross section versus $m_{\nino}$.  The recent DD limit by the 
COUPP experiment appears at least two orders of magnitude above the theory 
predictions of all of the models.  The probes of the SD cross section are 
actually better constrained by ID experiments.  The ID limits from ICECUBE and 
its projected DeepCore limit are shown in this figure, and the cross 
section is that of winos interacting with solar Hydrogen.  Though DeepCore will 
be able to access portions of parameter space, it will not be capable of 
reaching any of the the models' lower limits.
\section{%
  Indirect Detection Rates for the AMSB Models
  \label{sec:idrates}
}%
In this section the Indirect Detection (ID) rates for mAMSB, HCAMSB, and \ino
are given.  We consider the indirect detection of WIMPs from their annihilation in the solar core via neutrino telescopes, and we also consider gamma ray and 
antimatter fluxes originating in halo WIMP annihilations.  These fluxes depend 
(quadratically) on the assumed density profile of the galaxy, and the results 
of two profiles, isothermal and Navarro-Frenk-White \cite{Navarro:1995iw}, are 
given.  Most halo models are in near-agreement at the Earth's position ($\sim$ 
8 kpc from the galactic center), however they differ widely at the inner 
parsecs of the galaxy.  This translates into large uncertainties in the gamma 
fluxes from these regions.  As mentioned in Section \ref{sec:dmexp}, antimatter 
signals tend to have fewer uncertainties because they should originate closer 
to earth.\\%
\indent%
This section is broken into three parts.  In the first subsection we consider 
the \mam model and calculate ID rates by varying $\mnot$ and $\tanb$ and 
analyze the interactions that lead to these rates.  In the following subsection 
we do the same for HCAMSB, but this time we vary $\alpha$ and $\tanb$.  And 
finally, in the last subsection we show the calculated ID rates while varying 
the remaining parameter, $\mhf$, for the mAMSB, HCAMSB, and the \ino models.
\subsection{%
  Indirect Wino Detection for \mam
  \label{subsec:idmamsb}
}%
\begin{figure}[t]
  \begin{center}
    \includegraphics[trim= 0cm 0cm 0cm 12cm,clip,width=1\textwidth]
      {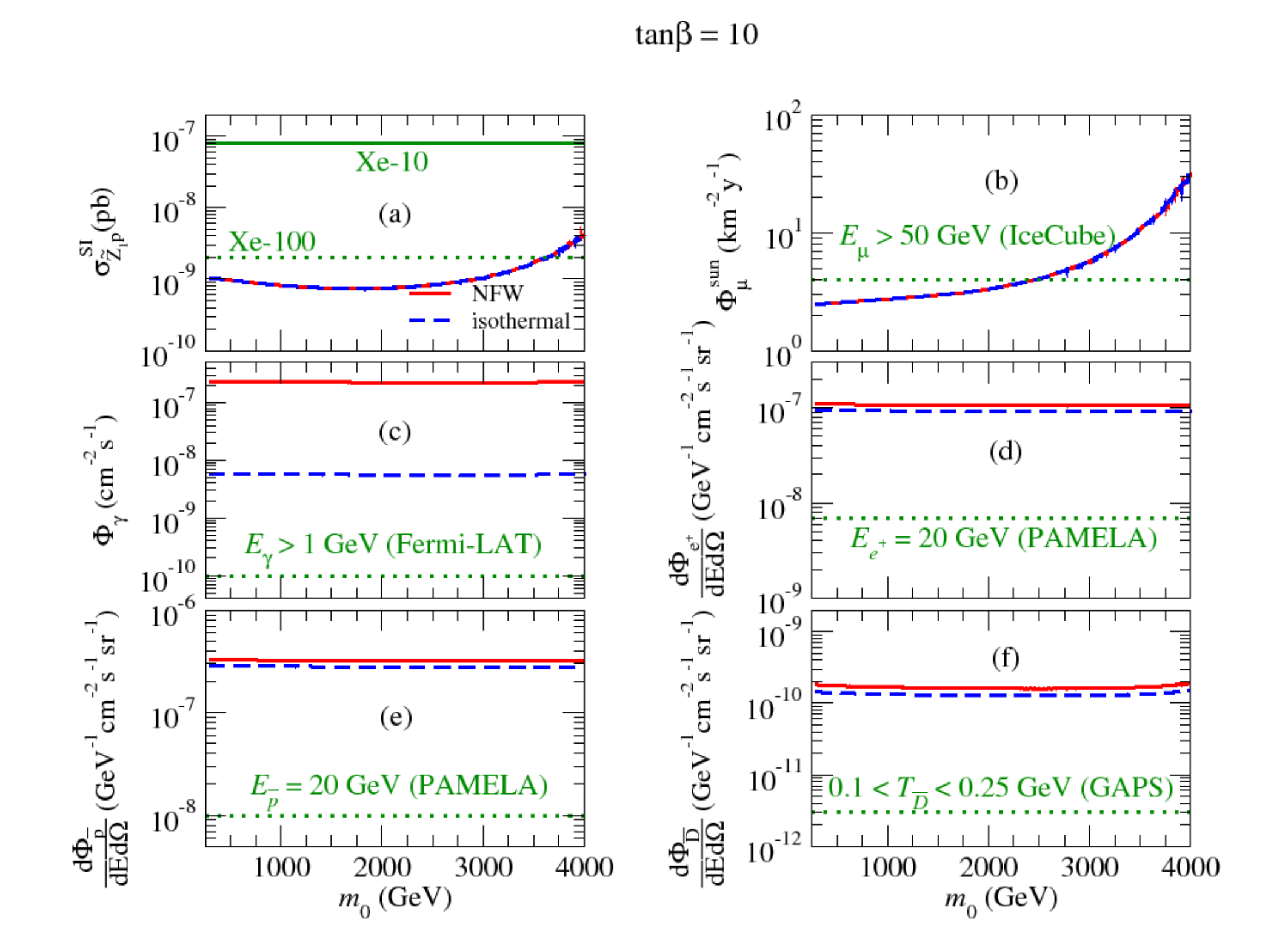}
    \caption{%
	Wino CDM direct detection (a) and indirect detection (b)--(f) rates 
	versus $\mnot$ in mAMSB for $\mhf=50$ TeV, $\tanb$=10, and $\mu>0$.
    }%
    \label{fig:id_mam10}
  \end{center}
\end{figure}

\begin{figure}[t]
  \begin{center}
    \includegraphics[trim= 0cm 0cm 0cm 12cm,clip,width=1\textwidth]
      {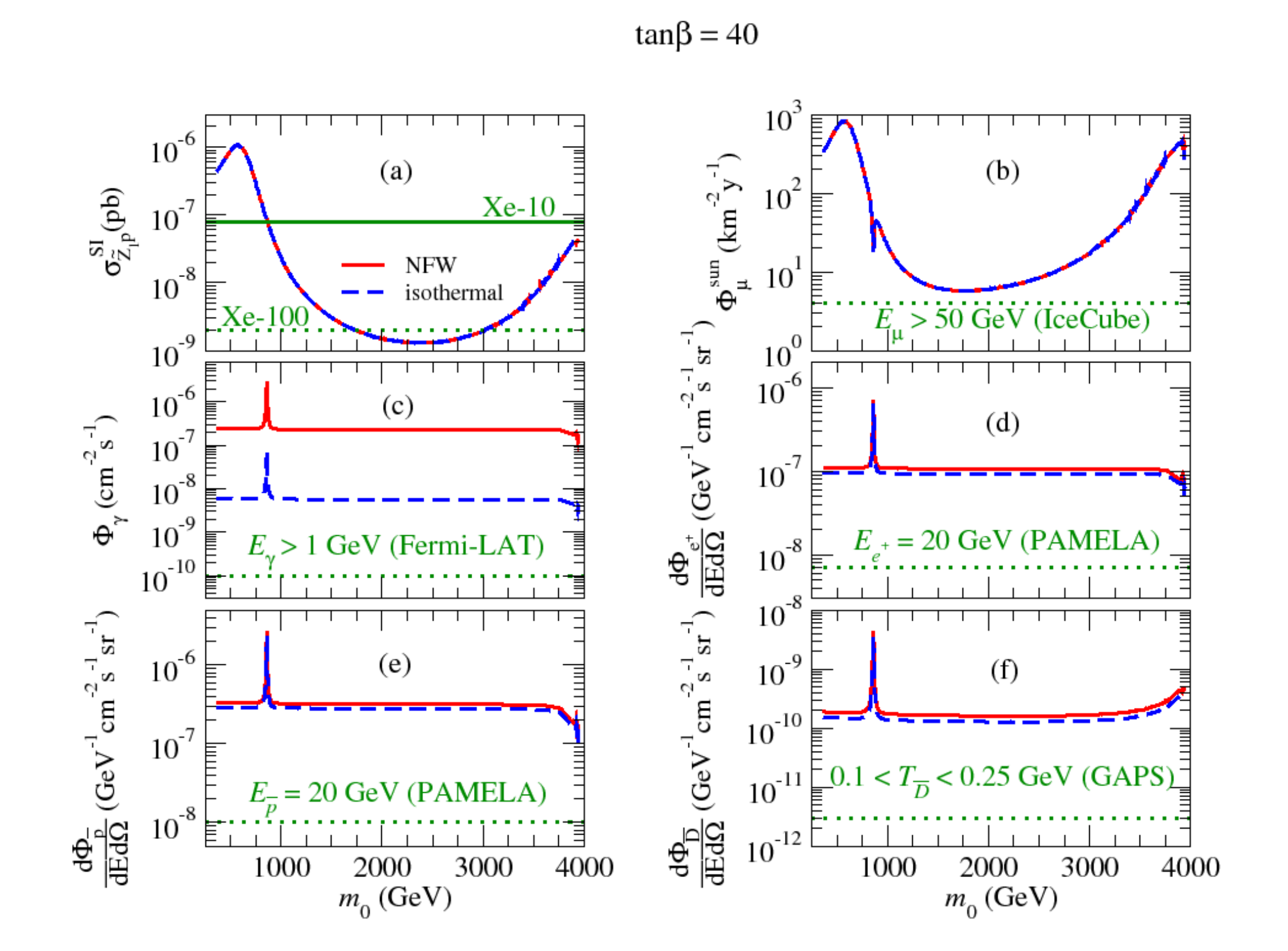}
    \caption{%
	Wino CDM direct detection (a) and indirect detection (b)--(f) rates 
	versus $\mnot$ in mAMSB for $\mhf=50$ TeV, $\tanb$=40, and $\mu>0$.
    }%
    \label{fig:id_mam40}
  \end{center}
\end{figure}

We begin the analysis with mAMSB.  We choose $\mhf =$ 50 TeV and $\tanb = 10$, 
which gives a wino-like neutralino with mass $m_{\nino} \simeq 144$ GeV.  All 
of the ID rates for these parameters, while varying $\mnot$, are shown in 
Figure \ref{fig:id_mam10} and each is discussed in turn.\\%
\indent%
In Figure \ref{fig:id_mam10} (a) we show the SI DD cross section for 
comparison, along with approximate reaches for \xen-10 and \xen-100.  For higher
$\mnot$, the value of $|\mu|$ drops, much like it does in mSUGRA when the 
focus-point region is approached.  Thus, when $\mnot$ is large, $\nino$ picks 
up a higgsino component that enhances the SI DD cross section since the 
relevant vertex is Higgs-higgsino-gaugino.  For $\mnot \gsim 3500$ GeV, it is 
seen that the $\nino$ becomes directly observable for \xen-100.\\%
\indent%
Figure \ref{fig:id_mam10} (b) shows the first ID rate which is the flux of muons
from the solar direction.  To calculate the flux we used the Isajet/DarkSusy 
interface (see Chapter \ref{chap:intro} for descriptions of these programs) and 
require $E_{\mu} > 50$ GeV as required for ICECUBE.  When $\nino$s annihilate 
in the core of the Sun they produce $\nu_{\mu}$s that travel to Earth and 
convert in water/ice to muons via the charged-current interaction.  $\nino$s 
are captured by the Sun in the first place through its spin-dependent
interactions with Hydrogen that is mainly sensitive to $Z^{*}$ exchange.  Thus, 
for larger values of $\mnot$ where the neutralino has more higgsino content, 
the flux is expected to be larger due to the $Z$-$\nino$-$\nino$ coupling.\\%
\indent%
In Figure \ref{fig:id_mam10} (c) we show the expected flux of gamma rays from 
the galactic core with $E_{\gamma} > 1$ GeV as required by the Fermi Gamma-ray
Space Telescope (FGST) experiment.  The gamma ray fluxes are due to neutralino 
annihilations and depend mainly on the $\nino\nino\rightarrow W^{+}W^{-}$ 
annihilation cross section which occurs via chargino exchange.  This is 
followed by $W\rightarrow q\bar{q}'\rightarrow$
$ \ \pi^{0}\rightarrow\gamma\gamma$ to produce the gammas.  The rates do 
not change with $\mnot$ because $\nino$s remain wino-like and the chargino
mass does not change significantly.  As expected, the type of profile plays a 
significant role in gamma detection, and the predictions for the two differ by 
over an order of magnitude.  Still, in both cases the fluxes exceed the 
approximate reach of FGST.\\%
\indent%
Figures \ref{fig:id_mam10} (d)-(f) show antimatter fluxes for $e^{+}$, 
$\bar{p}$, and $\bar{D}$ respectively that come from halo annihilations.  For 
$e^{+}$ and $\bar{p}$ there are observable rates predicted for Pamela 
\cite{Pearce:2002ef}, and $\bar{D}$ observables rates might be seen by GAPS 
\cite{Mori:2001dv}.  These elevated rates reflect the $wino-wino$ annihilation 
into $W^{+}W^{-}$ cross section.\\%
\indent%
Now the same calculations are done and plotted in Figure \ref{fig:id_mam40} 
with $\mhf=50$ TeV and $\mu>0$, but this time $\tanb$ is increased 40.  The SI
DD cross section in frame (a) has the usual enhancement due to increased 
higgsino content of $\nino$ at larger $\mnot$ values.  Additionally, there is
an enhancement for low $\mnot$ values as well.  As described in Section 
\ref{sec:dmexp}, $\sigma^{SI}_{\nino p}$ depends on Higgs and squark exchange 
diagrams between $\nino$ and quarks and through loops via Higgs exchange 
between $\nino$ and gluons.  Now that $\tanb$ is higher, the heavy Higgs mass 
is quite light, e.g., for $\mnot\sim$ 600 GeV the heavy Higgs mass is 
$m_{H}\sim$ 152 GeV.  This results in a huge DD cross section at low $\mnot$, 
and regions with $\mnot \lsim 900$ GeV and $\tanb=40$ are already excluded by 
DD WIMP searches.\\%
\indent%
Related to this, Figure \ref{fig:id_mam40} (b) shows an increase in the muon 
flux for low $\mnot$ and $\tanb=40$.  Usually the {\it spin-dependent} cross 
section is considered to be the most important part of $\nino$ capture at high 
$\tanb$ in the sun because the SI cross section receives minimal enhancement 
from the target (H and He) masses.  However, because of the lighter Higgs 
effects of the last paragraph at low $\mnot$, the SI cross section plays a 
significant role in $\nino$ capture and hence muon detection on Earth.  Of 
course, this region is already excluded by DD experiments.  We also note the 
interesting \quotes{anti-resonance} effect in $\Phi_{\mu}$ for $\mnot \sim 850$ 
GeV.  Here, $m_{A}\sim 2m_{\nino}$, and on the Higgs resonance 
$\nino\nino\rightarrow b\bar{b}$ is the dominant annihilation mode.  The 
$b$-decays modes produce a softer distribution of $\nu_{\mu}$s than from vector 
bosons leading to fewer muons passing the $E_{\mu}>50$ GeV requirement, and 
hence the sharp dip in flux.  Finally, at large $\mnot$ we again see an 
increase in muon flux due to the increased higgsino component of $\nino$.\\%
\indent%
Finally, gamma and antimatter fluxes versus $\mnot$ are shown in Figure 
\ref{fig:id_mam40} in frames (c) - (f). The rates are again flat due to the 
constant $\nino\nino$ annihilation rate to vector bosons.  The exception occurs 
at $\mnot\sim 850$ GeV where the $A$-resonance enhances the annihilation rate
\cite{Baer:2003bp}.  At high $\mnot$, $\pos$ and $\pbar$ fluxes drop slightly 
because $\nino\nino$ annihilations to $b\bbar$ become prominent relative to the 
usual vector boson products.  Subsequent decays lead to softer $\pos$ and 
$\pbar$, and hence less detectability above the $E_{\pos,\pbar}=20$ GeV 
threshold.  In frame (f), the $\Dbar$ rate actually increases with $\mnot$ 
since the softer $\Dbar$s produced are still in the low energy range of 
detection.
\subsection{%
  Indirect Wino Detection for \hca
  \label{subsec:idhcamsb}
}%
\begin{figure}[t]
  \begin{center}
    \includegraphics[trim= 0cm 0cm 0cm 12cm,clip,width=1\textwidth]
	{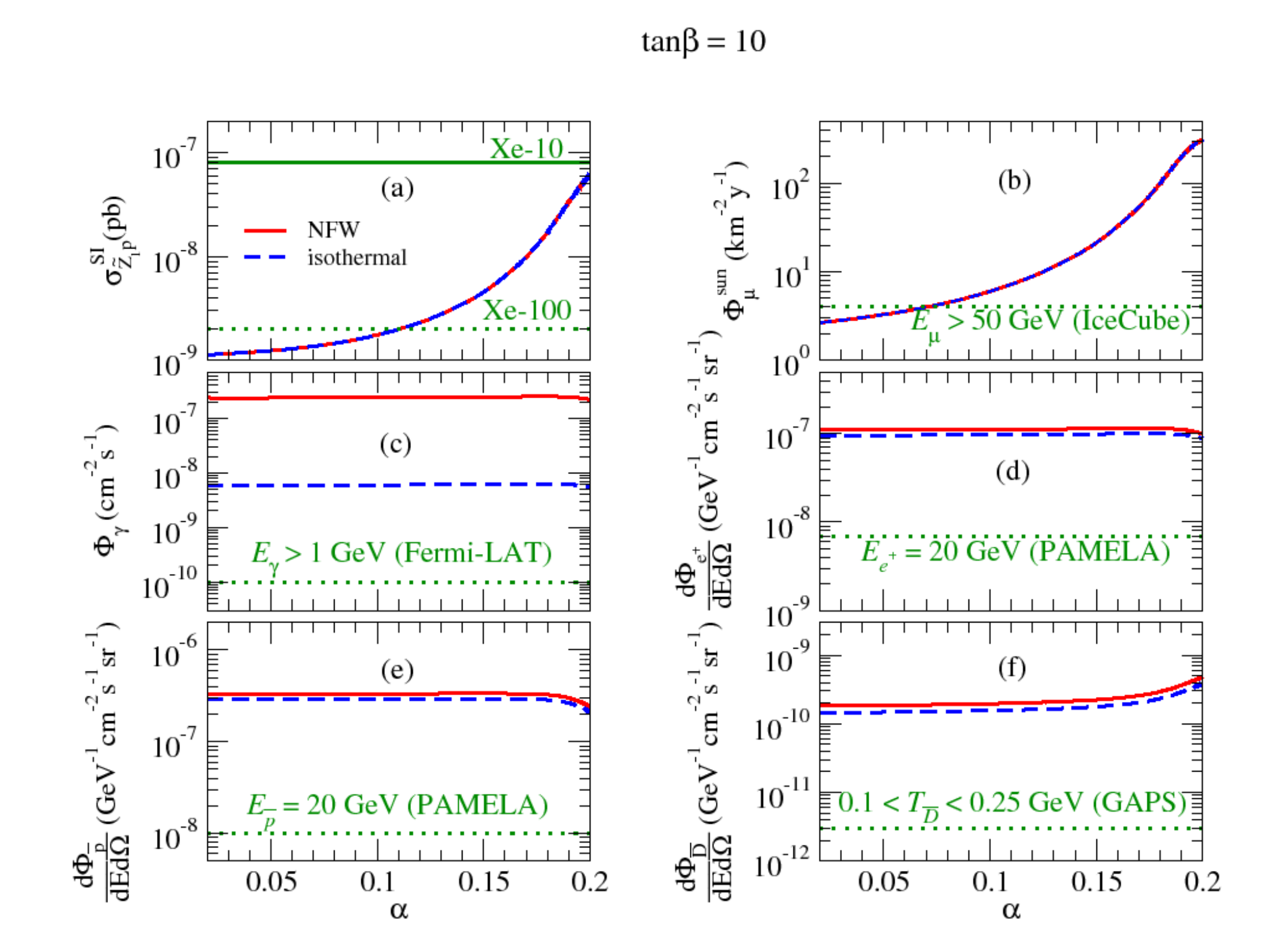}
    \caption{%
	Wino CDM direct detection (a) and indirect detection (b)--(f) rates 
	versus $\alpha$ in \hca for $\mhf=50$ TeV, $\tanb$=0, and $\mu>0$.
    }%
    \label{fig:id_hca10}
  \end{center}
\end{figure}

\begin{figure}[t]
  \begin{center}
    \includegraphics[trim= 0cm 0cm 0cm 12cm,clip,width=1\textwidth]
	{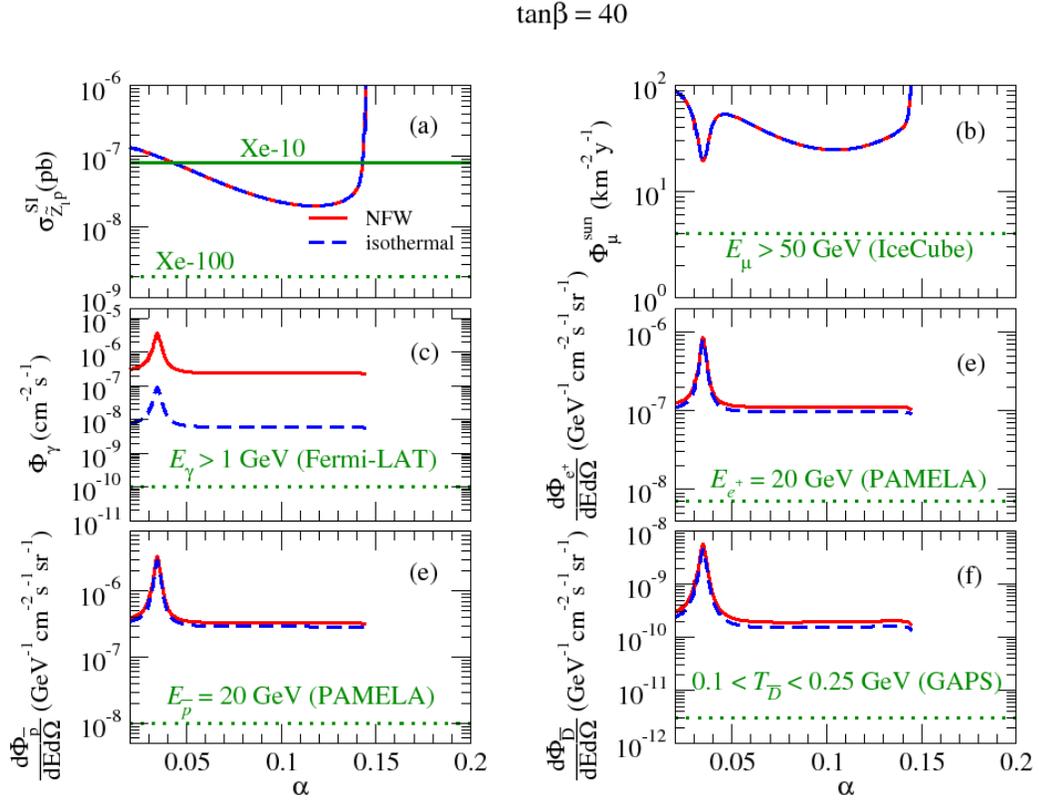}
    \caption{%
	Wino CDM direct detection (a) and indirect detection (b)--(f) rates 
	versus $\alpha$ in \hca for $\mhf=50$ TeV, $\tanb$=40, and $\mu>0$.
    }%
    \label{fig:id_hca40}
  \end{center}
\end{figure}

In this subsection results similar to those of the last subsection are shown 
for the \hca model for $\mhf=50$ TeV,  but now we vary the $\alpha$ parameter.  
It was seen in Chapter \ref{chap:hcamsb} that $\alpha\sim0$ corresponds to pure 
anomaly mediation, while larger $\alpha$ values give increasing $\mb$ at the 
GUT scale.  This affects the RGEs of sparticle masses through their hypercharge 
numbers, and this leads to a left-right split sparticle spectrum at the weak 
scale.  This effect, combined with the effect of a large t-quark Yukawa 
coupling, leads to light, dominantly left top squark state $\spart{t}{1}$ at 
large $\alpha$.  Raising $\alpha$ also leads to a diminished $|\mu|$ value, and 
this results in $\nino$ acquiring more of a higgsino component.\\%
\indent%
Figure \ref{fig:id_hca10} (a) shows the SI cross section for direct detection.  
Reminiscent of $\mnot$ in mAMSB, the cross section increases for larger values 
of $\alpha$, and this is, of course, due to the larger higgsino component of 
$\nino$.\\%
\indent%
In Figure \ref{fig:id_hca10} (b) is the muon flux versus $\alpha$ for the \hca
model.  At low $\alpha$ the flux is quite low, but increases considerable at 
high $\alpha$.  The higgsino component at high $\alpha$ leads to unsuppressed
couplings to $Z$.  Since the $\nino$--nucleon SD cross section, is mainly 
sensitive to $Z^{*}$-exchange, a higher flux of muons are seen on the Earth.\\%
\indent%
In Figure \ref{fig:id_hca10}, in frames (c)-(f) are the plots of the gamma ray 
flux and antimatter fluxes.  The high rates are due to the large $wino-wino$
$\rightarrow VV$ annihilation cross sections and are relatively flat with 
$\alpha$.  At large $\alpha$ the annihilation to $b\bbar$ states is enhanced 
leading to smaller detection rates for $\pos$ and $\pbar$ (softer spectrum does 
not pass experimental threshold) and slightly increased $\Dbar$ (by soft 
energy requirements).\\%
\indent%
In Figure \ref{fig:id_hca40}, we perform the same calculations as in Figure 
\ref{fig:id_hca10}, but now $\tanb$ is chosen to be 40.  In frame (a), the 
spin-independent DD cross section is enhanced relative to the $\tanb=10$ case 
because of the now much lighter, heavy and pseudoscalar Higgs ($H$ and $A$) and 
the increased $b$-quark Yukawa coupling.  As $\alpha$ increases, $m_{H}$ 
increases and the rate diminishes until the highest $\alpha$ values are 
reached.  Here the presence of the higgsino component of $\nino$ enhances the 
cross section.  As in the \mam case, the large SI enhancement leads to greater 
solar capture rate of $\nino$s, which in turn leads to greater muon fluxes on 
earth, and this is seen in frame (b).\\%
\indent%
In Figure \ref{fig:id_hca40} (b), we see the same type of anti-resonance effect 
as in Figure \ref{fig:id_mam40}, where near $2m_{\nino}\sim m_{A}$, 
annihilation to heavy fermion pairs dominate over the usual $VV$ pairs 
($V=W^{\pm}, \ Z$), and lead to a softer distribution of neutrinos.  In frames 
(c)-(f) are the halo annihilation rates for \hca with $\tanb=40$.  These rates 
are generally flat with changing $\alpha$, and do not suffer an increase as 
compared with low $\tanb$, since $wino-wino\rightarrow VV$ still dominates the 
annihilation rate.  The exception again occurs near the A-resonance, and halo 
annihilation is enhanced by pseudoscalar Higgs exchange.
\subsection{%
  Indirect Detection Rates for mAMSB, HCAMSB, and \ino versus $\mhf$
}%
\begin{figure}[t]
  \begin{center}
    \includegraphics[trim= 0cm 0cm 0cm 12cm,clip,width=1\textwidth]
	{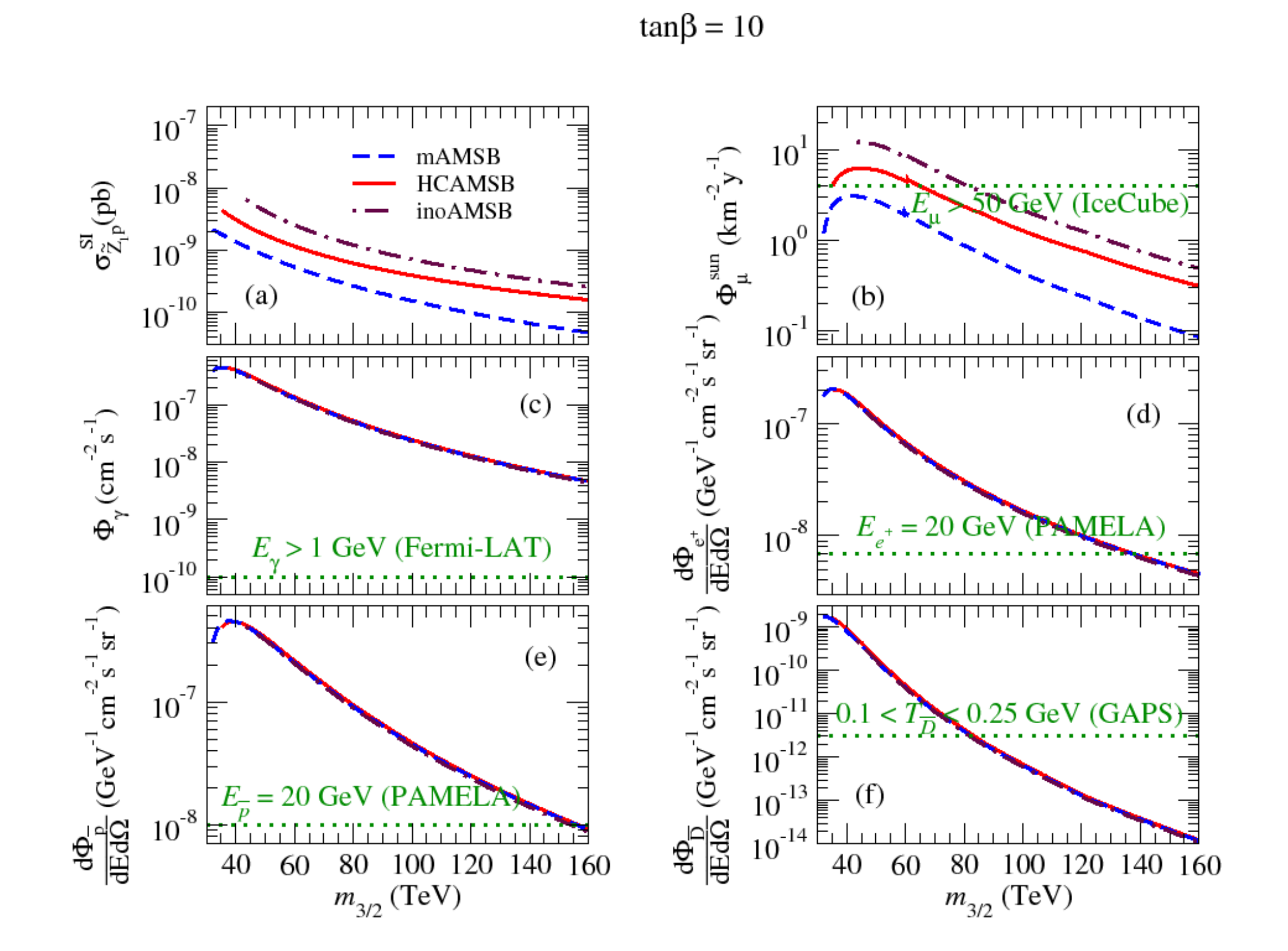}
    \caption{%
	DD and ID rates for wino CDM in mAMSB, HCAMSB, and \ino models versus 
	$\mhf$, for $\tanb=10$, and $\mu>0$.  For mAMSB, $\mnot=1$ TeV, and for 
	HCAMSB, $\alpha=0.1$.  In these plots we adopt the NFW DM halo profile.
    }%
    \label{fig:id_3models10}
  \end{center}
\end{figure}

\begin{figure}[t]
  \begin{center}
    \includegraphics[trim= 0cm 0cm 0cm 12cm,clip,width=1\textwidth]
	{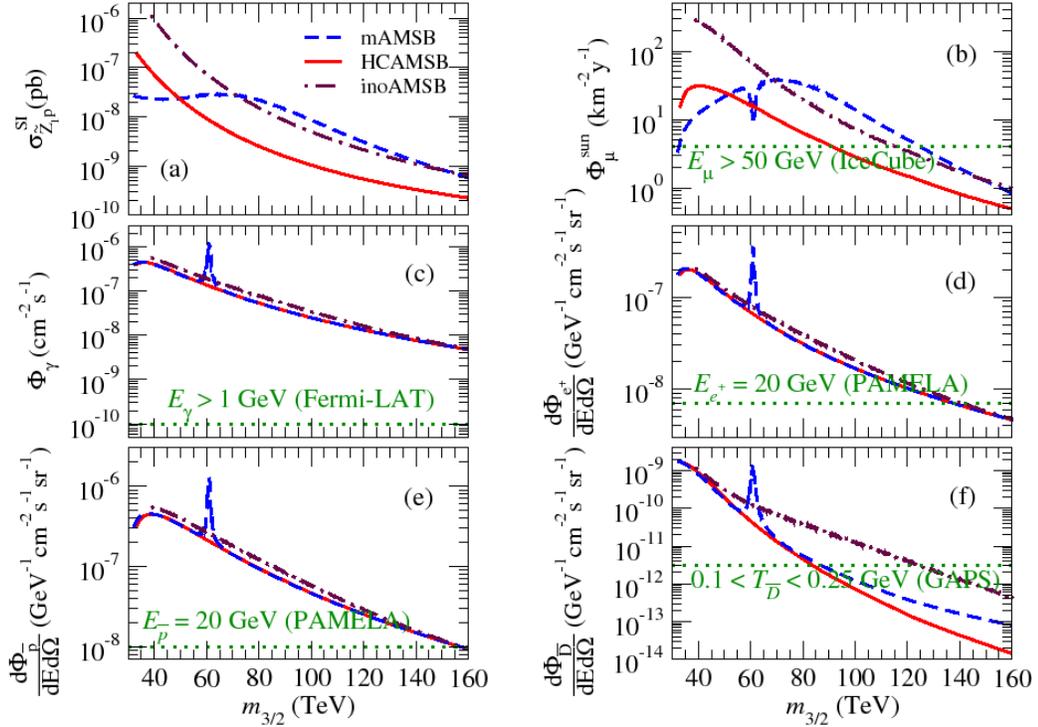}
    \caption{%
	DD and ID rates for wino CDM in mAMSB, HCAMSB, and \ino models versus 
	$\mhf$, for $\tanb=40$, and $\mu>0$.  For mAMSB, $\mnot=1$ TeV, and for 
	HCAMSB, $\alpha=0.1$.  In these plots we adopt the NFW DM halo profile.
    }%
    \label{fig:id_3models40}
  \end{center}
\end{figure}

In this subsection we explore the remaining parameter of our AMSB models: $\mhf$.
While $\mhf$ is roughly bounded from below by LEP2 constraints, in principle 
there is no upper bound for the parameter.  In general, increasing the 
gravitino mass increases all sparticle mass as can be seen for example in Table
\ref{tab:inospectrum} for the \ino case.  Thus we expect a diminution of rates 
toward higher $\mhf$ values due to decreased interactions.  Also, in all ID 
plots we assume the standard NFW halo profile.\\%
\indent%
In Figure \ref{fig:id_3models10} we show direct and indirect wino DM detection
rates versus $\mhf$ for all three models considered in this thesis, with 
$\tanb=10$ and $\mu>0$.  For the comparison, in the \hca case the choice 
$\alpha=0.1$ is made, and for \mam we take $\mnot=1$ TeV.\\%
\indent%
The spin-independent cross section is shown in frame (a) of Figure 
\ref{fig:id_3models10} for all models.  The larger \ino cross section is due in 
part to the smaller $\mu$ values for a given $\mhf$, which enhances $\nino$ 
scattering via Higgs exchange made possible by the Higgs-higgsino-gaugino 
vertex.\\%
\indent%
Figure \ref{fig:id_3models10} (b) shows the muon flux with the minimum energy 
of $E_{\mu}>50$ GeV required for detectability at IceCube.  Due to the relative 
values of $\mu$ in all three models, \ino yields the highest rates while \mam 
yields the lowest.  Since \ino generally has the lowest $|\mu|$ values, its $
\nino$ can interact more strongly through $Z^{*}$ exchange (see previous two 
subsections).  This enhances the SD cross section, which eventually leads to 
greater $\mu$ flux on earth.  The rough reach of IceCube is shown, and the 
figure indicates that the lower values of $\mhf$ may be accessible in 
$\nu_{\mu}\rightarrow\mu$ searches.\\%
\indent%
The ID rates for $\gamma$,$\pos$,$\pbar$, and $\Dbar$ are shown in 
Figure \ref{fig:id_3models10}, frames (c)-(f).  The rates for all models are 
nearly identical in each case for varying $\mhf$.  This is due to the dominance 
of $\nino\nino$$\rightarrow VV$ halo annihilations, which mainly depend on the 
gaugino (wino) component of $\nino$.  Rough reaches for Fermi-LAT, Pamela, and 
GAPS are shown for reference, and high wino annihilations should yield 
observable signals.\\%
\indent%
Increasing $\tanb$ to 40, the same rates are shown versus $\mhf$ in Figure 
\ref{fig:id_3models40}.  As in the previous subsections, the SI cross section in
frame (a) is enhanced relative to $\tanb=10$ due to the simultaneous decrease 
in the Higgs mass and increase in $b$-quark Yukawa coupling.  Because \ino has
both the smallest $m_{H}$ and smallest $|\mu|$ values of all models for low 
$\mhf$, this model shows the largest SI cross section.  As $\mhf$ increases, 
$m_{H}$ increases for \ino and \hca, but actually decreases for \mam.  Thus for
$\mhf\gsim 75$ TeV, the \mam model yields the highest value of 
$\sigma^{SI}(\nino p)$.\\%
\indent%
Figure \ref{fig:id_3models40}, frame (b), shows the muon flux, which is 
elevated relative to the $\tanb=10$ case, again due to the increased SI cross 
section of frame (a).  At low $\mhf$ the \ino model shows the highest flux due 
to its low $\mu$ value which enhances the $Z^{*}$ exchange in $\nino$-$q$ 
scattering.  Around $\mhf\sim 60$ TeV, the mass of A is near twice the $\nino$ 
mass, and resonant $\nino$ annihilations to $b\bbar$ final states dominate over 
$VV$.  This diminishes the flux of detectable muons (see previous two 
subsections for further discussion).  For higher $\mhf$ values, beyond the 
resonance, the $\mam$ model yields the highest muon flux due to its elevated 
value of $\sigma^{SI}(\nino p)$.\\%
\indent%
Figure \ref{fig:id_3models40}, frames (c)-(f), show the gamma and antimatter 
fluxes versus $\mhf$.  Relatively little change in the ID rates are seen in 
going from $\tanb=10$ to $\tanb=40$, since $wino-wino\rightarrow VV$ dominates 
the annihilations.  An exception, of course, is due to A-resonance enhancement 
of halo annihilations, where $2m_{\nino}\sim m_{A}$.  And $\Dbar$ rates appear 
enhanced for \ino and \mam at large $\mhf$ because in these cases the 
$\nino\nino$$\rightarrow$$b\bbar$ annihilation rate, which does receive $\tanb$
enhancement, contributes to the detection of rather low energy $\Dbar$s.\\%
\section{%
  \label{sec:dmsummary}
  Summary
}%
In this Chapter we have investigated aspects of cold dark matter in three models
of anomaly mediation: mAMSB, HCAMSB and inoAMSB. Typically, each gives rise to a
wino-like lightest neutralino, unless very high values of $m_0$ (for mAMSB) or
$\alpha$ (for HCAMSB) are used, in which case the $\nino$ becomes a mixed
wino-higgsino state.  In this class of models with a wino-like $\nino$, the 
thermal abundance of neutralino CDM is well below measured values, unless 
$m_{\nino}\gsim 1300$ GeV. We discuss four ways to reconcile the predicted 
abundance of CDM with experiment:%
\begin{enumerate}%
\item enhanced neutralino production via scalar field ({\it e.g.} moduli) 
      decay, 
\item enhanced neutralino production via gravitino decay, where gravitinos may 
      arise thermally, or by moduli or inflaton decay, 
\item enhanced neutralino production via heavy axino decay, and 
\item neutralino decay to axinos, where the bulk of CDM comes from a mixture of 
      vacuum mis-alignment produced axions and thermally produced axinos.
\end{enumerate}%
Cases 1 and 2 should lead to a situation where all of CDM is comprised of wino-
like WIMPs; they will be very hard, perhaps impossible, to tell apart.  Case 3 
would contain a mixture of axion and wino-like WIMP CDM. It is a scenario where
it is possible that both a WIMP and an axion could be detected. Case 4 predicts 
no direct or indirect detection of WIMPs, but a possible detection of relic 
axions.  It is important to note that more than one of these mechanisms may 
occur at once: for instance, we may gain additional neutralino production in 
the early universe from moduli, gravitino and axino decay all together.\\%
\indent%
In Sections \ref{sec:ddrates} and \ref{sec:idrates}, we presented rates for 
direct and indirect detection of relic wino-like WIMPs. The SI direct detection 
cross sections are bounded from below. Ultimately, ton-scale noble liquid or 
SuperCDMS experiments should probe out to $m_{\nino}\sim 500$ GeV, which would
exceed the 100 $\invfb$ reach of LHC; a non-observation of signal would put 
enormous stress on AMSB-like models as new physics. We also evaluated SD direct 
detection: current experiments have little reach for AMSB-like models, although 
IceCube DeepCore and possibly COUPP upgrades may probe more deeply.\\%
\indent%
WIMP indirect detection rates for all three AMSB models were also presented. 
The IceCube experiment has some reach for WIMPs from AMSB models, especially at 
high $\tanb$ or when the $\nino$ picks up a higgsino component. We noted an 
interesting inverse resonance effect in the muon flux detection rate, caused by 
transition from solar core annihilations to $VV$ states, to annihilations to 
mainly $b\bar{b}$ states. The detection of $\gamma$s, $e^+$s, $\bar{p}$s and 
$\bar{D}$s are all elevated in AMSB-like models compared to mSUGRA 
\cite{Baer:2003bp}, due to the high rate for $\nino\nino\to VV$ annihilation in 
the galactic halo. The results do depend on the assumed halo profile, 
especially for $\gamma$ ray detection in the direction of the galactic core. 
Generally, if a signal is seen in the $\pos$ channel, then one ought to be seen 
in the $\bar{p}$ channel, and ultimately in the $\gamma$, $\Dbar$ (if/when GAPS 
flies) or direct detection channel. In addition, a sparticle production signal 
should ultimately be seen at the LHC, at least for $m_{\gl}\lsim 2400$ GeV, once 100 $\invfb$ of integrated luminosity is accrued.\\%
\indent%
As a final remark, we should note here that the dark matter detection signals 
all provide complementary information to that which will be provided by the CERN
LHC.  At the LHC, each model-- mAMSB, HCAMSB and inoAMSB-- will provide a rich 
assortment of gluino and squark cascade decay signals which will include 
multi-jet plus multi-lepton plus missing $E_T$ events. In all cases, the 
wino-like lightest neutralino state will be signaled by the well-known presence 
of highly ionizing tracks (HITs) from quasi-stable charginos with track length 
of order {\it centimeters}, before they decay to soft pions plus a $\nino$.  It 
was discussed in Chapter \ref{chap:inoamsb} that the three models should be 
distinguishable at the LHC by the differing opposite-sign/same flavor dilepton 
invariant mass distributions. In the case of mAMSB, with 
$m_{\sl_L}\simeq m_{\sl_R}$, we expect a single mass edge from 
$\ninos{2}\to$ $l\sl_{L,R} \to$ $l^{+}l^{-} \nino$ decay. In HCAMSB, the 
sleptons are rather heavy, and instead $\ninos{2}\to\nino Z$ occurs at a large 
rate, leading to a bump in $m(l^{+}l^{-})\sim M_Z$, upon a continuum 
distribution. In inoAMSB, with 
$m_{\ninos{2}}>m_{\sl_{L,R}}$, but with $\sl_L$ and $\sl_R$ split in mass
(due to different $U(1)_Y$ quantum numbers), a characteristic 
{\it double mass edge} is expected in the $m(l^{+}l^{-})$ invariant mass 
distribution.

\chapter{%
  Conclusion
  \label{chap:conclusion}
}%
We conclude by summarizing the main points of this thesis.  We have compared 
three model types, all having supergravity anomaly contributions (AMSB) to the 
soft parameters.  We began with the \mam model of Chapter \ref{chap:mamsb} 
which was the starting point of comparison for all other models.  This model 
features a wino-like LSP ($\nino$), a lightest chargino ($\cino$) with  mass 
nearly equal to that of the LSP, and a near left-right degeneracy in scalar 
masses.  Since $m_{\nino} \sim m_{\cino}$, we expect the presence of HITs (for 
all of the models) as described in Chapter \ref{chap:mamsb}. The second 
lightest neutralino, $\ninos{2}$, is bino-like and can be produced in LHC pp 
collisions and subsequently decays through 
$\ninos{2}\rightarrow \spart{l}{L/R}^{\pm}l$, while slepton decays produce the 
LSP and dilepton final states.  Since left- and right-handed states are 
degenerate, the OS $m(l^{\pm}l^{\mp})$ distribution for these final states 
results in a single (comparatively large) mass edge, as can be see in Figure 
\ref{fig:ino_mll}.\\
\indent
We also examined the Hypercharged Anomaly Mediated Supersymmetry-Breaking model,
which incorporates effects from a geometrically separated hidden sector.  The 
hidden sector communicates the supersymmetry-breaking to the visible sector via 
a combination of supergravity anomaly and a new $U(1)$ mediation.  The soft 
masses are those of mAMSB, but with an extra contribution to the bino mass 
parameterized as $\alpha$ in Equation (\ref{eqn:hca_bino}).  Because left- and 
right-handed particles have different hypercharge quantum numbers, the mass 
spectrum for this model is highly left-right split (in contrast to mAMSB).  The 
LSP is still a wino-like neutralino with near-degeneracy to the lightest 
chargino, and we again expect highly-ionizing track chargino events.  The 
second-lightest neutralino, however, is higgsino-like in this case due to 
heaviness of the bino soft mass, $M_{1}$.  The higgsino component in 
$\ninos{2}$ allows for decays to $Z$ and $\nino$.  Without intermediate slepton 
decays, the OS dilepton distribution appears as a smooth distribution with a 
peak around $m_Z$.\\
\indent
The last model(s) we examined originate in string theories with Calabi-Yao 
orientifold compactifications and moduli stabilization through fluxes and 
non-perturbative (KKLT) effects.  The supersymmetry-breaking is communicated to
the MSSM via gravitation.  The resulting soft contributions appear as a 
combination of anomaly mediation and gaugino mediation, and we have labeled 
these parameters space points \quotes{inoAMSB}.  In these models, the 
AMSB contributions are only present for gaugino masses, $M_{1}, M_{2}$, and 
$M_{3}$.  But like gaugino mediation, the soft scalar and trilinear coupling are
suppressed, $i.e.$, $\mnot \sim \anot \sim 0$ at $M_{string}$.  We expect for 
this class of models to yield HITs, as usual for AMSB models.  We also expect 
with a double edge structure in the $m(l^{\pm}l^{\mp})$ distribution.  This is
because, like in mAMSB, $\ninos{2}$ is bino-like, but the sleptons here are 
left-right split leading to two distinct mass edges, instead of just one as in
the mAMSB framework (again, see Figure \ref{fig:ino_mll}).\\
\indent%
Finally, we explored the Dark Matter predictions for all of these models.  It 
is well-known that the neutralino LSPs of AMSB models annihilate and 
co-annihilate too efficiently to account for DM relic density of the universe.  
However, if any of these models are the correct description of nature, there 
are likely other $non$-$thermal$ sources of LSP-production that can serve to 
increase the value of the relic density.  Assuming these mechanism provide the 
correct DM abundance, we calculated direct and various indirect detection rates 
for a multitude of experiments for every model.  For the direct detection, we 
saw that CDMS data has already excluded some regions of the parameters spaces 
of these models.  It was also striking to find that the cross sections were all 
bounded from below (Figure \ref{fig:si_xs}), leading to a XENON 1 Ton reach that far 
exceeded the reach of the LHC; the extreme example being the \ino reach
$\mg \sim$ 6.2 TeV compared to $\mg \lsim 2.6$ TeV for 100 $\invfb$ of LHC 
data.\\
\indent%
For indirect detection, we found that the ICECUBE experiment provided some 
reach for AMSB models, particularly in cases with high $\tanb$, or in parameter 
space regions where $\nino$ has significant higgsino content (high $\mnot$ or 
high $\alpha$).  The detection elevated rates of gammas, positrons, 
antiprotons, and antideuterons in AMSB models is due to high rate of 
$\nino \nino \rightarrow VV$ annihilations.  We noted interesting inverse 
resonance effects in the muon flux.  We also observed how indirect detection 
rates depend on the assumed galactic DM profile.

\clearpage
\addcontentsline{toc}{chapter}{Bibliography}
\bibliographystyle{hunsrt}
\bibliography{bibliography}

\end{document}